\PassOptionsToPackage{dvipsnames,table}{xcolor}
\documentclass[acmtog,screen,nonacm]{acmart}

\AtBeginDocument{}
\setcopyright{none}
\copyrightyear{2026}
\acmYear{2026}
\usepackage{multirow}
\usepackage{placeins}
\usepackage{booktabs}
\usepackage{xcolor}
\usepackage[normalem]{ulem}

\usepackage{graphicx}
\usepackage{wrapfig}
\usepackage{pifont}
\usepackage{enumitem}
\usepackage{tabularx}
\usepackage{array}
\usepackage{graphbox}
\usepackage{float}
\usepackage{balance}

\begin{document}

\title{Compact Representation of Mipmapped SVBRDFs via Shared Gaussians}
\author{Fengdi Zhang}
\email{fengdizhang@tencent.com}
\affiliation{%
  \institution{Tencent}
  \country{China}}
\affiliation{%
  \institution{Tsinghua University}
  \country{China}}

\author{Haocheng Ren}
\email{dukeren@tencent.com}
\affiliation{%
  \institution{Tencent}
  \country{China}}

\author{Qing Luo}
\email{summerqluo@tencent.com}
\affiliation{%
  \institution{Tencent}
  \country{China}}

\author{Yaqing Li}
\email{fredli@tencent.com}
\affiliation{%
  \institution{Tencent}
  \country{China}}

\author{Jibing Lou}
\email{lollylou@tencent.com}
\affiliation{%
  \institution{Tencent}
  \country{China}}

\author{Hongwei Li}
\authornote{Corresponding author.}
\email{hongweixli@tencent.com}
\affiliation{%
  \institution{Tencent}
  \country{China}}

\renewcommand{\shortauthors}{Zhang et al.}

\begin{abstract}

Spatially-varying BRDFs (SVBRDFs) are central to material representation in computer graphics, but their high-resolution, multi-channel, mipmapped textures impose a substantial storage burden. Existing compression methods face a fundamental trade-off: block-based compression provides random access and hardware-friendly decoding but exploits redundancy only within local blocks; image codecs offer strong rate-distortion performance but are not designed for direct real-time texture access; and neural texture compression achieves high compression ratios but requires neural inference during decoding, which introduces additional runtime overhead, especially on mobile platforms. We present \emph{Gaussian Texture Compression (GTC)}, a compact 2D Gaussian-based representation for mipmapped SVBRDF texture stacks that delivers high-quality compression with flexible rate-distortion trade-offs. Our method is based on a key observation that there are two dominant sources of redundancy in such data: across mip levels and across material maps. Both share a common underlying structure: the same spatial support is reused, with only level- or map-specific information attached. This property naturally suits 2D Gaussians, since each Gaussian explicitly separates its spatial footprint from the values it carries, allowing the footprint to be shared while the values vary per level and per map. Building on this property, GTC shares Gaussians along both redundancy dimensions and is trained via a progressive optimization pipeline. Experiments show that GTC achieves higher reconstruction quality and lower memory usage than ASTC, the industry-standard GPU texture compression format, while supporting random-access, non-neural decoding suitable for real-time rendering.
\end{abstract}


\begin{teaserfigure}
  \includegraphics[width=\textwidth]{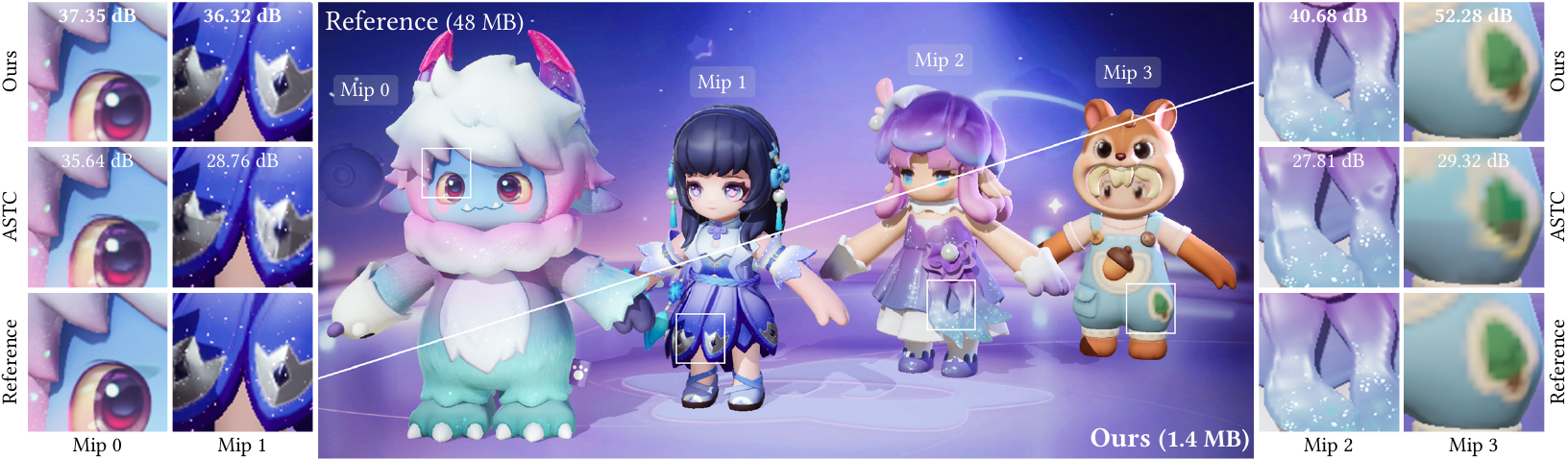}
  \caption{A production-game scene rendered with mipmapped SVBRDFs. From front to back, the four characters use mip levels 0, 1, 2, and 3, respectively. Insets compare our method, ASTC \(12{\times}12\), and the reference at the corresponding mip levels. Our method improves PSNR by 6.75 dB over ASTC while using 22.5\% less memory.}
  \Description{A production-game scene comparing GTC, ASTC, and reference SVBRDF renderings at four mip levels.}
  \label{fig:teaser}
\end{teaserfigure}

\maketitle

\section{Introduction}
Spatially-varying bidirectional reflectance distribution functions (SVBRDFs) are an essential method for representing materials in computer graphics. Conventionally, an SVBRDF is stored as a set of 2D images, i.e., textures, each encoding a different material property of the underlying shading model (e.g., Cook-Torrance). To ensure high-quality rendering, these SVBRDF maps are authored at high resolutions, which introduces a practical storage challenge. Consider modern AAA mobile game titles as examples. They typically include tens of thousands of high-resolution SVBRDF materials, accounting for approximately 60\% of total game assets and more than 10 GB in general. This makes every download and update a considerable burden for players and renders on-device storage a bottleneck for game adoption. Moreover, every texture needs to be expanded into a mipmap~\cite{williams1983pyramidal} to eliminate aliasing artifacts when viewed at a distance. Mipmapping adds approximately one-third extra storage on top of the base texture, thus exacerbating the storage problem. Consequently, achieving a compact representation of SVBRDF maps has become an urgent priority for both the gaming industry and real-time rendering community, representing a critical research challenge that demands immediate attention.

Block compression techniques~\cite{delp1979btc, iourcha1999s3tc, microsoft2024bc, nystad2012astc, strom2005etc1, strom2007etc2} represent the industry standard for addressing storage challenges owing to their support for random access on GPUs. However, their compression ratios remain limited as they only explore compression possibilities within a block. Conversely, in the multimedia field, image and video codecs~\cite{alakuijala2019jpegxl, chen2018overview} have demonstrated promising compression efficiency by detecting data redundancy across entire images, but they lack efficient GPU decoding and are thus unsuitable for real-time rendering applications.

Recent advances in AI have spurred neural texture compression methods~\cite{mueller2022instant, ntc2023siggraph, shin2024, Weinreich2024Real}. These approaches compress images into latent representations and use neural decoders to achieve high compression ratios. However, their decoding speed limits real-time rendering adoption, especially on mobile devices. Inspired by view synthesis advances, from NeRF~\cite{mildenhall2020nerf} to 3DGS~\cite{kerbl20233dgaussiansplattingrealtime}, several recent studies have explored representing 2D images using sets of 2D Gaussians~\cite{zhang2024imagegs, zhang2024gaussianimage1000fpsimage, zhu2025large}. These methods achieve image quality comparable to classical signal-processing codecs while maintaining decoding speeds exceeding 1\,000 FPS~\cite{zhang2024gaussianimage1000fpsimage} on commodity GPUs without requiring specialized tensor-processing hardware.

Following this line of work, a natural question arises: whether 2D Gaussians can represent SVBRDF maps. Two challenges emerge. First, adjacent levels in mipmaps share substantial low-frequency content; a na\"ive strategy that encodes each mip level independently would waste many Gaussians on redundant information. Second, SVBRDF maps typically contain three or more textures with cross-texture redundancies. For example, salient edges in albedo maps often correspond to those in normal maps, but these correlations are not uniform across all texture types (e.g., AO maps may lack such correspondences). Using a long feature vector~\cite{zhang2024imagegs} can waste storage when a Gaussian is useful only for a subset of material maps.

In this paper, we introduce GTC, a new 2DGS-based representation for SVBRDF maps that delivers high-quality compression with flexible rate-distortion trade-offs. Our key observation is that the two redundancies identified above share a common underlying structure: in both cases, the same spatial support is reused, with only level- or map-specific information layered on top. This makes 2D Gaussians particularly well-suited, as they \textit{explicitly decouple reusable spatial support} from such specific information, enabling the shared structure to be encoded once and reused. Guided by this insight, we exploit 2D Gaussians along both redundancy dimensions. To address the cross-mipmap redundancy, we design a \emph{coarse-to-fine} 2DGS reconstruction scheme that progressively adds Gaussians to capture residual high-frequency details at finer levels, while reusing coarse Gaussians as a shared low-frequency base. To address cross-map redundancy within an SVBRDF texture stack, we let multiple material maps share the same Gaussian geometry while assigning map-specific feature channels, ensuring that common spatial structures are encoded once and reused across maps.

Experiments demonstrate that GTC achieves a high compression ratio and fast decoding while preserving rendering quality. Compared to previous works in the same domain, by reducing the total number of Gaussians required to represent a full mipmap hierarchy, our approach yields a smaller memory footprint than Image-GS~\cite{zhang2024imagegs}. Compared with ASTC~\cite{nystad2012astc}, the industry-standard GPU texture compression format, our method delivers over $50\%$ memory savings at the same level of visual quality, while additionally supporting content-adaptive allocation of detail.

In summary, our main contributions are as follows:
\begin{itemize}[leftmargin=2.1em]
\item a compact representation of mipmapped SVBRDF maps based on shared Gaussians.
\item a progressive optimization scheme that enables flexible rate--quality control and content-adaptive Gaussian allocation.
\item a non-neural, random-access decoding method that operates efficiently on GPUs.
\end{itemize}

\section{Related Work}
\subsection{Image Representation via 2D Gaussians}
Many conventional image compression methods concentrate mainly on minimizing the amount of storage needed. JPEG-like methods~\cite{wallace1991jpeg} employ entropy coding, cosine transforms, and wavelets~\cite{antonini1992image} to attain strong compression performance. However, at low bitrates they often introduce visible artifacts such as discoloration. More recent formats such as AVIF~\cite{aomedia2025avif} and JPEG-XL~\cite{alakuijala2019jpegxl} have been designed to incorporate human visual perception models in order to enhance visual quality. Nevertheless, their relatively slow decoding makes them impractical for material textures, especially for multi-channel textures~\cite{ntc2023siggraph}.

Recent research demonstrates that Gaussian splatting is effective not only for 3D scene representation~\cite{kerbl20233dgaussiansplattingrealtime}, but also for 2D image compression and representation. GaussianImage~\cite{zhang2024gaussianimage1000fpsimage} models images using 2D Gaussians, achieving roughly the same compression ratio as JPEG at the same image quality. Building on a similar principle, Image-GS~\cite{zhang2024imagegs} introduces content-aware initialization and optimization, and extends to texture encoding. To further improve the fidelity of large images, multi-level Gaussian representations have been proposed~\cite{zhu2025large}, where two layers of 2D Gaussians are used to separately encode low-frequency regions and high-frequency details. These studies highlight the strong potential of 2D Gaussian-based representations. Our method is an extension of this line of work, formulating SVBRDF texture maps with a shared set of 2D Gaussians.

\subsection{Texture Compression}
While general-purpose image codecs prioritize offline compression efficiency, texture compression~(TC) in real-time rendering must additionally guarantee rapid hardware decoding, per-block random access, and low GPU memory bandwidth. These constraints have driven a family of block-based methods that encode independent pixel blocks at fixed bitrates. Delp and Mitchell~\cite{delp1979btc} introduced block truncation coding~(BTC), which represents each small grayscale block with two values and a per-pixel selector between them. Campbell et~al.~\cite{campbell1986ccc} extended BTC to color images via a lookup-table palette, and the S3 Texture Compression family~\cite{iourcha1999s3tc} generalized this scheme by storing two quantized color endpoints per block and deriving additional palette colors through interpolation. Standardized in DirectX as BC1--BC7~\cite{microsoft2024bc}, these variants added support for alpha channels, normal maps, and HDR content, and remain the dominant TC standard on desktop GPUs. On mobile platforms, Ericsson Texture Compression (ETC1/ETC2)~\cite{strom2005etc1,strom2007etc2} replaces endpoint storage with a single base color modulated by a trained per-block offset table and is deployed on billions of devices via OpenGL\,ES. ASTC~\cite{nystad2012astc} further generalizes block-based TC with variable block sizes, flexible bit allocation, and unified support for LDR, HDR, and 3D volumetric textures. Despite these advances, the reliance on fixed block partitioning used throughout this family prohibits content-adaptive allocation and only allows fixed compression ratios, limiting how far bitrates can be reduced while still maintaining visual quality.

\subsection{Neural Material Representations}
Due to high compactness of implicit neural representations, neural methods have been increasingly applied to material representation. A number of works focus on compressing SVBRDF texture maps while retaining the conventional shading pipeline. Implicit neural representations~\cite{sitzmann2020siren,tancik2020fourier,dupont2021coin,dupont2022coinpp} can overfit a compact coordinate network to encode texture data, but their per-pixel decoding cost---typically thousands of multiply-accumulate operations---prohibits real-time use. Hybrid approaches reduce this cost by pairing a lightweight neural decoder with an explicit spatial structure such as an adaptive coordinate grid~\cite{martel2021acorn}, a multi-resolution hash table~\cite{mueller2022instant}, or a learned codebook~\cite{takikawa2022variable}. Building on the hash-grid paradigm, Neural Texture Compression~(NTC)~\cite{ntc2023siggraph} encodes multi-channel material textures and their mipmap chains with random-access capability, while Weinreich et~al.~\cite{Weinreich2024Real} and Fujieda and Harada~\cite{shin2024} encode block-compressed textures (BC) into neural features, making them more suitable for real-time rendering pipeline. These methods push compression ratios beyond what traditional block-based schemes can achieve, yet they still rely on a neural decoder at inference time, support only a few fixed compression ratios, and employ uniform feature grids that cannot adapt allocation to spatially varying detail.

Going further, several methods encode the complete material appearance into an implicit neural representation, removing the need for explicit texture maps altogether. Neural Layered BRDFs~\cite{Fan2022NLBRDF} and MetaLayer~\cite{guo2023metalayer} learn compact latent representations of layered reflectance that replace expensive Monte Carlo layer simulation, while NeuMIP~\cite{kuznetsov2021neumip} maps query coordinates and view-light directions directly to shaded color at multiple resolutions. These models operate within offline Monte Carlo rendering pipelines and are not designed for real-time use. More recently, Zeltner et~al.~\cite{zeltner2024neural} and Xu et~al.~\cite{xu2025neural} bring neural materials to interactive frame rates via a compact shading network and quantized inference. However, these implicit neural materials sacrifice editability, require non-trivial engine integration, and remain too costly for low-end GPUs.

\section{Method}
\subsection{Overview}
\label{sec:overview}

\begin{figure}[t]
  \centering
  \includegraphics[width=\linewidth]{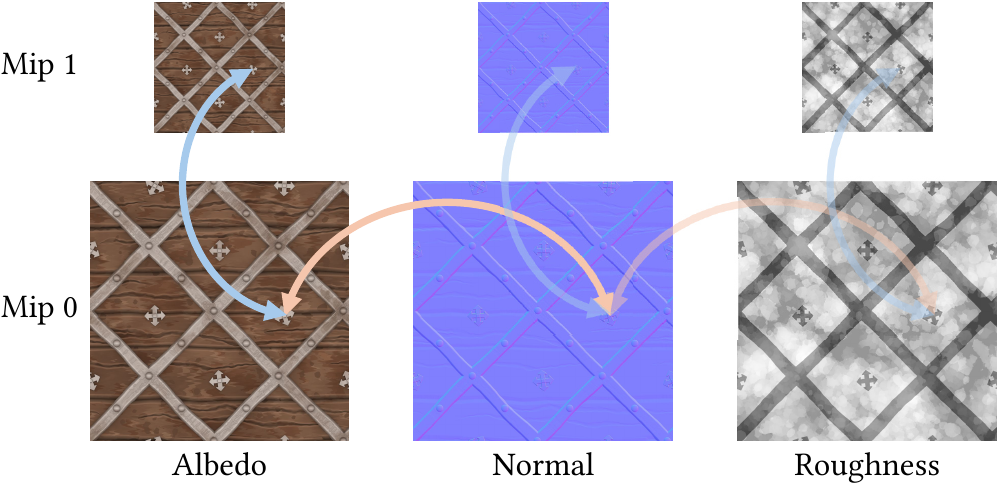}
  \caption{Redundancy in a standard SVBRDF asset. Blue arrows indicate shared structure across mip levels, and orange arrows indicate shared structure across material maps.}
  \label{fig:svbrdf_sample}
\end{figure}

Our objective is to design a highly compact representation for SVBRDF maps that drastically reduces storage requirements while maintaining reasonable decoding performance. A typical SVBRDF asset consists of multiple groups of maps, each representing different SVBRDF parameters. Moreover, each map forms a pyramid structure called \textit{mipmap}, ranging from full resolution to a single pixel. As can be seen in Figure~\ref{fig:svbrdf_sample}, there are two main types of redundancy across these dimensions: First, coarser mip levels are downsampled versions of finer levels and therefore contain the same low-frequency spatial information. Second, even though different material maps correspond to distinct physical properties, they refer to the same underlying surface and consequently exhibit a shared spatial structure. The key to obtaining a highly efficient representation is to eliminate these redundancies.

\begin{figure*}[t]
  \centering
  \includegraphics[width=\linewidth]{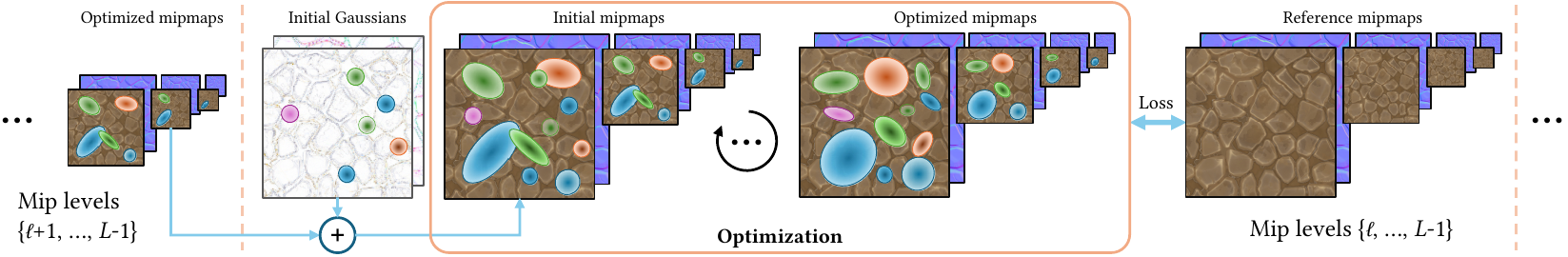}
  \caption{%
    Progressive optimization pipeline of GTC. We optimize from the coarsest level, $L-1$, to the finest level, $0$. At each level $\ell$, new Gaussians are initialized from the
    residual map, combined with coarser Gaussians whose labels lie in $\{\ell+1,\ldots,L-1\}$, and jointly optimized over the reconstructed mipmaps.
  }
  \label{fig:pipeline}
\end{figure*}

Our insight is that the spatial redundancy across mip levels offers substantial potential for reuse: a finer-level mipmap can be seen as a combination of the low frequency components from coarser levels and the high-frequency residual that remains at the current level. Based on this key observation, we develop our shared Gaussian representation (\S~\ref{sec:shared_representation}) and devise a progressive construction pipeline for SVBRDF mipmaps (\S~\ref{sec:progressive_optimization}) using shared Gaussians. Furthermore, by introducing a group-lasso regularizer and pruning, we reduce the redundancy caused by similar structures across material maps (\S~\ref{sec:regularization}). Finally, we introduce an efficient GPU random-access decoding algorithm for GTC maps (\S~\ref{sec:decoding}). Throughout this paper, we refer to the finest level as level 0 and the coarsest level as level $L - 1$.

\subsection{Representation via Shared Gaussians}
\label{sec:shared_representation}

Anisotropic 2D Gaussians are an efficient representation of images~\cite{zhang2024gaussianimage1000fpsimage, zhang2024imagegs}. Each Gaussian has the following trainable parameters:
\begin{equation}
\bigl(\boldsymbol{\mu}_i,\mathbf{s}_i,\theta_i,\mathbf{c}_i\bigr),
  \label{eq:base_gaussian_parameters}
\end{equation}
a center $\boldsymbol{\mu}_i$ defined in UV space, a two-dimensional support scale $\mathbf{s}_i=(s_{i,1},s_{i,2})^{\top}$, a rotation angle $\theta_i\in\mathbb{R}$ and a color vector $\mathbf{c}_i\in\mathbb{R}^{n}$. For each location $\mathbf{u}$, the weight of a Gaussian can be evaluated as
\begin{equation}
  G_i(\mathbf{u})
  =
  \exp\!\left(
    -\frac{1}{2}
    (\mathbf{u}-\boldsymbol{\mu}_i)^{\top}
    \boldsymbol{\Sigma}_i^{-1}
    (\mathbf{u}-\boldsymbol{\mu}_i)
  \right),
\end{equation}
where the covariance matrix $\boldsymbol{\Sigma}_i$ is parameterized by scale $\mathbf{s}_i$ and angle~$\theta_i$:
\begin{equation}
  \boldsymbol{\Sigma}_i
  =
  \mathbf{R}(\theta_i)\,
  \mathrm{diag}(s_{i,1}^{2},s_{i,2}^{2})\,
  \mathbf{R}(\theta_i)^{\top},
  \label{eq:2dgs_sigma}
\end{equation}
$\mathbf{R}(\theta_i)\in\mathbb{R}^{2\times 2}$ is the 2D rotation matrix with angle $\theta_i$, and $\mathrm{diag}(s_{i,1}^{2},s_{i,2}^{2})$ is the diagonal matrix formed from the squared scale parameters. An image $\mathbf{I}$ can then be rendered by summing the weighted colors:
\begin{equation}
  \hat{\mathbf{I}}(\mathbf{u})
  =
  \sum_{i=1}^{N}
  G_i(\mathbf{u})\,\mathbf{c}_i.
  \label{eq:single_image_render}
\end{equation}
Rather than representing a single image as in previous methods~\cite{zhang2024gaussianimage1000fpsimage, zhang2024imagegs}, our representation must encode a complete set of SVBRDF mipmaps. Thus, each Gaussian is associated with a specific level of the mipmap hierarchy. For a mipmap that has $L$ levels, we introduce a Level-of-Detail (LoD) label $\ell_i\in\{0,\ldots,L-1\}$ to identify its corresponding mip level. Furthermore, the color vector $\mathbf{c}_i$ is replaced by a feature vector $\mathbf{f}_i\in\mathbb{R}^{C}$ that stores the concatenated channels of all SVBRDF parameter maps, such as base color, normal, roughness and metallic.
In summary, the parameters of a single Gaussian in our representation are expressed as
\begin{equation}
  \bigl(\boldsymbol{\mu}_i,\mathbf{s}_i,\theta_i,\mathbf{f}_i,\ell_i\bigr).
  \label{eq:gaussian_parameters}
\end{equation}
With these new parameters, we update the rendering rule from Eq.~\ref{eq:single_image_render}~to
\begin{equation}
  \hat{\mathbf{I}}^{(l)}(\mathbf{u})
  =
  \sum_{i:\,l \le \ell_i\le L-1}
  G_i(\mathbf{u})\,\mathbf{f}_i.
  \label{eq:lod_render}
\end{equation}
A specific texture $\hat{\mathbf{I}}^{(l)}$ at mip level $l$ is calculated by activating all Gaussians with $\ell_i\ge l$ and accumulating their contributions. Although Gaussians are associated with specific levels, coarser-level Gaussians contribute to all finer levels. Thus finer levels do not need to maintain separate Gaussians for low-frequency content. By sharing these Gaussians, the overall number of Gaussians is therefore reduced.

\subsection{Progressive Optimization Pipeline}
\label{sec:progressive_optimization}

Building upon the representation discussed above, we introduce our optimization pipeline in this section. Under the hierarchical rendering rule in Eq.~\ref{eq:lod_render}, Gaussians introduced at coarser levels are reused by all finer levels, while newly inserted Gaussians at finer levels mainly explain the remaining high-frequency residual. We therefore optimize the representation level by level from the coarsest mip to the finest mip, which stabilizes the shared low-frequency basis before adding finer details.

Figure~\ref{fig:pipeline} illustrates our progressive optimization pipeline. At each level $\ell$ to be constructed, we first initialize new Gaussians from a same-resolution residual error map (\S~\ref{sec:gs_init}) under a specified Gaussian number budget (\S~\ref{sec:gs_budget}). These Gaussians have $\ell_i=\ell$ and capture details that are not explained by coarser levels. They are then optimized jointly with all previous coarser Gaussians and carried forward to the next finer level. To improve optimization efficiency, instead of evaluating every mip level at every iteration, we sample a mip level $l$ uniformly from the levels already covered by the current shared Gaussians, $\{\ell,\ldots,L-1\}$. The loss function used for each sampled level is:

\begin{equation}
  \begin{aligned}
    \mathcal{L}^{(l)} ={}&
    \lambda_{\mathrm{L1}}
    \left\lVert
      \hat{\mathbf{I}}^{(l)} - \mathbf{I}^{(l)}
    \right\rVert_1
    +
    \lambda_{\mathrm{ssim}}
    \bigl(1-\mathrm{SSIM}(\hat{\mathbf{I}}^{(l)},\mathbf{I}^{(l)})\bigr) \\
    &+
    \lambda_{\mathrm{reg}}\mathcal{L}_\mathrm{reg},
  \end{aligned}
  \label{eq:loss_total}
\end{equation}
where $\lambda_{\mathrm{L1}}$, $\lambda_{\mathrm{ssim}}$, and $\lambda_{\mathrm{reg}}$ are scalar weights for the $L_1$ term, the SSIM term, and the regularization term (\S~\ref{sec:regularization}), respectively. The regularization term encourages the sharing of Gaussians across the multi-channel SVBRDF maps. We discuss these essential techniques in the following sections.

\subsubsection{Gaussian Initialization}
\label{sec:gs_init}
Gaussians are known to be highly sensitive to their initial parameters~\cite{kerbl20233dgaussiansplattingrealtime, zhang2024imagegs}. Therefore careful initialization warrants particular attention. Because we use Gaussians to represent residual information at each level, we initialize new Gaussians from the residual distribution directly.
For $l<L-1$, let the coarser-level prediction at level $l$ be
\begin{equation}
  \hat{\mathbf{I}}^{(l)}_{\mathrm{coarse}}(\mathbf{u})
  =
  \sum_{i:\,l+1\le \ell_i\le L-1}
  G_i(\mathbf{u})\,\mathbf{f}_i.
\end{equation}
We compute a residual map as
\begin{equation}
  r^{(l)}(\mathbf{u})
  =
  \left\|
    \mathbf{I}^{(l)}(\mathbf{u})-\hat{\mathbf{I}}^{(l)}_{\mathrm{coarse}}(\mathbf{u})
  \right\|_{2}, \qquad l < L - 1.
  \label{eq:residual_map}
\end{equation}
Spatial regions with larger residuals receive more Gaussians, directing new primitives to content that remains under-explained by the coarser level.
At the coarsest level, the sampling procedure is unnecessary; we initialize one Gaussian to represent $\mathbf{I}^{(L-1)}(\mathbf{u})$.

For each newly inserted Gaussian, given the sampled center $\boldsymbol{\mu}_i$, its feature vector $\mathbf{f}_i$ is initialized to the
residual value $\mathbf{I}^{(l)}(\boldsymbol{\mu}_i)-\hat{\mathbf{I}}^{(l)}_{\mathrm{coarse}}(\boldsymbol{\mu}_i)$. The rotation $\theta_i$ is initialized to zero, and the scale parameters $\mathbf{s}_i$ are set using predefined level-dependent scales so that newly inserted Gaussians have a comparable texel footprint across levels.

\subsubsection{Gaussian Number Budget}
\label{sec:gs_budget}

Given a maximum Gaussian budget $N_{\max}$, we must distribute it across mip levels under two key constraints: (1) coarse levels require sufficient Gaussians to maintain quality as they are reused by all finer levels and cover large pixel footprints; (2) finer levels must receive at least as many Gaussians as coarser levels to capture their richer high-frequency detail. The total budget is bounded by the cumulative pixel count across all levels.

Given these constraints, we formulate a heuristic allocation strategy to distribute the Gaussian budget across all mip levels, demonstrating consistent performance in our experiments. We allocate Gaussian budgets to each level progressively from coarse to fine. Starting from a remaining budget $B^{(L-1)}=N_{\max}$, the target budget of level $l$ is
\begin{equation}
  N_{\mathrm{bud}}^{(l)}
  =
  \min\!\left(
    H^{(l)}W^{(l)},
    \left\lfloor \frac{B^{(l)}}{l+1} \right\rfloor
  \right),
  \qquad
  B^{(l-1)} = B^{(l)} - N_{\mathrm{bud}}^{(l)},
  \label{eq:lod_budget}
\end{equation}
where $L$ is the number of mip levels, and $B^{(l)}$ is the remaining budget before allocating level $l$. This allocation scheme respects the cap on the number of texels at coarse levels while leaving progressively larger budgets for finer levels, where more Gaussians are needed to capture high-frequency details.

\subsubsection{Regularization and Pruning}
\label{sec:regularization}
Ideally every Gaussian would be reused across all material maps. However, because the specifics of the material maps differ, each channel depends on its own particular subset of Gaussians. In this situation, only limited redundancy can be removed by the shared representation. Furthermore, if numerous Gaussians possess zero feature values, the total storage of a Gaussian representation may exceed that of the original image. To tackle these challenges, we propose a regularizer on Gaussian features to encourage the sharing of Gaussians across channels. The $\mathcal{L}_{\mathrm{reg}}$ term in Eq.~\ref{eq:loss_total} is defined as
\begin{equation}
  \begin{aligned}
\mathcal{L}_\mathrm{reg} = \frac{1}{\sqrt{C}}
    \sum_{i:\,l\le\ell_i\le L-1}
    \lVert\mathbf{f}_i\rVert_2.
      \end{aligned}
  \label{eq:regularization}
\end{equation}
It is a group-lasso regularizer~\cite{yuan2006model} on the $C$-dimensional feature vector carried by each visible Gaussian. This regularizer penalizes scenarios wherein individual maps develop nearly separate Gaussians and encourages reuse of the same Gaussian across multiple material maps whenever possible. In practice, the group-lasso term suppresses Gaussians whose overall contribution is weak, thereby enhancing the stability of subsequent Gaussian pruning operations.

During optimization, we prune Gaussians whose feature vector entries are all below a specified threshold. This pruning step is applied every 100 iterations throughout the optimization process. Pruning not only reduces the total optimization time but also yields a more compact representation with fewer Gaussians.

After optimization, we employ adaptive quantization to obtain a compact representation. Please refer to the supplementary document for details.

\subsection{Random-Access Decoding}
\label{sec:decoding}
We propose a two-stage random-access decoding method for reconstructing SVBRDF maps. Our approach extends the per-pixel linked list technique~\cite{Thibieroz2011PPLL} by partitioning the image domain into a regular grid of tiles, where each tile maintains a list of Gaussians whose footprints intersect it. In the preparation stage, we launch one compute thread per Gaussian. Each thread retrieves the corresponding Gaussian parameters from the compressed representation, computes its axis-aligned bounding box, and appends the Gaussian identifier to the lists of all tiles it overlaps. In the splatting stage, reconstructing a texel requires only determining its enclosing tile and traversing the associated Gaussian list to accumulate color contributions from all overlapping Gaussians. Since the preparation stage executes only once, this two-stage design enables efficient random access to individual texels without requiring decompression of the entire map.

\section{Results}
\label{sec:results}
\subsection{Experimental Setup}

\begin{figure*}[t]
  \centering
  \includegraphics[width=\textwidth]{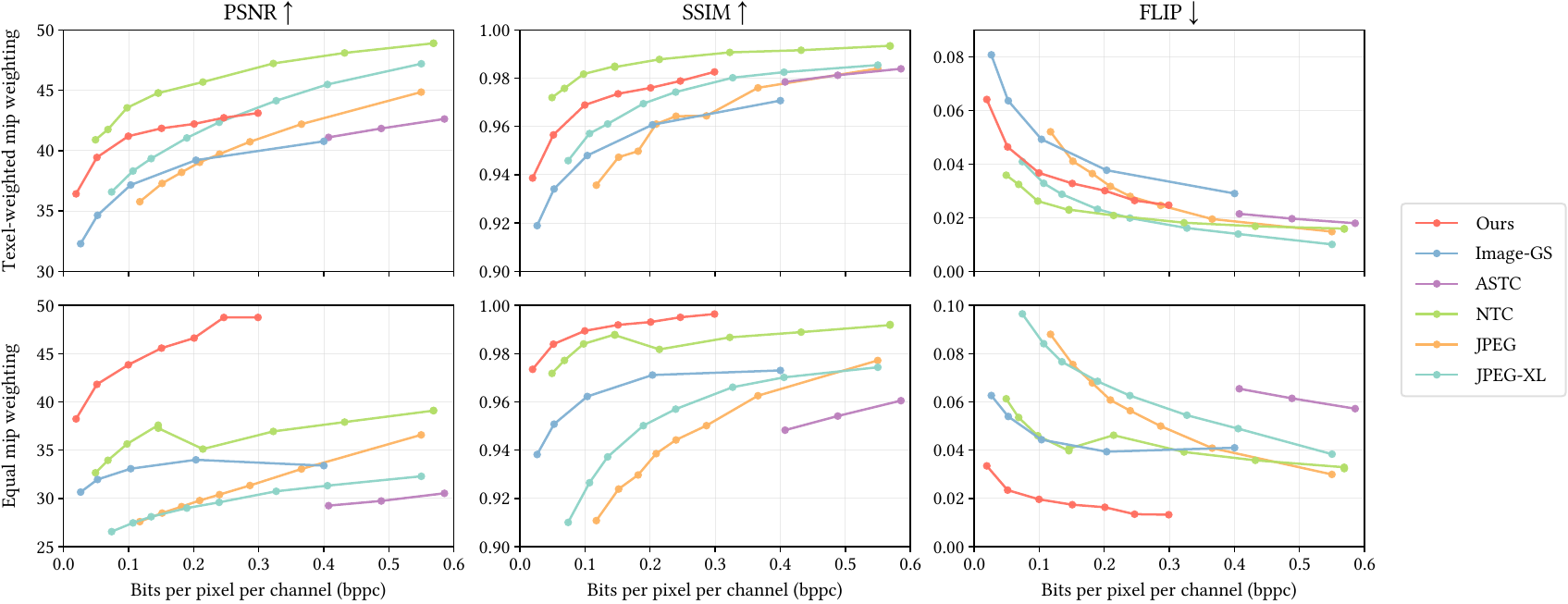}
  \caption{Rate--distortion comparison on the evaluation dataset. Rows show texel-weighted and equal-mip aggregation; columns report PSNR, SSIM, and FLIP over all material maps and mip levels. Lower FLIP is better.}
  \label{fig:svbrdf-rd-metrics}
\end{figure*}

\begin{table*}[t]
  \centering
  \scriptsize
  \setlength{\tabcolsep}{3.0pt}
  \renewcommand{\arraystretch}{1.08}
  \caption{Texel-weighted storage-grouped comparison on the evaluation dataset among non-neural random-access representations. Methods are grouped by bppc; lower bppc and FLIP are better, and higher PSNR and SSIM are better.}
  \label{tab:svbrdf-bppc-matched}
  \resizebox{\textwidth}{!}{%
  \begin{tabular}{lccccccccccccccc}
    \toprule
     & \multicolumn{3}{c}{$\sim$0.02 bppc} & \multicolumn{3}{c}{$\sim$0.05 bppc} & \multicolumn{3}{c}{$\sim$0.1 bppc} & \multicolumn{3}{c}{$\sim$0.2 bppc} & \multicolumn{3}{c}{$\sim$0.35 bppc} \\
    \cmidrule(lr){2-4} \cmidrule(lr){5-7} \cmidrule(lr){8-10} \cmidrule(lr){11-13} \cmidrule(lr){14-16}
    Method & Ours & Image-GS & ASTC & Ours & Image-GS & ASTC & Ours & Image-GS & ASTC & Ours & Image-GS & ASTC & Ours & Image-GS & ASTC \\
    bppc & \textbf{0.019} & 0.026 & - & \textbf{0.051} & 0.052 & - & \textbf{0.100} & 0.103 & - & \textbf{0.201} & 0.204 & - & \textbf{0.299} & 0.400 & 0.407 \\
    \midrule
    PSNR $\uparrow$ & \textbf{36.42} & 32.29 & - & \textbf{39.44} & 34.64 & - & \textbf{41.19} & 37.15 & - & \textbf{42.21} & 39.21 & - & \textbf{43.11} & 40.77 & 41.10 \\
    SSIM $\uparrow$ & \textbf{0.9386} & 0.9189 & - & \textbf{0.9565} & 0.9341 & - & \textbf{0.9689} & 0.9480 & - & \textbf{0.9760} & 0.9607 & - & \textbf{0.9826} & 0.9707 & 0.9784 \\
    FLIP $\downarrow$ & \textbf{0.0641} & 0.0807 & - & \textbf{0.0464} & 0.0636 & - & \textbf{0.0367} & 0.0492 & - & \textbf{0.0301} & 0.0377 & - & 0.0247 & 0.0290 & \textbf{0.0214} \\
    \bottomrule
  \end{tabular}
  }
\end{table*}

\paragraph{Dataset.}
We evaluate on a collection of 20 SVBRDF texture sets, consisting of two subsets: the \emph{atlas} subset, which contains 10 UV-atlas assets selected from TexVerse~\cite{zhang2025texverse}, and the \emph{material} subset that contains 10 PBR material assets from FreeStylized~\cite{freestylized2026}, a public collection of game-ready textures. Each texture stack has a complete mipmap pyramid, with the highest-resolution level at $2048\times2048$. Each material also includes at least three SVBRDF maps, including base color, normal, and packed maps such as ambient occlusion, roughness, metallic or height. Additional mipmap generation details are provided in the supplementary document.

\paragraph{Metrics and aggregation.}
We report PSNR~\cite{huynh2008psnr}, SSIM~\cite{wang2004ssim}, and FLIP~\cite{andersson2020flip} to evaluate reconstruction quality. All metrics are computed between the reconstructed texture maps and the corresponding ground-truth mipmap levels. Since mipmap evaluation can be biased by the number of texels at each level, we report two complementary aggregation protocols. \emph{Texel-weighted} aggregation weights every texel equally and therefore emphasizes the highest-resolution levels. \emph{Equal-mip} aggregation gives each mip level the same weight and therefore exposes reconstruction errors in coarse mip levels that would otherwise be hidden by texel count. Unless otherwise specified, scores are aggregated over all evaluated material maps and mipmap levels in the dataset.  Storage is reported as bits per pixel per channel (bppc), computed over the dataset.

\paragraph{Baselines.}
We compare our method with several strong baselines, including JPEG~\cite{wallace1991jpeg}, JPEG-XL~\cite{alakuijala2019jpegxl}, ASTC~\cite{nystad2012astc}, NTC~\cite{ntc2023siggraph}, and Image-GS~\cite{zhang2024imagegs}. JPEG, JPEG-XL, and ASTC are applied to individual texture images. NTC and our method compress the material stack jointly and explicitly model texture mipmaps. We use official ASTC encoder~\cite{astcenc} with the \textit{exhaustive} preset to ensure its best quality.

To make a fair comparison with Image-GS, we use a mip-aware Gaussian budget allocation rather than a naive single-budget baseline. Specifically, we specify a Gaussian budget for the finest mip level and reduce this budget progressively toward coarser levels using a preset decay factor. This gives finer mips more Gaussians while assigning progressively fewer Gaussians to coarser mips. Experiments show that a decay factor of $1.5$ gives the best overall reconstruction quality across the evaluation set. Full settings are reported in the supplementary document.

\paragraph{Optimization.}
We optimize Eq.~\ref{eq:loss_total} with Adam~\cite{kingma2014adam}, using learning rates of $5\times10^{-4}$ for Gaussian centers and features, and $2\times10^{-3}$ for Gaussian scale parameters and rotations. The loss uses $\lambda_{\mathrm{L1}}=1.0$, $\lambda_{\mathrm{ssim}}=0.1$, and $\lambda_{\mathrm{reg}}=10^{-7}$. We use a pruning threshold of $3\times10^{-4}$ in our experiments.

\subsection{Quantitative Results}

Figure~\ref{fig:svbrdf-rd-metrics} summarizes the rate-distortion behavior of all methods. Among the non-neural baselines, our method consistently achieves the best quality-memory trade-off, particularly under equal-mip aggregation. This advantage stems from our method delivering more plausible reconstruction quality at low mip levels. NTC emerges as the strongest competitor in pure rate-distortion quality, especially under texel-weighted aggregation where the highest-resolution levels dominate the metric. When storage budgets are relaxed, entropy-encoding methods such as JPEG and JPEG-XL can achieve better quality too. This advantage arises because these methods can allocate nonlinear capacity to exploit global high-frequency redundancy through either learning-based or signal-analytic approaches. Our method performs slightly below NTC in these scenarios and outperforms at coarser mip levels (see Fig.~\ref{fig:main-mipmap-error-inset-fantasy-character}). Moreover NTC requires a neural decoder at inference time, while entropy-encoding methods do not support GPU-friendly decoding or random access.

Table~\ref{tab:svbrdf-bppc-matched} gives a closer texel-weighted bppc-matched comparison among fast non-neural representations. Against the Image-GS baseline, our method improves PSNR by $\geq3.0$ dB at $0.02$, $0.05$, $0.1$, and $0.2$ bppc. The perceptual improvement is also consistent: FLIP is reduced by roughly $20$--$27\%$ at these points. At a higher bppc range, our method reaches $43.11$ dB and $0.9826$ SSIM at $0.299$ bppc, while ASTC reaches $41.10$ dB and $0.9784$ SSIM at $0.407$ bppc. Our method provides higher PSNR and SSIM while using $26.5\%$ less storage. ASTC obtains a slightly lower FLIP with a substantially larger memory footprint.

\subsection{Qualitative Results}

Figure~\ref{fig:main-qual-comparison} shows representative qualitative comparisons. Each example shows base color, normal, and packed material-property maps at specific mip levels. Our method preserves sharp semantic boundaries in base color maps, maintains coherent high-frequency structures in normal maps, and avoids introducing obvious block-discontinuity artifacts in packed property maps. The benefit is most visible at coarse mip levels. Note that our method achieves this quality with the smallest bppc among all competitors.

We also show mip-level renderings for our method and ASTC in Fig.~\ref{fig:teaser}.  Comparisons across the full mipmap pyramid for different methods are presented in Fig.~\ref{fig:main-mipmap-error-inset-fantasy-character}, while a bitrate sweep of our method is shown in Fig.~\ref{fig:main-gtc-sweep-brick-wall-battle-axe}. Additional qualitative comparisons are provided in the supplementary document.

\subsection{Ablation Study}

\begin{table}[t]
  \centering
  \footnotesize
  \setlength{\tabcolsep}{3.0pt}
  \renewcommand{\arraystretch}{1.08}
  \caption{Ablation study over the evaluation dataset. Values are mean \(\pm\) standard deviation; ``eq'' and ``texel'' denote equal-mip and texel-weighted aggregation, respectively.}
  \vspace{-5pt}
  \label{tab:gtc-ablation}
  \resizebox{\columnwidth}{!}{%
  \begin{tabular}{lccccc}
    \toprule
    Variant &
    bppc $\downarrow$ &
    PSNR$_{\mathrm{eq}}$ $\uparrow$ &
    FLIP$_{\mathrm{eq}}$ $\downarrow$ &
    PSNR$_{\mathrm{texel}}$ $\uparrow$ &
    FLIP$_{\mathrm{texel}}$ $\downarrow$ \\
    \midrule
    Uniform init &
    $0.114{\pm}0.016$ & $50.04{\pm}4.14$ & $0.012{\pm}0.005$ &
    $45.40{\pm}4.27$ & $0.021{\pm}0.008$ \\
    Uniform LoD budget &
    $\mathbf{0.080{\pm}0.013}$ & $48.52{\pm}4.23$ & $0.015{\pm}0.005$ &
    $45.09{\pm}3.92$ & $0.026{\pm}0.008$ \\
    w/o group-lasso &
    $0.172{\pm}0.025$ & $48.85{\pm}5.36$ & $0.015{\pm}0.007$ &
    $44.73{\pm}3.87$ & $0.023{\pm}0.006$ \\
    w/o sparse pruning &
    $0.200{\pm}0.017$ & $49.77{\pm}5.40$ & $0.012{\pm}0.006$ &
    $45.87{\pm}3.58$ & $0.021{\pm}0.004$ \\
    Full &
    $0.132{\pm}0.041$ & $\mathbf{51.74{\pm}3.73}$ & $\mathbf{0.011{\pm}0.004}$ &
    $\mathbf{46.68{\pm}3.76}$ & $\mathbf{0.020{\pm}0.007}$ \\
    \bottomrule
  \end{tabular}}
\end{table}
Table~\ref{tab:gtc-ablation} shows ablation studies on the technical components of our method. Replacing residual-guided Gaussian initialization with uniform initialization reduces equal-mip PSNR by $1.70$ dB and texel-weighted PSNR by $1.28$ dB. A uniform LoD budget induces more compact results, however it gives the lowest equal-mip quality and the largest FLIP errors, indicating the effectiveness of our budget allocation strategy. Removing the group-lasso regularizer or pruning drastically increases bppc while worsening PSNR and FLIP, confirming that these components help improve the compactness of our shared Gaussians. Our full pipeline gives the best quality among all variants.

\section{Limitations and Future Work}
\begin{figure}[!h]
  \centering
  \footnotesize
  \setlength{\tabcolsep}{2pt}
  \begin{tabular}{@{}*{3}{c}@{}}
    \includegraphics[width=0.32\linewidth]{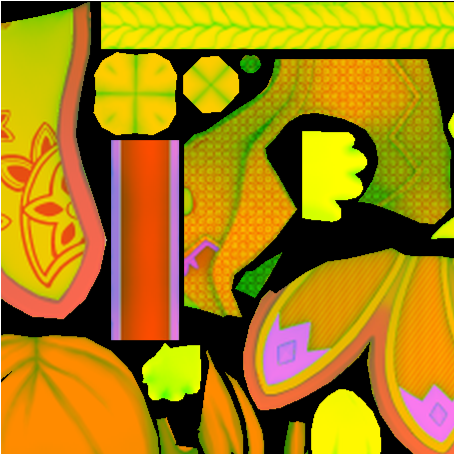} &
    \includegraphics[width=0.32\linewidth]{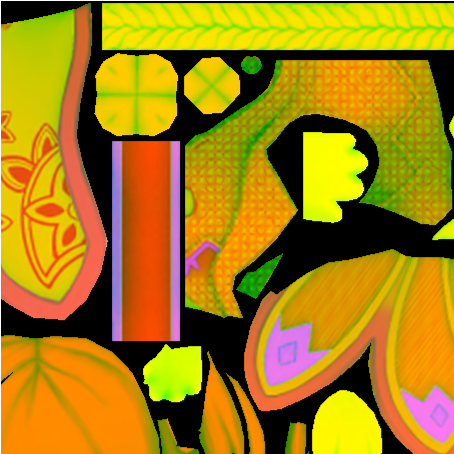} &
    \includegraphics[width=0.32\linewidth]{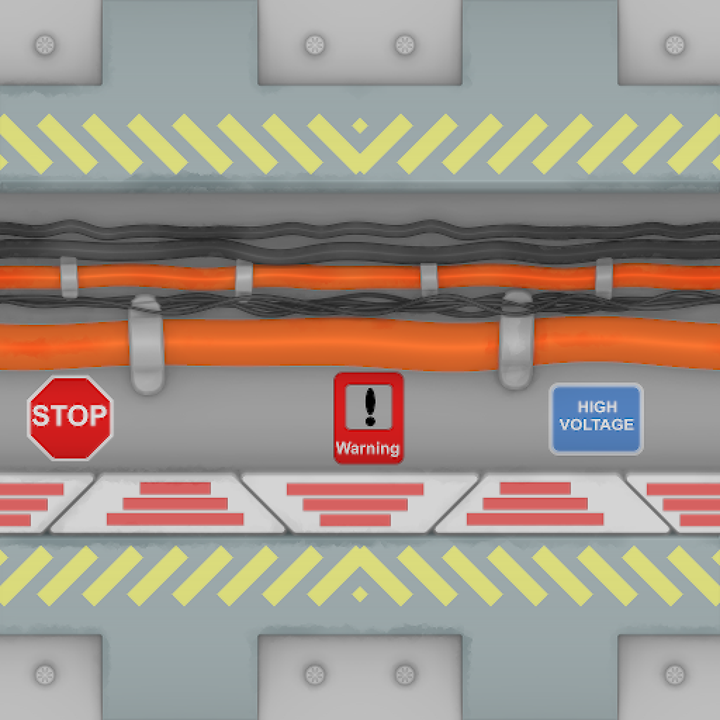} \\
    (a) Reference & (b) Ours & (c) Repeated patterns \\
  \end{tabular}
  \caption{Illustration of our limitations. Better solutions for (a, b) very high-frequency stochastic patterns and (c) repeated patterns remain future work.}
  \label{fig:failed-case}
  \vspace{-1em}
\end{figure}

First, very high-frequency stochastic patterns and thin repetitive structures are still difficult to represent with a finite number of Gaussian primitives. Fig.~\ref{fig:failed-case} (a,b) shows results for the roughness and metallic textures of a cloth material. Our method achieves high quality on the smoother left region, but the cloth pattern on the right becomes noticeably blurrier than the reference when zoomed in. Promising extensions include compactly supported kernels~\cite{thomas2025splinesplat}, deformable or anisotropic splatting kernels~\cite{huang2025drk,hsu2026a2tg}, edge-aware splats~\cite{chelani2025edgegaussians}, and hybrid Gaussian-wavelet representations~\cite{saragadam2023wire}, which may reduce boundary bleeding and allocate primitives more effectively near high-frequency structures.

Second, SVBRDF maps often contain repeated patterns, as shown in Fig.~\ref{fig:failed-case} (c). Although Gaussian-based representations can adaptively allocate primitives to local structures, they do not explicitly exploit long-range repetition and may require many similar Gaussians in repetitive regions. Prior work has reduced such redundancy by reorganizing UV layouts while preserving random access~\cite{luo2023texture,knodt2025texture}.

We view this as a promising direction for future work.

Finally, our method depends on per-asset optimization which may become a bottleneck for large material repositories. In line with recent efforts to speed up 3DGS reconstruction~\cite{Sharp2025iclr}, an appealing avenue for future work is to employ a pretrained feed-forward network to provide an initial yet good Gaussian configuration to speed up the optimization process.

\section{Conclusion}
We have presented Gaussian Texture Compression (GTC), a compact 2D Gaussian-based representation for mipmapped SVBRDF texture stacks. By observing the redundancies across mip levels and between different material maps, we employ shared 2D Gaussians as a unified spatial basis, yielding an efficient representation. To achieve high efficiency, we propose a progressive optimization scheme alongside an optimized non-neural decoder. We envision GTC as a step toward a new direction in texture compression, demonstrating that explicit, lightweight primitives can jointly exploit the extensive redundancy present within SVBRDF assets, bridging the gap between the efficiency of traditional block-based formats and the strong compression performance of neural approaches.

\balance
\bibliographystyle{unsrtnat}
\bibliography{references}

\clearpage
\begingroup
\setkeys{Gin}{interpolate=false}

\newcommand{\mainQualFontSize}{\footnotesize}
\newcommand{\mainQualMetricGap}{1pt}
\newcommand{\mainQualMetricTopGap}{27pt}
\newcommand{\mainQualRowGap}{27pt}
\newcommand{\mainQualMipLabelW}{0.032\textwidth}
\newcommand{\mainQualCellW}{0.106\textwidth}
\newcommand{\mainQualVCenter}[1]{\raisebox{-0.5\height}{#1}}
\newcommand{\mainQualImg}[2]{\mainQualVCenter{\includegraphics[width=\mainQualCellW]{figs/qual_subfigs/#1/#2.png}}}
\newcommand{\mainQualMip}[2]{\mainQualVCenter{\makebox[\mainQualMipLabelW][r]{\begin{tabular}{@{}r@{}}\textbf{#1}\\#2\end{tabular}}}}
\newcommand{\mainQualStream}[1]{\mainQualVCenter{\rotatebox{90}{\textbf{#1}}}}
\newcommand{\mainQualRow}[3]{#2 & \mainQualImg{#1}{#3_gt} & \mainQualImg{#1}{#3_gt_zoom} & \mainQualImg{#1}{#3_gtc} & \mainQualImg{#1}{#3_image_gs} & \mainQualImg{#1}{#3_astc} & \mainQualImg{#1}{#3_ntc} & \mainQualImg{#1}{#3_jpeg} & \mainQualImg{#1}{#3_jxl}}
\newcommand{\mainQualMetric}[3]{\vtop{\hbox{\begin{tabular}{@{}c@{}}\textbf{#1}\\[\mainQualMetricGap]#2\\[\mainQualMetricGap]#3\end{tabular}}}}
\newcommand{\mainQualMetricName}[1]{\vtop{\hbox{\begin{tabular}{@{}c@{}}\makebox[\mainQualCellW][c]{\textbf{#1}}\\[\mainQualMetricGap]\makebox[\mainQualCellW][c]{}\\[\mainQualMetricGap]\makebox[\mainQualCellW][c]{}\end{tabular}}}}
\newcommand{\mainQualSep}{\noalign{\vskip 2pt}\hline\noalign{\vskip 2pt}}

\begin{figure*}[p!]
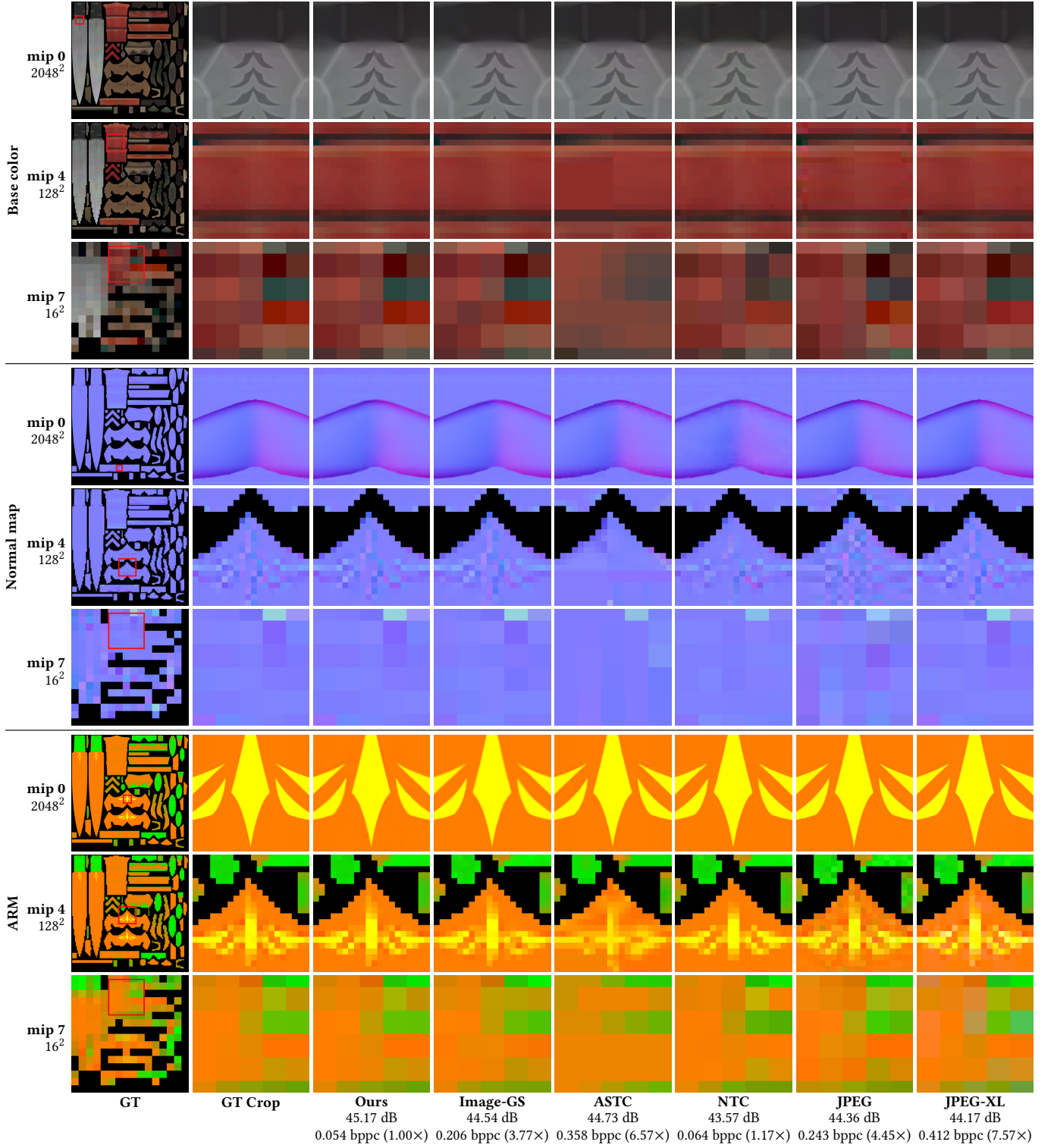

  \centering
  \vspace{-1.5ex}
  \mainQualFontSize
  \setlength{\tabcolsep}{0.7pt}
  \renewcommand{\arraystretch}{0.78}
  \makebox[\textwidth][c]{\resizebox{\textwidth}{!}{%
  \begin{tabular}{@{}c@{\hspace{5pt}}c@{\hspace{3pt}}*{7}{c@{\hspace{2pt}}}c@{}}
    & \mainQualRow{sword_25941}{\mainQualMip{mip 0}{2048$^2$}}{bc_m0} \\[\mainQualRowGap]
    \mainQualStream{Base color} & \mainQualRow{sword_25941}{\mainQualMip{mip 4}{128$^2$}}{bc_m4} \\[\mainQualRowGap]
    & \mainQualRow{sword_25941}{\mainQualMip{mip 7}{16$^2$}}{bc_m7} \\
    \mainQualSep
    & \mainQualRow{sword_25941}{\mainQualMip{mip 0}{2048$^2$}}{n_m0} \\[\mainQualRowGap]
    \mainQualStream{Normal map} & \mainQualRow{sword_25941}{\mainQualMip{mip 4}{128$^2$}}{n_m4} \\[\mainQualRowGap]
    & \mainQualRow{sword_25941}{\mainQualMip{mip 7}{16$^2$}}{n_m7} \\
    \mainQualSep
    & \mainQualRow{sword_25941}{\mainQualMip{mip 0}{2048$^2$}}{arm_m0} \\[\mainQualRowGap]
    \mainQualStream{ARM} & \mainQualRow{sword_25941}{\mainQualMip{mip 4}{128$^2$}}{arm_m4} \\[\mainQualRowGap]
    & \mainQualRow{sword_25941}{\mainQualMip{mip 7}{16$^2$}}{arm_m7} \\[\mainQualMetricTopGap]
    \multicolumn{2}{c}{} &
    \mainQualMetricName{GT} &
    \mainQualMetricName{GT Crop} &
    \mainQualMetric{Ours}{45.17 dB}{0.054 bppc (1.00$\times$)} &
    \mainQualMetric{Image-GS}{44.54 dB}{0.206 bppc (3.77$\times$)} &
    \mainQualMetric{ASTC}{44.73 dB}{0.358 bppc (6.57$\times$)} &
    \mainQualMetric{NTC}{43.57 dB}{0.064 bppc (1.17$\times$)} &
    \mainQualMetric{JPEG}{44.36 dB}{0.243 bppc (4.45$\times$)} &
    \mainQualMetric{JPEG-XL}{44.17 dB}{0.412 bppc (7.57$\times$)} \\
  \end{tabular}%
  }}
  \caption{Qualitative comparison on the \textit{Sword} SVBRDF stack. Reported PSNR values use texel-weighted aggregation over the full mipmap pyramid. ARM denotes ambient occlusion, roughness, and metallic channels. GTC preserves details at low bitrate while avoiding block artifacts and unstable coarse-mip~reconstruction.}
  \label{fig:main-qual-comparison}
\end{figure*}

\clearpage

\newcommand{\mipQualFontSize}{\scriptsize}
\newcommand{\mipQualHGap}{1pt}
\newcommand{\mipQualVGap}{14.8pt}
\newcommand{\mipQualSepGap}{1pt}
\newcommand{\mipQualStreamMethodGap}{0pt}
\newcommand{\mipQualMethodImageGap}{2pt}
\newcommand{\mipQualTableTargetW}{1.0\textwidth}

\newcommand{\mipQualHeaderGap}{1pt}
\newcommand{\mipQualRowGap}{\mipQualVGap}
\newcommand{\mipQualStreamGap}{\mipQualSepGap}
\newcommand{\mipQualColGap}{\mipQualHGap}
\newcommand{\mipQualMethodLabelW}{0.075\textwidth}
\newcommand{\mipQualCellW}{0.062\textwidth}
\newcommand{\mipQualVCenter}[1]{\raisebox{\dimexpr\depth/2-\height/2\relax}{#1}}
\newcommand{\mipQualCellStrut}{}
\newcommand{\mipQualImg}[1]{\mipQualCellStrut\mipQualVCenter{\includegraphics[width=\mipQualCellW]{#1}}}
\newcommand{\mipQualMip}[2]{\mipQualVCenter{\makebox[\mipQualCellW][c]{\begin{tabular}{@{}c@{}}\textbf{#1}\\#2\end{tabular}}}}
\newcommand{\mipQualMethod}[2]{%
  \mipQualVCenter{\makebox[\mipQualMethodLabelW][r]{%
    \if\relax\detokenize{#2}\relax
      \textbf{#1}%
    \else
      \begin{tabular}{@{}r@{}}\textbf{#1}\\#2\end{tabular}%
    \fi
  }}%
}
\newcommand{\mipQualStream}[1]{\smash{\mipQualVCenter{\rotatebox{90}{\textbf{#1}}}}}
\newcommand{\mipQualSep}{\noalign{\vskip \mipQualStreamGap}\hline\noalign{\vskip \mipQualStreamGap}}

\newcommand{\oursSweepFontSize}{\scriptsize}
\newcommand{\oursSweepMetricGap}{1pt}
\newcommand{\oursSweepMetricTopGap}{26.7pt}
\newcommand{\oursSweepRowGap}{26.7pt}
\newcommand{\oursSweepMipLabelW}{0.032\textwidth}
\newcommand{\oursSweepCellW}{0.105\textwidth}
\newcommand{\oursSweepTableTargetW}{1.0\textwidth}
\newcommand{\oursSweepVCenter}[1]{\raisebox{-0.5\height}{#1}}
\newcommand{\oursSweepImg}[1]{\oursSweepVCenter{\includegraphics[width=\oursSweepCellW]{#1}}}
\newcommand{\oursSweepMip}[2]{\oursSweepVCenter{\makebox[\oursSweepMipLabelW][r]{\begin{tabular}{@{}r@{}}\textbf{#1}\\#2\end{tabular}}}}
\newcommand{\oursSweepStream}[1]{\oursSweepVCenter{\rotatebox{90}{\textbf{#1}}}}
\newcommand{\oursSweepMetric}[2]{%
  \vtop{\hbox{\begin{tabular}{@{}c@{}}
    \makebox[\oursSweepCellW][c]{\textbf{#1}}\\[\oursSweepMetricGap]
    \makebox[\oursSweepCellW][c]{#2}
  \end{tabular}}}%
}
\newcommand{\oursSweepMetricName}[1]{%
  \vtop{\hbox{\begin{tabular}{@{}c@{}}
    \makebox[\oursSweepCellW][c]{\textbf{#1}}\\[\oursSweepMetricGap]
    \makebox[\oursSweepCellW][c]{}
  \end{tabular}}}%
}
\newcommand{\oursSweepSep}{\noalign{\vskip 2pt}\hline\noalign{\vskip 2pt}}

\begin{figure*}[p!]
  \centering
  \captionsetup{aboveskip=2pt,belowskip=2pt}
  \begin{minipage}{\textwidth}
  \centering
  \mipQualFontSize
  \setlength{\tabcolsep}{0pt}
  \renewcommand{\arraystretch}{1.0}
  \makebox[\textwidth][c]{\resizebox{\mipQualTableTargetW}{!}{%
  \begin{tabular}{@{}c@{\hspace{\mipQualStreamMethodGap}}c@{\hspace{\mipQualMethodImageGap}}*{11}{c@{\hspace{\mipQualColGap}}}c@{}}
    \multicolumn{2}{c}{} & \mipQualMip{mip 0}{2048$^2$} & \mipQualMip{mip 1}{1024$^2$} & \mipQualMip{mip 2}{512$^2$} & \mipQualMip{mip 3}{256$^2$} & \mipQualMip{mip 4}{128$^2$} & \mipQualMip{mip 5}{64$^2$} & \mipQualMip{mip 6}{32$^2$} & \mipQualMip{mip 7}{16$^2$} & \mipQualMip{mip 8}{8$^2$} & \mipQualMip{mip 9}{4$^2$} & \mipQualMip{mip 10}{2$^2$} & \mipQualMip{mip 11}{1$^2$} \\[\mipQualHeaderGap]
      & \mipQualMethod{GT}{} & \mipQualImg{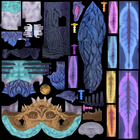} & \mipQualImg{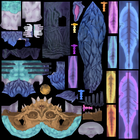} & \mipQualImg{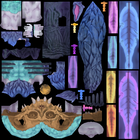} & \mipQualImg{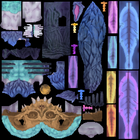} & \mipQualImg{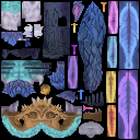} & \mipQualImg{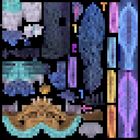} & \mipQualImg{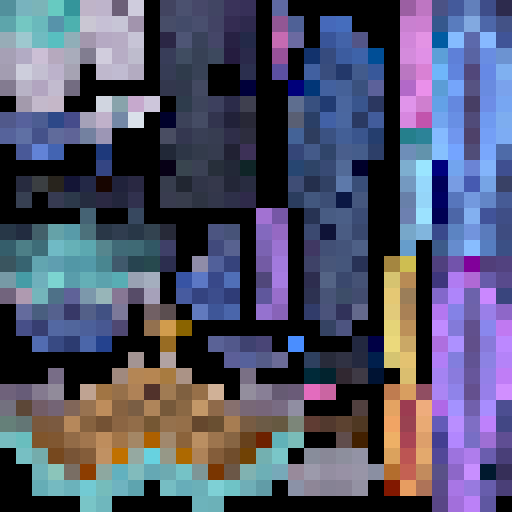} & \mipQualImg{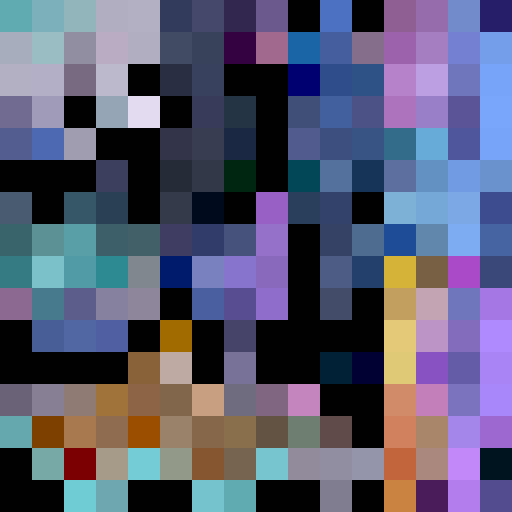} & \mipQualImg{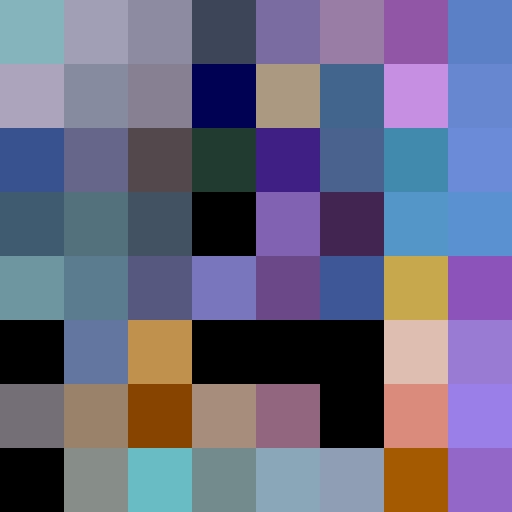} & \mipQualImg{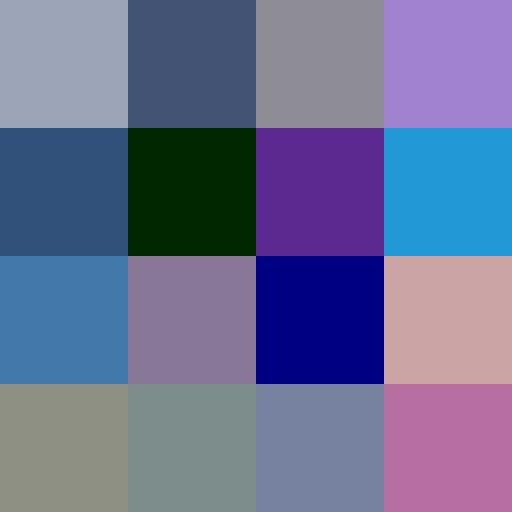} & \mipQualImg{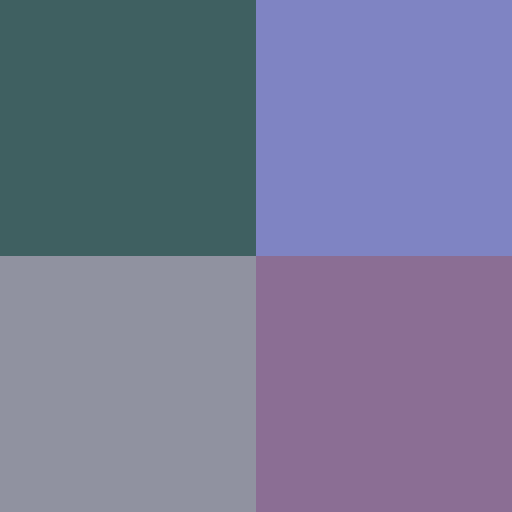} & \mipQualImg{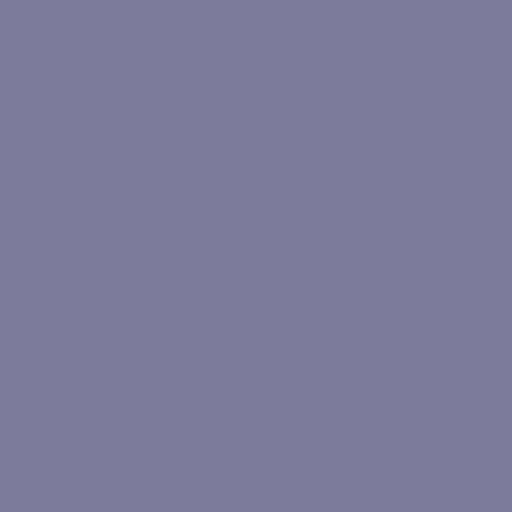} \\[\mipQualRowGap]
      & \mipQualMethod{Ours}{0.210 bppc} & \mipQualImg{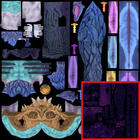} & \mipQualImg{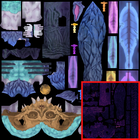} & \mipQualImg{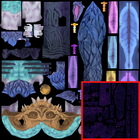} & \mipQualImg{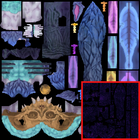} & \mipQualImg{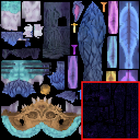} & \mipQualImg{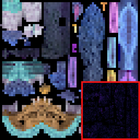} & \mipQualImg{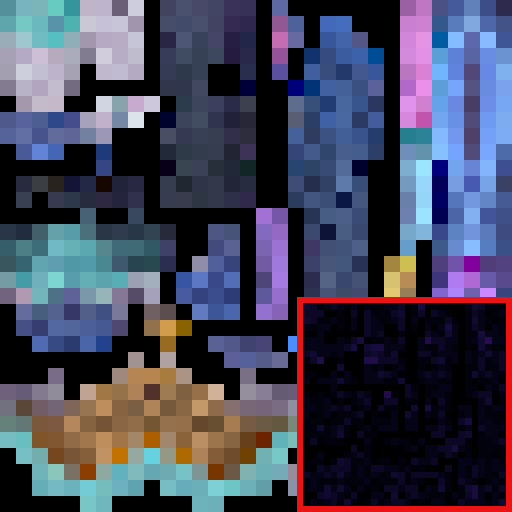} & \mipQualImg{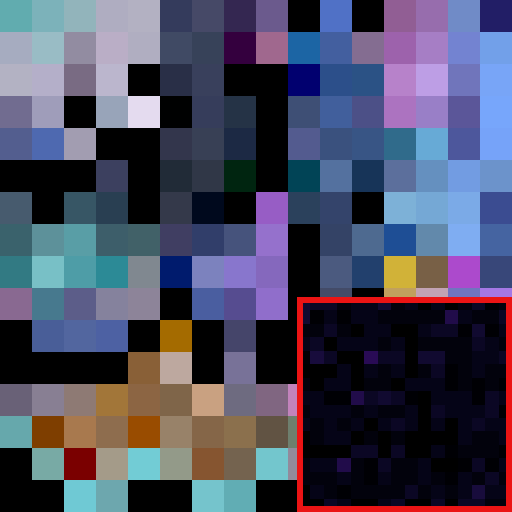} & \mipQualImg{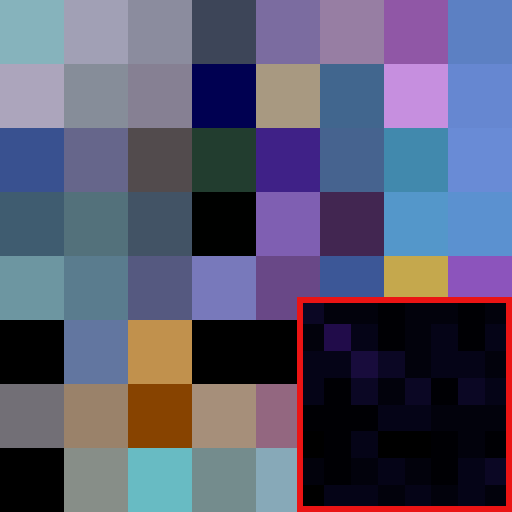} & \mipQualImg{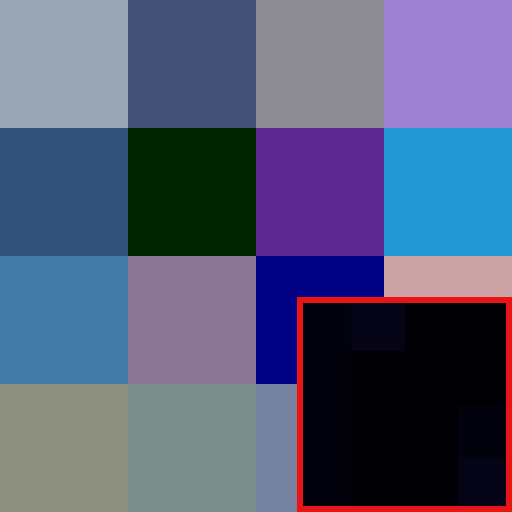} & \mipQualImg{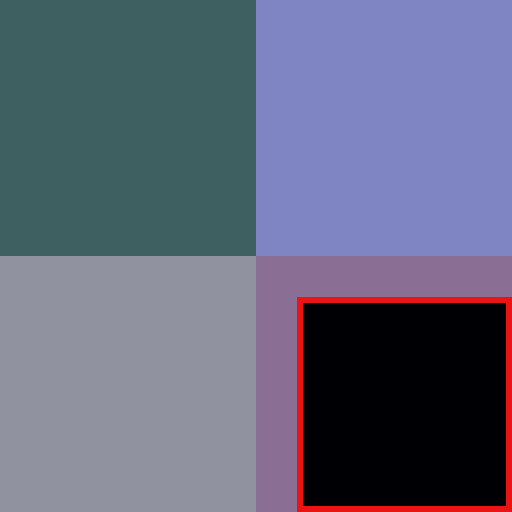} & \mipQualImg{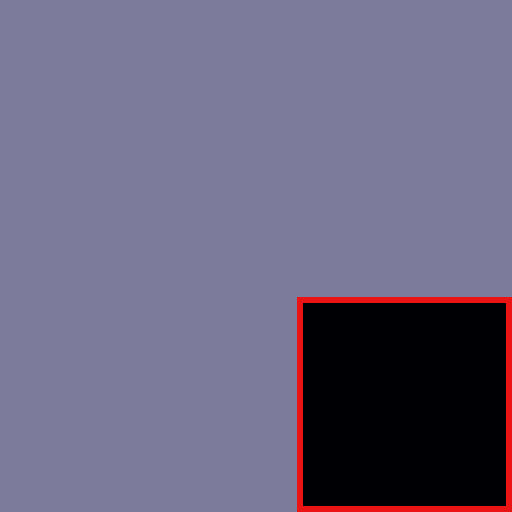} \\[\mipQualRowGap]
    \mipQualStream{Base color} & \mipQualMethod{Image-GS}{0.445 bppc} & \mipQualImg{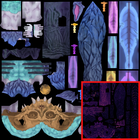} & \mipQualImg{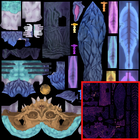} & \mipQualImg{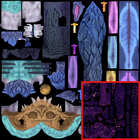} & \mipQualImg{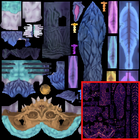} & \mipQualImg{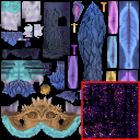} & \mipQualImg{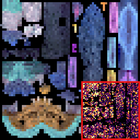} & \mipQualImg{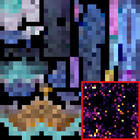} & \mipQualImg{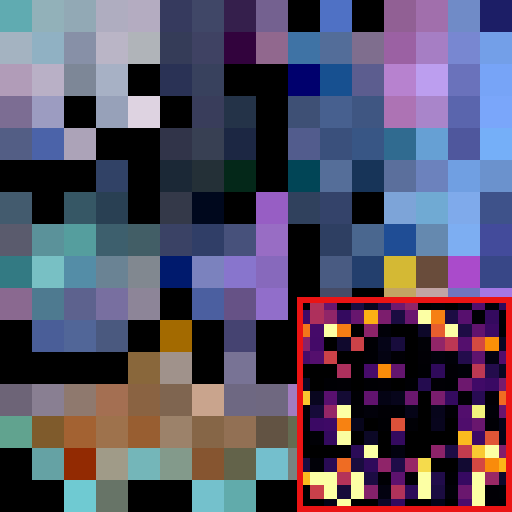} & \mipQualImg{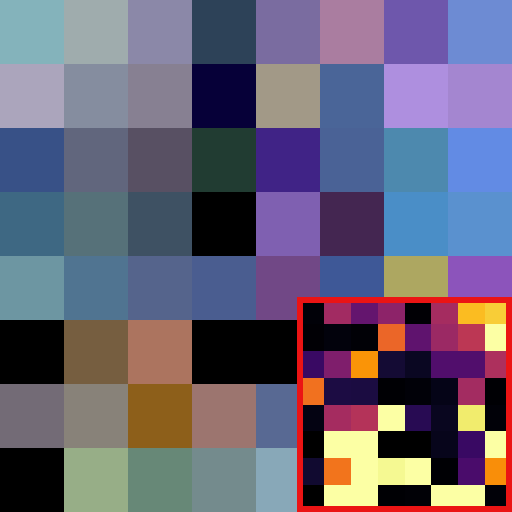} & \mipQualImg{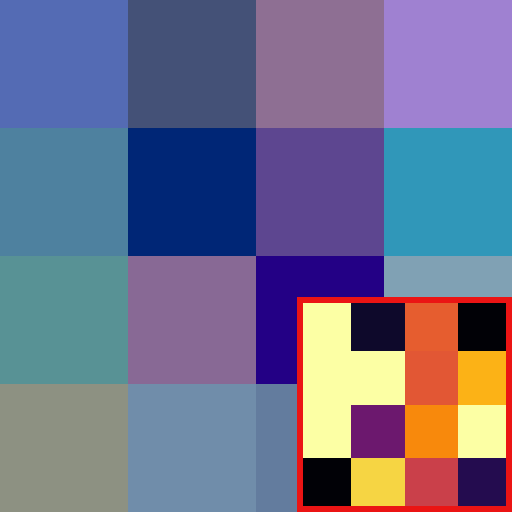} & \mipQualImg{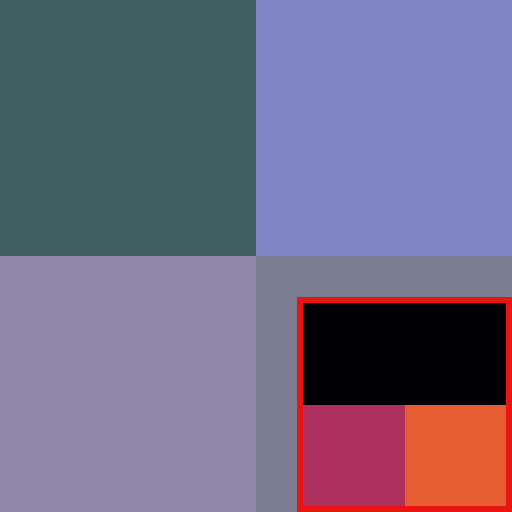} & \mipQualImg{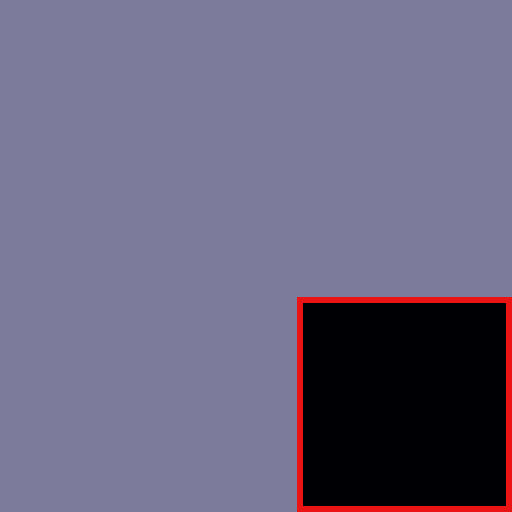} \\[\mipQualRowGap]
      & \mipQualMethod{ASTC}{0.384 bppc} & \mipQualImg{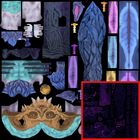} & \mipQualImg{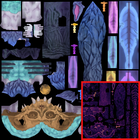} & \mipQualImg{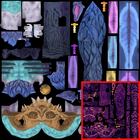} & \mipQualImg{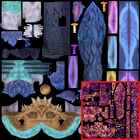} & \mipQualImg{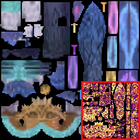} & \mipQualImg{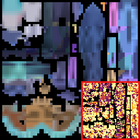} & \mipQualImg{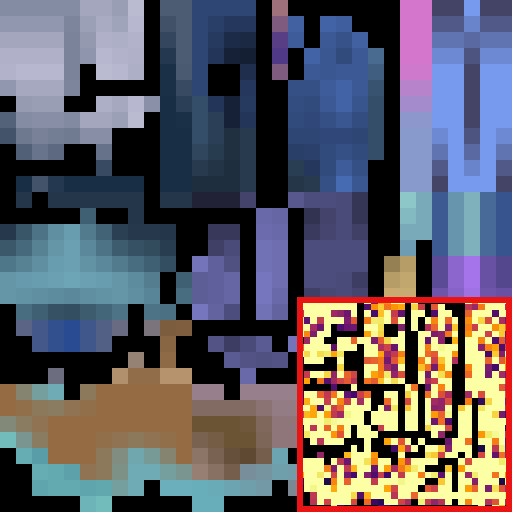} & \mipQualImg{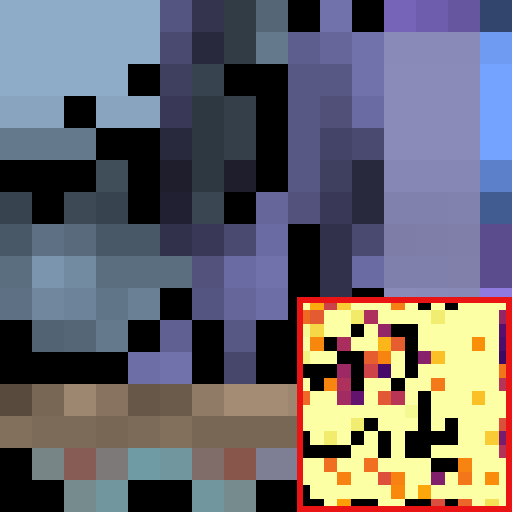} & \mipQualImg{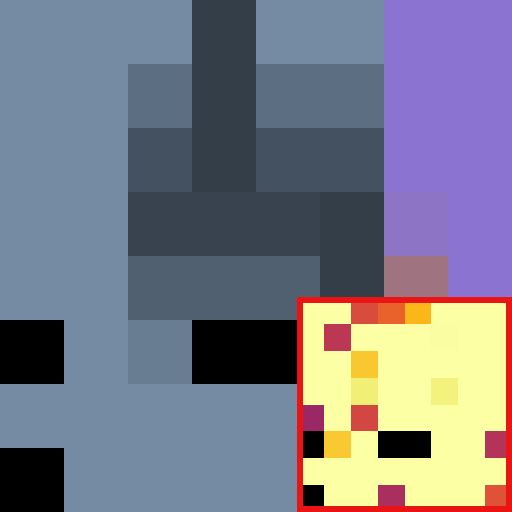} & \mipQualImg{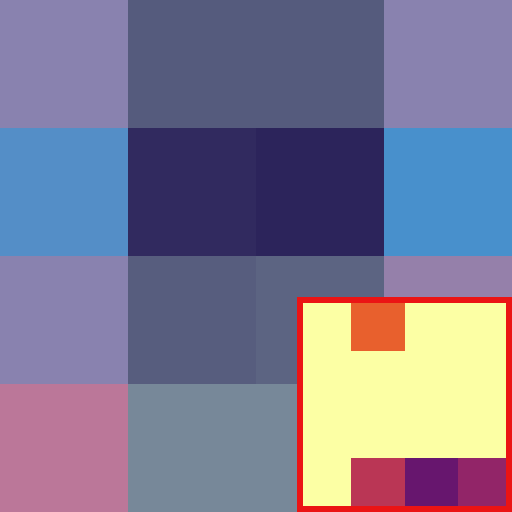} & \mipQualImg{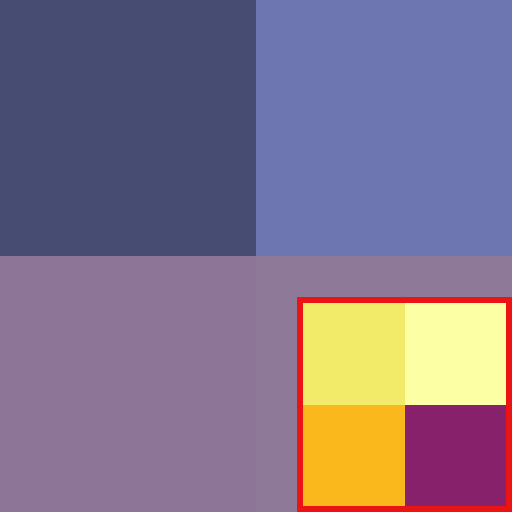} & \mipQualImg{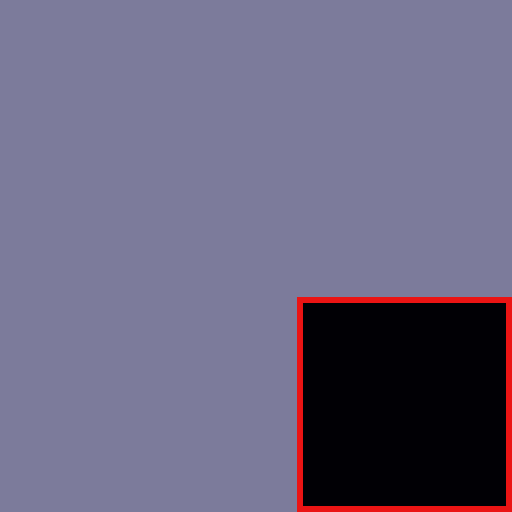} \\[\mipQualRowGap]
      & \mipQualMethod{NTC}{0.290 bppc} & \mipQualImg{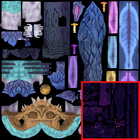} & \mipQualImg{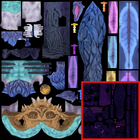} & \mipQualImg{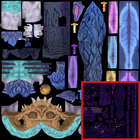} & \mipQualImg{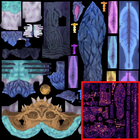} & \mipQualImg{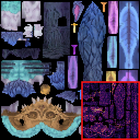} & \mipQualImg{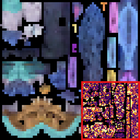} & \mipQualImg{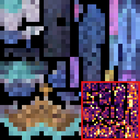} & \mipQualImg{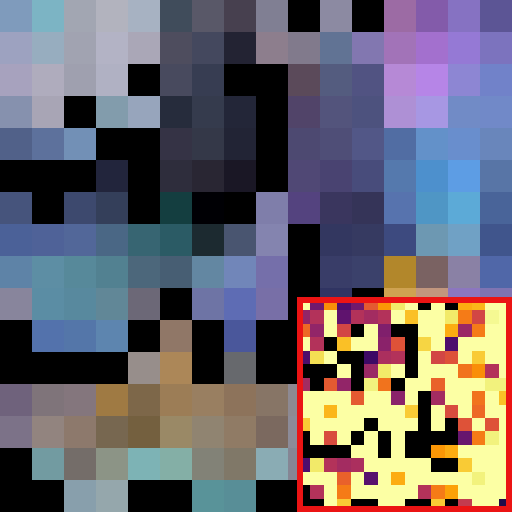} & \mipQualImg{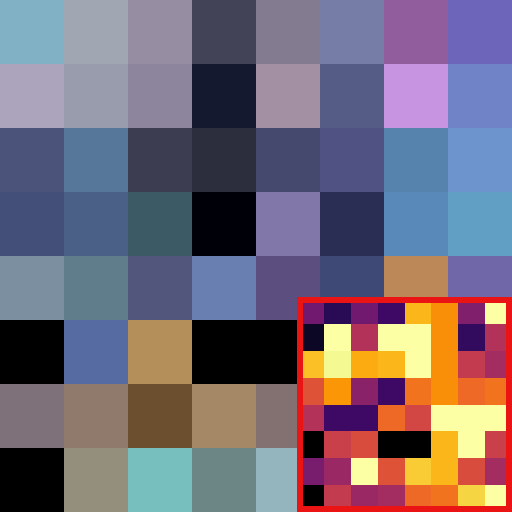} & \mipQualImg{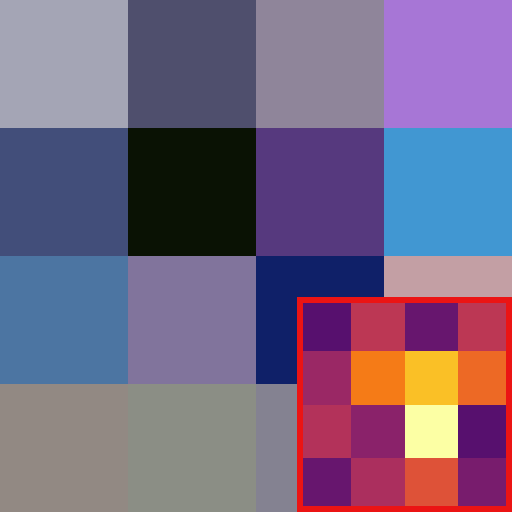} & \mipQualImg{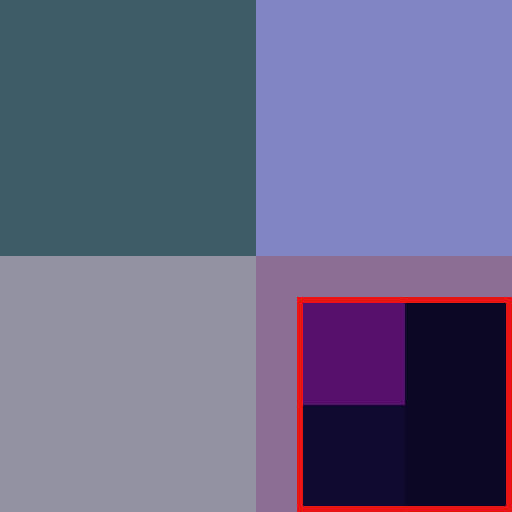} & \mipQualImg{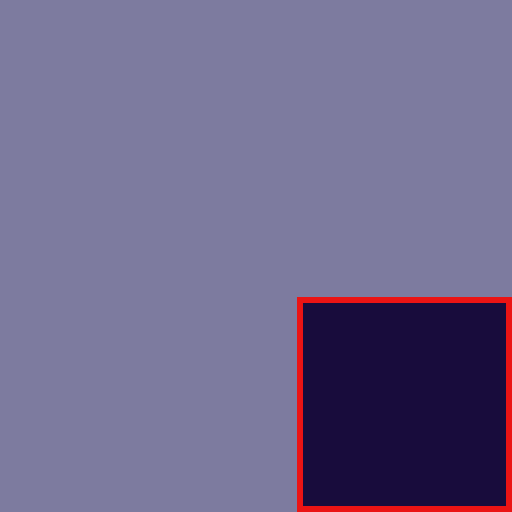} \\
    \mipQualSep
      & \mipQualMethod{GT}{} & \mipQualImg{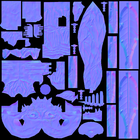} & \mipQualImg{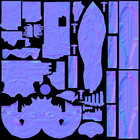} & \mipQualImg{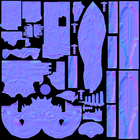} & \mipQualImg{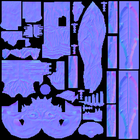} & \mipQualImg{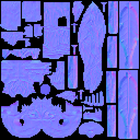} & \mipQualImg{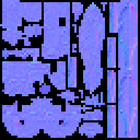} & \mipQualImg{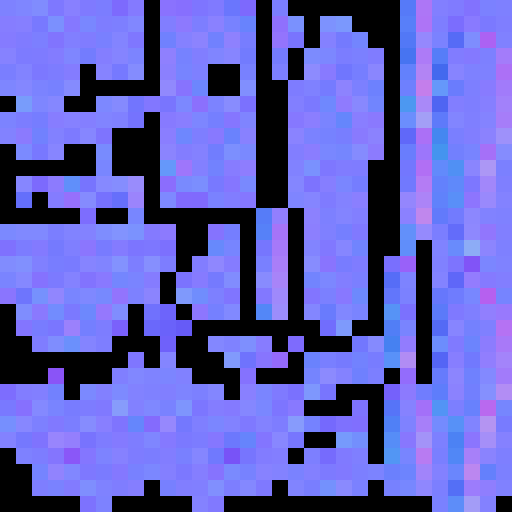} & \mipQualImg{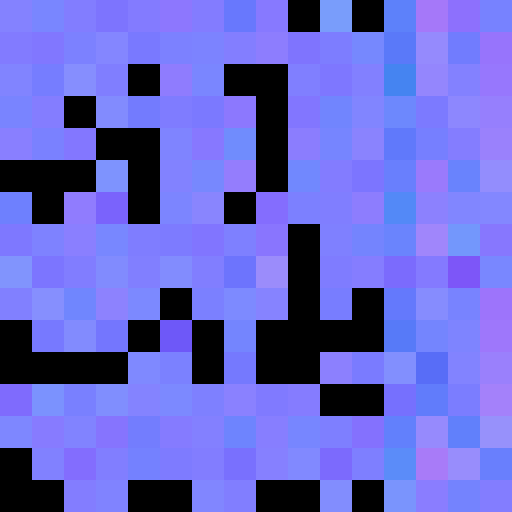} & \mipQualImg{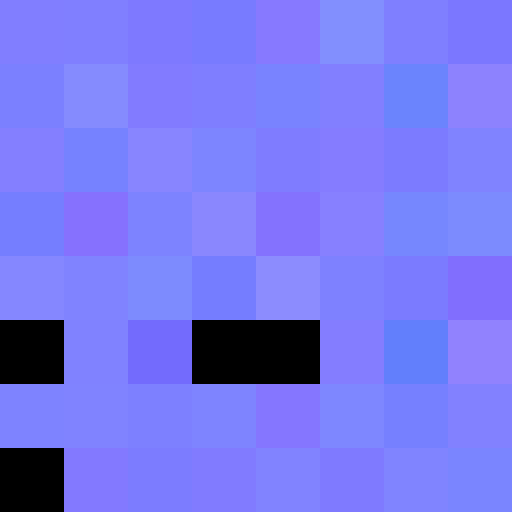} & \mipQualImg{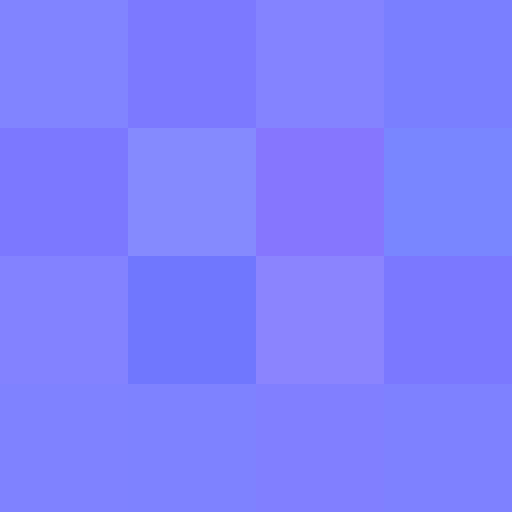} & \mipQualImg{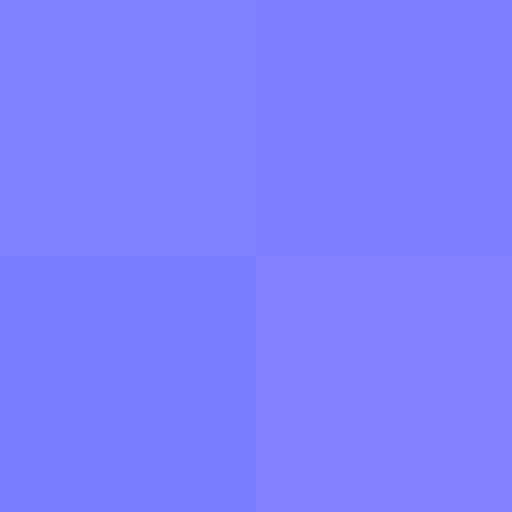} & \mipQualImg{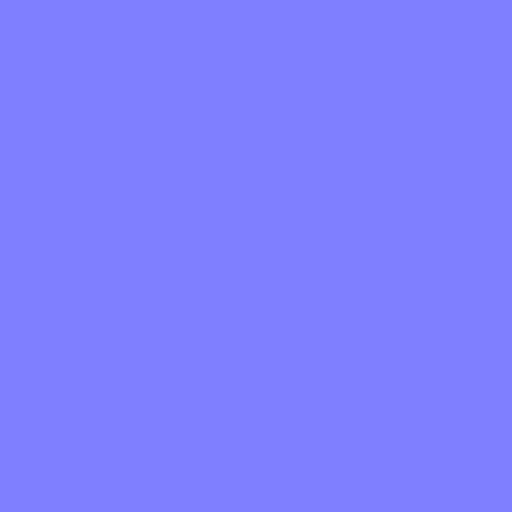} \\[\mipQualRowGap]
      & \mipQualMethod{Ours}{0.210 bppc} & \mipQualImg{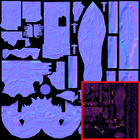} & \mipQualImg{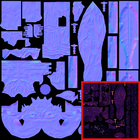} & \mipQualImg{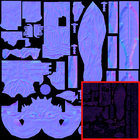} & \mipQualImg{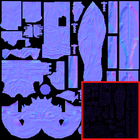} & \mipQualImg{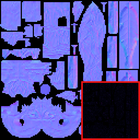} & \mipQualImg{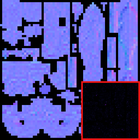} & \mipQualImg{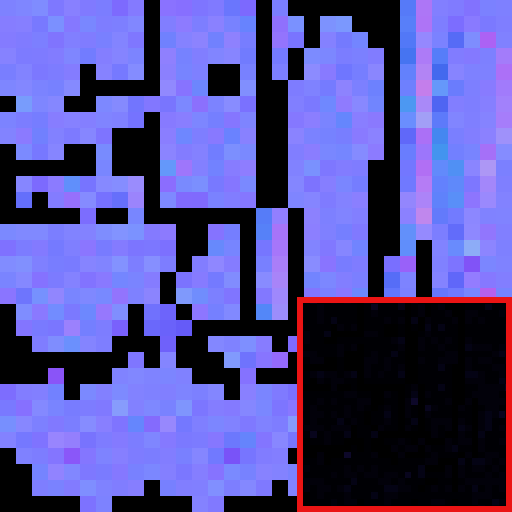} & \mipQualImg{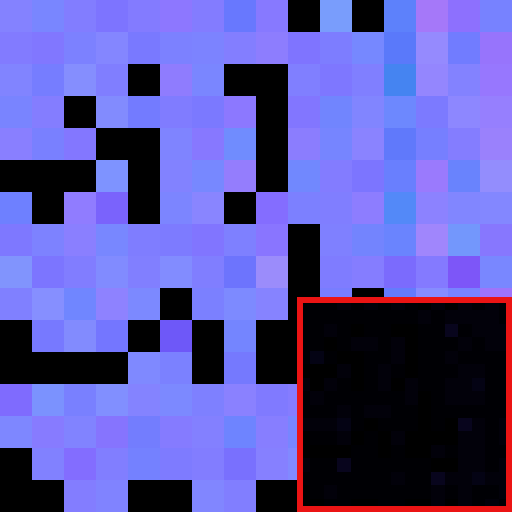} & \mipQualImg{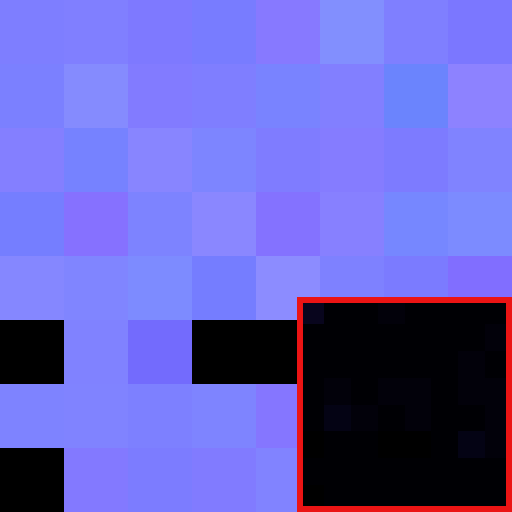} & \mipQualImg{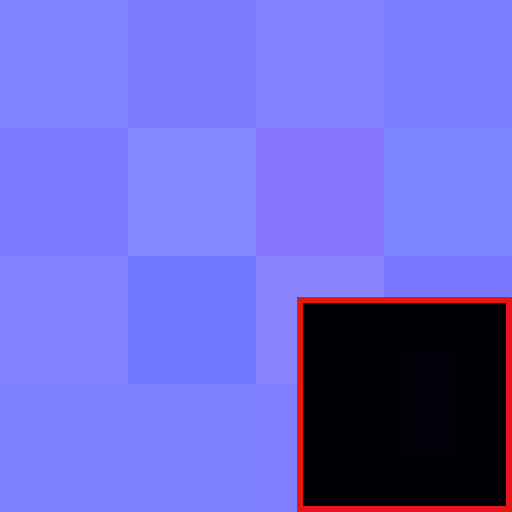} & \mipQualImg{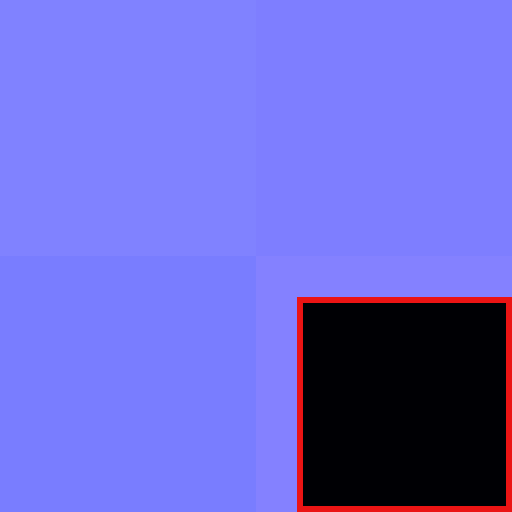} & \mipQualImg{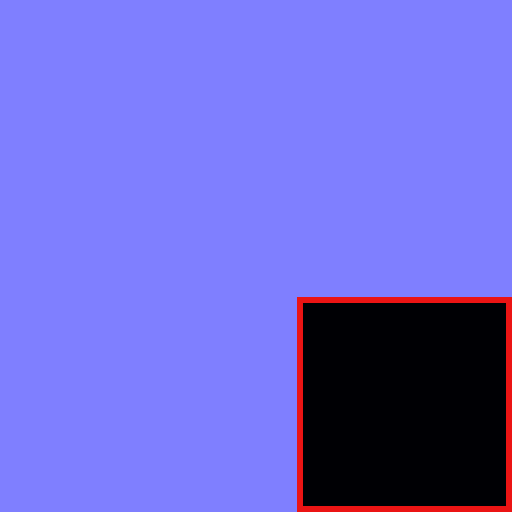} \\[\mipQualRowGap]
    \mipQualStream{Normal map} & \mipQualMethod{Image-GS}{0.445 bppc} & \mipQualImg{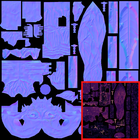} & \mipQualImg{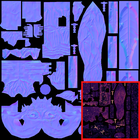} & \mipQualImg{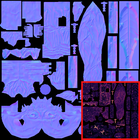} & \mipQualImg{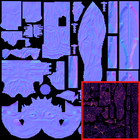} & \mipQualImg{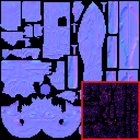} & \mipQualImg{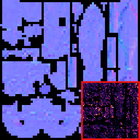} & \mipQualImg{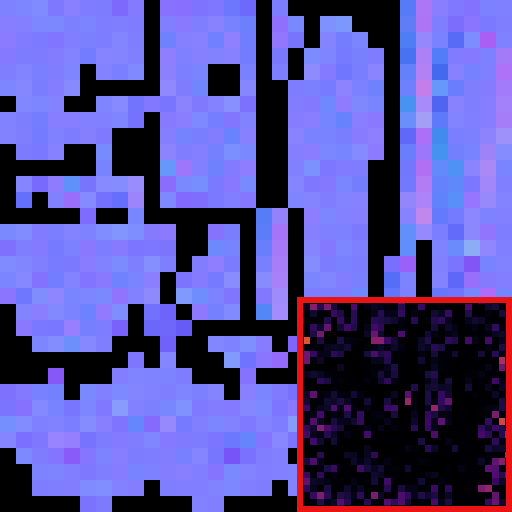} & \mipQualImg{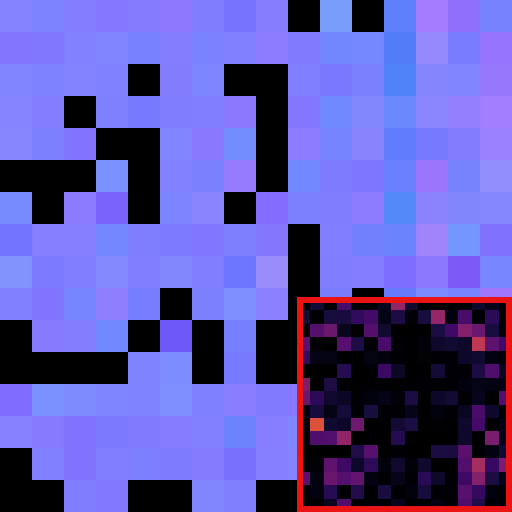} & \mipQualImg{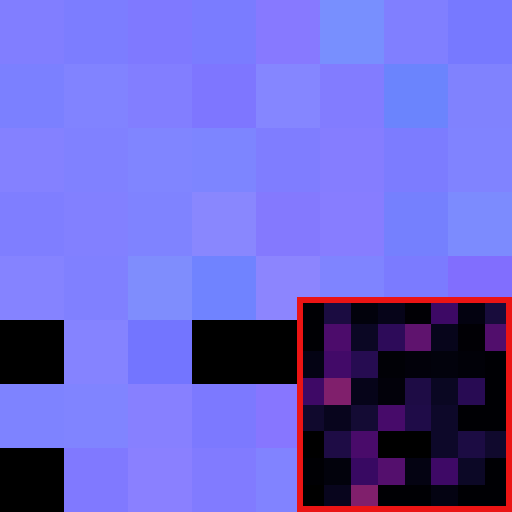} & \mipQualImg{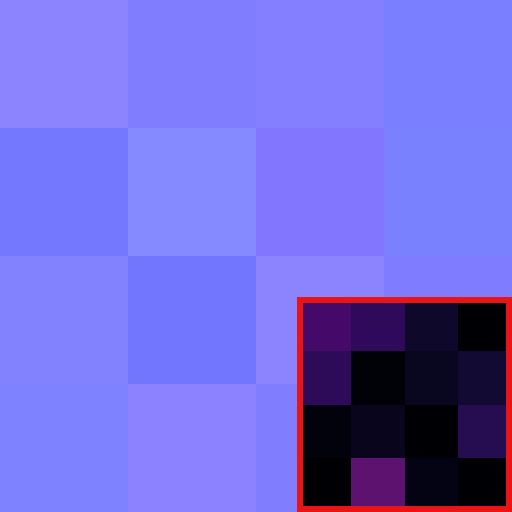} & \mipQualImg{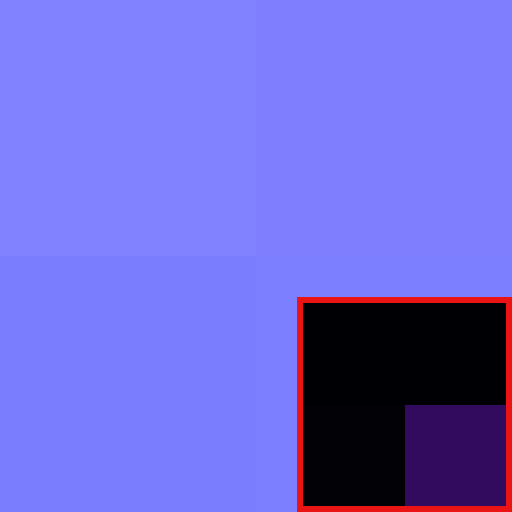} & \mipQualImg{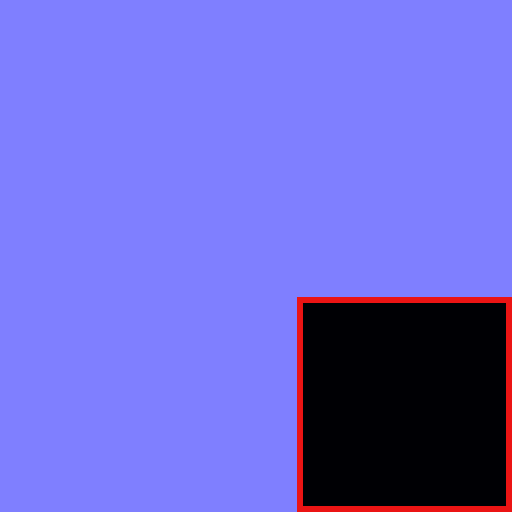} \\[\mipQualRowGap]
      & \mipQualMethod{ASTC}{0.384 bppc} & \mipQualImg{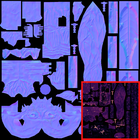} & \mipQualImg{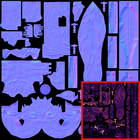} & \mipQualImg{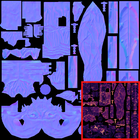} & \mipQualImg{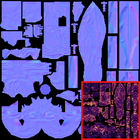} & \mipQualImg{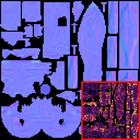} & \mipQualImg{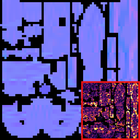} & \mipQualImg{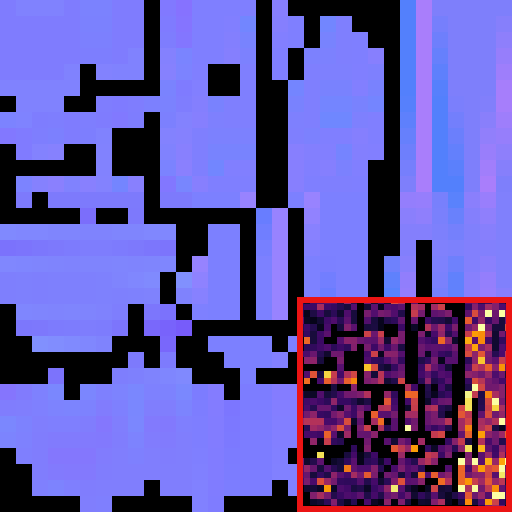} & \mipQualImg{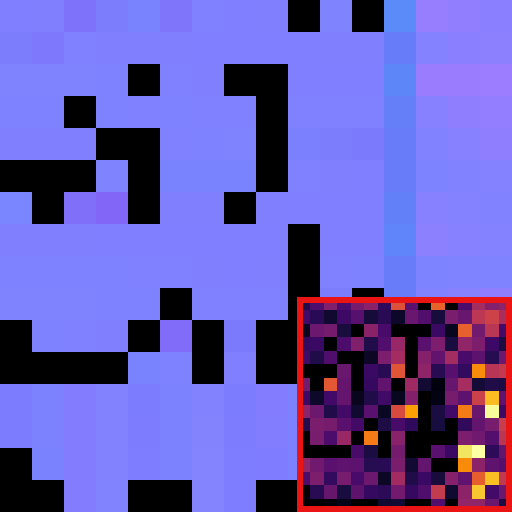} & \mipQualImg{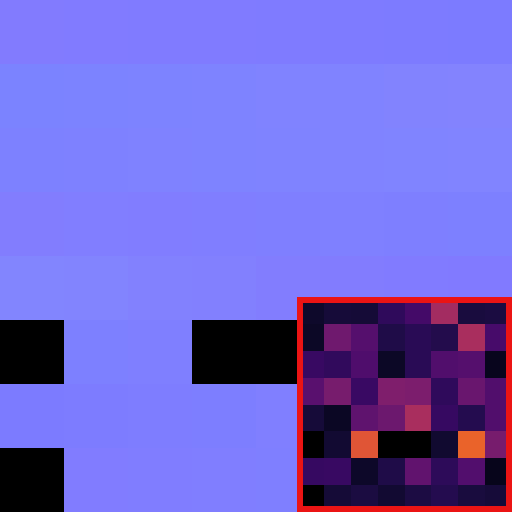} & \mipQualImg{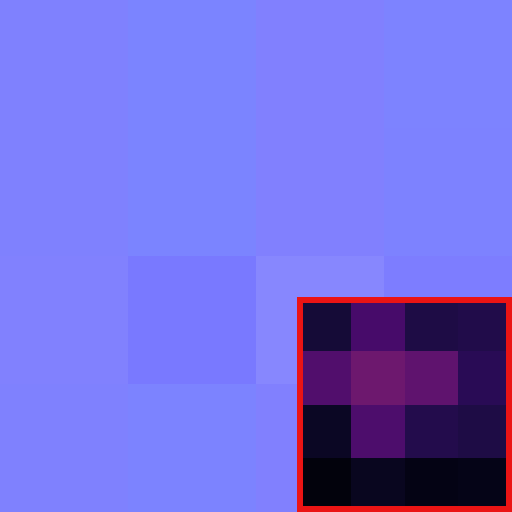} & \mipQualImg{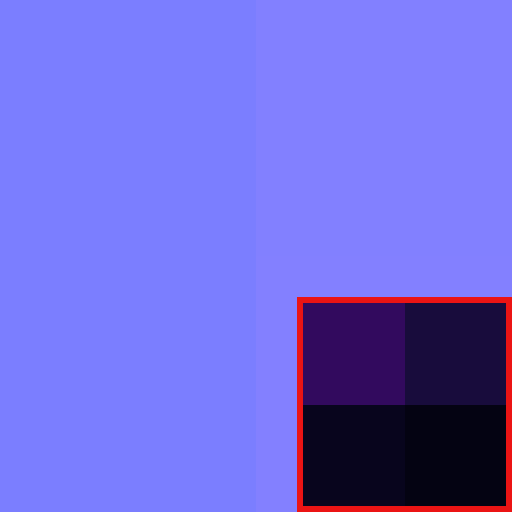} & \mipQualImg{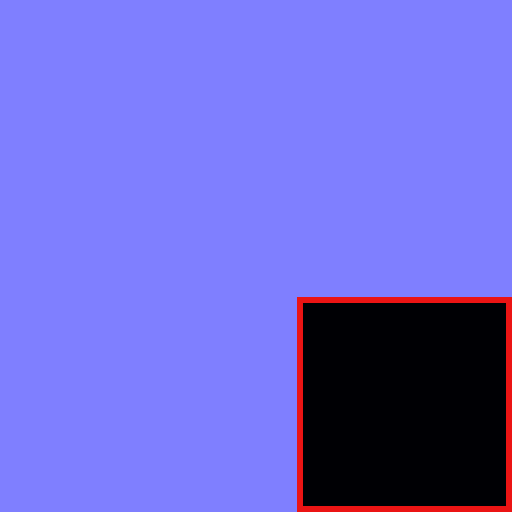} \\[\mipQualRowGap]
      & \mipQualMethod{NTC}{0.290 bppc} & \mipQualImg{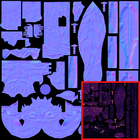} & \mipQualImg{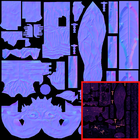} & \mipQualImg{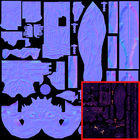} & \mipQualImg{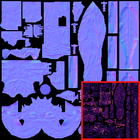} & \mipQualImg{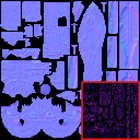} & \mipQualImg{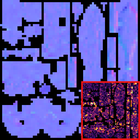} & \mipQualImg{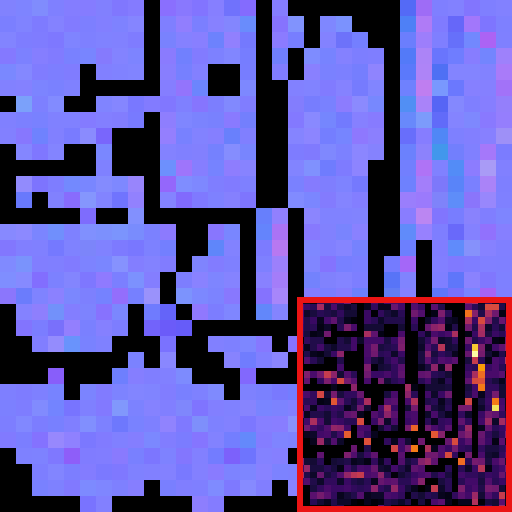} & \mipQualImg{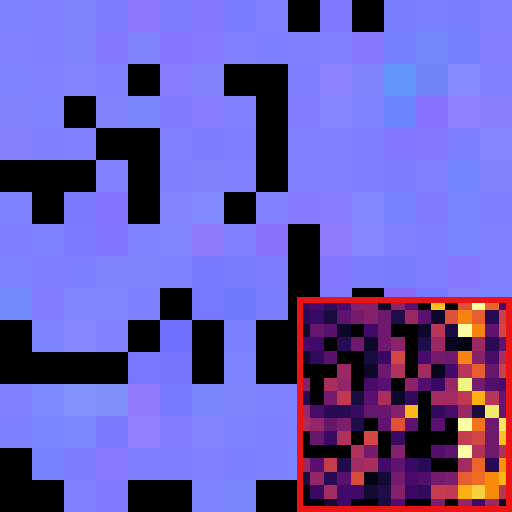} & \mipQualImg{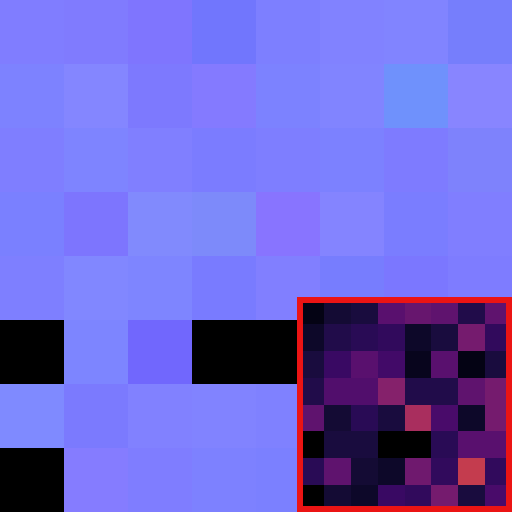} & \mipQualImg{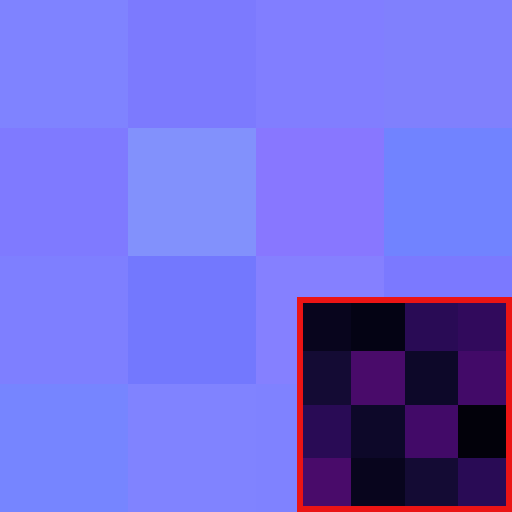} & \mipQualImg{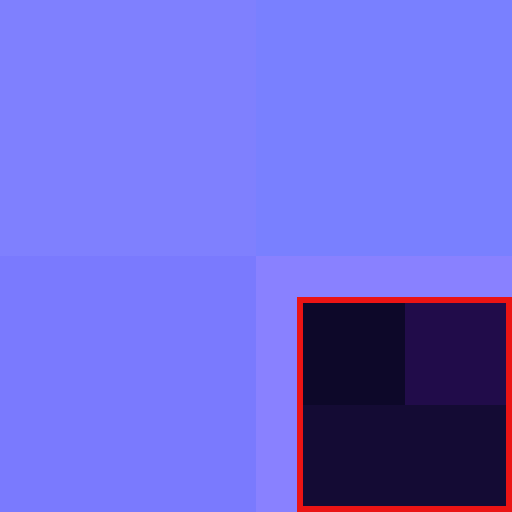} & \mipQualImg{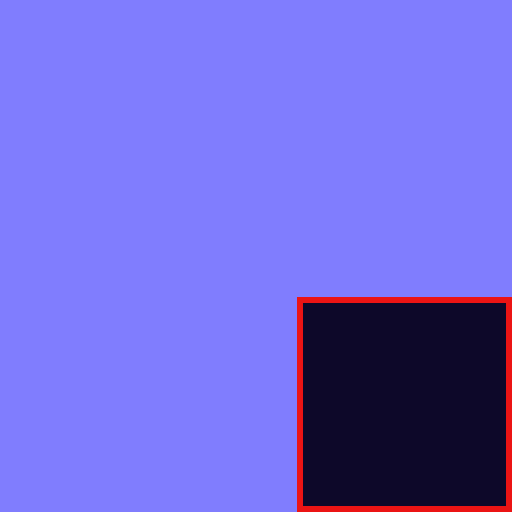} \\
  \end{tabular}%
  }}
  \caption{Mipmap reconstruction comparison on the \textit{Fantasy Character} SVBRDF stack. Rows show the reference and reconstructed mipmap pyramids, with insets showing per-pixel error maps against the reference. GTC delivers favorable rate-distortion performance, particularly at coarser levels.}
  \label{fig:main-mipmap-error-inset-fantasy-character}
  \par\vspace{7pt}
  \oursSweepFontSize
  \setlength{\tabcolsep}{0.7pt}
  \renewcommand{\arraystretch}{0.78}
  \makebox[\textwidth][c]{\resizebox{\oursSweepTableTargetW}{!}{%
  \begin{tabular}{@{}c@{\hspace{2pt}}c@{\hspace{3pt}}*{4}{c@{\hspace{2pt}}}@{\hspace{4pt}}c@{\hspace{2pt}}c@{\hspace{3pt}}*{3}{c@{\hspace{2pt}}}c@{}}
     & \oursSweepMip{mip 0}{2048$^2$} & \oursSweepImg{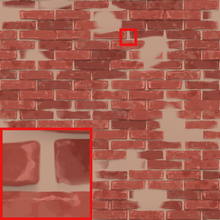} & \oursSweepImg{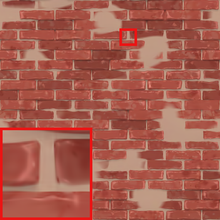} & \oursSweepImg{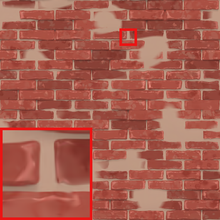} & \oursSweepImg{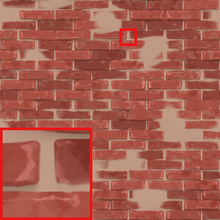} & & \oursSweepMip{mip 0}{2048$^2$} & \oursSweepImg{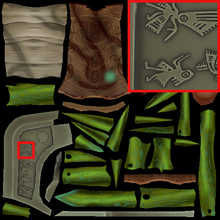} & \oursSweepImg{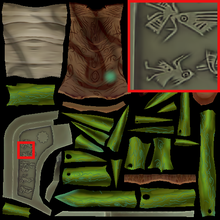} & \oursSweepImg{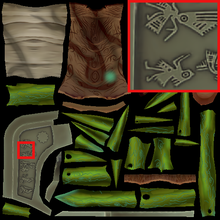} & \oursSweepImg{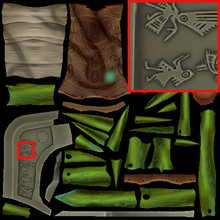} \\[\oursSweepRowGap]
    \oursSweepStream{Base color} & \oursSweepMip{mip 4}{128$^2$} & \oursSweepImg{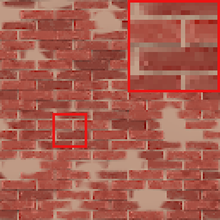} & \oursSweepImg{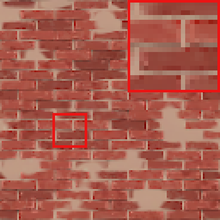} & \oursSweepImg{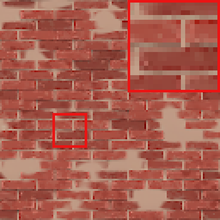} & \oursSweepImg{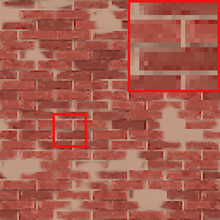} & \oursSweepStream{Base color} & \oursSweepMip{mip 4}{128$^2$} & \oursSweepImg{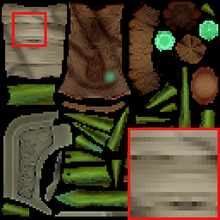} & \oursSweepImg{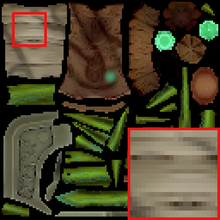} & \oursSweepImg{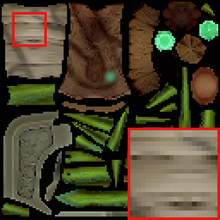} & \oursSweepImg{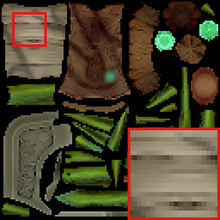} \\[\oursSweepRowGap]
     & \oursSweepMip{mip 7}{16$^2$} & \oursSweepImg{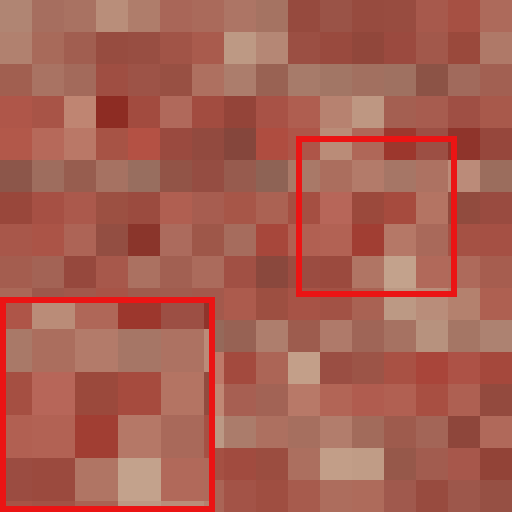} & \oursSweepImg{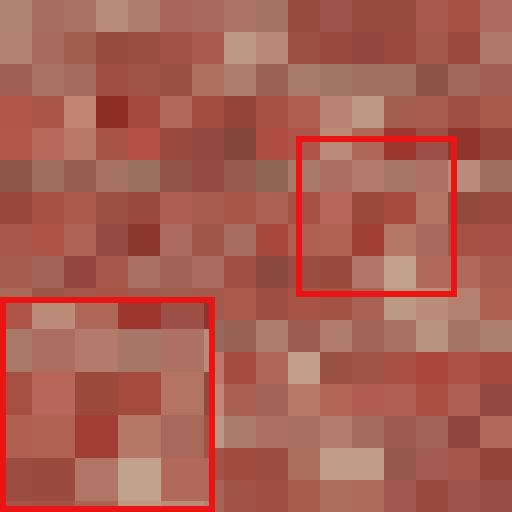} & \oursSweepImg{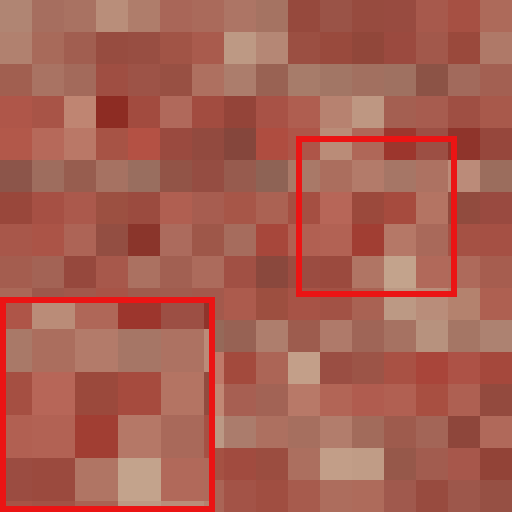} & \oursSweepImg{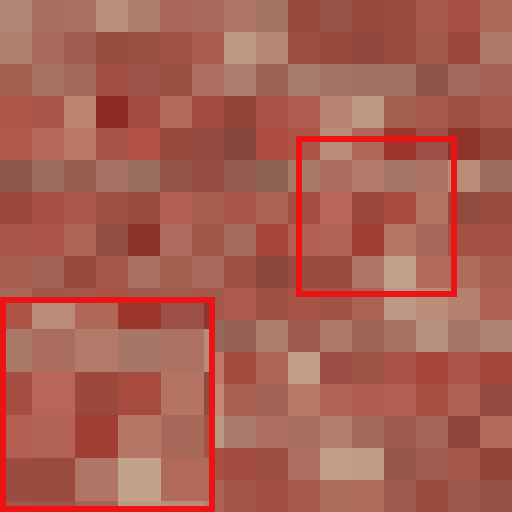} & & \oursSweepMip{mip 7}{16$^2$} & \oursSweepImg{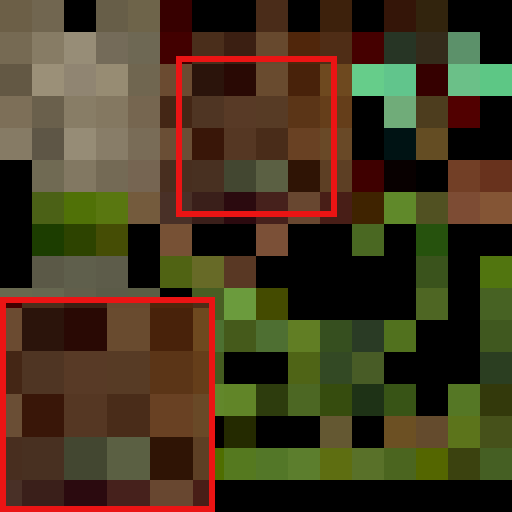} & \oursSweepImg{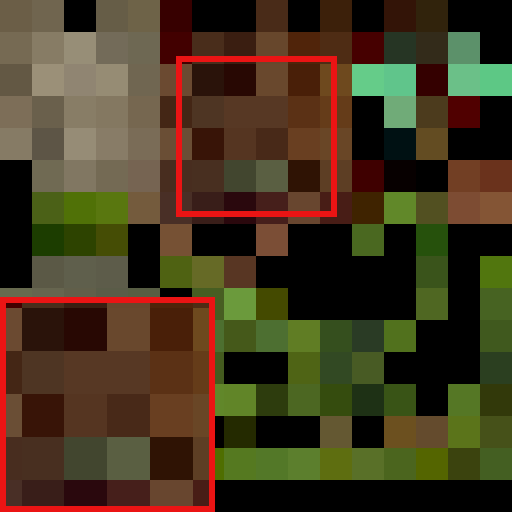} & \oursSweepImg{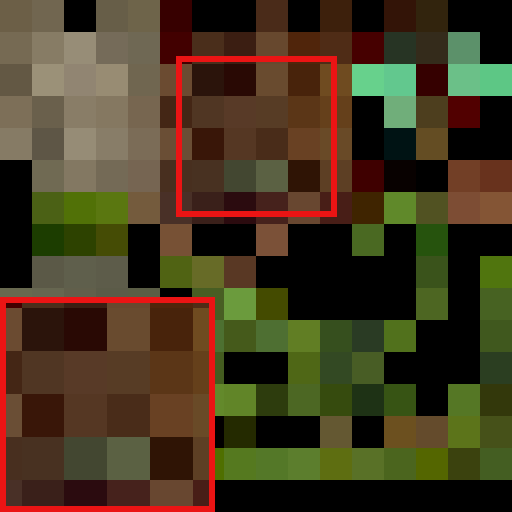} & \oursSweepImg{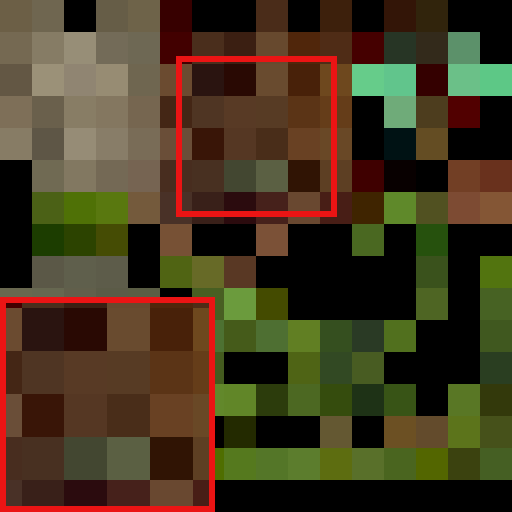} \\[\oursSweepMetricTopGap]
    \multicolumn{2}{c}{} &
    \oursSweepMetricName{GT} &
    \oursSweepMetric{Ours}{0.056 bppc} &
    \oursSweepMetric{Ours}{0.111 bppc} &
    \oursSweepMetric{Ours}{0.248 bppc} &
    \multicolumn{2}{c}{} &
    \oursSweepMetricName{GT} &
    \oursSweepMetric{Ours}{0.050 bppc} &
    \oursSweepMetric{Ours}{0.104 bppc} &
    \oursSweepMetric{Ours}{0.135 bppc} \\
  \end{tabular}%
  }}
  \caption{Bitrate sweep for our method on the base color maps of \textit{Brick Wall} (left) and \textit{Battle Axe} (right). Higher bitrates recover progressively finer details.}
  \label{fig:main-gtc-sweep-brick-wall-battle-axe}
  \end{minipage}
\end{figure*}

\endgroup
\clearpage

\section*{Supplementary Material}
\setcounter{section}{0}
\setcounter{subsection}{0}
\setcounter{equation}{0}
\setcounter{figure}{0}
\setcounter{table}{0}
\renewcommand{\thesection}{S\arabic{section}}
\renewcommand{\thesubsection}{\thesection.\arabic{subsection}}
\renewcommand{\theequation}{S\arabic{equation}}
\renewcommand{\thefigure}{S\arabic{figure}}
\renewcommand{\thetable}{S\arabic{table}}
\renewcommand{\theHsection}{supp.S\arabic{section}}
\renewcommand{\theHsubsection}{supp.S\arabic{section}.\arabic{subsection}}
\renewcommand{\theHequation}{supp.S\arabic{equation}}
\renewcommand{\theHfigure}{supp.S\arabic{figure}}
\renewcommand{\theHtable}{supp.S\arabic{table}}

\section{Implementation Details}
\subsection{Mipmap Generation}

All ground-truth mipmap pyramids are generated offline with NVIDIA Texture Tools Exporter~\cite{nvidiaTextureToolsExporter,nvidiaTextureToolsExporterBlog2020}. 
For each texture stack, the finest stored mip level has resolution $2048\times2048$, and each subsequent level halves both spatial dimensions until the final $1\times1$ level.

\subsection{Benchmark Settings}

\paragraph{GTC}
We evaluate seven rate settings with maximum Gaussian budgets of
\(10{,}000\), \(25{,}941\), \(50{,}000\), \(87{,}381\), \(150{,}000\), \(283{,}989\), and \(500{,}000\).

\paragraph{Image-GS}
Each mip level is compressed independently. For each level \(l\), all material channels are concatenated into a single joint texture stack and optimized as one Image-GS model. The Gaussian budget is \(N_{\mathrm{IGS}}^{(l)}=\mathrm{clamp}(\mathrm{round}(N_0/r^l),1,H^{(l)}W^{(l)})\), where \(l=0\) is the finest level, \(N_0\in\{2048,4096,8192,16384,32768\}\), and \(r=1.5\). All other parameters use official defaults.

\paragraph{ASTC}
We use ASTC Encoder v5.3.0 with the LDR-linear profile and the exhaustive search preset. We evaluate block sizes \(\{4{\times}4,5{\times}4,\\5{\times}5,6{\times}5,6{\times}6,8{\times}5,8{\times}6,8{\times}8,10{\times}5,10{\times}6,10{\times}8,10{\times}10,12{\times}10,12{\times}12\}\). All other parameters use official defaults.

\paragraph{NTC}
We use the official NVIDIA RTXNTC command-line tool with official JSON manifests. We sweep its bits-per-pixel rate-control setting over \(\{1,1.5,2,3,4,5,6,8,12,16,20\}\). All other parameters use official defaults.

\paragraph{JPEG}
We use the libjpeg-turbo encoder with a quality sweep over \(\{20,30,40,50,60,70,80,90,95,100\}\). We enable optimized coding, progressive mode, arithmetic coding, integer DCT, \(1{\times}1\) sampling for RGB maps, grayscale mode for scalar maps, and 12-bit precision for inputs above 8 bits when applicable. All other parameters use official defaults.

\paragraph{JPEG-XL}
We use the official libjxl encoder with a distance sweep over \(\{0,0.25,0.5,0.75,1,1.5,2,3,4,6\}\). We set the effort level to 10 and disable JPEG-lossless mode. All other parameters use official defaults.

\subsection{Optimization Schedule}

\paragraph{Phase 1: progressive optimization and pruning.}
We concatenate all material channels at every mip level and start from the coarsest $1\times1$ level with one Gaussian.
For subsequent levels, the per-level Gaussian budget is derived from the user-specified maximum Gaussian budget using the Gaussian number budget rule described in the main paper.
Every 100 iterations, we deactivate feature channels with $|f_{i,c}|<3\times 10^{-4}$ and remove Gaussians whose channels have all become inactive.
The current level is treated as saturated when fewer than 20 Gaussians are pruned at a checkpoint; the optimizer then advances to the next finer level.
Once all levels have been introduced, feature regularization and pruning are disabled, the optimizer is reset, all learning rates are divided by 10, and refinement begins.

\paragraph{Phase 2: fixed-set refinement.}
The Gaussian set is fixed, and only the parameters continue to be optimized.
We keep the best checkpoint according to texel-weighted aggregated PSNR over the mipmap pyramid.
If this score does not improve for 5000 iterations, the best checkpoint is restored, all learning rates are divided by another factor of 10, and quantization-aware refinement begins.

\paragraph{Phase 3: quantization-aware refinement.}
Quantization (Sec.~\ref{sec:fixed_bit_quantization}) is enabled in the forward pass, the optimizer is reset, and training continues with quantization-aware fine-tuning.
This stage terminates when the PSNR fails to improve for 1000 iterations.

\subsection{Balancing Updates Across Mip Levels}

At each iteration, one mip level is sampled uniformly.
With our indexing convention, level 0 is the finest level and level $L-1$ is the coarsest level.
A Gaussian with LoD label $\ell_i$ is visible at a sampled level $l$ when $l\le\ell_i$, which occurs with probability $(\ell_i+1)/L$.
Coarser Gaussians are therefore updated more often because they are shared by more levels.
To balance the effective update magnitude, after each Adam step we rescale the update of the full parameter vector $\boldsymbol{\varphi}_i$ of Gaussian $i$ by
\begin{equation}
  m_{\ell_i} = \frac{1}{\ell_i+1},
  \qquad
  \boldsymbol{\varphi}_i
  \leftarrow
  \boldsymbol{\varphi}_i^{\mathrm{old}}
  +
  m_{\ell_i}
  \bigl(
    \boldsymbol{\varphi}_i^{\mathrm{adam}}
    -
    \boldsymbol{\varphi}_i^{\mathrm{old}}
  \bigr),
  \label{eq:lr_impl}
\end{equation}
where $\boldsymbol{\varphi}_i^{\mathrm{old}}$ denotes the parameter vector before the Adam step and $\boldsymbol{\varphi}_i^{\mathrm{adam}}$ denotes the parameter vector after the Adam step.
For numerical stability, the implementation stores inverse Gaussian scales and clamps them from below by $10^{-4}$.

\subsection{Quantization}
\label{sec:fixed_bit_quantization}

The deployment format uses frozen uniform quantizers initialized from the current parameters.
We use one global quantizer for Gaussian centers, one global quantizer for rotations, one LoD-specific quantizer for Gaussian scales at each LoD, and one LoD-specific quantizer for features at each LoD.
Before quantization, rotations are wrapped to the canonical $\pi$-periodic interval $[-\pi/2,\pi/2)$, and Gaussian scales are mapped to the $\log_2$ domain.

For a parameter group $\mathbf{x}$ quantized with $b$ bits, let $\mathbf{x}_{\min}$ and $\mathbf{x}_{\max}$ denote its elementwise minimum and maximum over the current state.
We initialize the frozen quantizer with
\begin{equation}
\begin{aligned}
  \boldsymbol{\Delta}
  &=
  \frac{\mathbf{x}_{\max}-\mathbf{x}_{\min}}{2^{b}-1},
  \qquad
  \mathbf{o}=\mathbf{x}_{\min}, \\
  \mathbf{q}
  &=
  \mathrm{clip}\!\left(
    \left\lfloor
      \frac{\mathbf{x}-\mathbf{o}}{\boldsymbol{\Delta}}
    \right\rceil,
    0,
    2^{b}-1
  \right),
  \qquad
  \hat{\mathbf{x}}=\mathbf{q}\odot\boldsymbol{\Delta}+\mathbf{o},
\end{aligned}
\label{eq:quant_impl}
\end{equation}
where $\boldsymbol{\Delta}$ is the quantization step size, $\mathbf{o}$ is the offset, $\mathbf{q}$ is the integer code, $\hat{\mathbf{x}}$ is the dequantized value used in the forward pass, and $\odot$ denotes element-wise multiplication.
During quantization-aware refinement, gradients through the quantizer are approximated with the straight-through estimator~\cite{bengio2013estimating}.

Bit widths for Gaussian centers, Gaussian scales, rotations, and features are selected by searching over $b\in[6,16]$ for each parameter group.
Candidate bit-width tuples are grouped and evaluated from smallest to largest estimated storage size.
We keep the first size group whose best candidate preserves the finest mip level within 0.5 dB of the restored checkpoint.

\newcommand{\additionalResultsHeading}{%
  \vspace*{-1.0\baselineskip}
  \section{Additional Results}
}
\setkeys{Gin}{interpolate=false}

\newcommand{\qualFontSize}{\footnotesize}
\newcommand{\qualMetricGap}{1pt}
\newcommand{\qualMetricTopGap}{27pt}
\newcommand{\qualRowGap}{27pt}
\newcommand{\qualMipLabelW}{0.032\textwidth}
\newcommand{\qualCellW}{0.106\textwidth}
\newcommand{\qualTableTargetW}{\linewidth}
\newcommand{\qualVCenter}[1]{\raisebox{-0.5\height}{#1}}
\newcommand{\qualImg}[2]{\qualVCenter{\includegraphics[width=\qualCellW]{figs/qual_subfigs/#1/#2.png}}}
\newcommand{\qualMip}[2]{\qualVCenter{\makebox[\qualMipLabelW][r]{\begin{tabular}{@{}r@{}}\textbf{#1}\\#2\end{tabular}}}}
\newcommand{\qualStream}[1]{\qualVCenter{\rotatebox{90}{\textbf{#1}}}}
\newcommand{\qualRow}[3]{%
  #2 &
  \qualImg{#1}{#3_gt} &
  \qualImg{#1}{#3_gt_zoom} &
  \qualImg{#1}{#3_gtc} &
  \qualImg{#1}{#3_image_gs} &
  \qualImg{#1}{#3_astc} &
  \qualImg{#1}{#3_ntc} &
  \qualImg{#1}{#3_jpeg} &
  \qualImg{#1}{#3_jxl}%
}
\newcommand{\qualHead}{%
  & &
  \textbf{GT} &
  \textbf{GT Crop} &
  \textbf{Ours} &
  \textbf{Image-GS} &
  \textbf{ASTC} &
  \textbf{NTC} &
  \textbf{JPEG} &
  \textbf{JPEG-XL} \\
}
\newcommand{\qualMetric}[3]{%
  \vtop{\hbox{\begin{tabular}{@{}c@{}}
    \textbf{#1}\\[\qualMetricGap]
    #2\\[\qualMetricGap]
    #3
  \end{tabular}}}%
}
\newcommand{\qualMetricName}[1]{%
  \vtop{\hbox{\begin{tabular}{@{}c@{}}
    \makebox[\qualCellW][c]{\textbf{#1}}\\[\qualMetricGap]
    \makebox[\qualCellW][c]{}\\[\qualMetricGap]
    \makebox[\qualCellW][c]{}
  \end{tabular}}}%
}
\newcommand{\qualEmptyMetric}{\vtop{\hbox{\makebox[\qualCellW]{}}}}
\newcommand{\qualSep}{\noalign{\vskip 2pt}\hline\noalign{\vskip 2pt}}
\providecommand{\additionalResultsHeading}{}

\begin{figure*}[t]
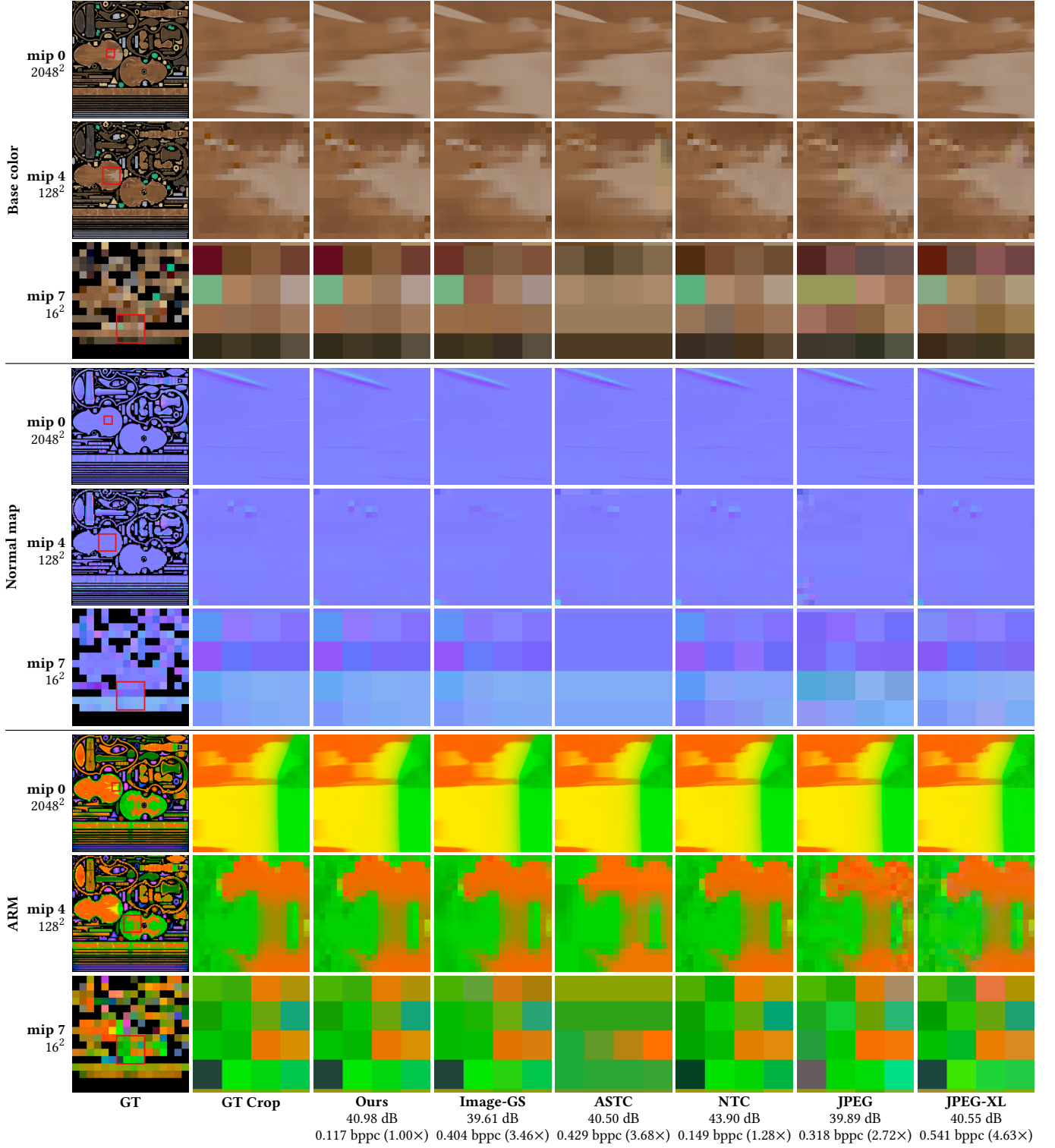

  \additionalResultsHeading
  \centering
  \qualFontSize
  \setlength{\tabcolsep}{0.7pt}
  \renewcommand{\arraystretch}{0.78}
  \resizebox{\qualTableTargetW}{!}{%
  \begin{tabular}{@{}c@{\hspace{5pt}}c@{\hspace{3pt}}*{7}{c@{\hspace{2pt}}}c@{}}
    & \qualRow{violin_87381}{\qualMip{mip 0}{2048$^2$}}{bc_m0} \\[\qualRowGap]
    \qualStream{Base color} & \qualRow{violin_87381}{\qualMip{mip 4}{128$^2$}}{bc_m4} \\[\qualRowGap]
    & \qualRow{violin_87381}{\qualMip{mip 7}{16$^2$}}{bc_m7} \\
    \qualSep
    & \qualRow{violin_87381}{\qualMip{mip 0}{2048$^2$}}{n_m0} \\[\qualRowGap]
    \qualStream{Normal map} & \qualRow{violin_87381}{\qualMip{mip 4}{128$^2$}}{n_m4} \\[\qualRowGap]
    & \qualRow{violin_87381}{\qualMip{mip 7}{16$^2$}}{n_m7} \\
    \qualSep
    & \qualRow{violin_87381}{\qualMip{mip 0}{2048$^2$}}{arm_m0} \\[\qualRowGap]
    \qualStream{ARM} & \qualRow{violin_87381}{\qualMip{mip 4}{128$^2$}}{arm_m4} \\[\qualRowGap]
    & \qualRow{violin_87381}{\qualMip{mip 7}{16$^2$}}{arm_m7} \\[\qualMetricTopGap]
    \multicolumn{2}{c}{} &
    \qualMetricName{GT} &
    \qualMetricName{GT Crop} &
    \qualMetric{Ours}{40.98 dB}{0.117 bppc (1.00$\times$)} &
    \qualMetric{Image-GS}{39.61 dB}{0.404 bppc (3.46$\times$)} &
    \qualMetric{ASTC}{40.50 dB}{0.429 bppc (3.68$\times$)} &
    \qualMetric{NTC}{43.90 dB}{0.149 bppc (1.28$\times$)} &
    \qualMetric{JPEG}{39.89 dB}{0.318 bppc (2.72$\times$)} &
    \qualMetric{JPEG-XL}{40.55 dB}{0.541 bppc (4.63$\times$)} \\
  \end{tabular}%
  }
  \caption{Qualitative comparison on the \textit{Violin} SVBRDF stack. Reported PSNR values use texel-weighted aggregation over the full mipmap pyramid. ARM denotes ambient occlusion, roughness, and metallic channels.}
  \label{fig:selected3-violin-87381}
\end{figure*}

\begin{figure*}[t]
  \centering
  \qualFontSize
  \setlength{\tabcolsep}{0.7pt}
  \renewcommand{\arraystretch}{0.78}
  \resizebox{\qualTableTargetW}{!}{%
  \begin{tabular}{@{}c@{\hspace{5pt}}c@{\hspace{3pt}}*{7}{c@{\hspace{2pt}}}c@{}}
    & \qualRow{cliff_rocks_87381}{\qualMip{mip 0}{2048$^2$}}{bc_m0} \\[\qualRowGap]
    \qualStream{Base color} & \qualRow{cliff_rocks_87381}{\qualMip{mip 4}{128$^2$}}{bc_m4} \\[\qualRowGap]
    & \qualRow{cliff_rocks_87381}{\qualMip{mip 7}{16$^2$}}{bc_m7} \\
    \qualSep
    & \qualRow{cliff_rocks_87381}{\qualMip{mip 0}{2048$^2$}}{n_m0} \\[\qualRowGap]
    \qualStream{Normal map} & \qualRow{cliff_rocks_87381}{\qualMip{mip 4}{128$^2$}}{n_m4} \\[\qualRowGap]
    & \qualRow{cliff_rocks_87381}{\qualMip{mip 7}{16$^2$}}{n_m7} \\
    \qualSep
    & \qualRow{cliff_rocks_87381}{\qualMip{mip 0}{2048$^2$}}{arh_m0} \\[\qualRowGap]
    \qualStream{ARH} & \qualRow{cliff_rocks_87381}{\qualMip{mip 4}{128$^2$}}{arh_m4} \\[\qualRowGap]
    & \qualRow{cliff_rocks_87381}{\qualMip{mip 7}{16$^2$}}{arh_m7} \\[\qualMetricTopGap]
    \multicolumn{2}{c}{} &
    \qualMetricName{GT} &
    \qualMetricName{GT Crop} &
    \qualMetric{Ours}{51.62 dB}{0.128 bppc (1.00$\times$)} &
    \qualMetric{Image-GS}{48.02 dB}{0.404 bppc (3.15$\times$)} &
    \qualMetric{ASTC}{51.19 dB}{0.597 bppc (4.65$\times$)} &
    \qualMetric{NTC}{52.47 dB}{0.151 bppc (1.18$\times$)} &
    \qualMetric{JPEG}{50.66 dB}{0.554 bppc (4.31$\times$)} &
    \qualMetric{JPEG-XL}{51.12 dB}{0.176 bppc (1.37$\times$)} \\
  \end{tabular}%
  }
  \caption{Qualitative comparison on the \textit{Cliff Rocks} SVBRDF stack. Reported PSNR values use texel-weighted aggregation over the full mipmap pyramid. ARH denotes ambient occlusion, roughness, and height channels.}
  \label{fig:selected3-cliff-rocks-87381}
\end{figure*}

\begin{figure*}[t]
  \centering
  \qualFontSize
  \setlength{\tabcolsep}{0.7pt}
  \renewcommand{\arraystretch}{0.78}
  \resizebox{\qualTableTargetW}{!}{%
  \begin{tabular}{@{}c@{\hspace{5pt}}c@{\hspace{3pt}}*{7}{c@{\hspace{2pt}}}c@{}}
    & \qualRow{sand_25941}{\qualMip{mip 0}{2048$^2$}}{bc_m0} \\[\qualRowGap]
    \qualStream{Base color} & \qualRow{sand_25941}{\qualMip{mip 4}{128$^2$}}{bc_m4} \\[\qualRowGap]
    & \qualRow{sand_25941}{\qualMip{mip 7}{16$^2$}}{bc_m7} \\
    \qualSep
    & \qualRow{sand_25941}{\qualMip{mip 0}{2048$^2$}}{n_m0} \\[\qualRowGap]
    \qualStream{Normal map} & \qualRow{sand_25941}{\qualMip{mip 4}{128$^2$}}{n_m4} \\[\qualRowGap]
    & \qualRow{sand_25941}{\qualMip{mip 7}{16$^2$}}{n_m7} \\
    \qualSep
    & \qualRow{sand_25941}{\qualMip{mip 0}{2048$^2$}}{arh_m0} \\[\qualRowGap]
    \qualStream{ARH} & \qualRow{sand_25941}{\qualMip{mip 4}{128$^2$}}{arh_m4} \\[\qualRowGap]
    & \qualRow{sand_25941}{\qualMip{mip 7}{16$^2$}}{arh_m7} \\[\qualMetricTopGap]
    \multicolumn{2}{c}{} &
    \qualMetricName{GT} &
    \qualMetricName{GT Crop} &
    \qualMetric{Ours}{49.58 dB}{0.054 bppc (1.00$\times$)} &
    \qualMetric{Image-GS}{47.63 dB}{0.104 bppc (1.92$\times$)} &
    \qualMetric{ASTC}{48.86 dB}{0.597 bppc (10.99$\times$)} &
    \qualMetric{NTC}{50.49 dB}{0.074 bppc (1.36$\times$)} &
    \qualMetric{JPEG}{49.01 dB}{0.293 bppc (5.40$\times$)} &
    \qualMetric{JPEG-XL}{48.11 dB}{0.081 bppc (1.49$\times$)} \\
  \end{tabular}%
  }
  \caption{Qualitative comparison on the \textit{Sand} SVBRDF stack. Reported PSNR values use texel-weighted aggregation over the full mipmap pyramid. ARH denotes ambient occlusion, roughness, and height channels.}
  \label{fig:selected3-sand-25941}
\end{figure*}

\clearpage
\setkeys{Gin}{interpolate=false}

\newcommand{\mipQualFontSize}{\scriptsize}
\newcommand{\mipQualHGap}{1pt}
\newcommand{\mipQualVGap}{14.5pt}
\newcommand{\mipQualSepGap}{1pt}
\newcommand{\mipQualStreamMethodGap}{0pt}
\newcommand{\mipQualMethodImageGap}{2pt}
\newcommand{\mipQualTableTargetW}{\linewidth}

\newcommand{\mipQualHeaderGap}{1pt}
\newcommand{\mipQualRowGap}{\mipQualVGap}
\newcommand{\mipQualStreamGap}{\mipQualSepGap}
\newcommand{\mipQualColGap}{\mipQualHGap}
\newcommand{\mipQualMethodLabelW}{0.075\textwidth}
\newcommand{\mipQualCellW}{0.062\textwidth}
\newcommand{\mipQualVCenter}[1]{\raisebox{\dimexpr\depth/2-\height/2\relax}{#1}}
\newcommand{\mipQualCellStrut}{}
\newcommand{\mipQualImg}[1]{\mipQualCellStrut\mipQualVCenter{\includegraphics[width=\mipQualCellW]{#1}}}
\newcommand{\mipQualMip}[2]{\mipQualVCenter{\makebox[\mipQualCellW][c]{\begin{tabular}{@{}c@{}}\textbf{#1}\\#2\end{tabular}}}}
\newcommand{\mipQualMethod}[2]{%
  \mipQualVCenter{\makebox[\mipQualMethodLabelW][r]{%
    \if\relax\detokenize{#2}\relax
      \textbf{#1}%
    \else
      \begin{tabular}{@{}r@{}}\textbf{#1}\\#2\end{tabular}%
    \fi
  }}%
}
\newcommand{\mipQualStream}[1]{\smash{\mipQualVCenter{\rotatebox{90}{\textbf{#1}}}}}
\newcommand{\mipQualSep}{\noalign{\vskip \mipQualStreamGap}\hline\noalign{\vskip \mipQualStreamGap}}

\clearpage
\begin{figure*}[p!]
  \centering
  \mipQualFontSize
  \setlength{\tabcolsep}{0pt}
  \renewcommand{\arraystretch}{1.0}
  \resizebox{\mipQualTableTargetW}{!}{%
  \begin{tabular}{@{}c@{\hspace{\mipQualStreamMethodGap}}c@{\hspace{\mipQualMethodImageGap}}*{11}{c@{\hspace{\mipQualColGap}}}c@{}}
    \multicolumn{2}{c}{} & \mipQualMip{mip 0}{2048$^2$} & \mipQualMip{mip 1}{1024$^2$} & \mipQualMip{mip 2}{512$^2$} & \mipQualMip{mip 3}{256$^2$} & \mipQualMip{mip 4}{128$^2$} & \mipQualMip{mip 5}{64$^2$} & \mipQualMip{mip 6}{32$^2$} & \mipQualMip{mip 7}{16$^2$} & \mipQualMip{mip 8}{8$^2$} & \mipQualMip{mip 9}{4$^2$} & \mipQualMip{mip 10}{2$^2$} & \mipQualMip{mip 11}{1$^2$} \\[\mipQualHeaderGap]
      & \mipQualMethod{GT}{} & \mipQualImg{figs/mipmap_qual_error_inset_subfigs/fantasy_character/basecolor/gt/gt_mip00.png} & \mipQualImg{figs/mipmap_qual_error_inset_subfigs/fantasy_character/basecolor/gt/gt_mip01.png} & \mipQualImg{figs/mipmap_qual_error_inset_subfigs/fantasy_character/basecolor/gt/gt_mip02.png} & \mipQualImg{figs/mipmap_qual_error_inset_subfigs/fantasy_character/basecolor/gt/gt_mip03.png} & \mipQualImg{figs/mipmap_qual_error_inset_subfigs/fantasy_character/basecolor/gt/gt_mip04.png} & \mipQualImg{figs/mipmap_qual_error_inset_subfigs/fantasy_character/basecolor/gt/gt_mip05.png} & \mipQualImg{figs/mipmap_qual_error_inset_subfigs/fantasy_character/basecolor/gt/gt_mip06.png} & \mipQualImg{figs/mipmap_qual_error_inset_subfigs/fantasy_character/basecolor/gt/gt_mip07.png} & \mipQualImg{figs/mipmap_qual_error_inset_subfigs/fantasy_character/basecolor/gt/gt_mip08.png} & \mipQualImg{figs/mipmap_qual_error_inset_subfigs/fantasy_character/basecolor/gt/gt_mip09.png} & \mipQualImg{figs/mipmap_qual_error_inset_subfigs/fantasy_character/basecolor/gt/gt_mip10.png} & \mipQualImg{figs/mipmap_qual_error_inset_subfigs/fantasy_character/basecolor/gt/gt_mip11.png} \\[\mipQualRowGap]
      & \mipQualMethod{Ours}{0.210 bppc} & \mipQualImg{figs/mipmap_qual_error_inset_subfigs/fantasy_character/basecolor/ours/ours_mip00.png} & \mipQualImg{figs/mipmap_qual_error_inset_subfigs/fantasy_character/basecolor/ours/ours_mip01.png} & \mipQualImg{figs/mipmap_qual_error_inset_subfigs/fantasy_character/basecolor/ours/ours_mip02.png} & \mipQualImg{figs/mipmap_qual_error_inset_subfigs/fantasy_character/basecolor/ours/ours_mip03.png} & \mipQualImg{figs/mipmap_qual_error_inset_subfigs/fantasy_character/basecolor/ours/ours_mip04.png} & \mipQualImg{figs/mipmap_qual_error_inset_subfigs/fantasy_character/basecolor/ours/ours_mip05.png} & \mipQualImg{figs/mipmap_qual_error_inset_subfigs/fantasy_character/basecolor/ours/ours_mip06.png} & \mipQualImg{figs/mipmap_qual_error_inset_subfigs/fantasy_character/basecolor/ours/ours_mip07.png} & \mipQualImg{figs/mipmap_qual_error_inset_subfigs/fantasy_character/basecolor/ours/ours_mip08.png} & \mipQualImg{figs/mipmap_qual_error_inset_subfigs/fantasy_character/basecolor/ours/ours_mip09.png} & \mipQualImg{figs/mipmap_qual_error_inset_subfigs/fantasy_character/basecolor/ours/ours_mip10.png} & \mipQualImg{figs/mipmap_qual_error_inset_subfigs/fantasy_character/basecolor/ours/ours_mip11.png} \\[\mipQualRowGap]
    \mipQualStream{Base color} & \mipQualMethod{Image-GS}{0.445 bppc} & \mipQualImg{figs/mipmap_qual_error_inset_subfigs/fantasy_character/basecolor/image_gs/image_gs_mip00.png} & \mipQualImg{figs/mipmap_qual_error_inset_subfigs/fantasy_character/basecolor/image_gs/image_gs_mip01.png} & \mipQualImg{figs/mipmap_qual_error_inset_subfigs/fantasy_character/basecolor/image_gs/image_gs_mip02.png} & \mipQualImg{figs/mipmap_qual_error_inset_subfigs/fantasy_character/basecolor/image_gs/image_gs_mip03.png} & \mipQualImg{figs/mipmap_qual_error_inset_subfigs/fantasy_character/basecolor/image_gs/image_gs_mip04.png} & \mipQualImg{figs/mipmap_qual_error_inset_subfigs/fantasy_character/basecolor/image_gs/image_gs_mip05.png} & \mipQualImg{figs/mipmap_qual_error_inset_subfigs/fantasy_character/basecolor/image_gs/image_gs_mip06.png} & \mipQualImg{figs/mipmap_qual_error_inset_subfigs/fantasy_character/basecolor/image_gs/image_gs_mip07.png} & \mipQualImg{figs/mipmap_qual_error_inset_subfigs/fantasy_character/basecolor/image_gs/image_gs_mip08.png} & \mipQualImg{figs/mipmap_qual_error_inset_subfigs/fantasy_character/basecolor/image_gs/image_gs_mip09.png} & \mipQualImg{figs/mipmap_qual_error_inset_subfigs/fantasy_character/basecolor/image_gs/image_gs_mip10.png} & \mipQualImg{figs/mipmap_qual_error_inset_subfigs/fantasy_character/basecolor/image_gs/image_gs_mip11.png} \\[\mipQualRowGap]
      & \mipQualMethod{ASTC}{0.384 bppc} & \mipQualImg{figs/mipmap_qual_error_inset_subfigs/fantasy_character/basecolor/astc/astc_mip00.png} & \mipQualImg{figs/mipmap_qual_error_inset_subfigs/fantasy_character/basecolor/astc/astc_mip01.png} & \mipQualImg{figs/mipmap_qual_error_inset_subfigs/fantasy_character/basecolor/astc/astc_mip02.png} & \mipQualImg{figs/mipmap_qual_error_inset_subfigs/fantasy_character/basecolor/astc/astc_mip03.png} & \mipQualImg{figs/mipmap_qual_error_inset_subfigs/fantasy_character/basecolor/astc/astc_mip04.png} & \mipQualImg{figs/mipmap_qual_error_inset_subfigs/fantasy_character/basecolor/astc/astc_mip05.png} & \mipQualImg{figs/mipmap_qual_error_inset_subfigs/fantasy_character/basecolor/astc/astc_mip06.png} & \mipQualImg{figs/mipmap_qual_error_inset_subfigs/fantasy_character/basecolor/astc/astc_mip07.png} & \mipQualImg{figs/mipmap_qual_error_inset_subfigs/fantasy_character/basecolor/astc/astc_mip08.png} & \mipQualImg{figs/mipmap_qual_error_inset_subfigs/fantasy_character/basecolor/astc/astc_mip09.png} & \mipQualImg{figs/mipmap_qual_error_inset_subfigs/fantasy_character/basecolor/astc/astc_mip10.png} & \mipQualImg{figs/mipmap_qual_error_inset_subfigs/fantasy_character/basecolor/astc/astc_mip11.png} \\[\mipQualRowGap]
      & \mipQualMethod{NTC}{0.290 bppc} & \mipQualImg{figs/mipmap_qual_error_inset_subfigs/fantasy_character/basecolor/ntc/ntc_mip00.png} & \mipQualImg{figs/mipmap_qual_error_inset_subfigs/fantasy_character/basecolor/ntc/ntc_mip01.png} & \mipQualImg{figs/mipmap_qual_error_inset_subfigs/fantasy_character/basecolor/ntc/ntc_mip02.png} & \mipQualImg{figs/mipmap_qual_error_inset_subfigs/fantasy_character/basecolor/ntc/ntc_mip03.png} & \mipQualImg{figs/mipmap_qual_error_inset_subfigs/fantasy_character/basecolor/ntc/ntc_mip04.png} & \mipQualImg{figs/mipmap_qual_error_inset_subfigs/fantasy_character/basecolor/ntc/ntc_mip05.png} & \mipQualImg{figs/mipmap_qual_error_inset_subfigs/fantasy_character/basecolor/ntc/ntc_mip06.png} & \mipQualImg{figs/mipmap_qual_error_inset_subfigs/fantasy_character/basecolor/ntc/ntc_mip07.png} & \mipQualImg{figs/mipmap_qual_error_inset_subfigs/fantasy_character/basecolor/ntc/ntc_mip08.png} & \mipQualImg{figs/mipmap_qual_error_inset_subfigs/fantasy_character/basecolor/ntc/ntc_mip09.png} & \mipQualImg{figs/mipmap_qual_error_inset_subfigs/fantasy_character/basecolor/ntc/ntc_mip10.png} & \mipQualImg{figs/mipmap_qual_error_inset_subfigs/fantasy_character/basecolor/ntc/ntc_mip11.png} \\
    \mipQualSep
      & \mipQualMethod{GT}{} & \mipQualImg{figs/mipmap_qual_error_inset_subfigs/fantasy_character/normal/gt/gt_mip00.png} & \mipQualImg{figs/mipmap_qual_error_inset_subfigs/fantasy_character/normal/gt/gt_mip01.png} & \mipQualImg{figs/mipmap_qual_error_inset_subfigs/fantasy_character/normal/gt/gt_mip02.png} & \mipQualImg{figs/mipmap_qual_error_inset_subfigs/fantasy_character/normal/gt/gt_mip03.png} & \mipQualImg{figs/mipmap_qual_error_inset_subfigs/fantasy_character/normal/gt/gt_mip04.png} & \mipQualImg{figs/mipmap_qual_error_inset_subfigs/fantasy_character/normal/gt/gt_mip05.png} & \mipQualImg{figs/mipmap_qual_error_inset_subfigs/fantasy_character/normal/gt/gt_mip06.png} & \mipQualImg{figs/mipmap_qual_error_inset_subfigs/fantasy_character/normal/gt/gt_mip07.png} & \mipQualImg{figs/mipmap_qual_error_inset_subfigs/fantasy_character/normal/gt/gt_mip08.png} & \mipQualImg{figs/mipmap_qual_error_inset_subfigs/fantasy_character/normal/gt/gt_mip09.png} & \mipQualImg{figs/mipmap_qual_error_inset_subfigs/fantasy_character/normal/gt/gt_mip10.png} & \mipQualImg{figs/mipmap_qual_error_inset_subfigs/fantasy_character/normal/gt/gt_mip11.png} \\[\mipQualRowGap]
      & \mipQualMethod{Ours}{0.210 bppc} & \mipQualImg{figs/mipmap_qual_error_inset_subfigs/fantasy_character/normal/ours/ours_mip00.png} & \mipQualImg{figs/mipmap_qual_error_inset_subfigs/fantasy_character/normal/ours/ours_mip01.png} & \mipQualImg{figs/mipmap_qual_error_inset_subfigs/fantasy_character/normal/ours/ours_mip02.png} & \mipQualImg{figs/mipmap_qual_error_inset_subfigs/fantasy_character/normal/ours/ours_mip03.png} & \mipQualImg{figs/mipmap_qual_error_inset_subfigs/fantasy_character/normal/ours/ours_mip04.png} & \mipQualImg{figs/mipmap_qual_error_inset_subfigs/fantasy_character/normal/ours/ours_mip05.png} & \mipQualImg{figs/mipmap_qual_error_inset_subfigs/fantasy_character/normal/ours/ours_mip06.png} & \mipQualImg{figs/mipmap_qual_error_inset_subfigs/fantasy_character/normal/ours/ours_mip07.png} & \mipQualImg{figs/mipmap_qual_error_inset_subfigs/fantasy_character/normal/ours/ours_mip08.png} & \mipQualImg{figs/mipmap_qual_error_inset_subfigs/fantasy_character/normal/ours/ours_mip09.png} & \mipQualImg{figs/mipmap_qual_error_inset_subfigs/fantasy_character/normal/ours/ours_mip10.png} & \mipQualImg{figs/mipmap_qual_error_inset_subfigs/fantasy_character/normal/ours/ours_mip11.png} \\[\mipQualRowGap]
    \mipQualStream{Normal map} & \mipQualMethod{Image-GS}{0.445 bppc} & \mipQualImg{figs/mipmap_qual_error_inset_subfigs/fantasy_character/normal/image_gs/image_gs_mip00.png} & \mipQualImg{figs/mipmap_qual_error_inset_subfigs/fantasy_character/normal/image_gs/image_gs_mip01.png} & \mipQualImg{figs/mipmap_qual_error_inset_subfigs/fantasy_character/normal/image_gs/image_gs_mip02.png} & \mipQualImg{figs/mipmap_qual_error_inset_subfigs/fantasy_character/normal/image_gs/image_gs_mip03.png} & \mipQualImg{figs/mipmap_qual_error_inset_subfigs/fantasy_character/normal/image_gs/image_gs_mip04.png} & \mipQualImg{figs/mipmap_qual_error_inset_subfigs/fantasy_character/normal/image_gs/image_gs_mip05.png} & \mipQualImg{figs/mipmap_qual_error_inset_subfigs/fantasy_character/normal/image_gs/image_gs_mip06.png} & \mipQualImg{figs/mipmap_qual_error_inset_subfigs/fantasy_character/normal/image_gs/image_gs_mip07.png} & \mipQualImg{figs/mipmap_qual_error_inset_subfigs/fantasy_character/normal/image_gs/image_gs_mip08.png} & \mipQualImg{figs/mipmap_qual_error_inset_subfigs/fantasy_character/normal/image_gs/image_gs_mip09.png} & \mipQualImg{figs/mipmap_qual_error_inset_subfigs/fantasy_character/normal/image_gs/image_gs_mip10.png} & \mipQualImg{figs/mipmap_qual_error_inset_subfigs/fantasy_character/normal/image_gs/image_gs_mip11.png} \\[\mipQualRowGap]
      & \mipQualMethod{ASTC}{0.384 bppc} & \mipQualImg{figs/mipmap_qual_error_inset_subfigs/fantasy_character/normal/astc/astc_mip00.png} & \mipQualImg{figs/mipmap_qual_error_inset_subfigs/fantasy_character/normal/astc/astc_mip01.png} & \mipQualImg{figs/mipmap_qual_error_inset_subfigs/fantasy_character/normal/astc/astc_mip02.png} & \mipQualImg{figs/mipmap_qual_error_inset_subfigs/fantasy_character/normal/astc/astc_mip03.png} & \mipQualImg{figs/mipmap_qual_error_inset_subfigs/fantasy_character/normal/astc/astc_mip04.png} & \mipQualImg{figs/mipmap_qual_error_inset_subfigs/fantasy_character/normal/astc/astc_mip05.png} & \mipQualImg{figs/mipmap_qual_error_inset_subfigs/fantasy_character/normal/astc/astc_mip06.png} & \mipQualImg{figs/mipmap_qual_error_inset_subfigs/fantasy_character/normal/astc/astc_mip07.png} & \mipQualImg{figs/mipmap_qual_error_inset_subfigs/fantasy_character/normal/astc/astc_mip08.png} & \mipQualImg{figs/mipmap_qual_error_inset_subfigs/fantasy_character/normal/astc/astc_mip09.png} & \mipQualImg{figs/mipmap_qual_error_inset_subfigs/fantasy_character/normal/astc/astc_mip10.png} & \mipQualImg{figs/mipmap_qual_error_inset_subfigs/fantasy_character/normal/astc/astc_mip11.png} \\[\mipQualRowGap]
      & \mipQualMethod{NTC}{0.290 bppc} & \mipQualImg{figs/mipmap_qual_error_inset_subfigs/fantasy_character/normal/ntc/ntc_mip00.png} & \mipQualImg{figs/mipmap_qual_error_inset_subfigs/fantasy_character/normal/ntc/ntc_mip01.png} & \mipQualImg{figs/mipmap_qual_error_inset_subfigs/fantasy_character/normal/ntc/ntc_mip02.png} & \mipQualImg{figs/mipmap_qual_error_inset_subfigs/fantasy_character/normal/ntc/ntc_mip03.png} & \mipQualImg{figs/mipmap_qual_error_inset_subfigs/fantasy_character/normal/ntc/ntc_mip04.png} & \mipQualImg{figs/mipmap_qual_error_inset_subfigs/fantasy_character/normal/ntc/ntc_mip05.png} & \mipQualImg{figs/mipmap_qual_error_inset_subfigs/fantasy_character/normal/ntc/ntc_mip06.png} & \mipQualImg{figs/mipmap_qual_error_inset_subfigs/fantasy_character/normal/ntc/ntc_mip07.png} & \mipQualImg{figs/mipmap_qual_error_inset_subfigs/fantasy_character/normal/ntc/ntc_mip08.png} & \mipQualImg{figs/mipmap_qual_error_inset_subfigs/fantasy_character/normal/ntc/ntc_mip09.png} & \mipQualImg{figs/mipmap_qual_error_inset_subfigs/fantasy_character/normal/ntc/ntc_mip10.png} & \mipQualImg{figs/mipmap_qual_error_inset_subfigs/fantasy_character/normal/ntc/ntc_mip11.png} \\
    \mipQualSep
      & \mipQualMethod{GT}{} & \mipQualImg{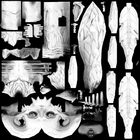} & \mipQualImg{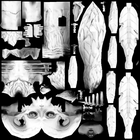} & \mipQualImg{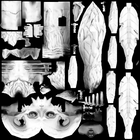} & \mipQualImg{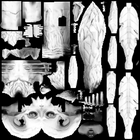} & \mipQualImg{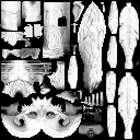} & \mipQualImg{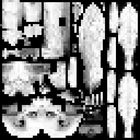} & \mipQualImg{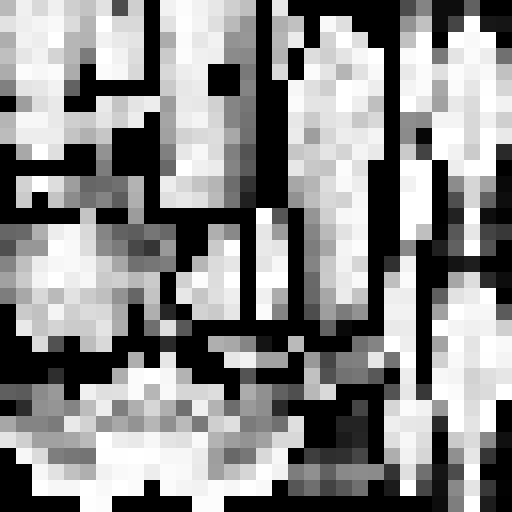} & \mipQualImg{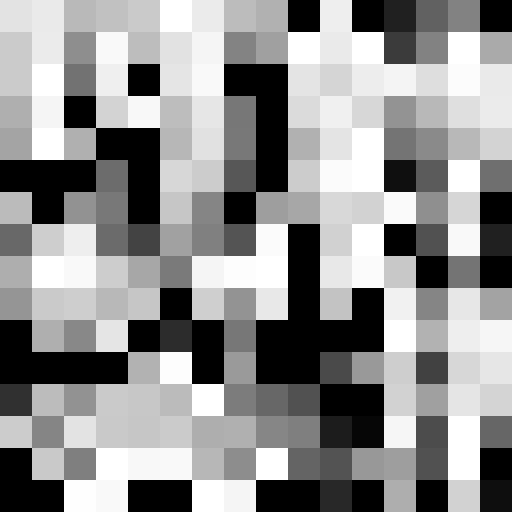} & \mipQualImg{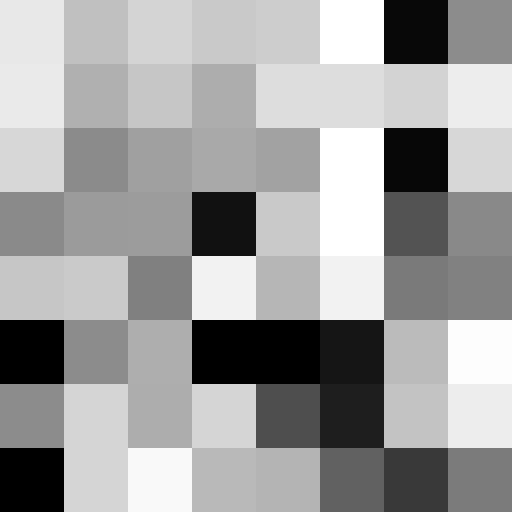} & \mipQualImg{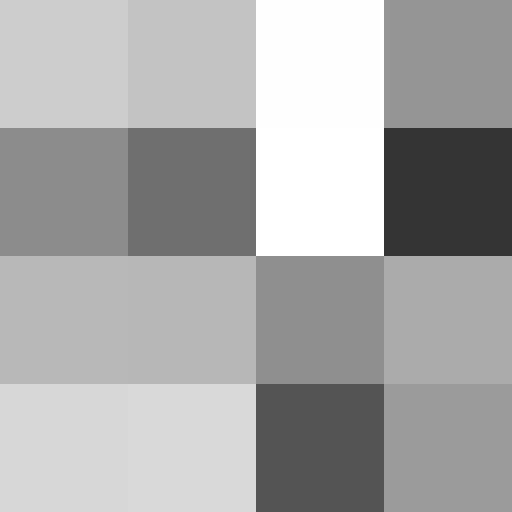} & \mipQualImg{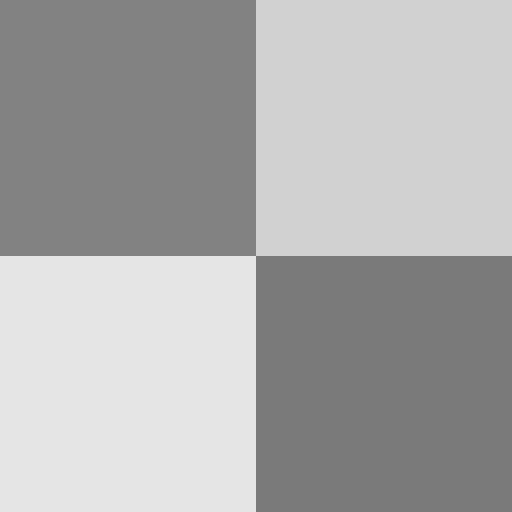} & \mipQualImg{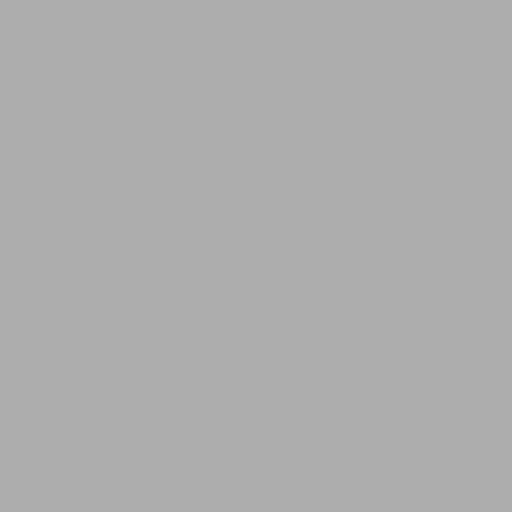} \\[\mipQualRowGap]
      & \mipQualMethod{Ours}{0.210 bppc} & \mipQualImg{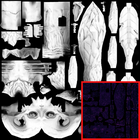} & \mipQualImg{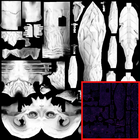} & \mipQualImg{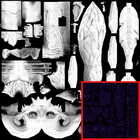} & \mipQualImg{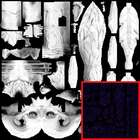} & \mipQualImg{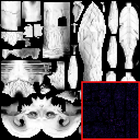} & \mipQualImg{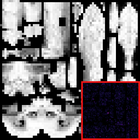} & \mipQualImg{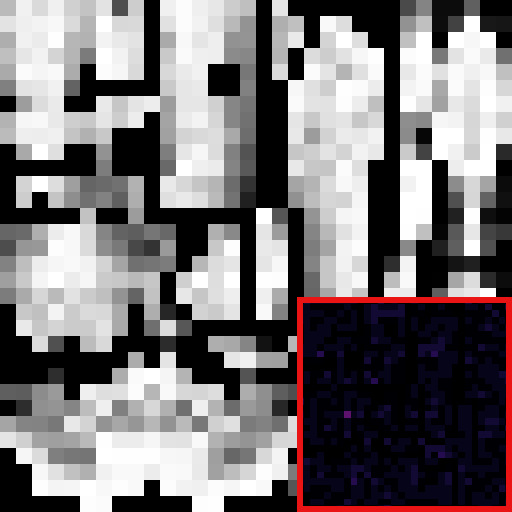} & \mipQualImg{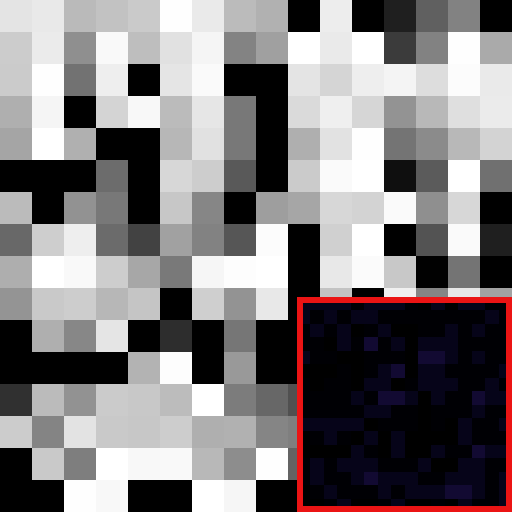} & \mipQualImg{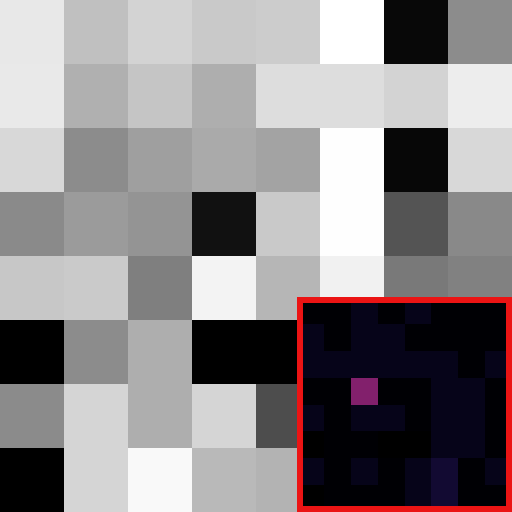} & \mipQualImg{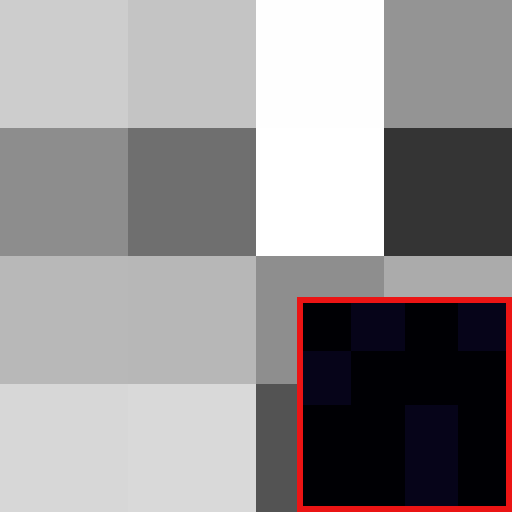} & \mipQualImg{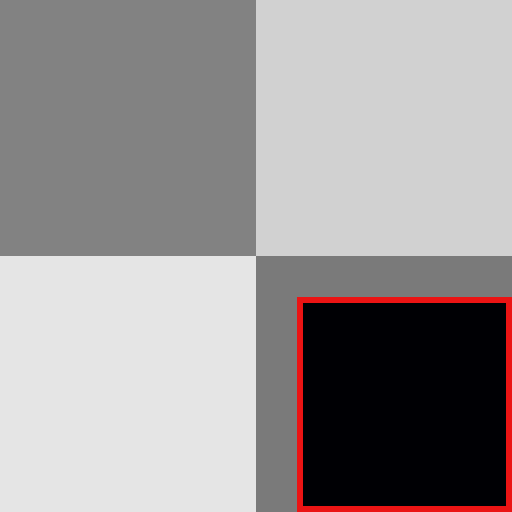} & \mipQualImg{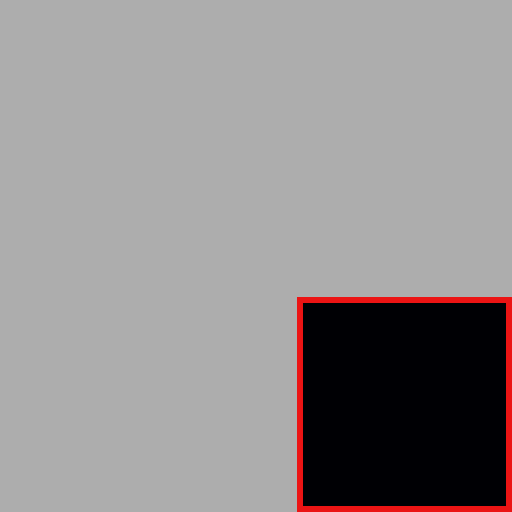} \\[\mipQualRowGap]
    \mipQualStream{AO} & \mipQualMethod{Image-GS}{0.445 bppc} & \mipQualImg{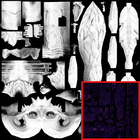} & \mipQualImg{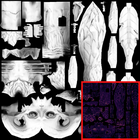} & \mipQualImg{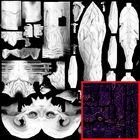} & \mipQualImg{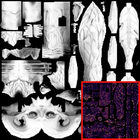} & \mipQualImg{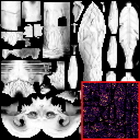} & \mipQualImg{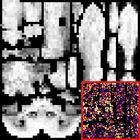} & \mipQualImg{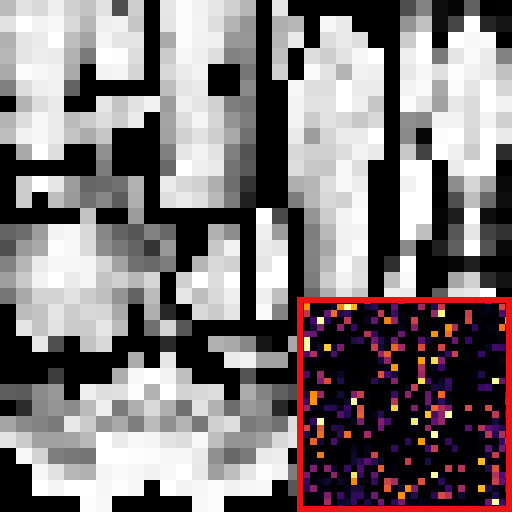} & \mipQualImg{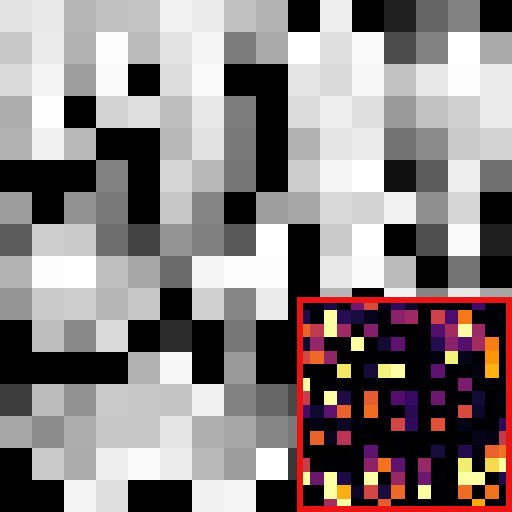} & \mipQualImg{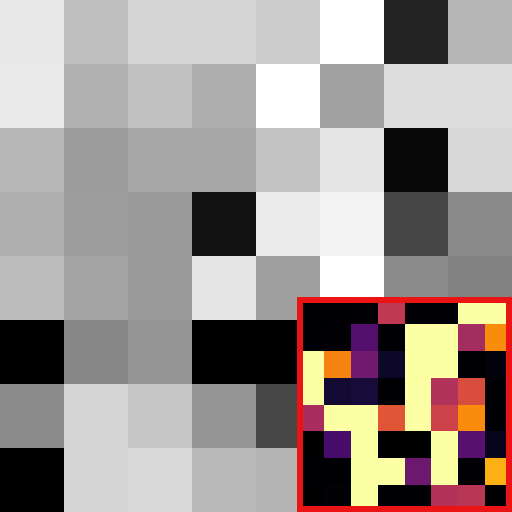} & \mipQualImg{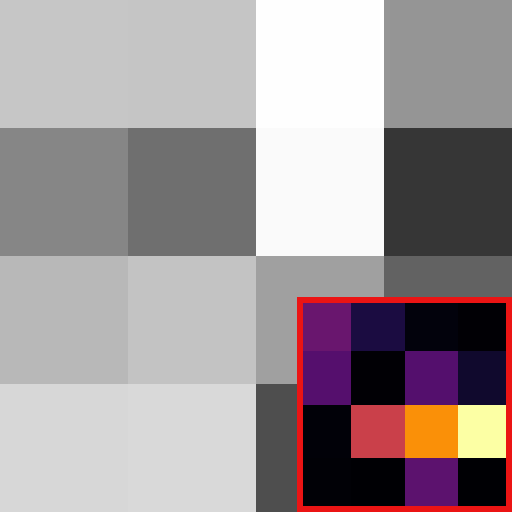} & \mipQualImg{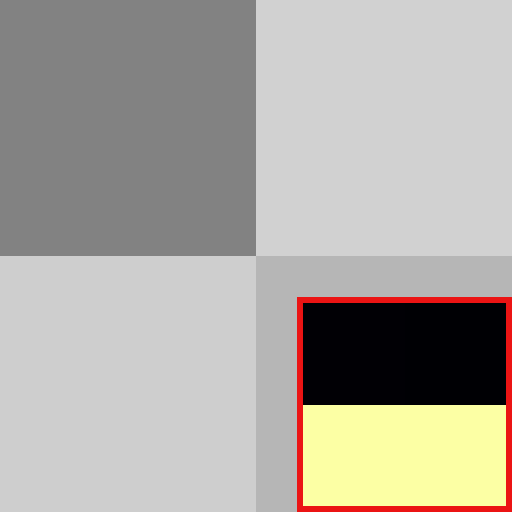} & \mipQualImg{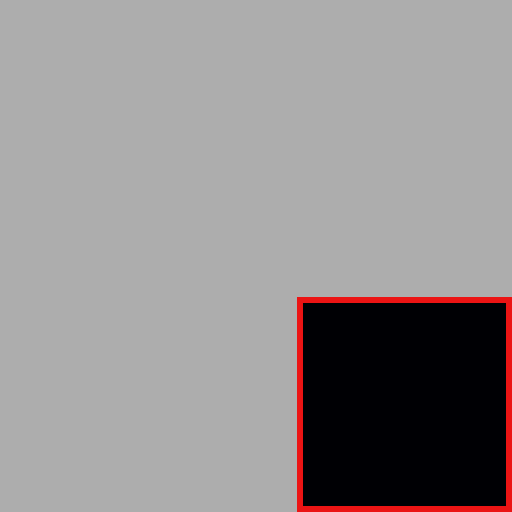} \\[\mipQualRowGap]
      & \mipQualMethod{ASTC}{0.384 bppc} & \mipQualImg{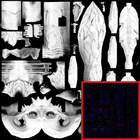} & \mipQualImg{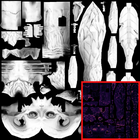} & \mipQualImg{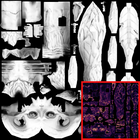} & \mipQualImg{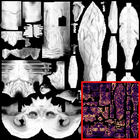} & \mipQualImg{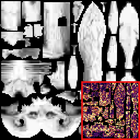} & \mipQualImg{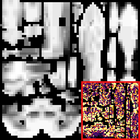} & \mipQualImg{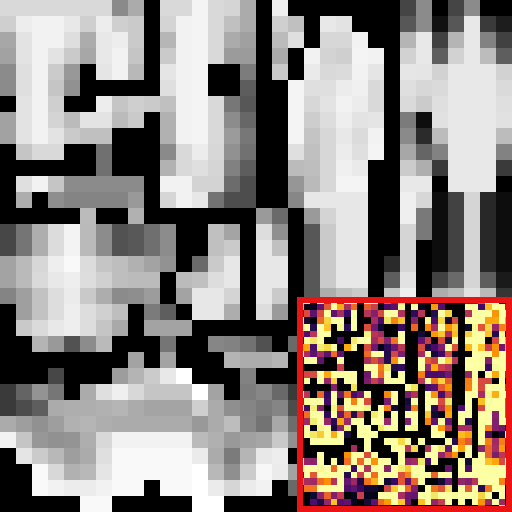} & \mipQualImg{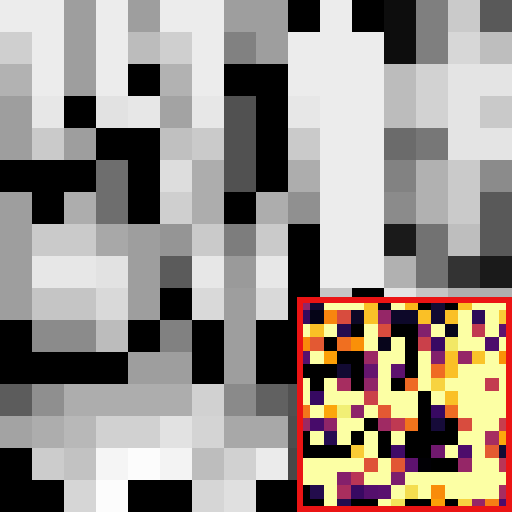} & \mipQualImg{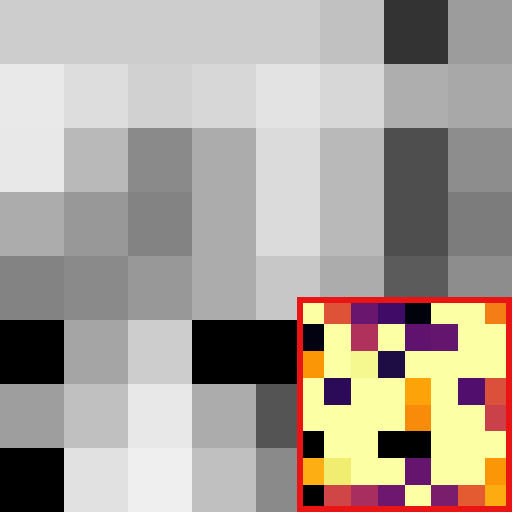} & \mipQualImg{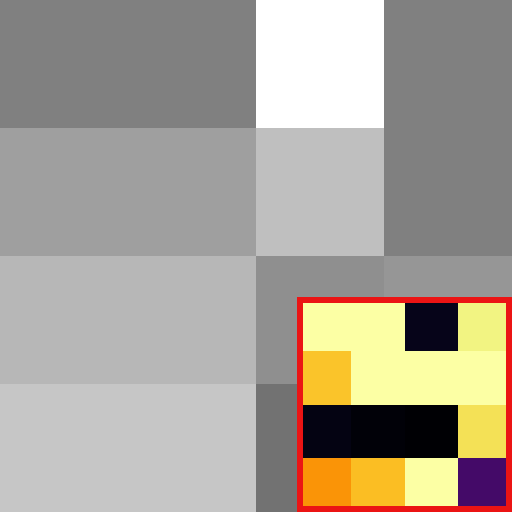} & \mipQualImg{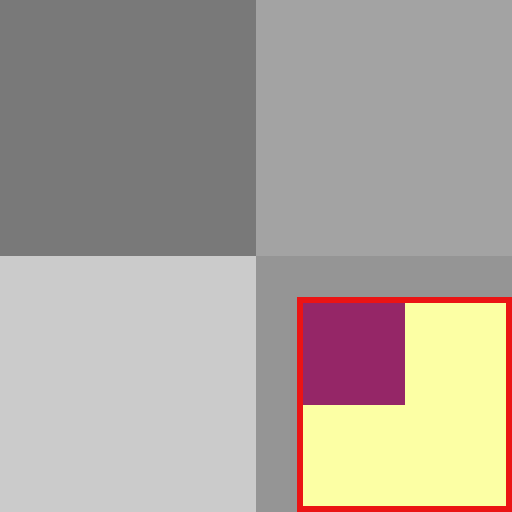} & \mipQualImg{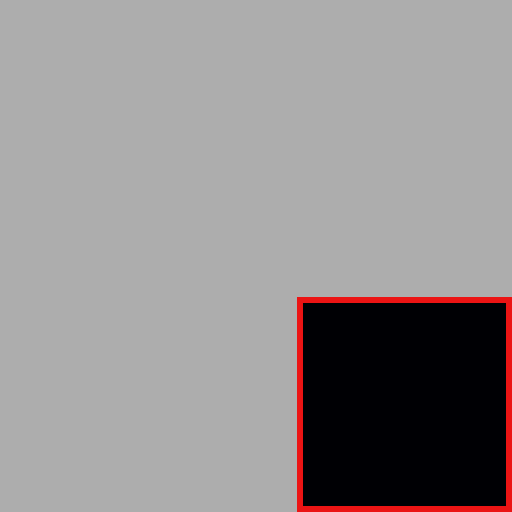} \\[\mipQualRowGap]
      & \mipQualMethod{NTC}{0.290 bppc} & \mipQualImg{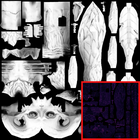} & \mipQualImg{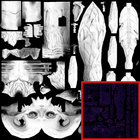} & \mipQualImg{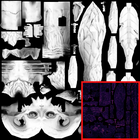} & \mipQualImg{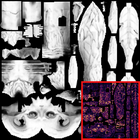} & \mipQualImg{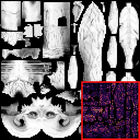} & \mipQualImg{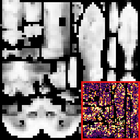} & \mipQualImg{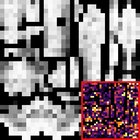} & \mipQualImg{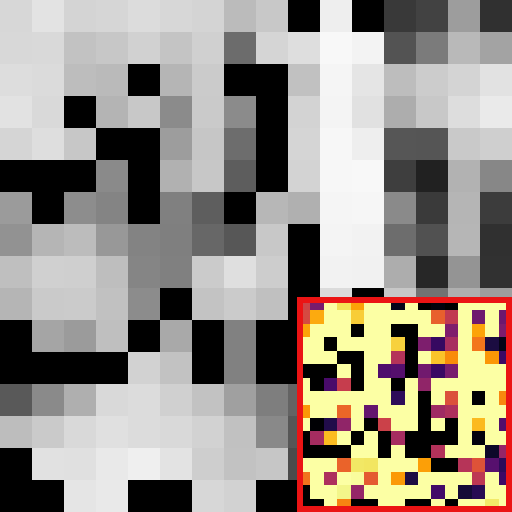} & \mipQualImg{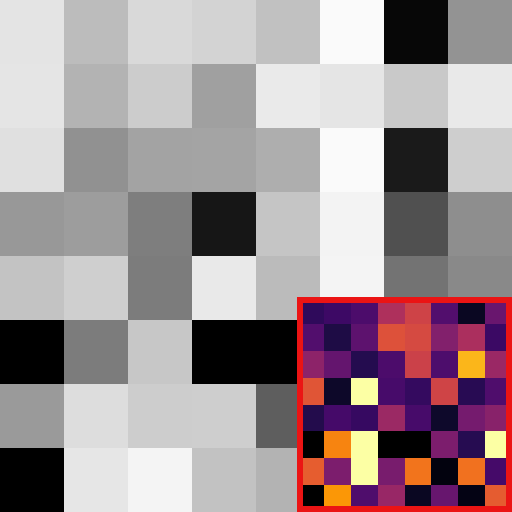} & \mipQualImg{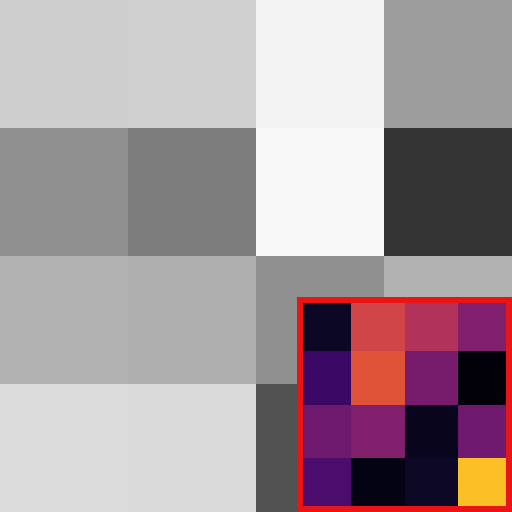} & \mipQualImg{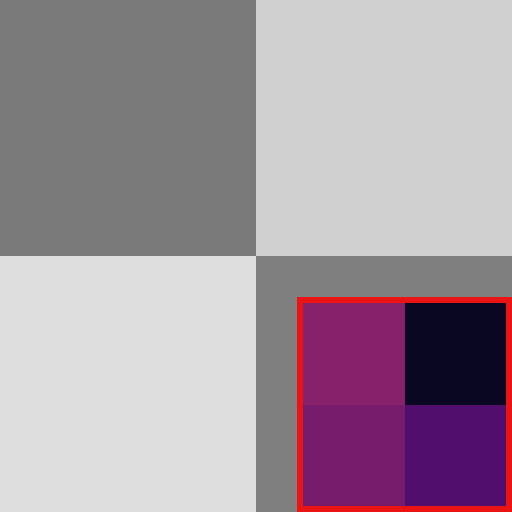} & \mipQualImg{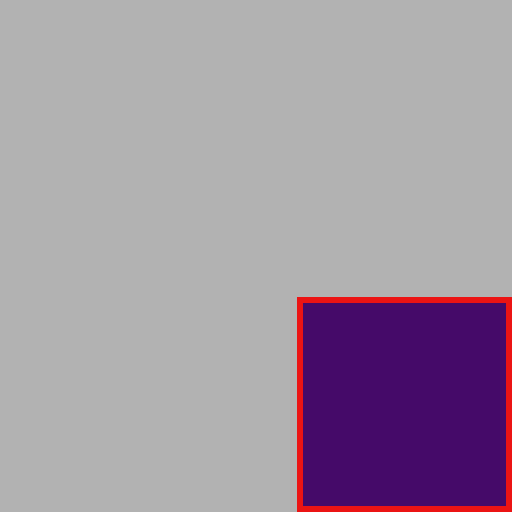} \\

  \end{tabular}%
  }
  \caption{Mipmap reconstruction comparison on the \textit{Fantasy Character} SVBRDF stack. Rows show the reference and reconstructed mipmap pyramids, with insets showing per-pixel error maps against the reference. AO denotes ambient occlusion.}
  \label{fig:mipmap-error-inset-fantasy-character}
\end{figure*}

\clearpage
\begin{figure*}[p!]
  \centering
  \mipQualFontSize
  \setlength{\tabcolsep}{0pt}
  \renewcommand{\arraystretch}{1.0}
  \resizebox{\mipQualTableTargetW}{!}{%
  \begin{tabular}{@{}c@{\hspace{\mipQualStreamMethodGap}}c@{\hspace{\mipQualMethodImageGap}}*{11}{c@{\hspace{\mipQualColGap}}}c@{}}
    \multicolumn{2}{c}{} & \mipQualMip{mip 0}{2048$^2$} & \mipQualMip{mip 1}{1024$^2$} & \mipQualMip{mip 2}{512$^2$} & \mipQualMip{mip 3}{256$^2$} & \mipQualMip{mip 4}{128$^2$} & \mipQualMip{mip 5}{64$^2$} & \mipQualMip{mip 6}{32$^2$} & \mipQualMip{mip 7}{16$^2$} & \mipQualMip{mip 8}{8$^2$} & \mipQualMip{mip 9}{4$^2$} & \mipQualMip{mip 10}{2$^2$} & \mipQualMip{mip 11}{1$^2$} \\[\mipQualHeaderGap]
      & \mipQualMethod{GT}{} & \mipQualImg{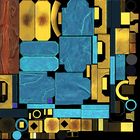} & \mipQualImg{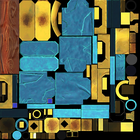} & \mipQualImg{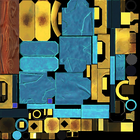} & \mipQualImg{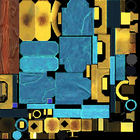} & \mipQualImg{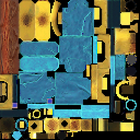} & \mipQualImg{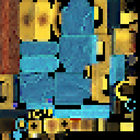} & \mipQualImg{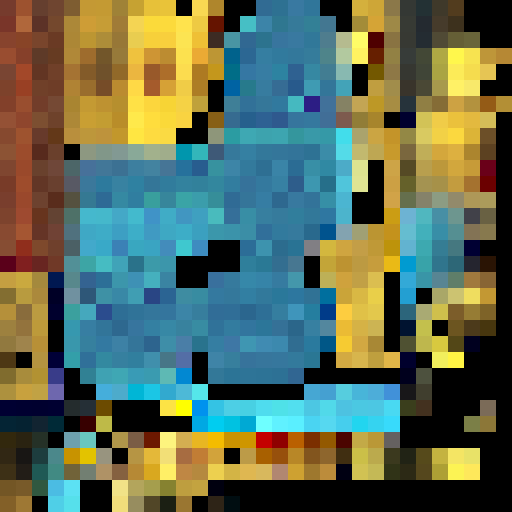} & \mipQualImg{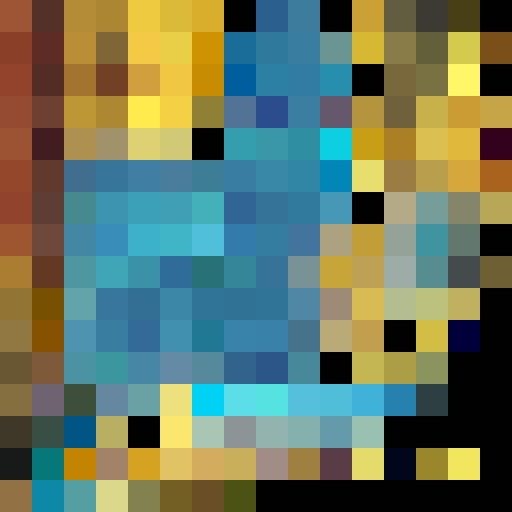} & \mipQualImg{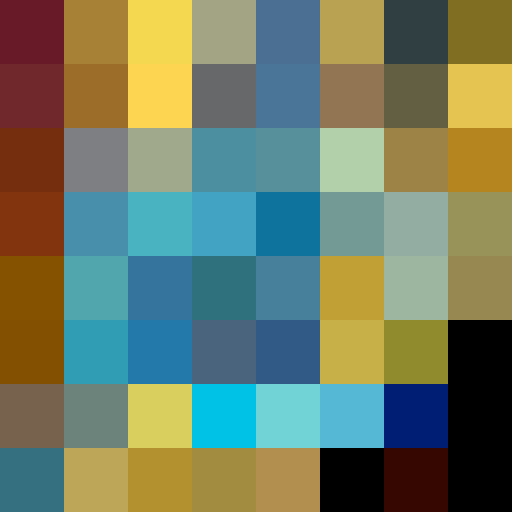} & \mipQualImg{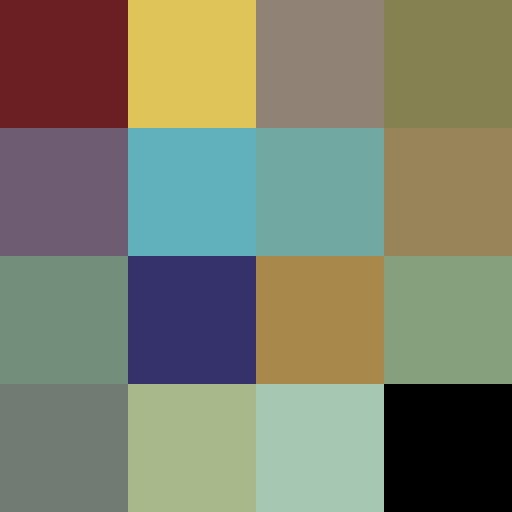} & \mipQualImg{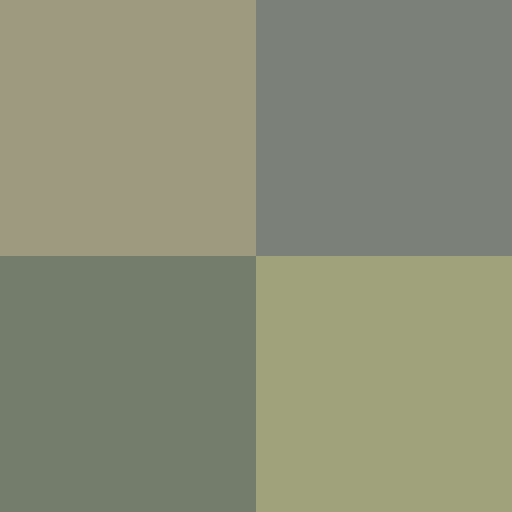} & \mipQualImg{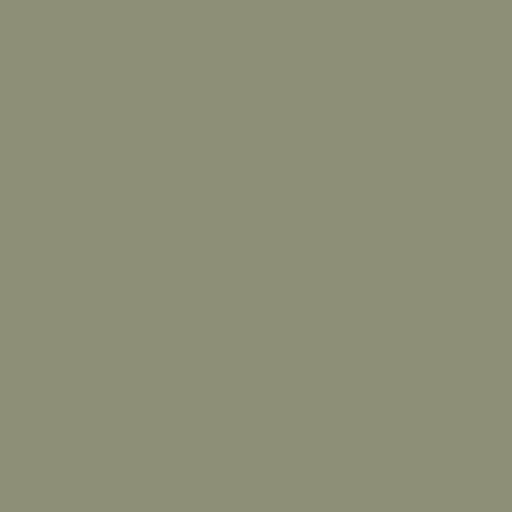} \\[\mipQualRowGap]
      & \mipQualMethod{Ours}{0.211 bppc} & \mipQualImg{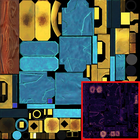} & \mipQualImg{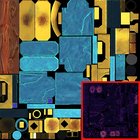} & \mipQualImg{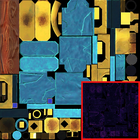} & \mipQualImg{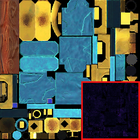} & \mipQualImg{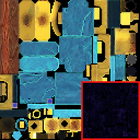} & \mipQualImg{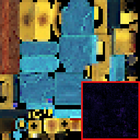} & \mipQualImg{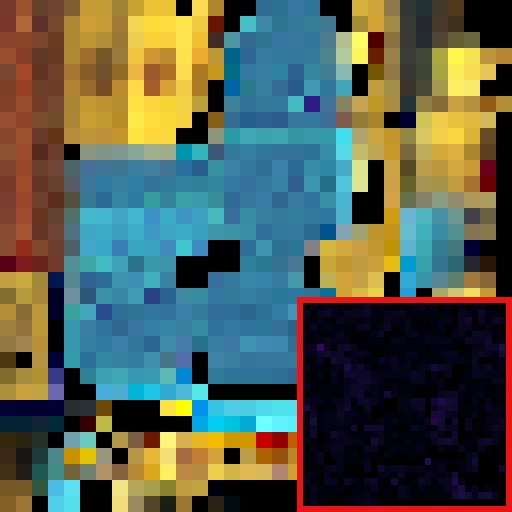} & \mipQualImg{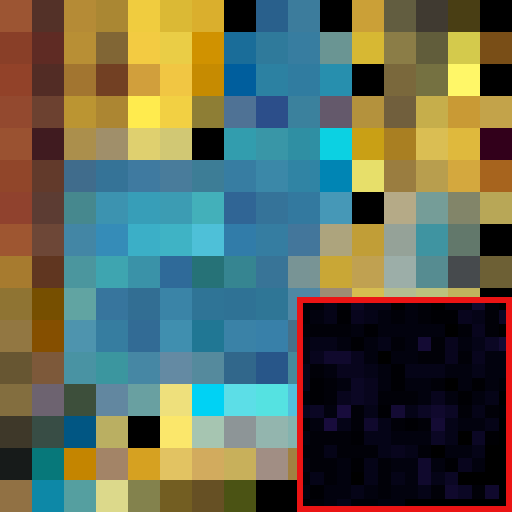} & \mipQualImg{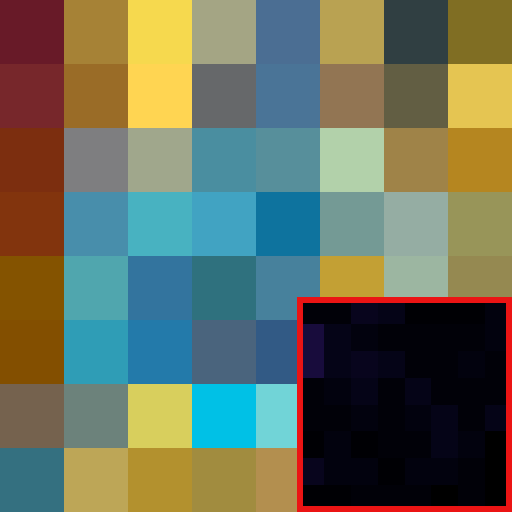} & \mipQualImg{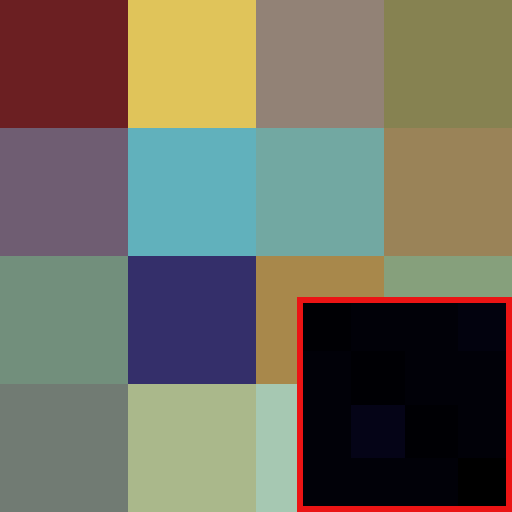} & \mipQualImg{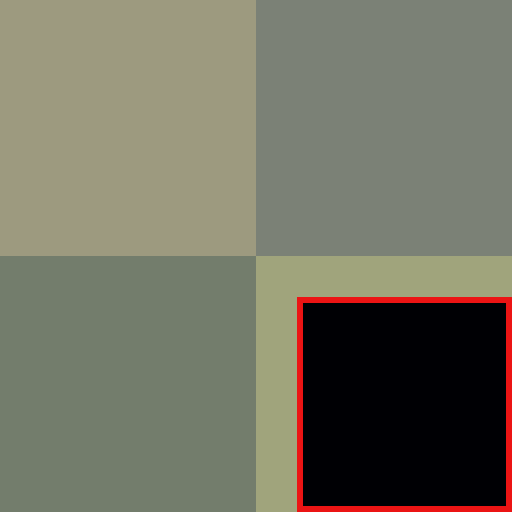} & \mipQualImg{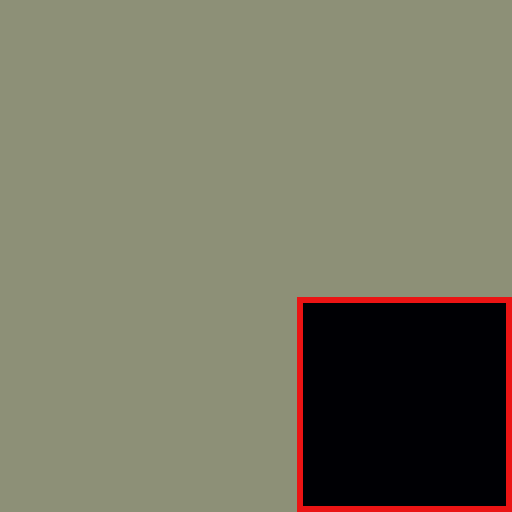} \\[\mipQualRowGap]
    \mipQualStream{Base color} & \mipQualMethod{Image-GS}{0.368 bppc} & \mipQualImg{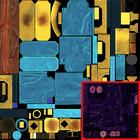} & \mipQualImg{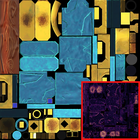} & \mipQualImg{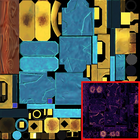} & \mipQualImg{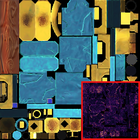} & \mipQualImg{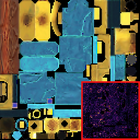} & \mipQualImg{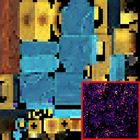} & \mipQualImg{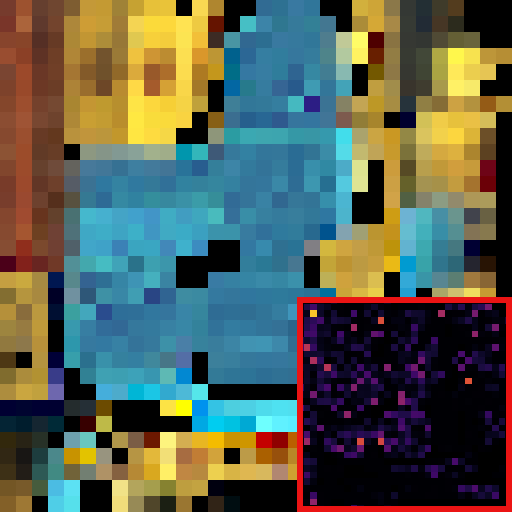} & \mipQualImg{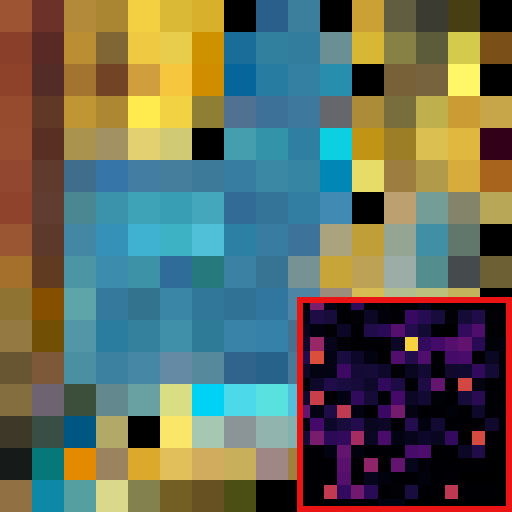} & \mipQualImg{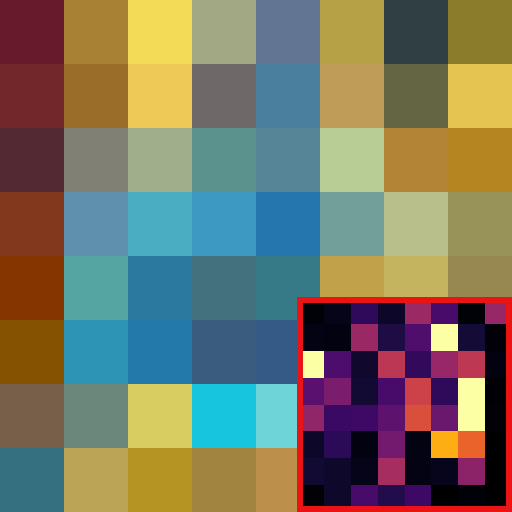} & \mipQualImg{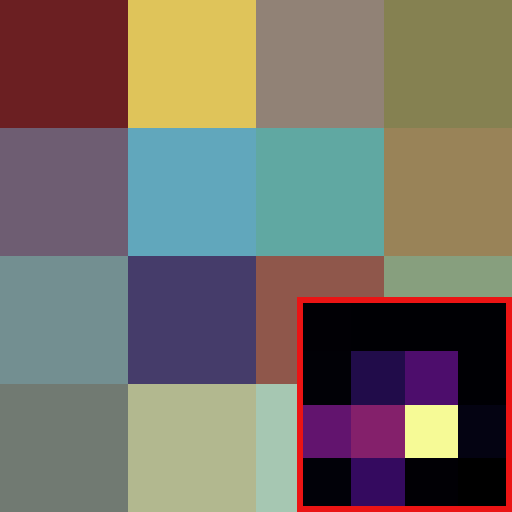} & \mipQualImg{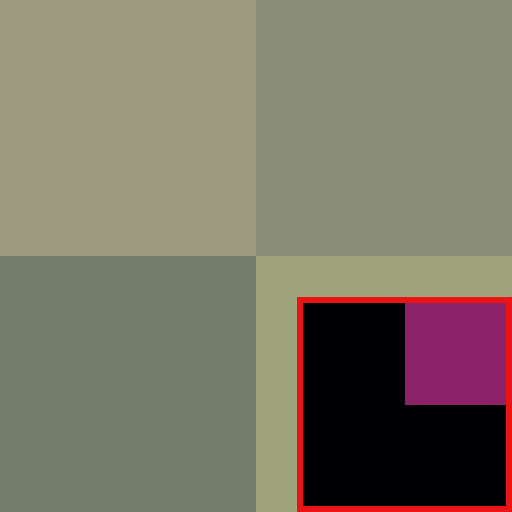} & \mipQualImg{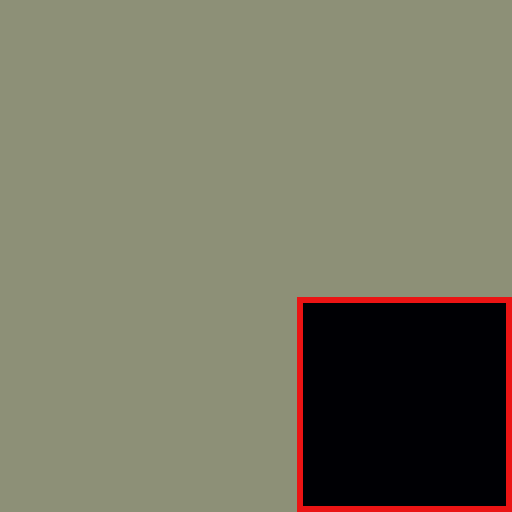} \\[\mipQualRowGap]
      & \mipQualMethod{ASTC}{0.429 bppc} & \mipQualImg{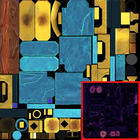} & \mipQualImg{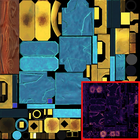} & \mipQualImg{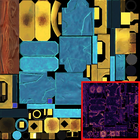} & \mipQualImg{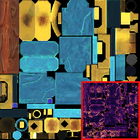} & \mipQualImg{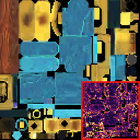} & \mipQualImg{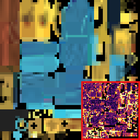} & \mipQualImg{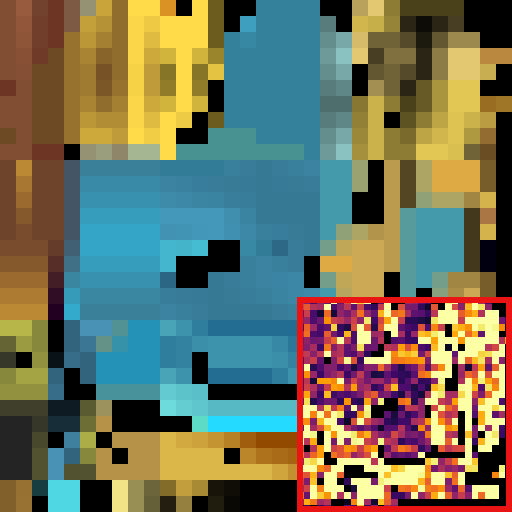} & \mipQualImg{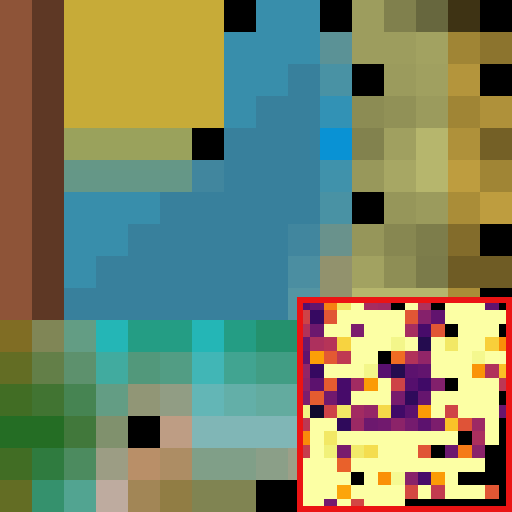} & \mipQualImg{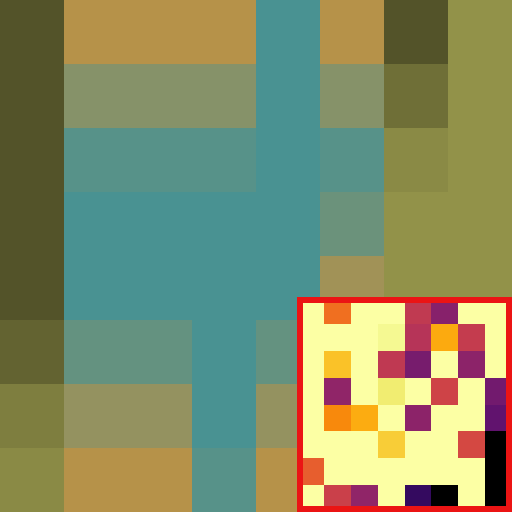} & \mipQualImg{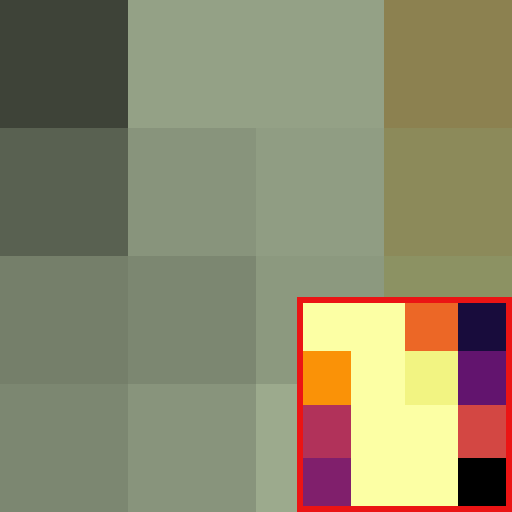} & \mipQualImg{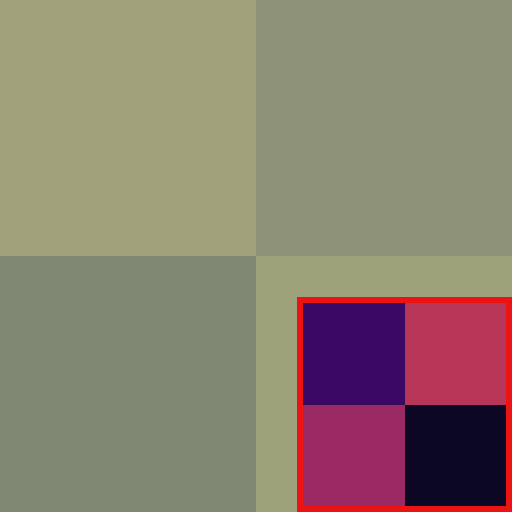} & \mipQualImg{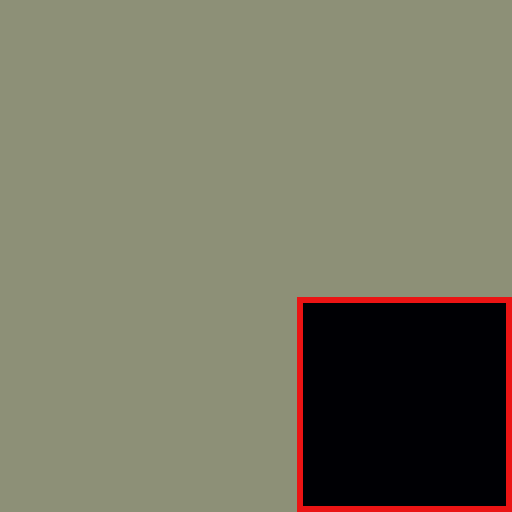} \\[\mipQualRowGap]
      & \mipQualMethod{NTC}{0.253 bppc} & \mipQualImg{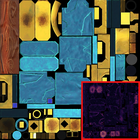} & \mipQualImg{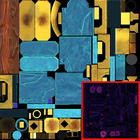} & \mipQualImg{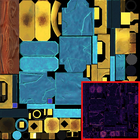} & \mipQualImg{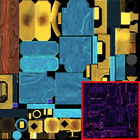} & \mipQualImg{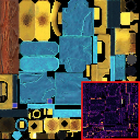} & \mipQualImg{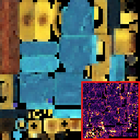} & \mipQualImg{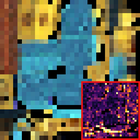} & \mipQualImg{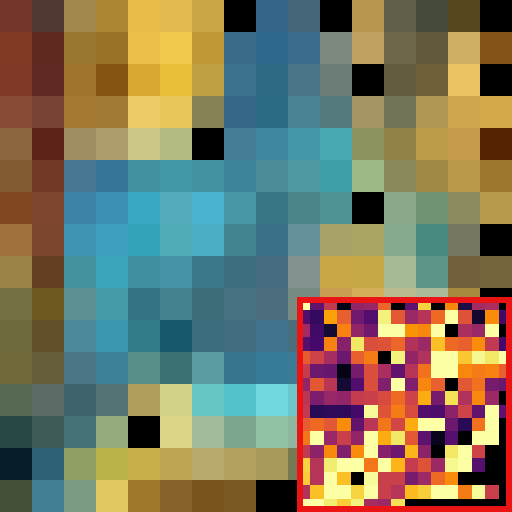} & \mipQualImg{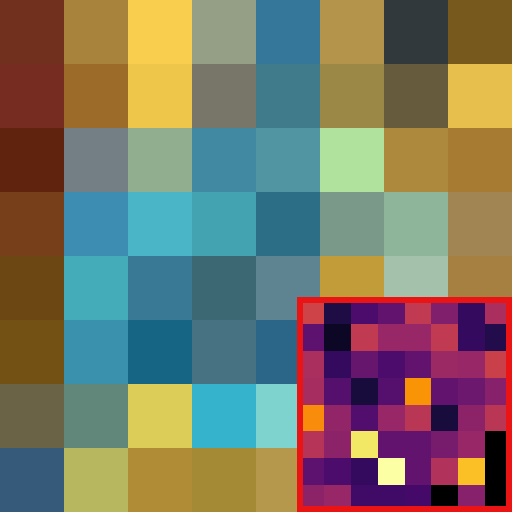} & \mipQualImg{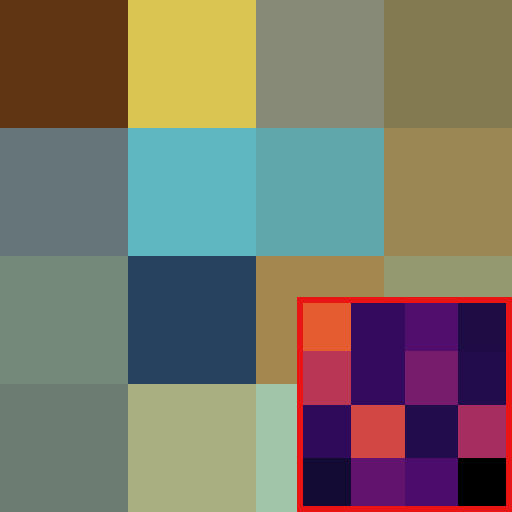} & \mipQualImg{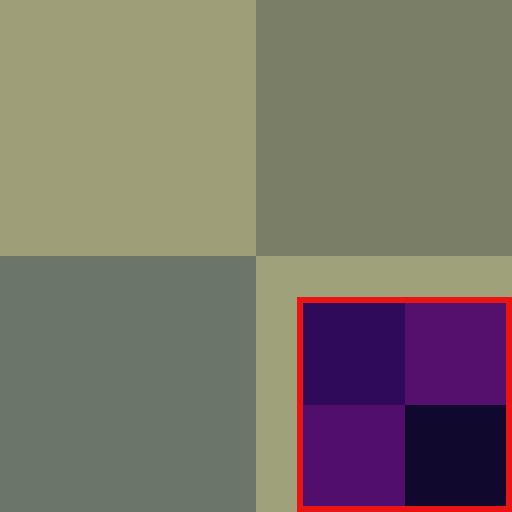} & \mipQualImg{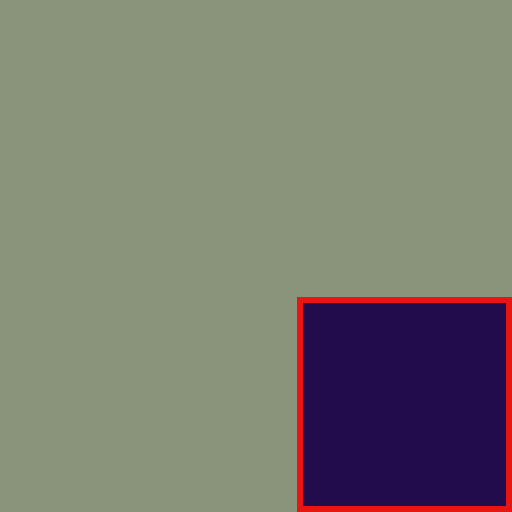} \\
    \mipQualSep
      & \mipQualMethod{GT}{} & \mipQualImg{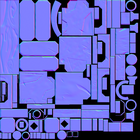} & \mipQualImg{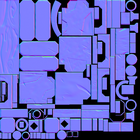} & \mipQualImg{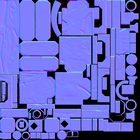} & \mipQualImg{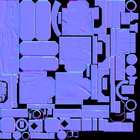} & \mipQualImg{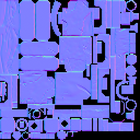} & \mipQualImg{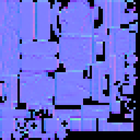} & \mipQualImg{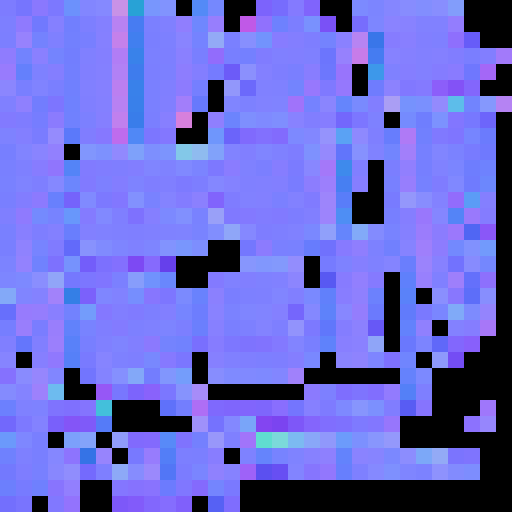} & \mipQualImg{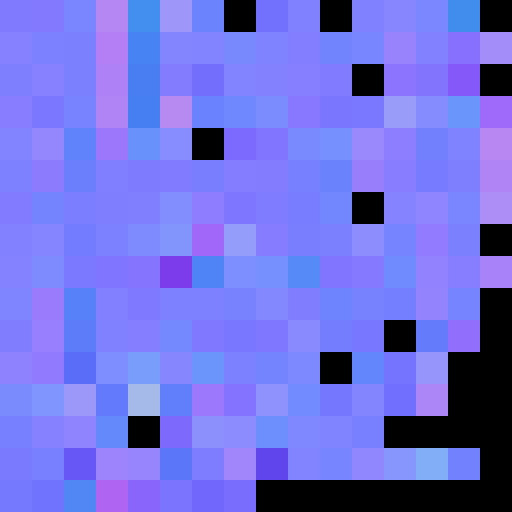} & \mipQualImg{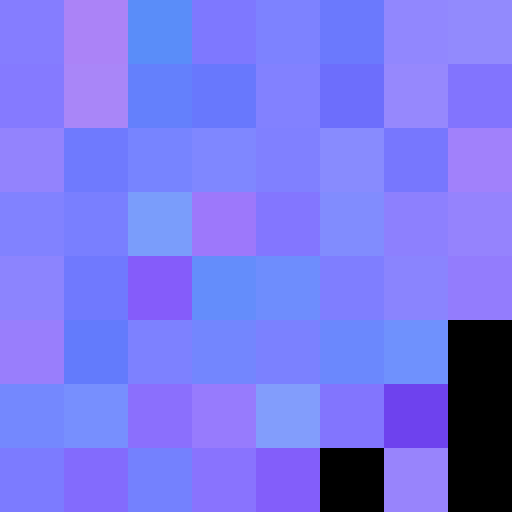} & \mipQualImg{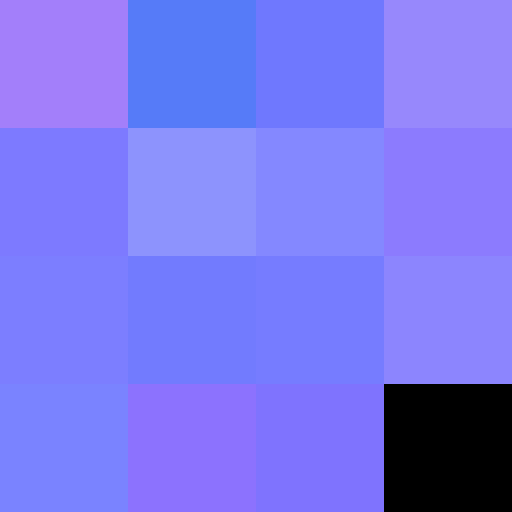} & \mipQualImg{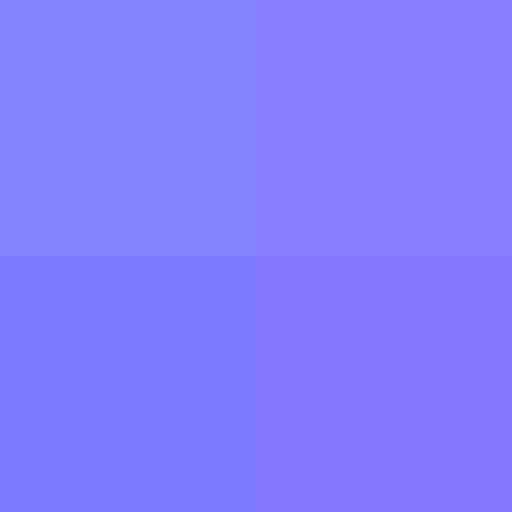} & \mipQualImg{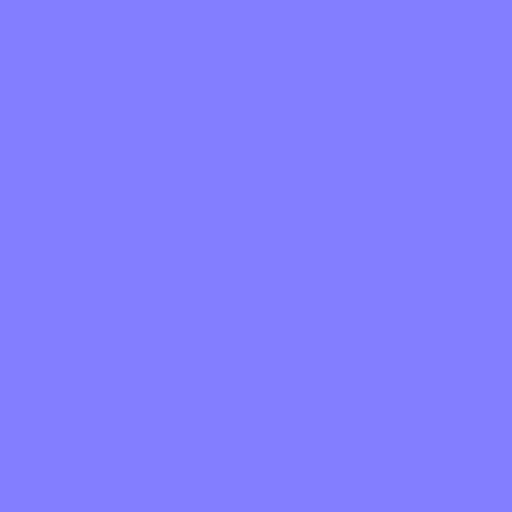} \\[\mipQualRowGap]
      & \mipQualMethod{Ours}{0.211 bppc} & \mipQualImg{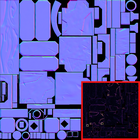} & \mipQualImg{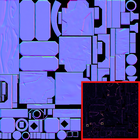} & \mipQualImg{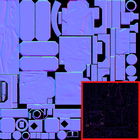} & \mipQualImg{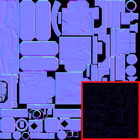} & \mipQualImg{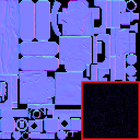} & \mipQualImg{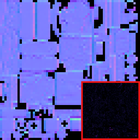} & \mipQualImg{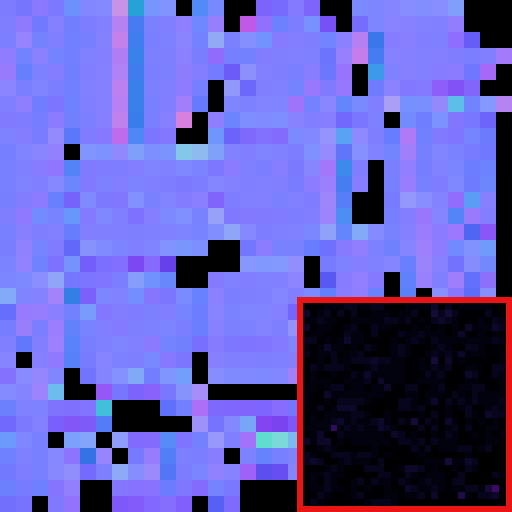} & \mipQualImg{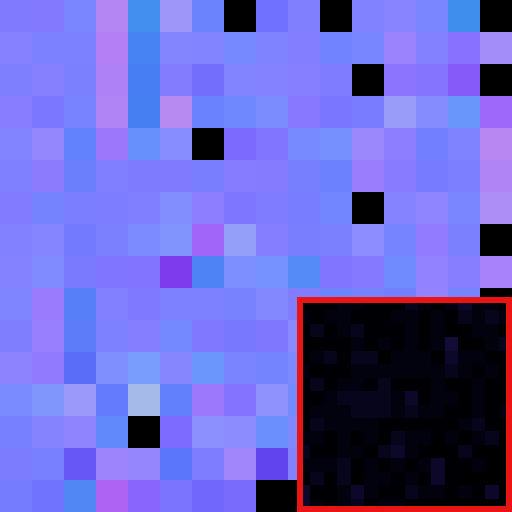} & \mipQualImg{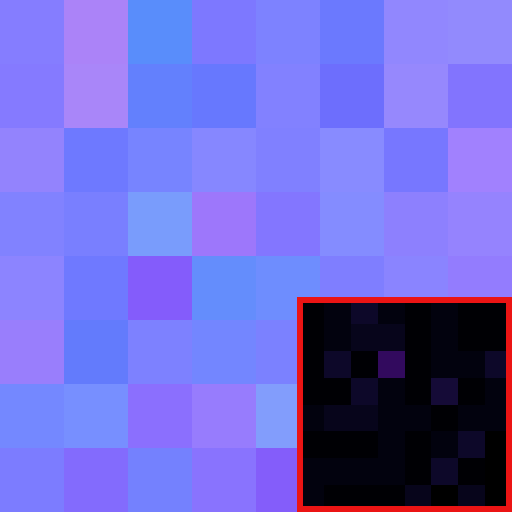} & \mipQualImg{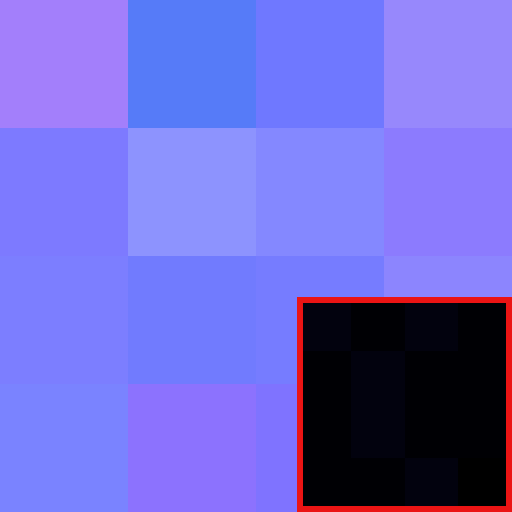} & \mipQualImg{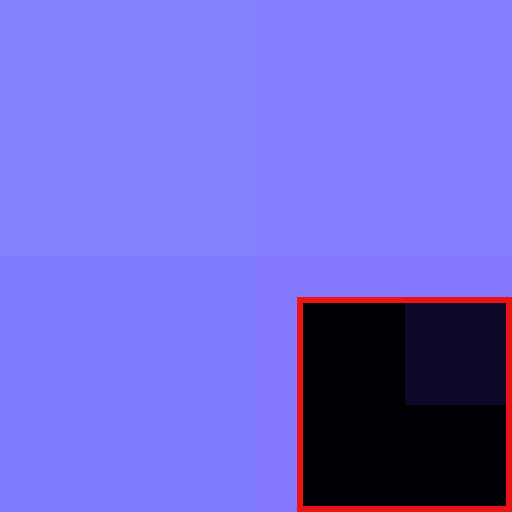} & \mipQualImg{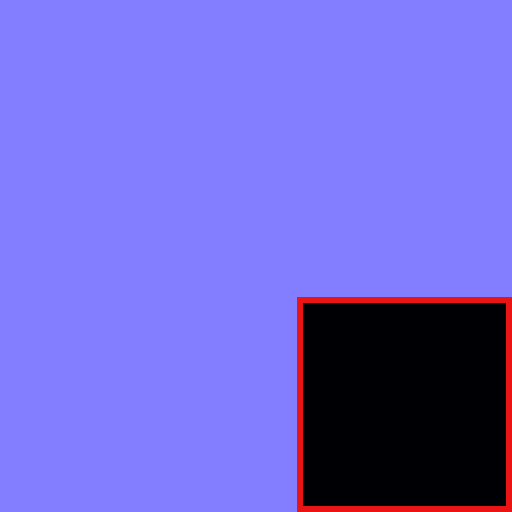} \\[\mipQualRowGap]
    \mipQualStream{Normal map} & \mipQualMethod{Image-GS}{0.368 bppc} & \mipQualImg{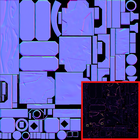} & \mipQualImg{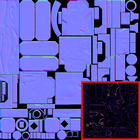} & \mipQualImg{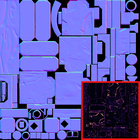} & \mipQualImg{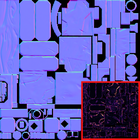} & \mipQualImg{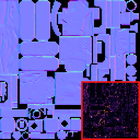} & \mipQualImg{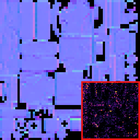} & \mipQualImg{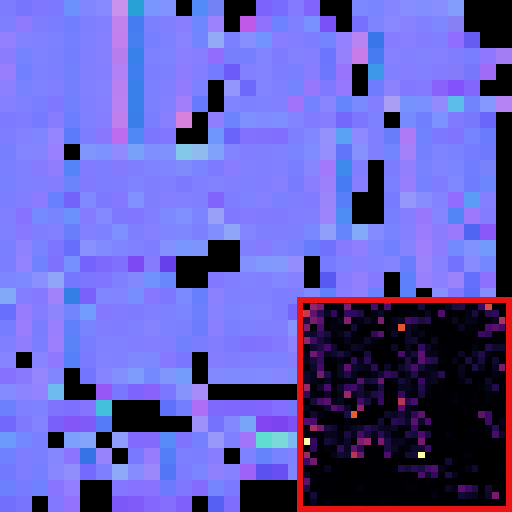} & \mipQualImg{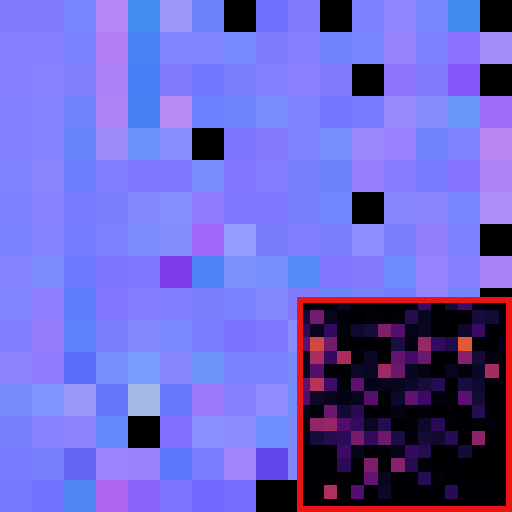} & \mipQualImg{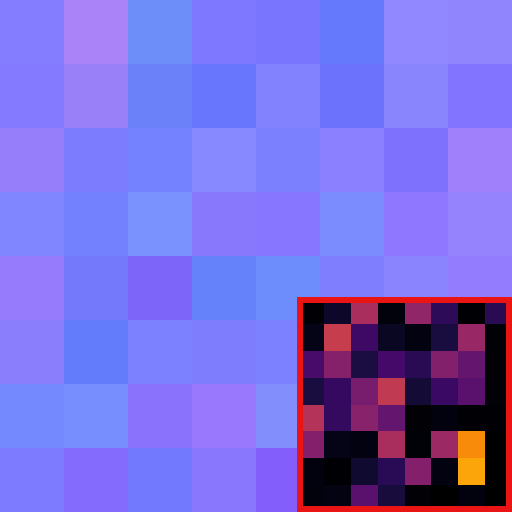} & \mipQualImg{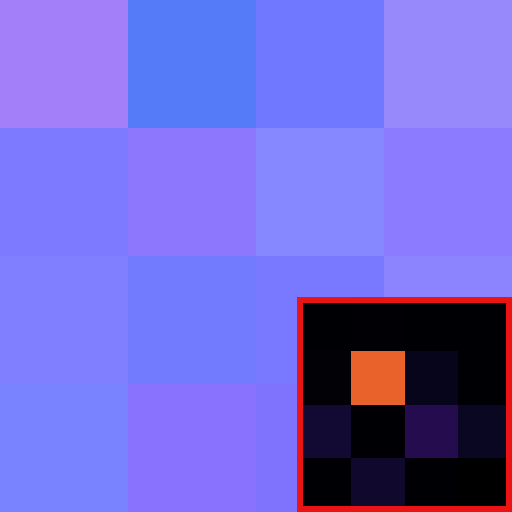} & \mipQualImg{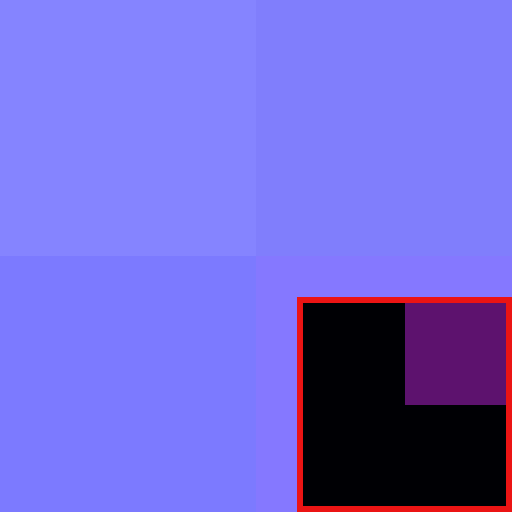} & \mipQualImg{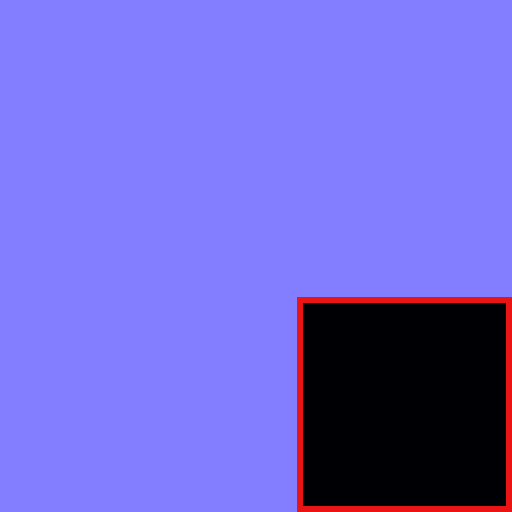} \\[\mipQualRowGap]
      & \mipQualMethod{ASTC}{0.429 bppc} & \mipQualImg{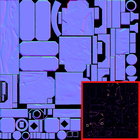} & \mipQualImg{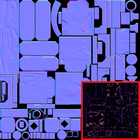} & \mipQualImg{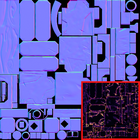} & \mipQualImg{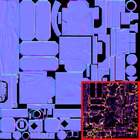} & \mipQualImg{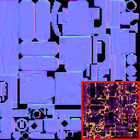} & \mipQualImg{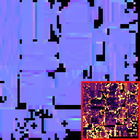} & \mipQualImg{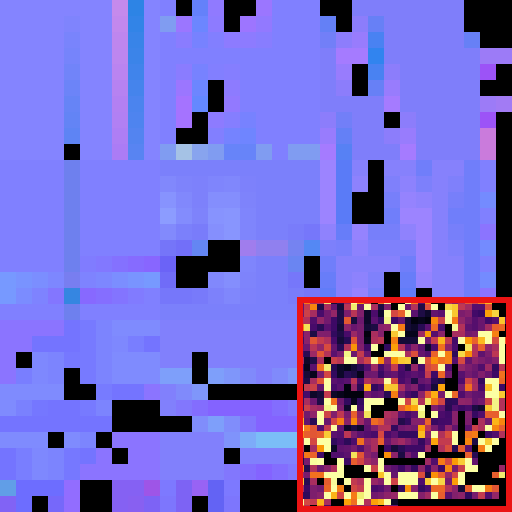} & \mipQualImg{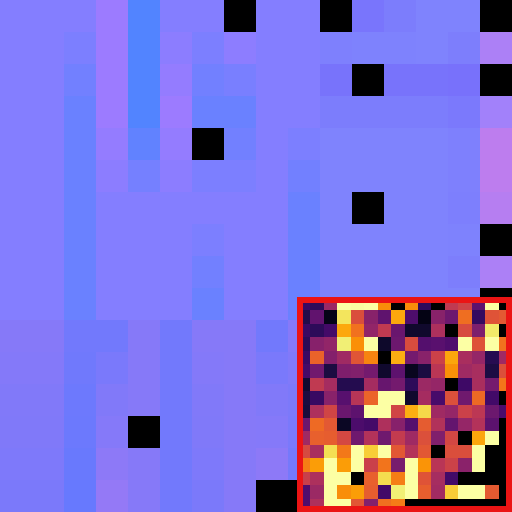} & \mipQualImg{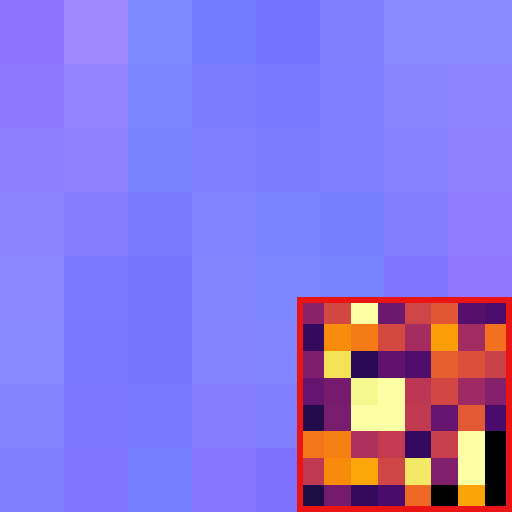} & \mipQualImg{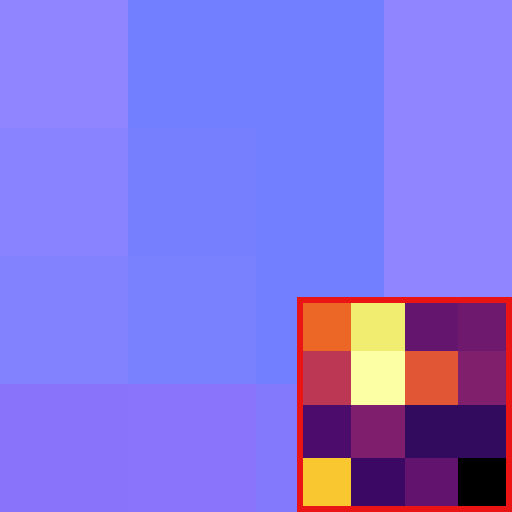} & \mipQualImg{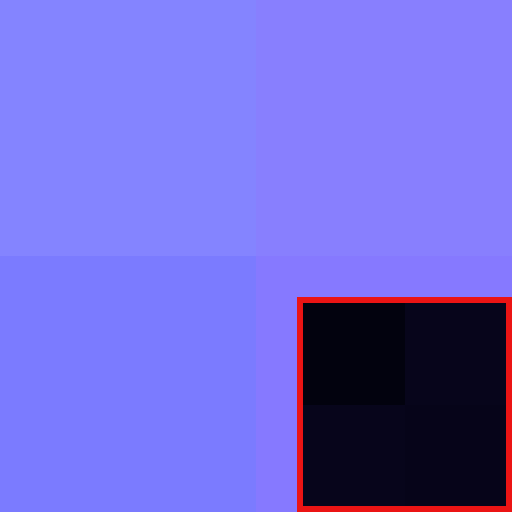} & \mipQualImg{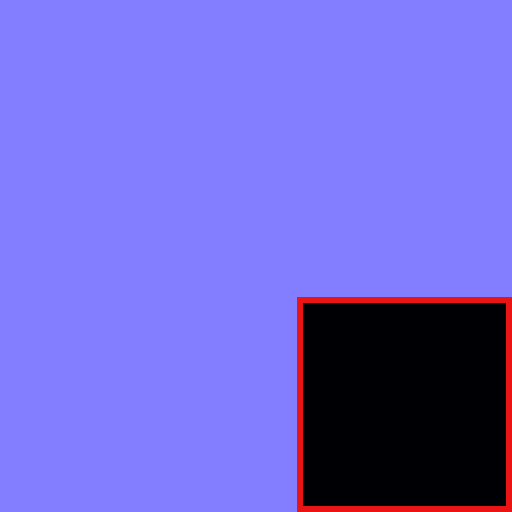} \\[\mipQualRowGap]
      & \mipQualMethod{NTC}{0.253 bppc} & \mipQualImg{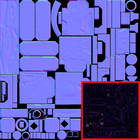} & \mipQualImg{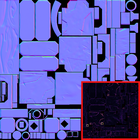} & \mipQualImg{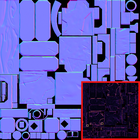} & \mipQualImg{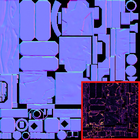} & \mipQualImg{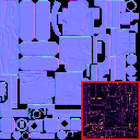} & \mipQualImg{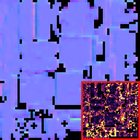} & \mipQualImg{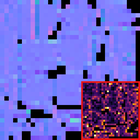} & \mipQualImg{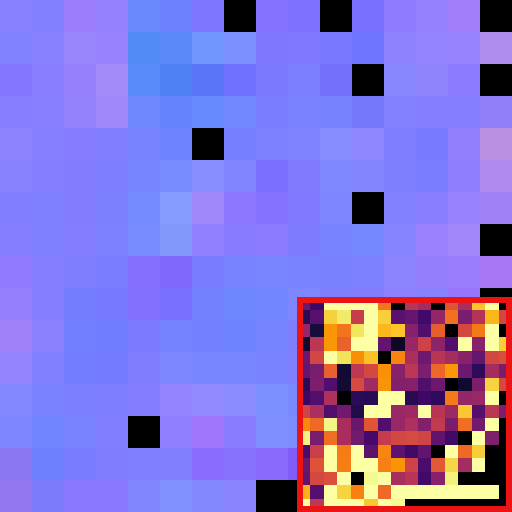} & \mipQualImg{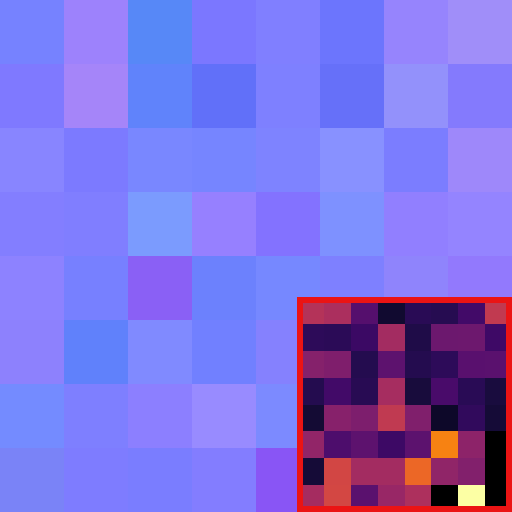} & \mipQualImg{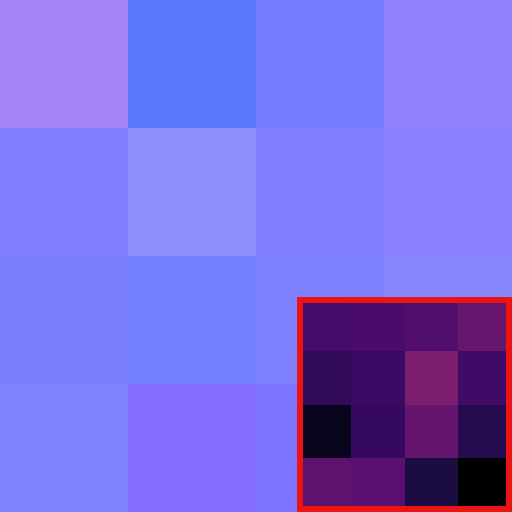} & \mipQualImg{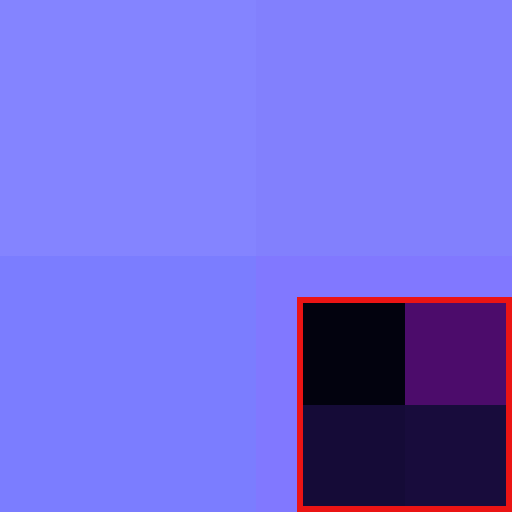} & \mipQualImg{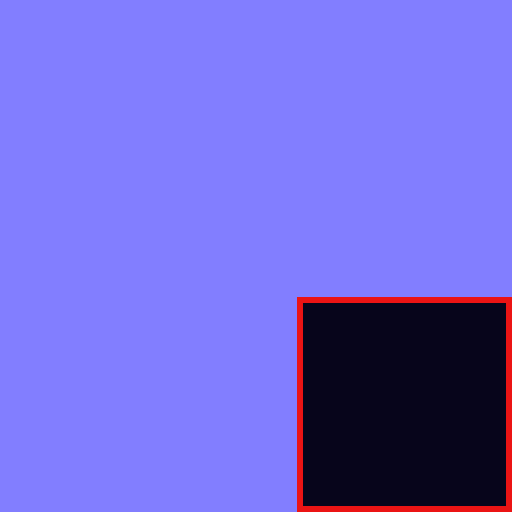} \\
    \mipQualSep
      & \mipQualMethod{GT}{} & \mipQualImg{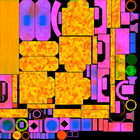} & \mipQualImg{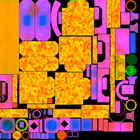} & \mipQualImg{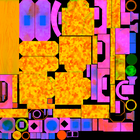} & \mipQualImg{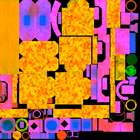} & \mipQualImg{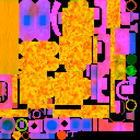} & \mipQualImg{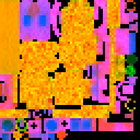} & \mipQualImg{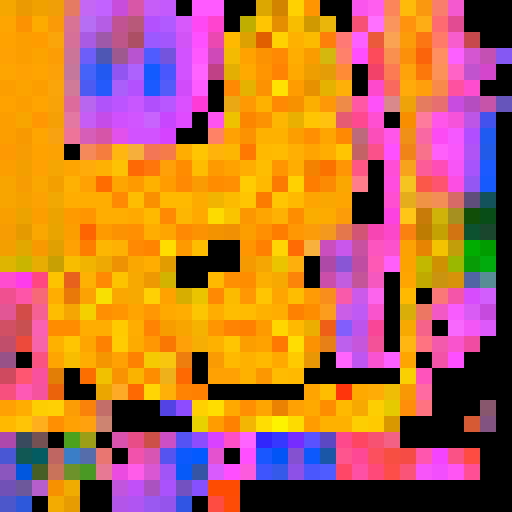} & \mipQualImg{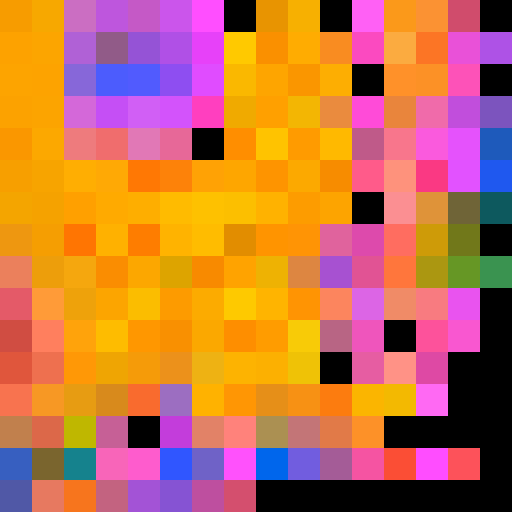} & \mipQualImg{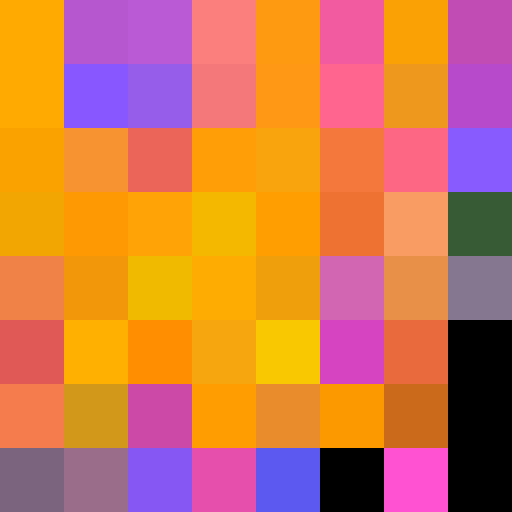} & \mipQualImg{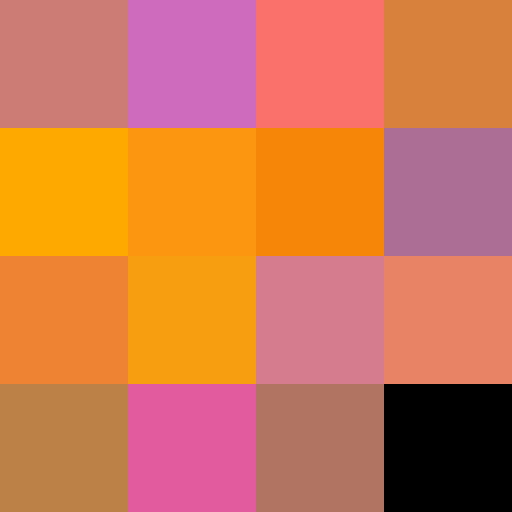} & \mipQualImg{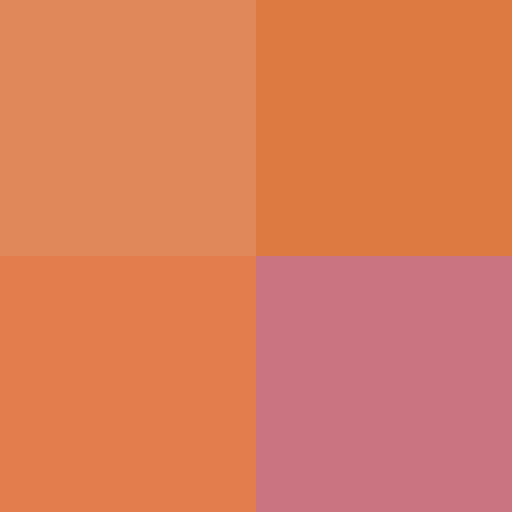} & \mipQualImg{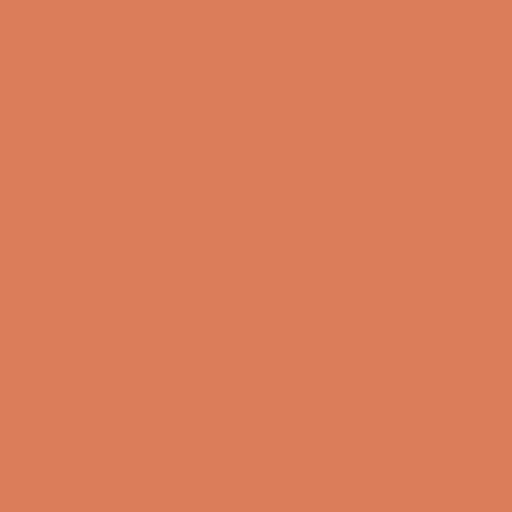} \\[\mipQualRowGap]
      & \mipQualMethod{Ours}{0.211 bppc} & \mipQualImg{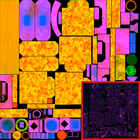} & \mipQualImg{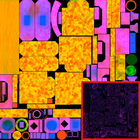} & \mipQualImg{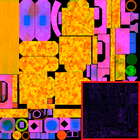} & \mipQualImg{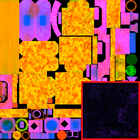} & \mipQualImg{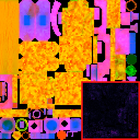} & \mipQualImg{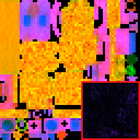} & \mipQualImg{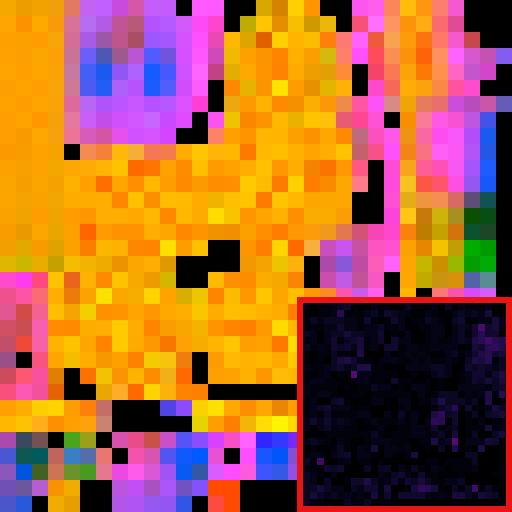} & \mipQualImg{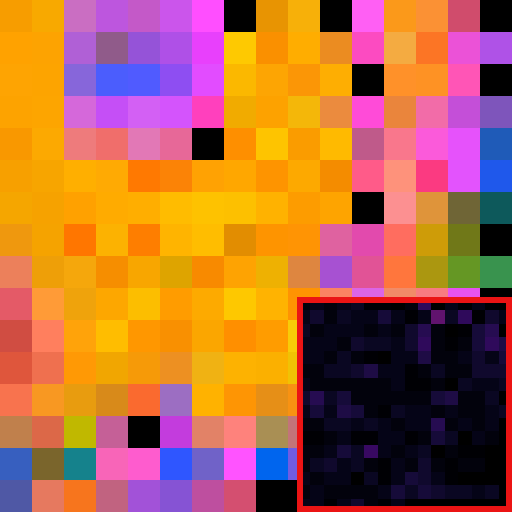} & \mipQualImg{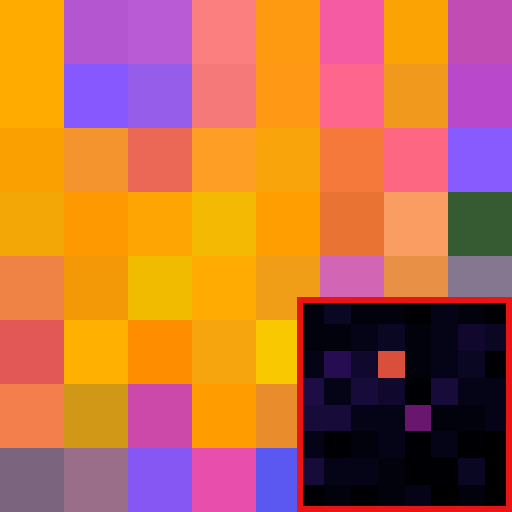} & \mipQualImg{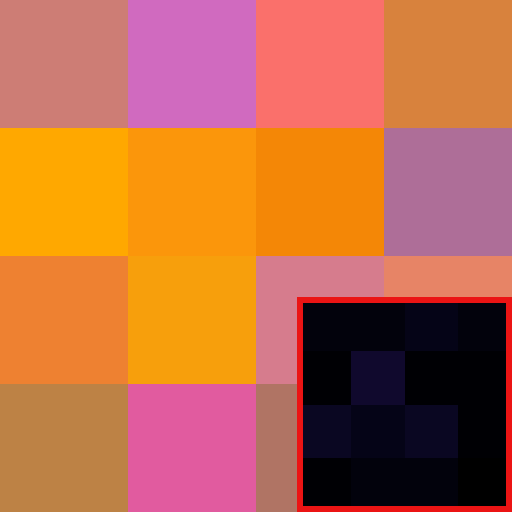} & \mipQualImg{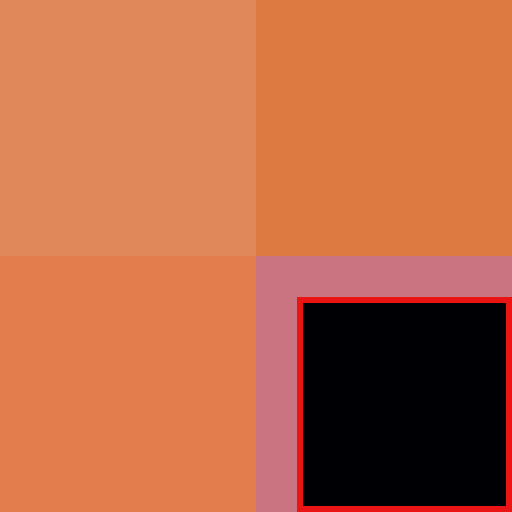} & \mipQualImg{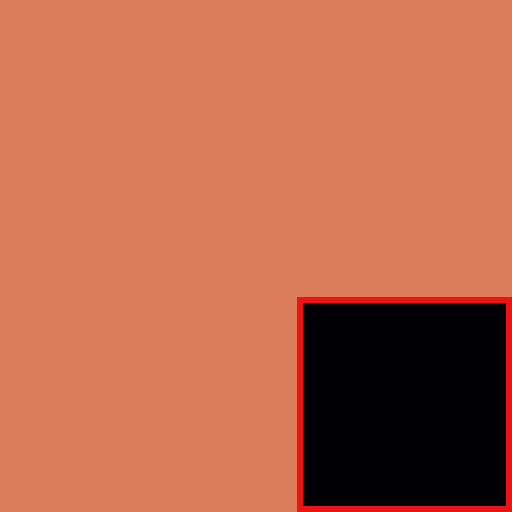} \\[\mipQualRowGap]
    \mipQualStream{ARM} & \mipQualMethod{Image-GS}{0.368 bppc} & \mipQualImg{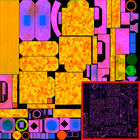} & \mipQualImg{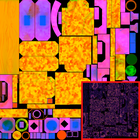} & \mipQualImg{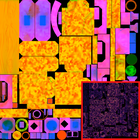} & \mipQualImg{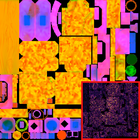} & \mipQualImg{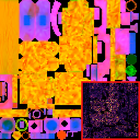} & \mipQualImg{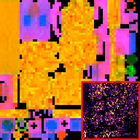} & \mipQualImg{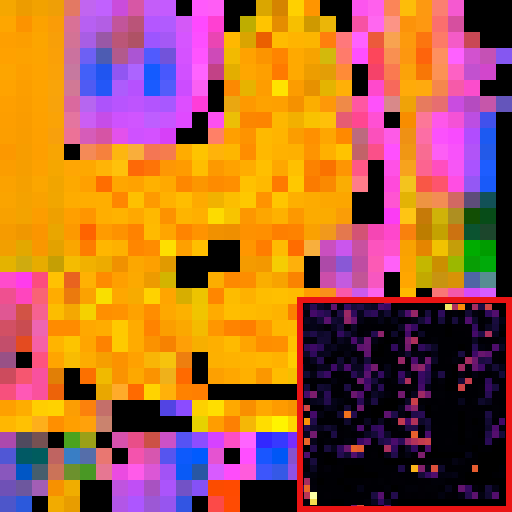} & \mipQualImg{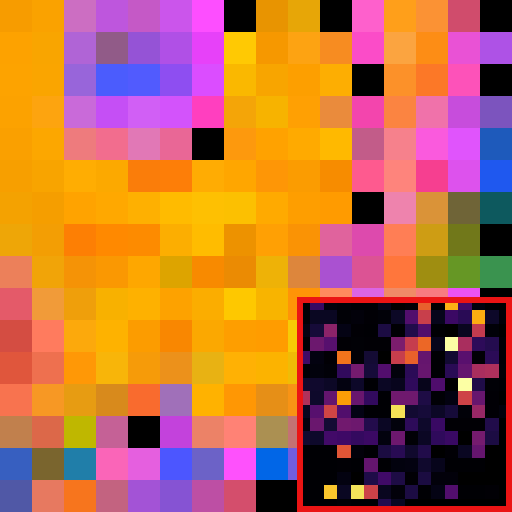} & \mipQualImg{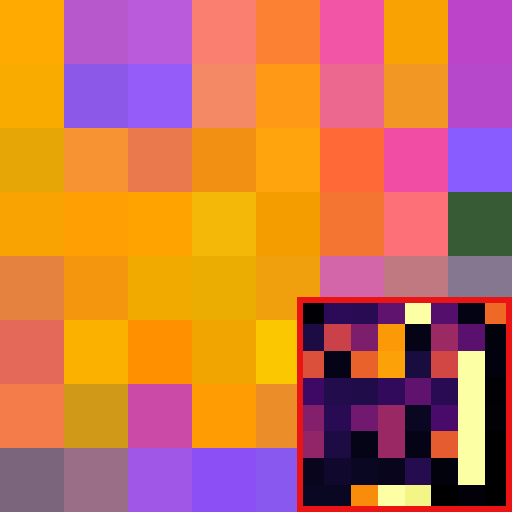} & \mipQualImg{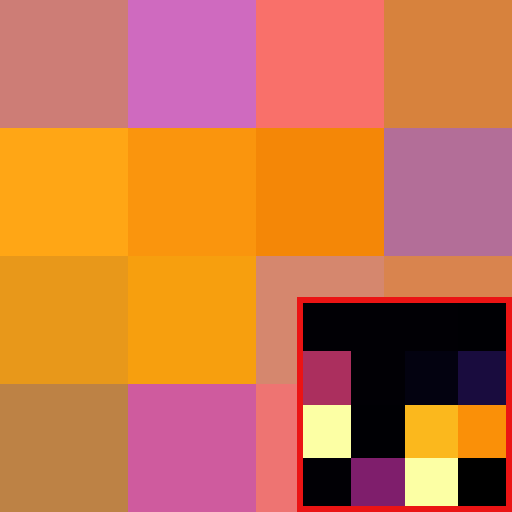} & \mipQualImg{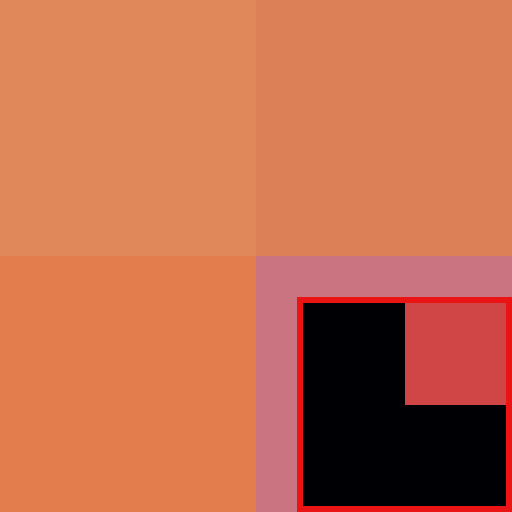} & \mipQualImg{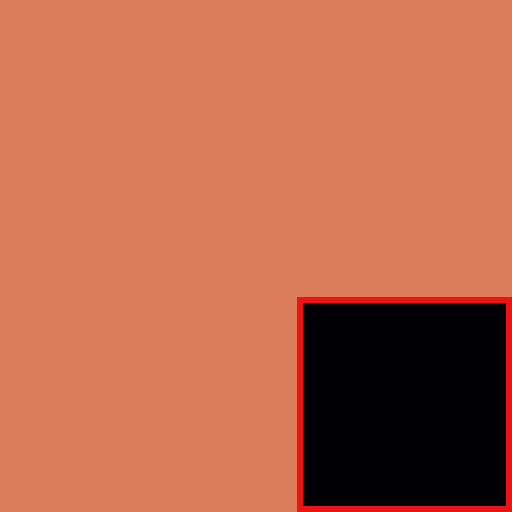} \\[\mipQualRowGap]
      & \mipQualMethod{ASTC}{0.429 bppc} & \mipQualImg{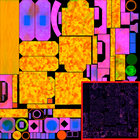} & \mipQualImg{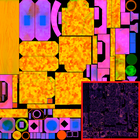} & \mipQualImg{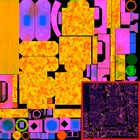} & \mipQualImg{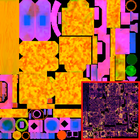} & \mipQualImg{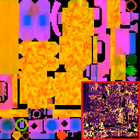} & \mipQualImg{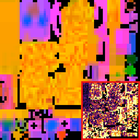} & \mipQualImg{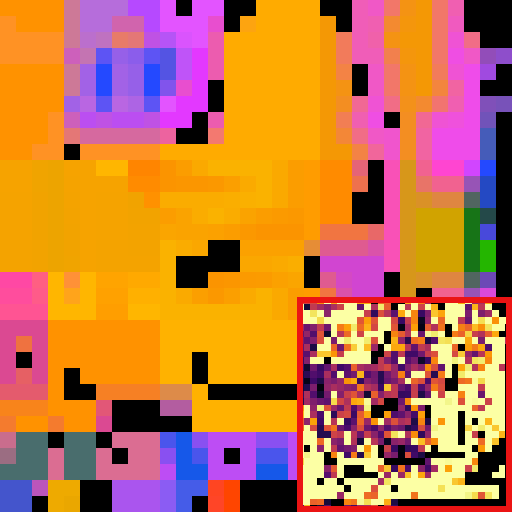} & \mipQualImg{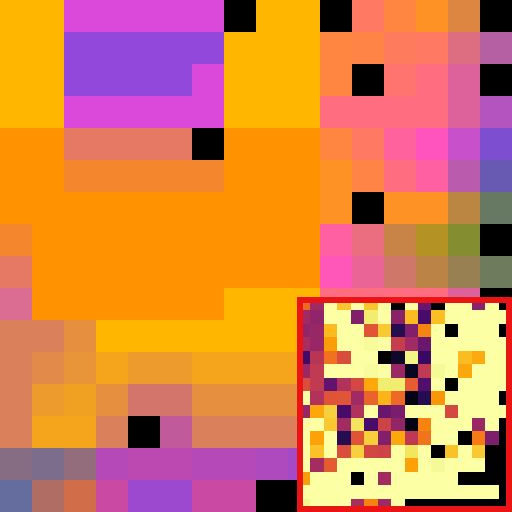} & \mipQualImg{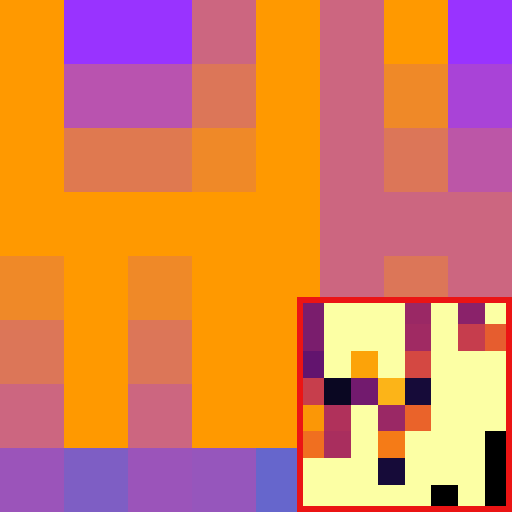} & \mipQualImg{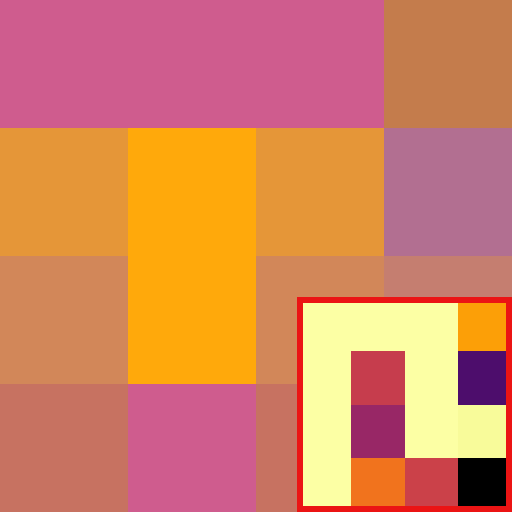} & \mipQualImg{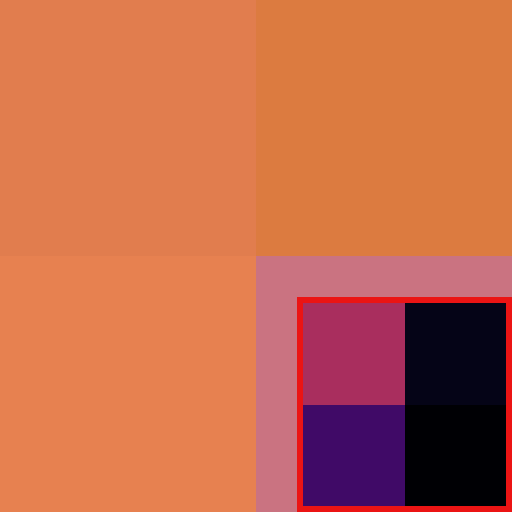} & \mipQualImg{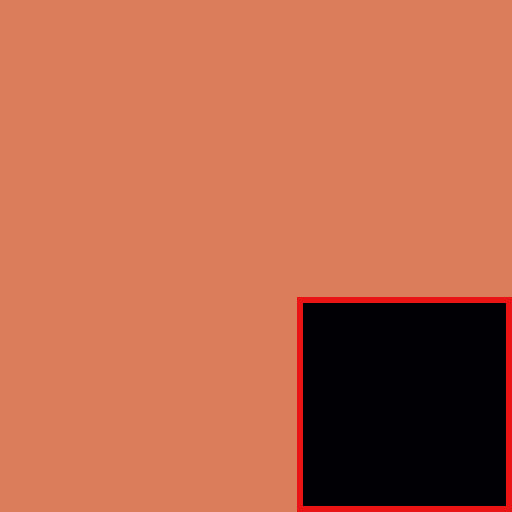} \\[\mipQualRowGap]
      & \mipQualMethod{NTC}{0.253 bppc} & \mipQualImg{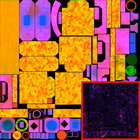} & \mipQualImg{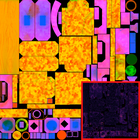} & \mipQualImg{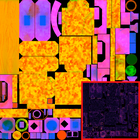} & \mipQualImg{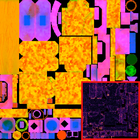} & \mipQualImg{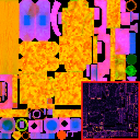} & \mipQualImg{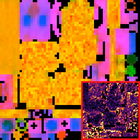} & \mipQualImg{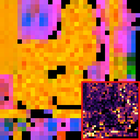} & \mipQualImg{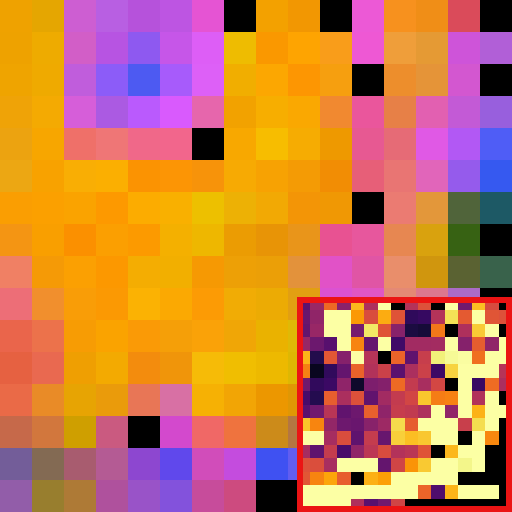} & \mipQualImg{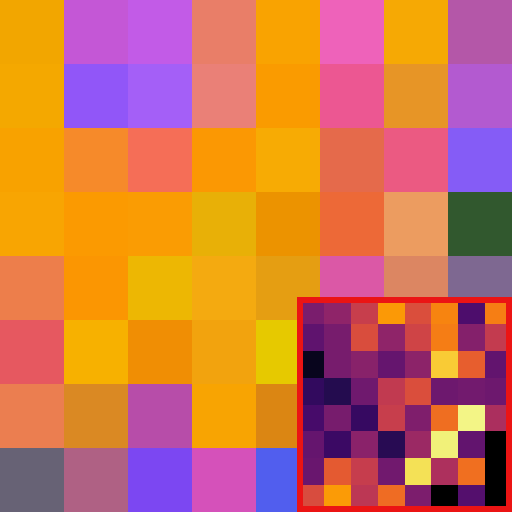} & \mipQualImg{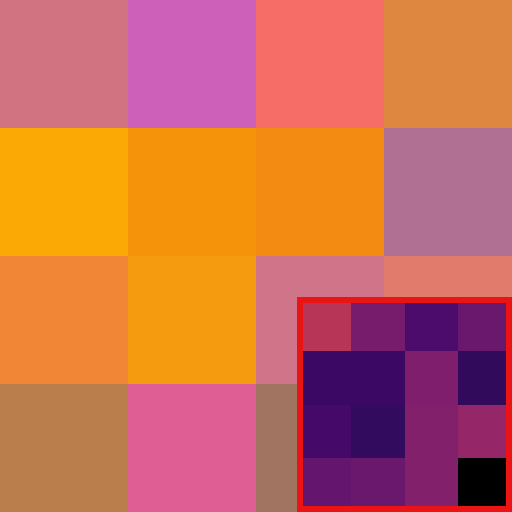} & \mipQualImg{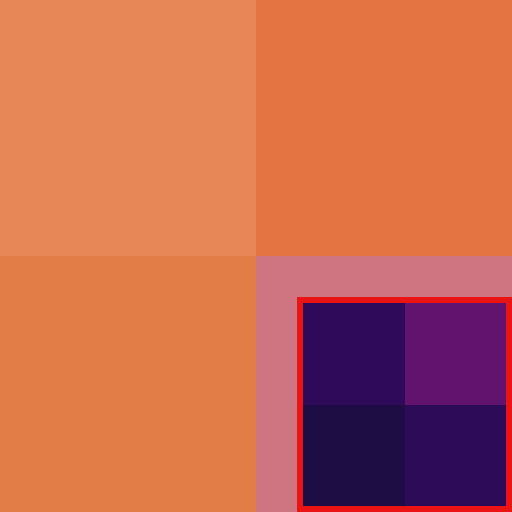} & \mipQualImg{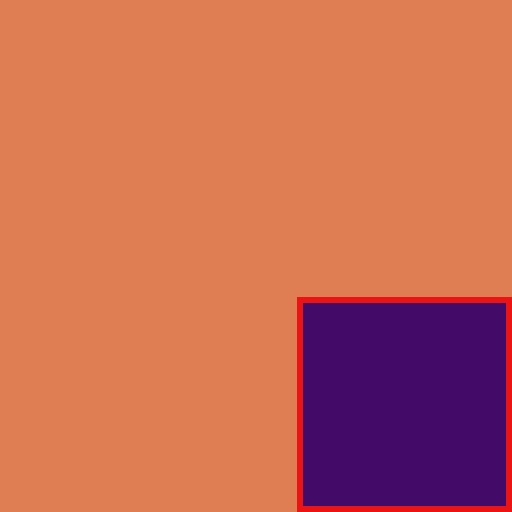} \\

  \end{tabular}%
  }
  \caption{Mipmap reconstruction comparison on the \textit{Hammer} SVBRDF stack. Rows show the reference and reconstructed mipmap pyramids, with insets showing per-pixel error maps against the reference. ARM denotes ambient occlusion, roughness, and metallic channels.}
  \label{fig:mipmap-error-inset-hammer}
\end{figure*}

\clearpage
\begin{figure*}[p!]
  \centering
  \mipQualFontSize
  \setlength{\tabcolsep}{0pt}
  \renewcommand{\arraystretch}{1.0}
  \resizebox{\mipQualTableTargetW}{!}{%
  \begin{tabular}{@{}c@{\hspace{\mipQualStreamMethodGap}}c@{\hspace{\mipQualMethodImageGap}}*{11}{c@{\hspace{\mipQualColGap}}}c@{}}
    \multicolumn{2}{c}{} & \mipQualMip{mip 0}{2048$^2$} & \mipQualMip{mip 1}{1024$^2$} & \mipQualMip{mip 2}{512$^2$} & \mipQualMip{mip 3}{256$^2$} & \mipQualMip{mip 4}{128$^2$} & \mipQualMip{mip 5}{64$^2$} & \mipQualMip{mip 6}{32$^2$} & \mipQualMip{mip 7}{16$^2$} & \mipQualMip{mip 8}{8$^2$} & \mipQualMip{mip 9}{4$^2$} & \mipQualMip{mip 10}{2$^2$} & \mipQualMip{mip 11}{1$^2$} \\[\mipQualHeaderGap]
      & \mipQualMethod{GT}{} & \mipQualImg{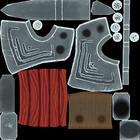} & \mipQualImg{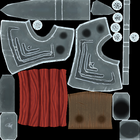} & \mipQualImg{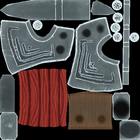} & \mipQualImg{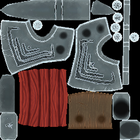} & \mipQualImg{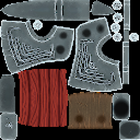} & \mipQualImg{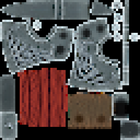} & \mipQualImg{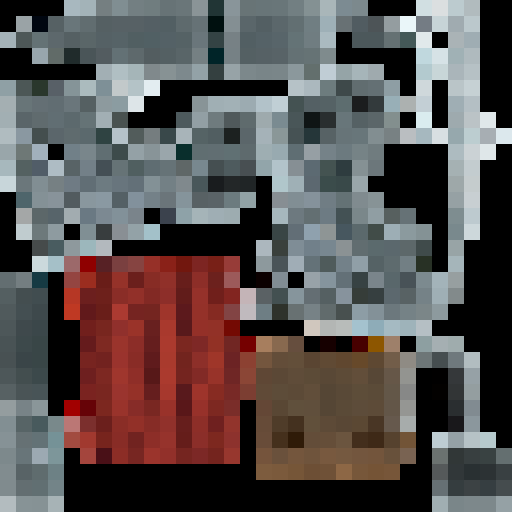} & \mipQualImg{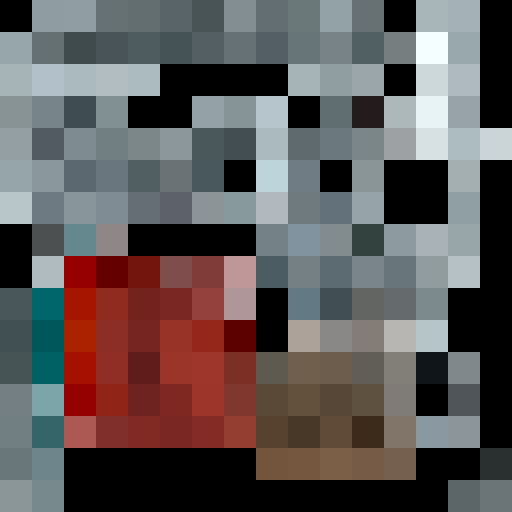} & \mipQualImg{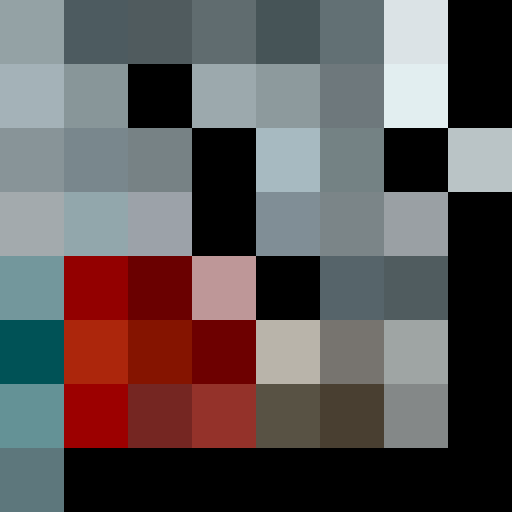} & \mipQualImg{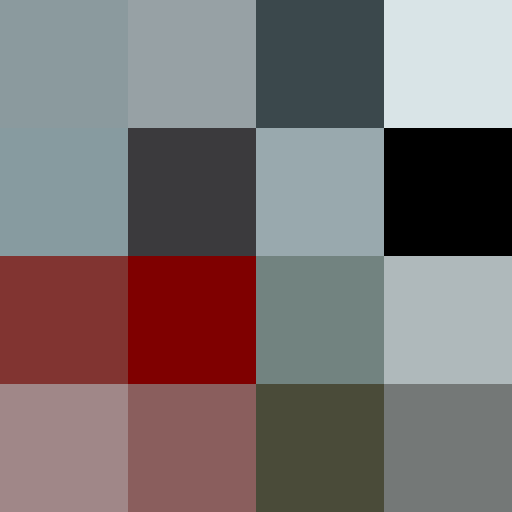} & \mipQualImg{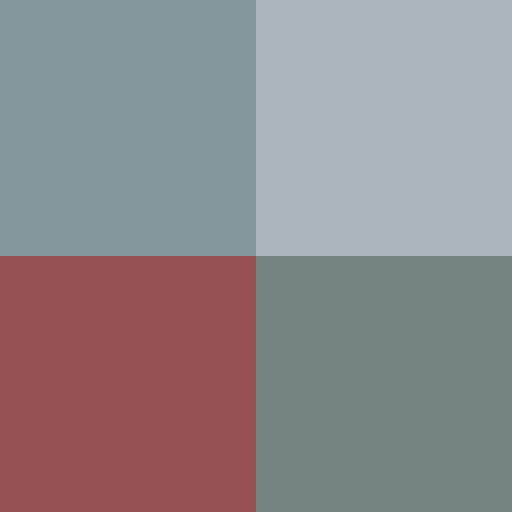} & \mipQualImg{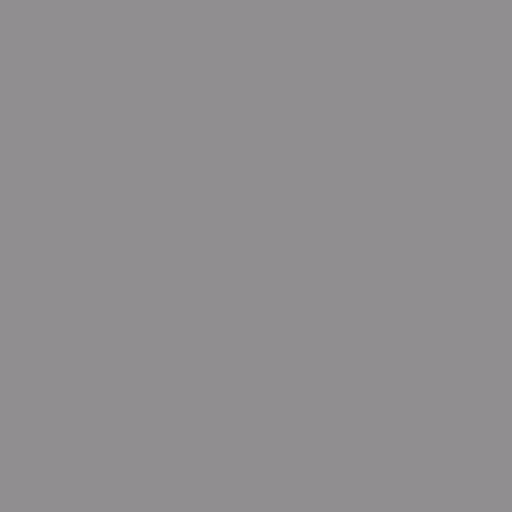} \\[\mipQualRowGap]
      & \mipQualMethod{Ours}{0.195 bppc} & \mipQualImg{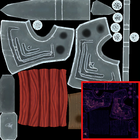} & \mipQualImg{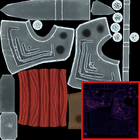} & \mipQualImg{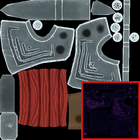} & \mipQualImg{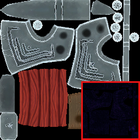} & \mipQualImg{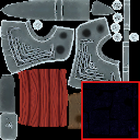} & \mipQualImg{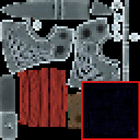} & \mipQualImg{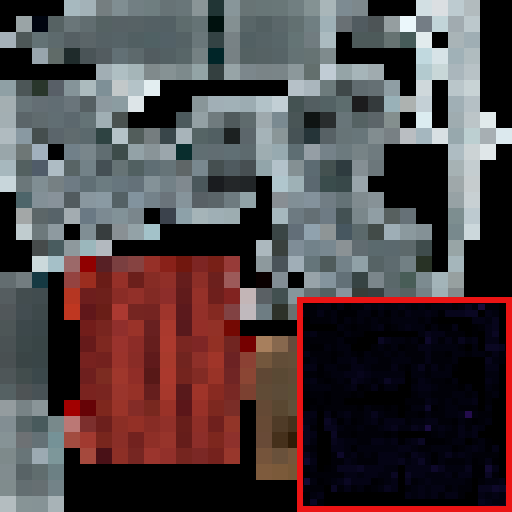} & \mipQualImg{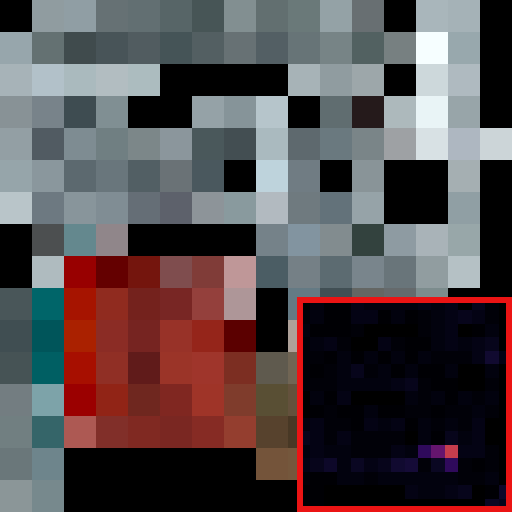} & \mipQualImg{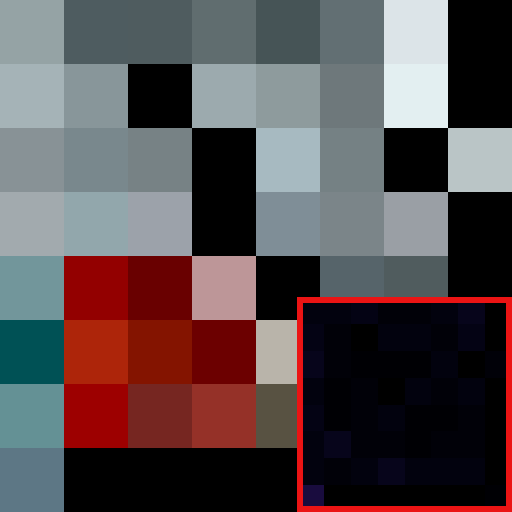} & \mipQualImg{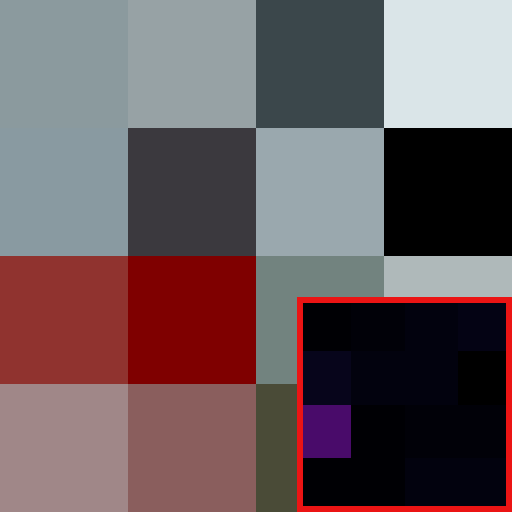} & \mipQualImg{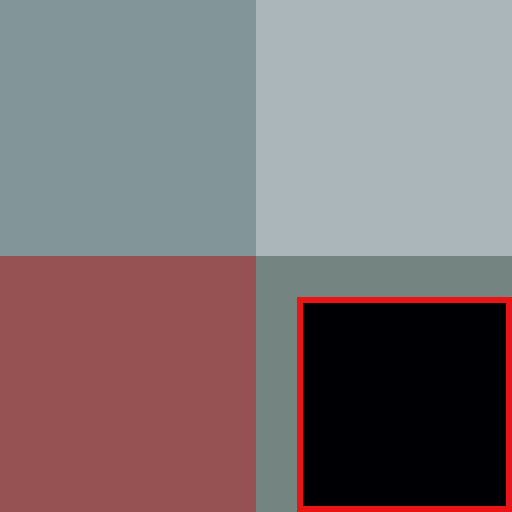} & \mipQualImg{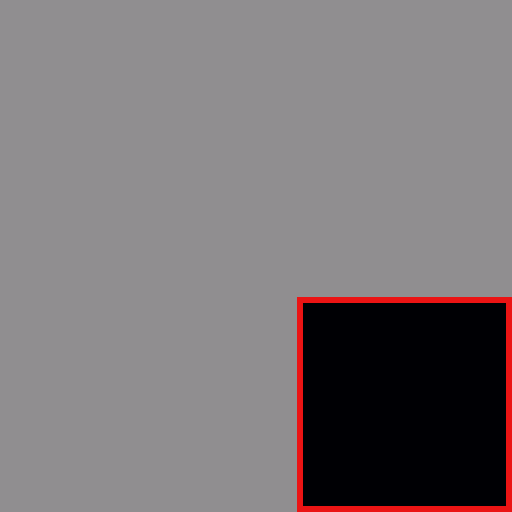} \\[\mipQualRowGap]
    \mipQualStream{Base color} & \mipQualMethod{Image-GS}{0.404 bppc} & \mipQualImg{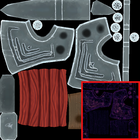} & \mipQualImg{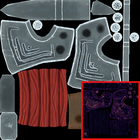} & \mipQualImg{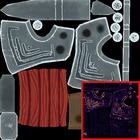} & \mipQualImg{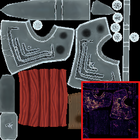} & \mipQualImg{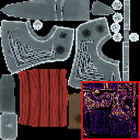} & \mipQualImg{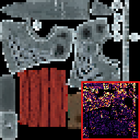} & \mipQualImg{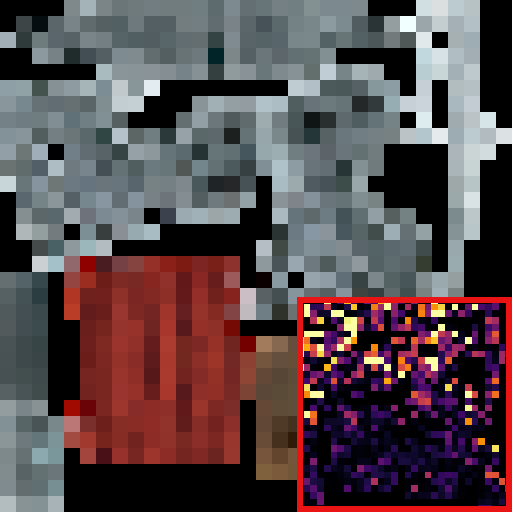} & \mipQualImg{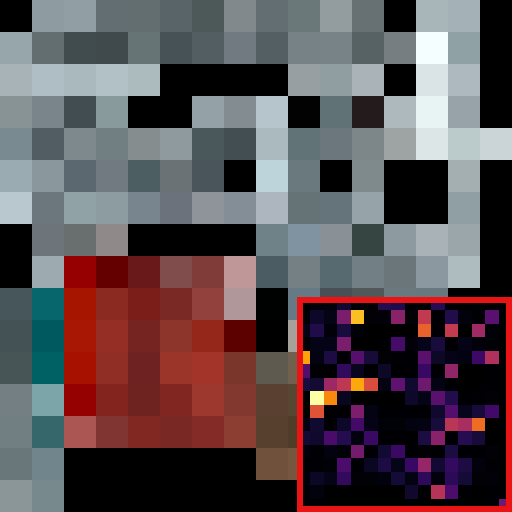} & \mipQualImg{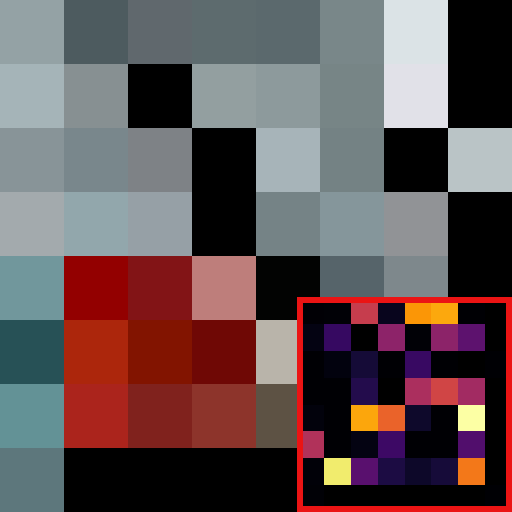} & \mipQualImg{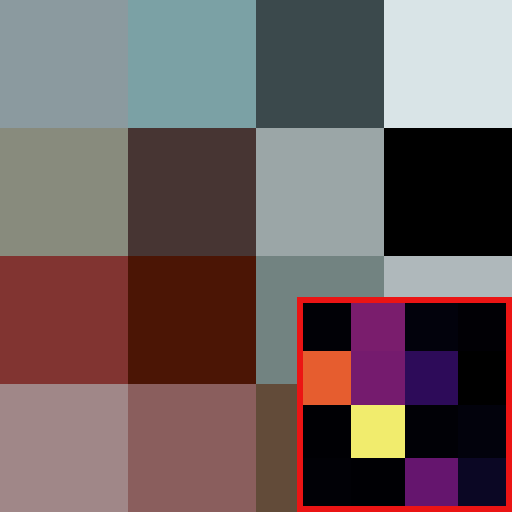} & \mipQualImg{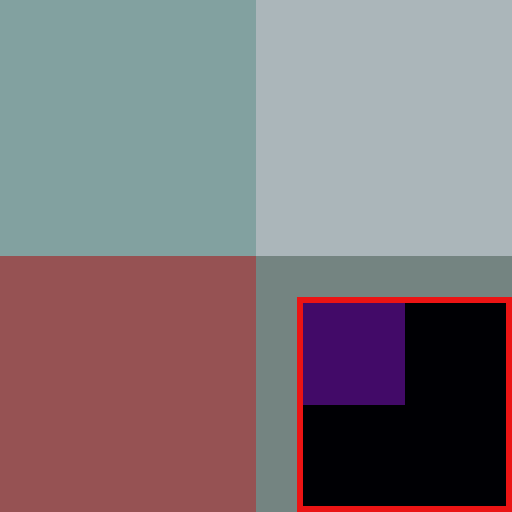} & \mipQualImg{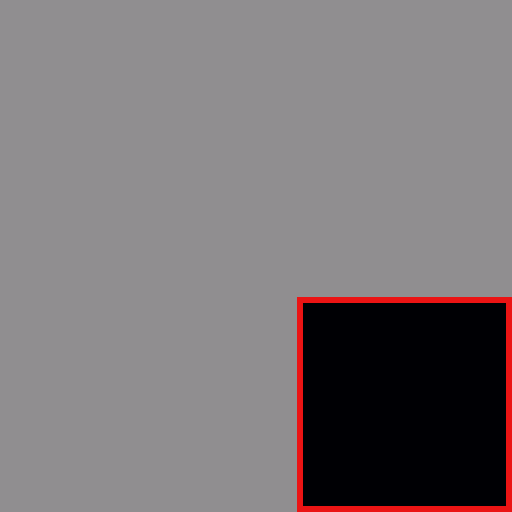} \\[\mipQualRowGap]
      & \mipQualMethod{ASTC}{0.429 bppc} & \mipQualImg{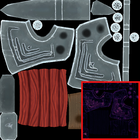} & \mipQualImg{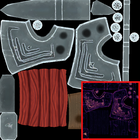} & \mipQualImg{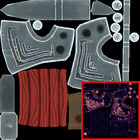} & \mipQualImg{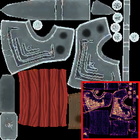} & \mipQualImg{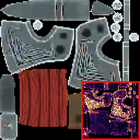} & \mipQualImg{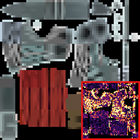} & \mipQualImg{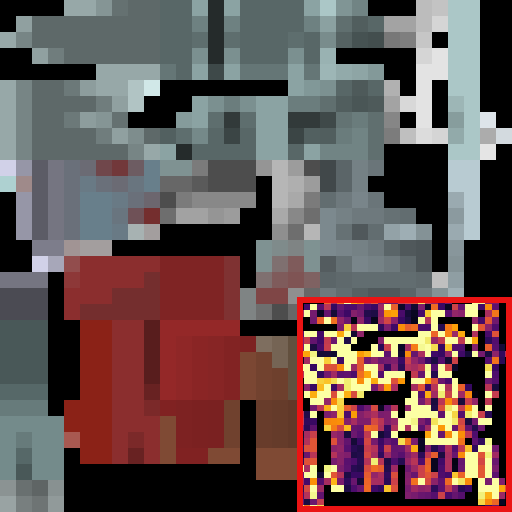} & \mipQualImg{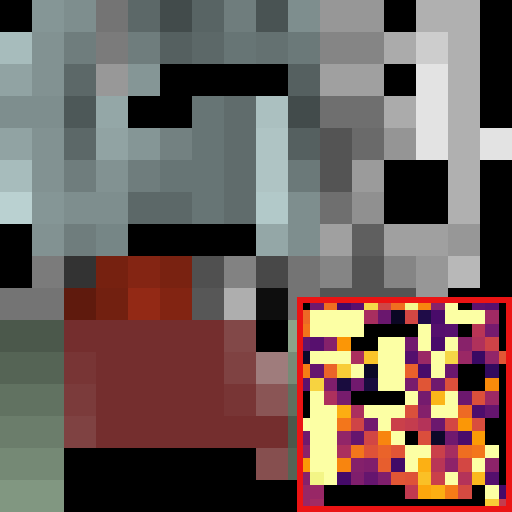} & \mipQualImg{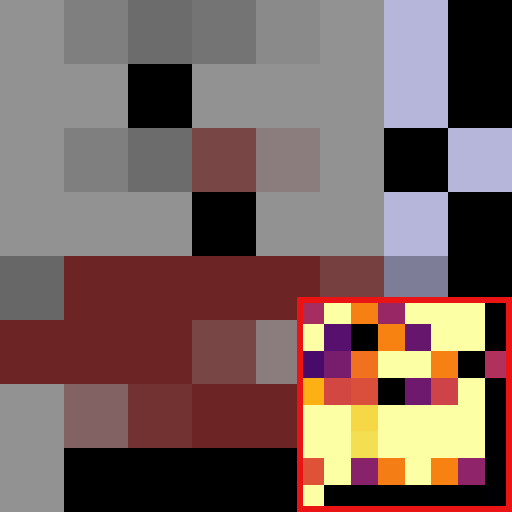} & \mipQualImg{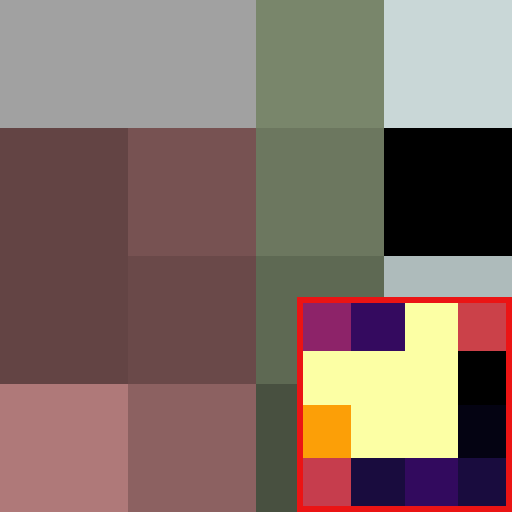} & \mipQualImg{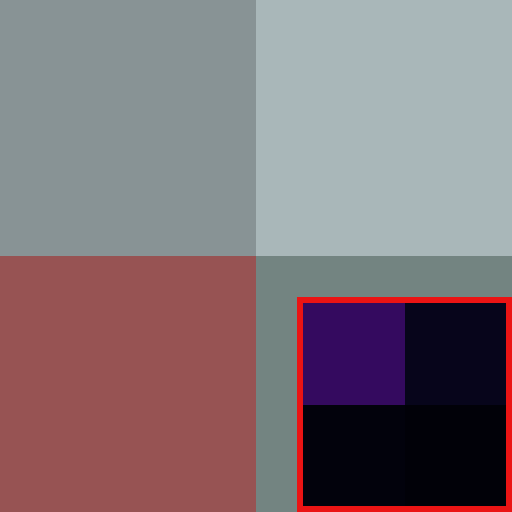} & \mipQualImg{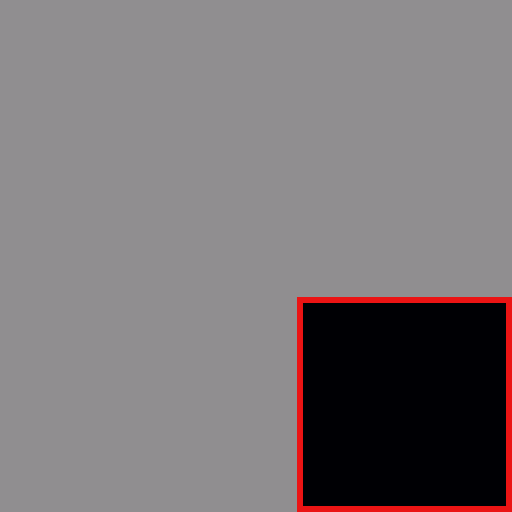} \\[\mipQualRowGap]
      & \mipQualMethod{NTC}{0.226 bppc} & \mipQualImg{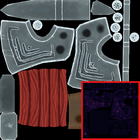} & \mipQualImg{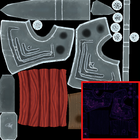} & \mipQualImg{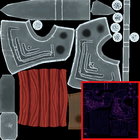} & \mipQualImg{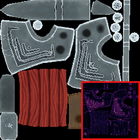} & \mipQualImg{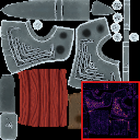} & \mipQualImg{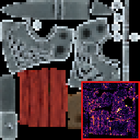} & \mipQualImg{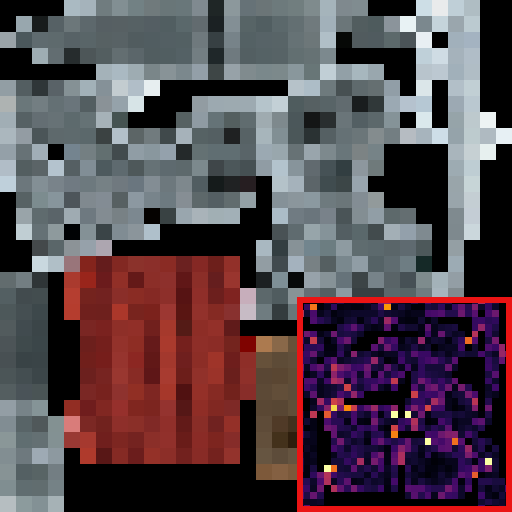} & \mipQualImg{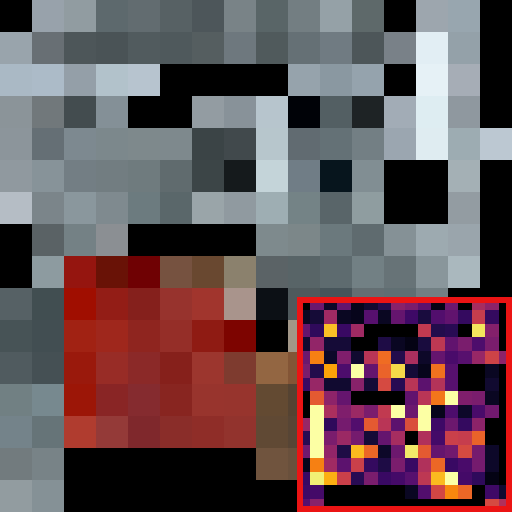} & \mipQualImg{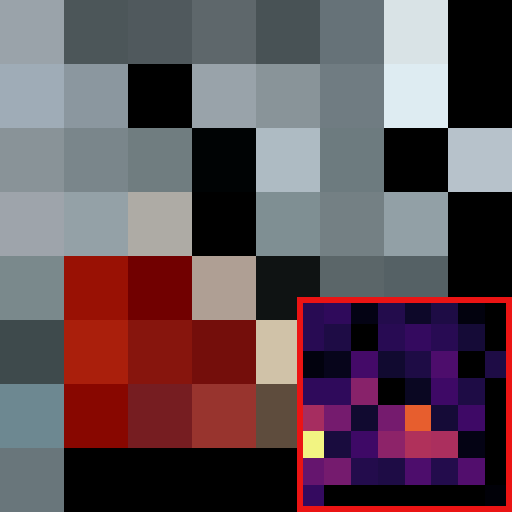} & \mipQualImg{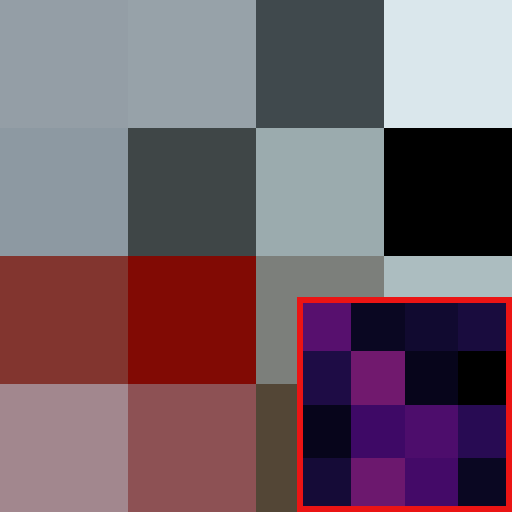} & \mipQualImg{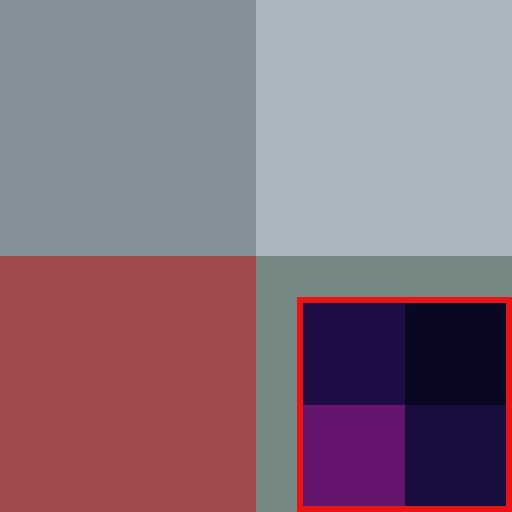} & \mipQualImg{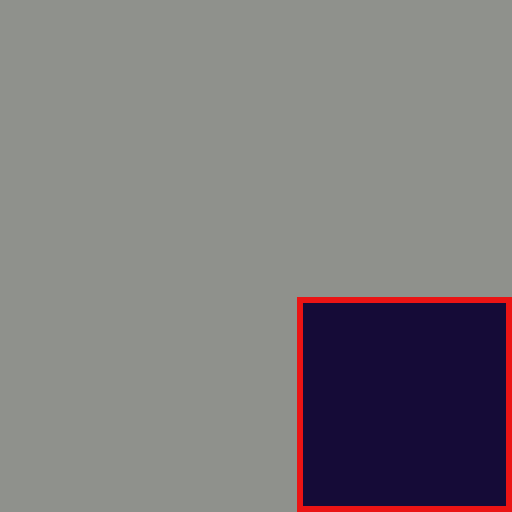} \\
    \mipQualSep
      & \mipQualMethod{GT}{} & \mipQualImg{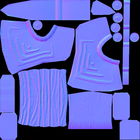} & \mipQualImg{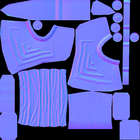} & \mipQualImg{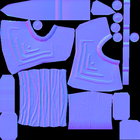} & \mipQualImg{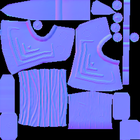} & \mipQualImg{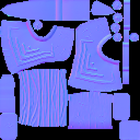} & \mipQualImg{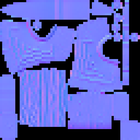} & \mipQualImg{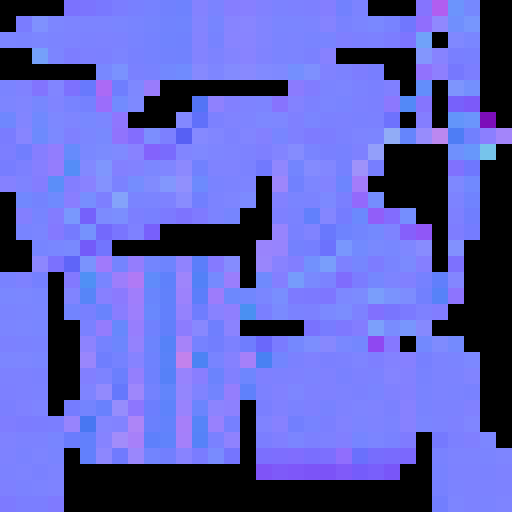} & \mipQualImg{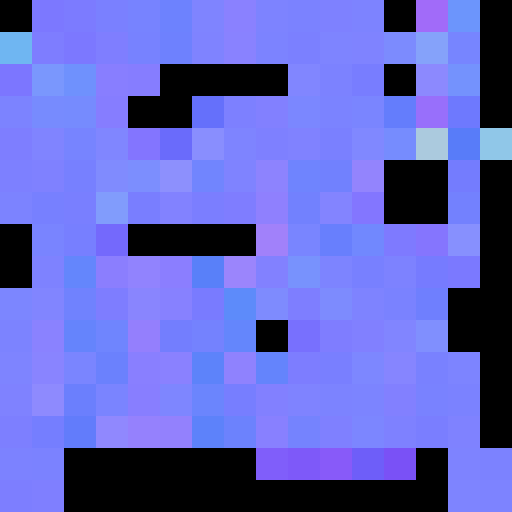} & \mipQualImg{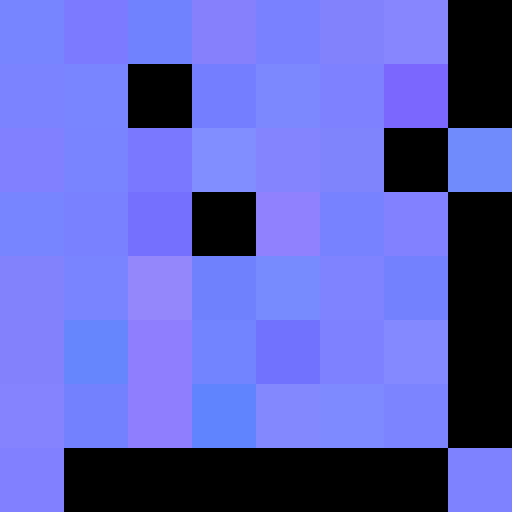} & \mipQualImg{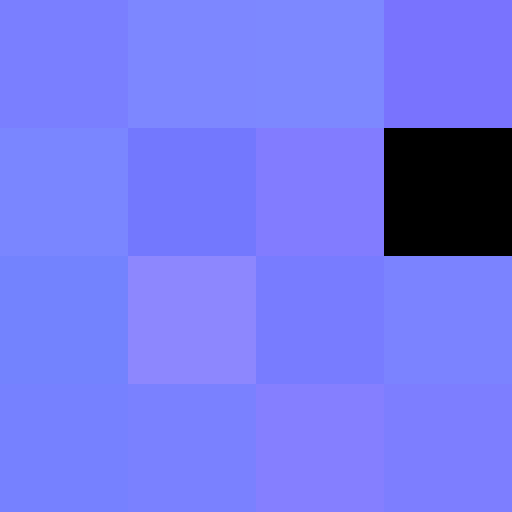} & \mipQualImg{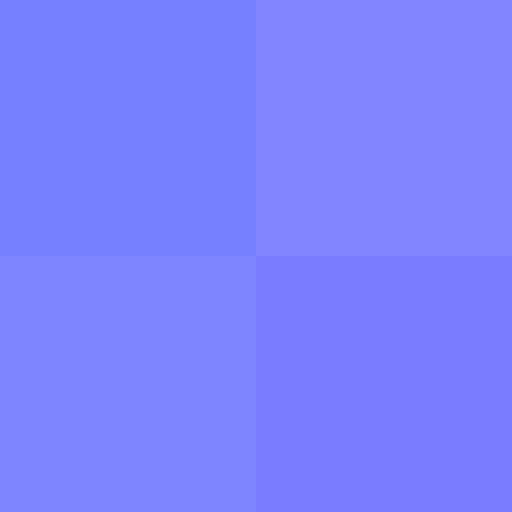} & \mipQualImg{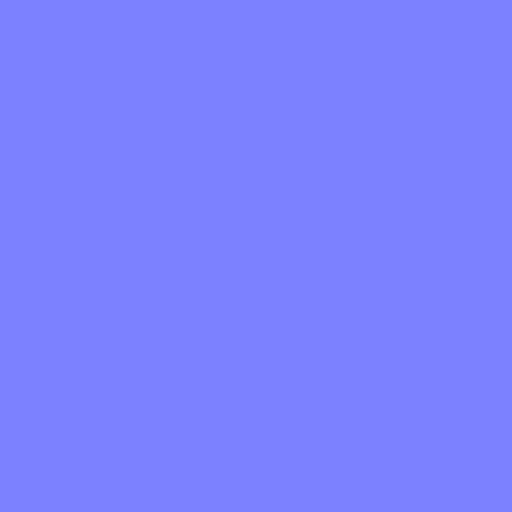} \\[\mipQualRowGap]
      & \mipQualMethod{Ours}{0.195 bppc} & \mipQualImg{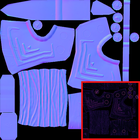} & \mipQualImg{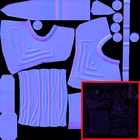} & \mipQualImg{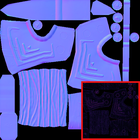} & \mipQualImg{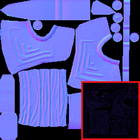} & \mipQualImg{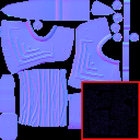} & \mipQualImg{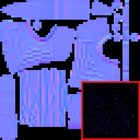} & \mipQualImg{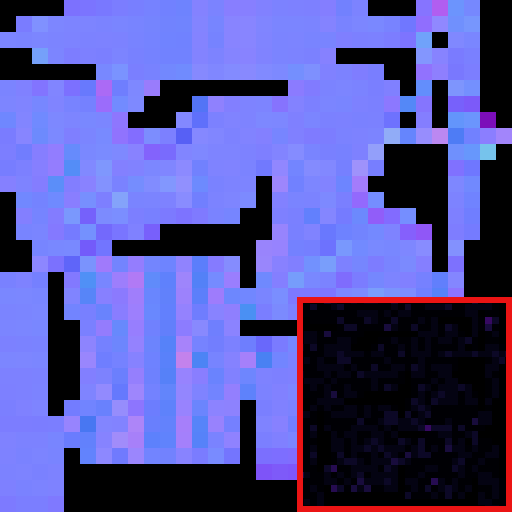} & \mipQualImg{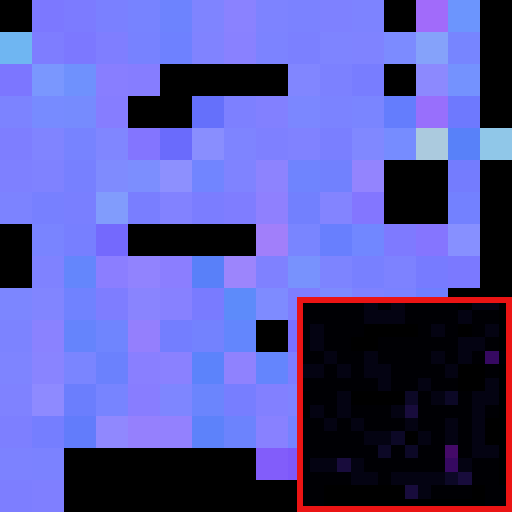} & \mipQualImg{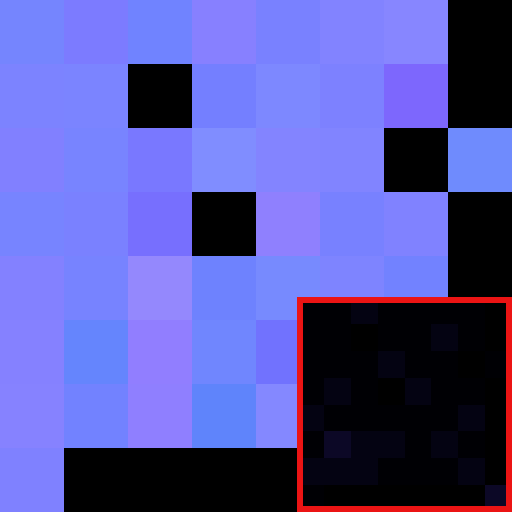} & \mipQualImg{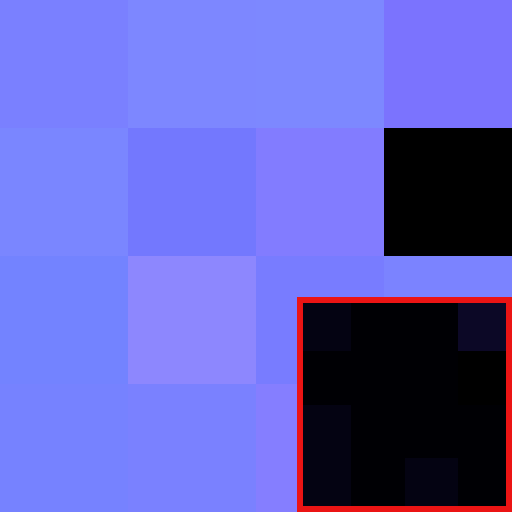} & \mipQualImg{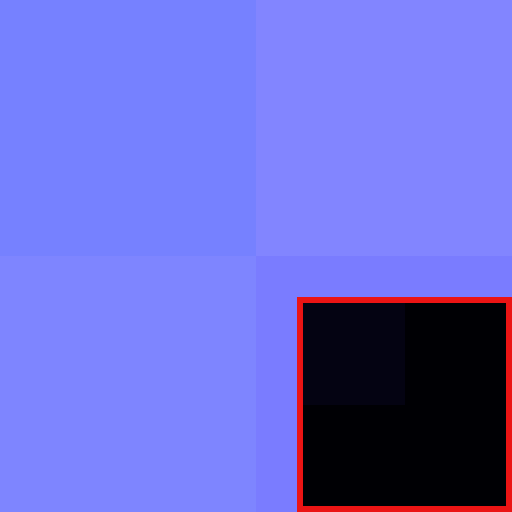} & \mipQualImg{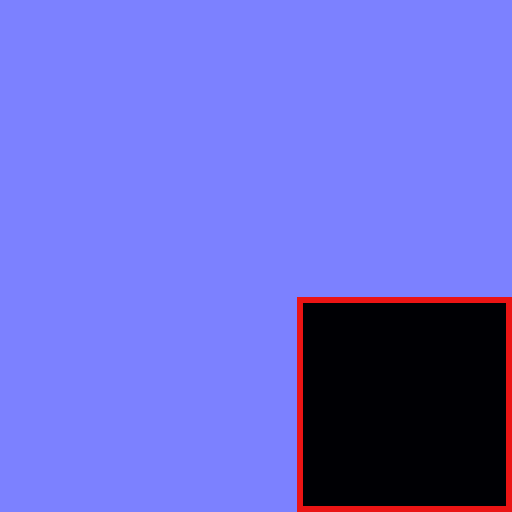} \\[\mipQualRowGap]
    \mipQualStream{Normal map} & \mipQualMethod{Image-GS}{0.404 bppc} & \mipQualImg{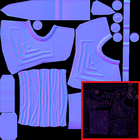} & \mipQualImg{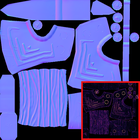} & \mipQualImg{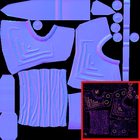} & \mipQualImg{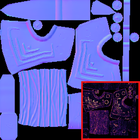} & \mipQualImg{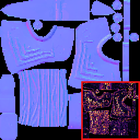} & \mipQualImg{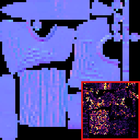} & \mipQualImg{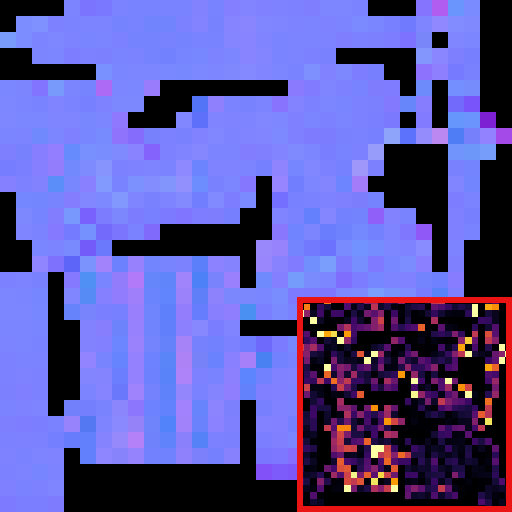} & \mipQualImg{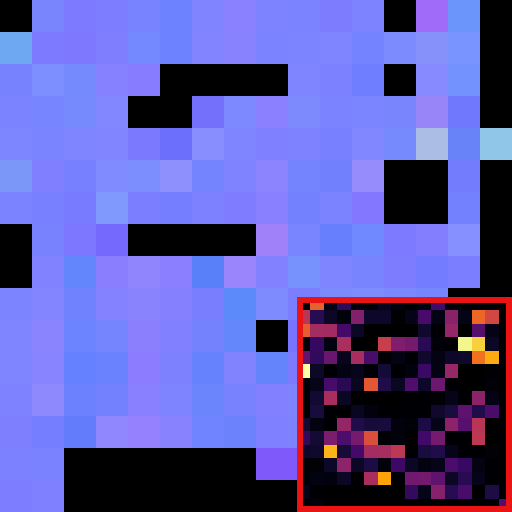} & \mipQualImg{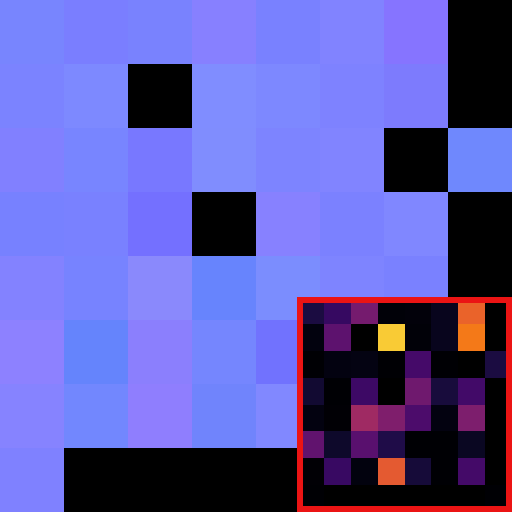} & \mipQualImg{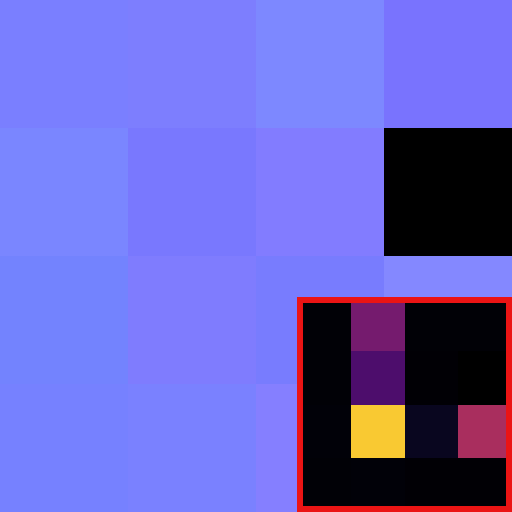} & \mipQualImg{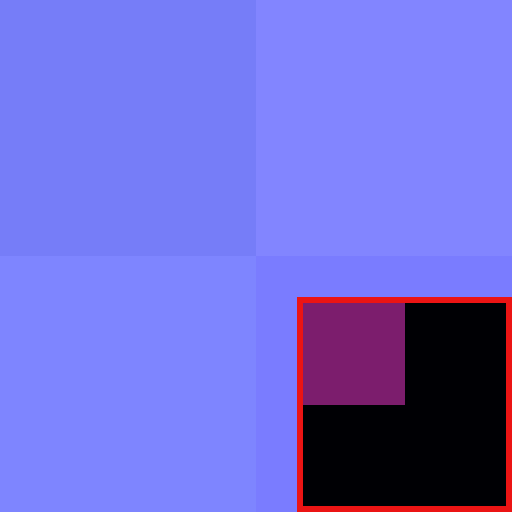} & \mipQualImg{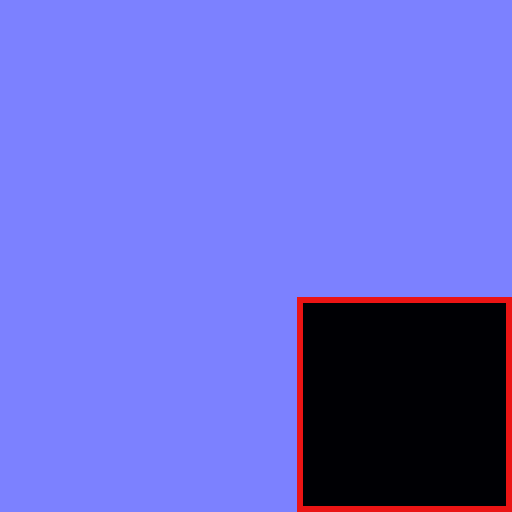} \\[\mipQualRowGap]
      & \mipQualMethod{ASTC}{0.429 bppc} & \mipQualImg{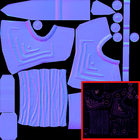} & \mipQualImg{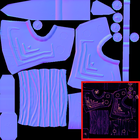} & \mipQualImg{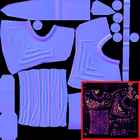} & \mipQualImg{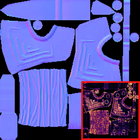} & \mipQualImg{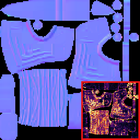} & \mipQualImg{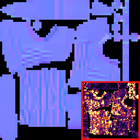} & \mipQualImg{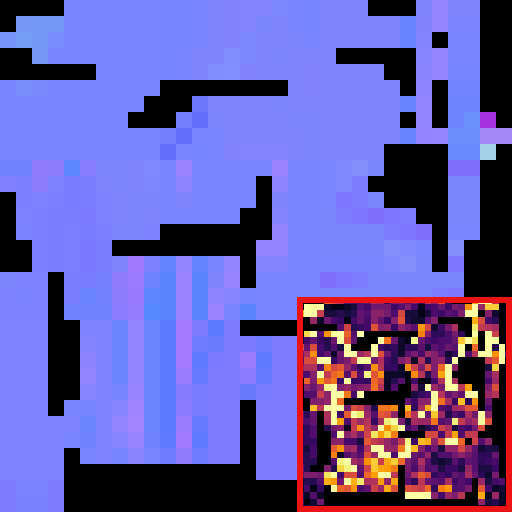} & \mipQualImg{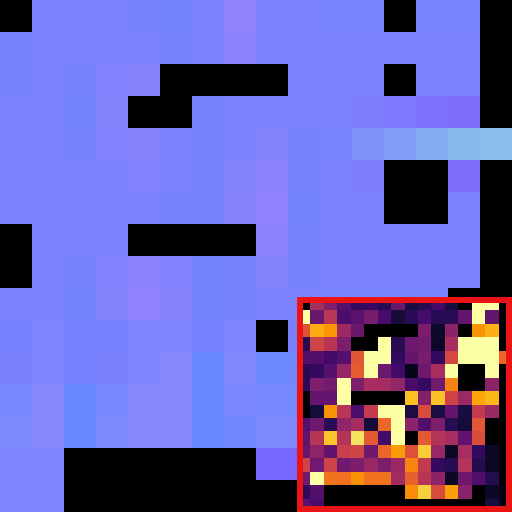} & \mipQualImg{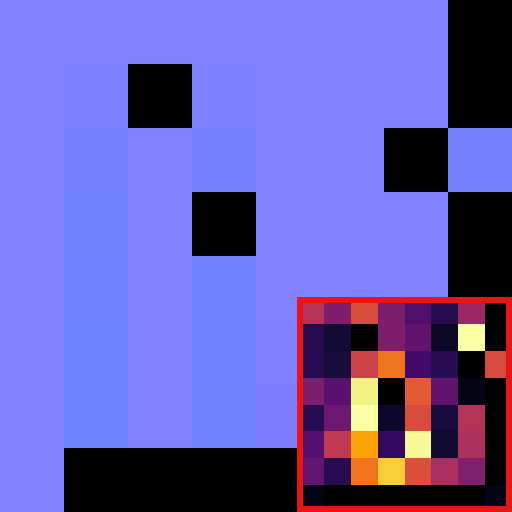} & \mipQualImg{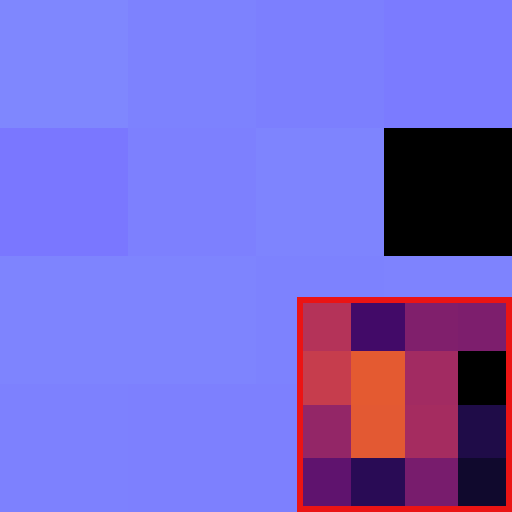} & \mipQualImg{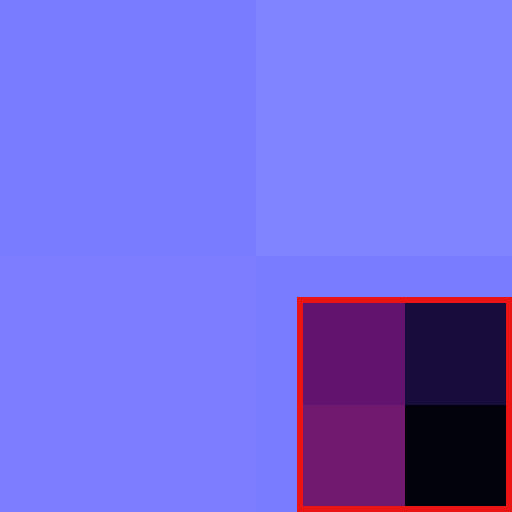} & \mipQualImg{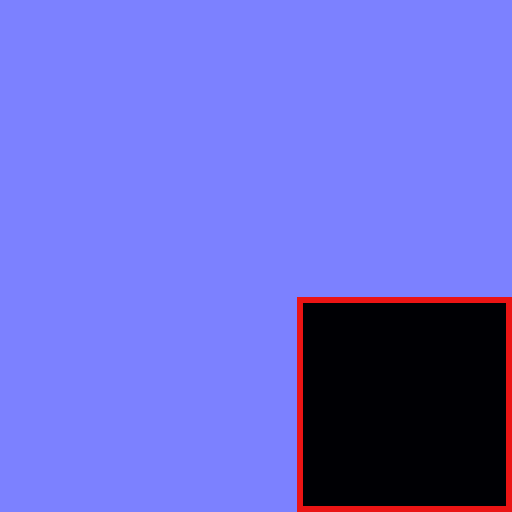} \\[\mipQualRowGap]
      & \mipQualMethod{NTC}{0.226 bppc} & \mipQualImg{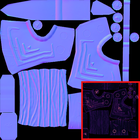} & \mipQualImg{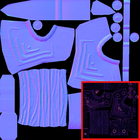} & \mipQualImg{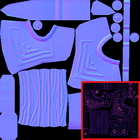} & \mipQualImg{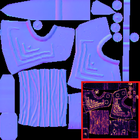} & \mipQualImg{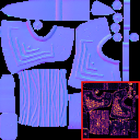} & \mipQualImg{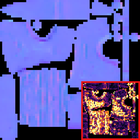} & \mipQualImg{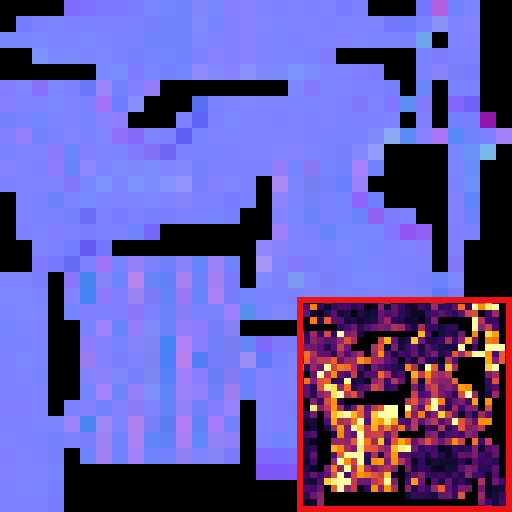} & \mipQualImg{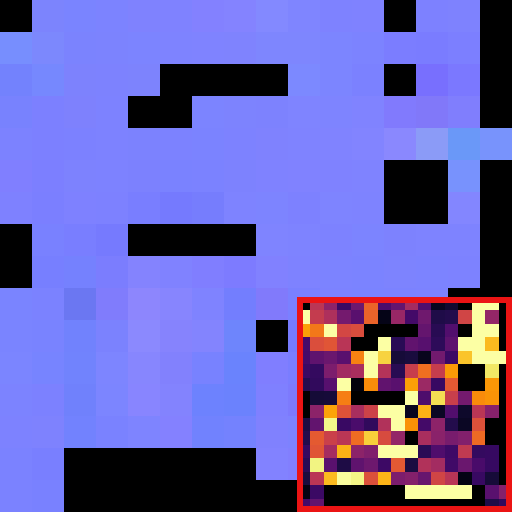} & \mipQualImg{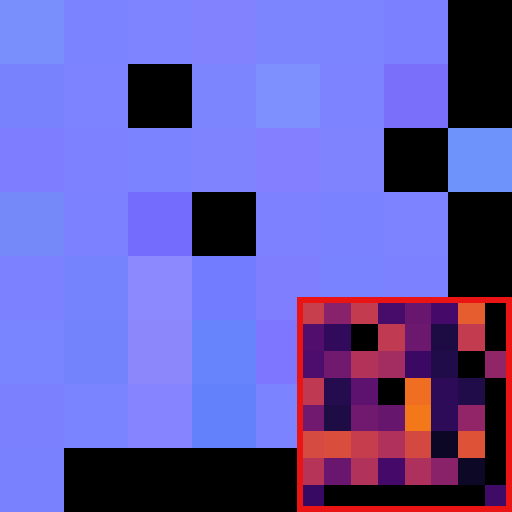} & \mipQualImg{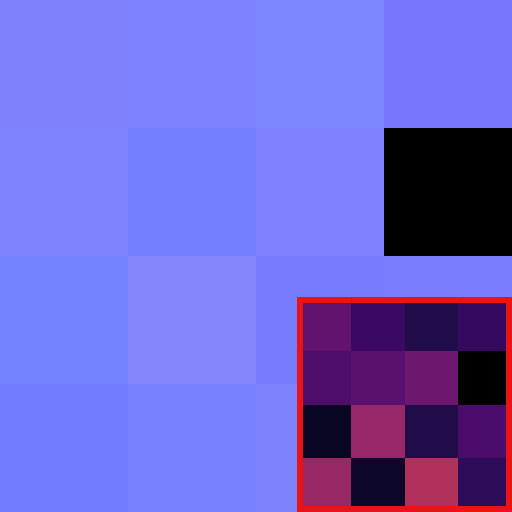} & \mipQualImg{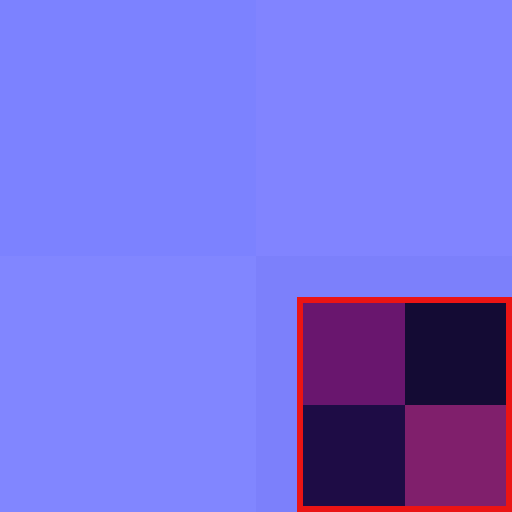} & \mipQualImg{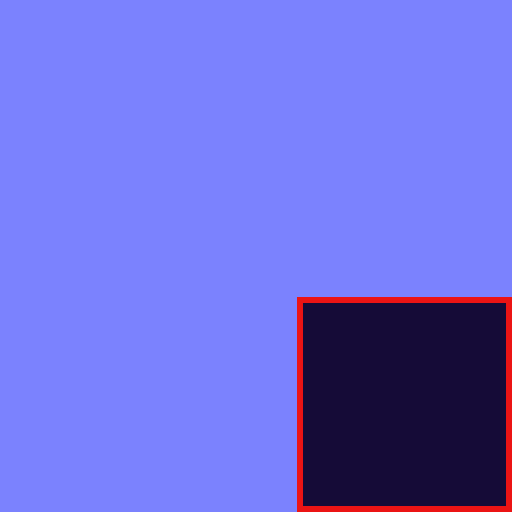} \\
    \mipQualSep
      & \mipQualMethod{GT}{} & \mipQualImg{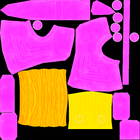} & \mipQualImg{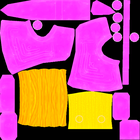} & \mipQualImg{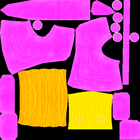} & \mipQualImg{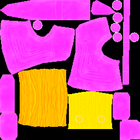} & \mipQualImg{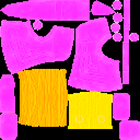} & \mipQualImg{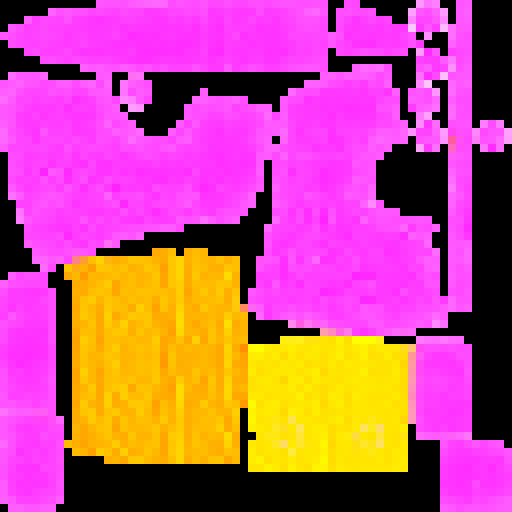} & \mipQualImg{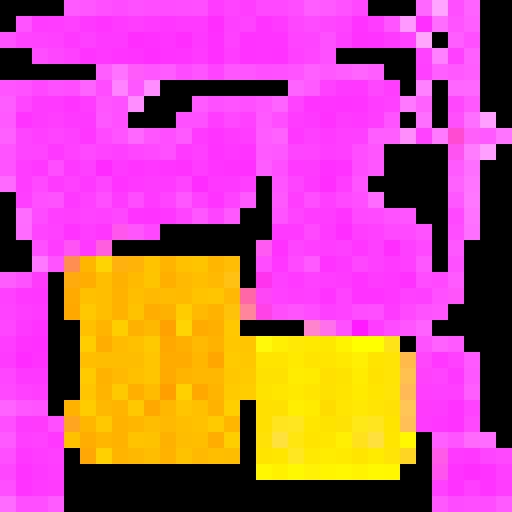} & \mipQualImg{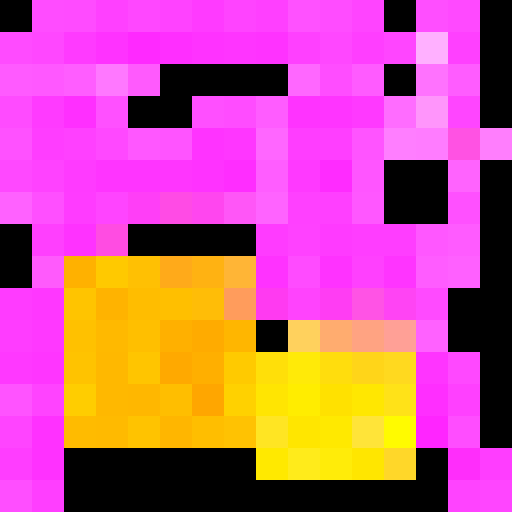} & \mipQualImg{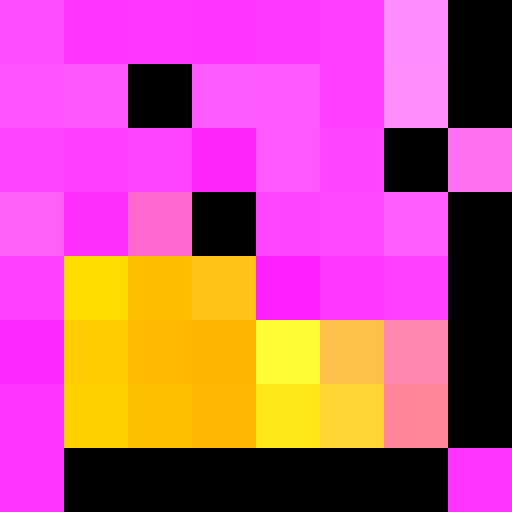} & \mipQualImg{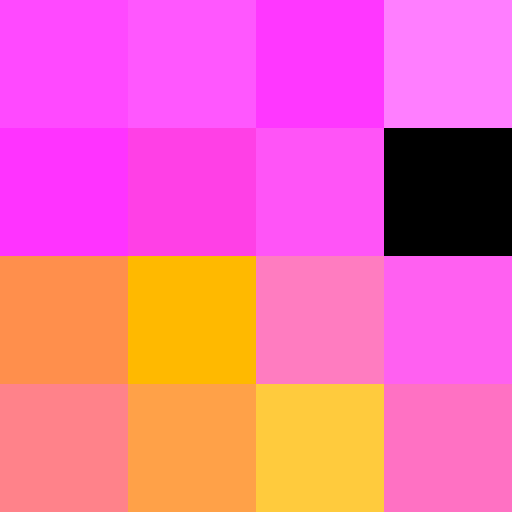} & \mipQualImg{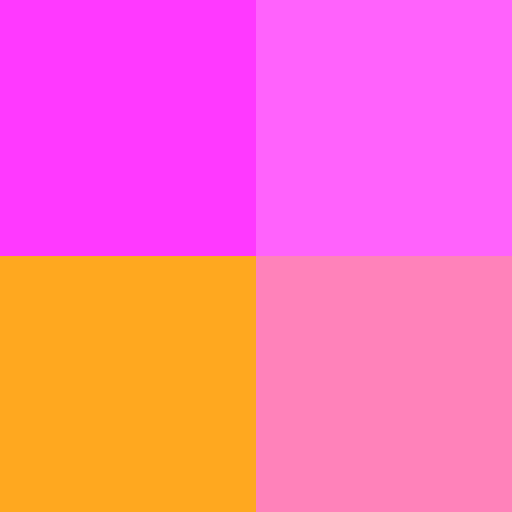} & \mipQualImg{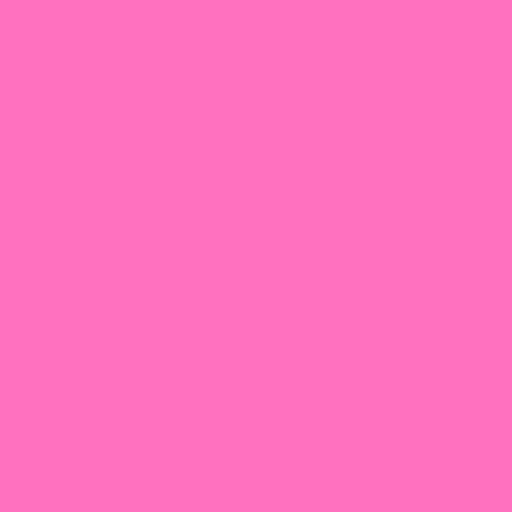} \\[\mipQualRowGap]
      & \mipQualMethod{Ours}{0.195 bppc} & \mipQualImg{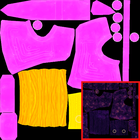} & \mipQualImg{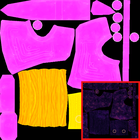} & \mipQualImg{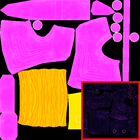} & \mipQualImg{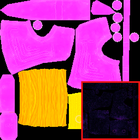} & \mipQualImg{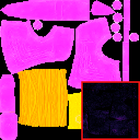} & \mipQualImg{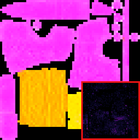} & \mipQualImg{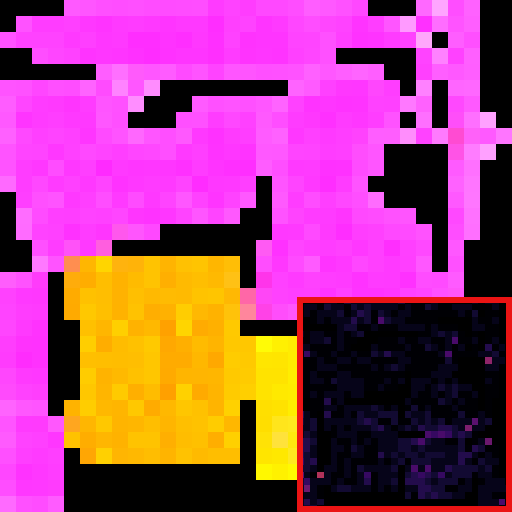} & \mipQualImg{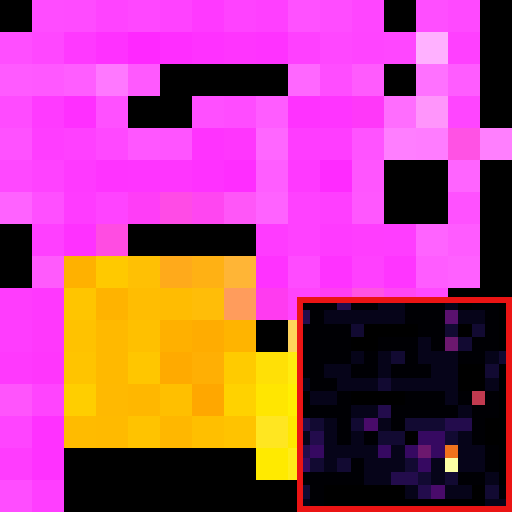} & \mipQualImg{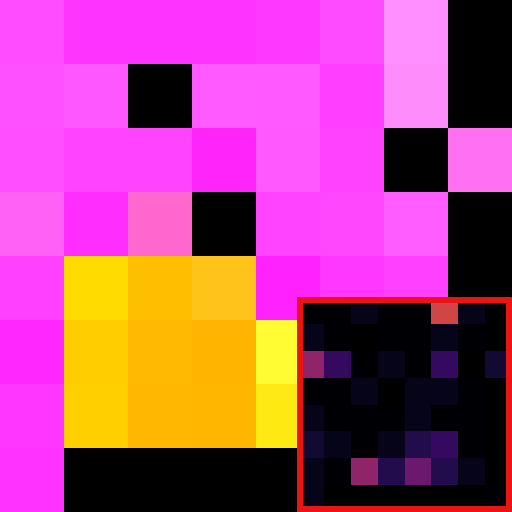} & \mipQualImg{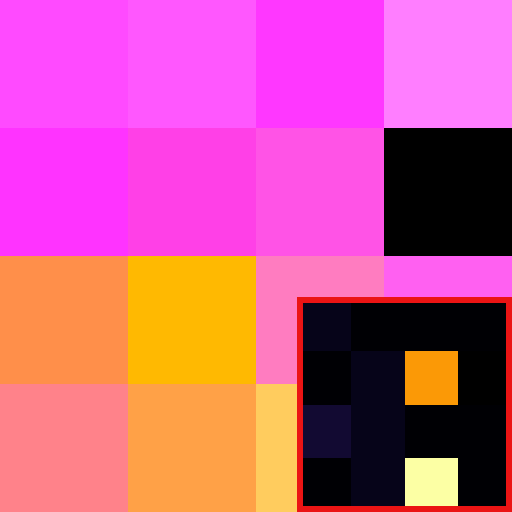} & \mipQualImg{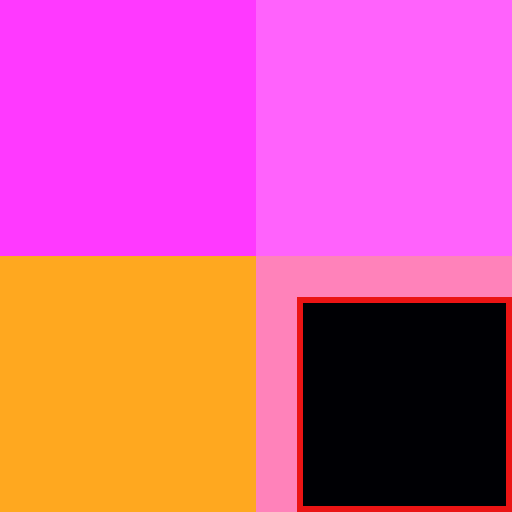} & \mipQualImg{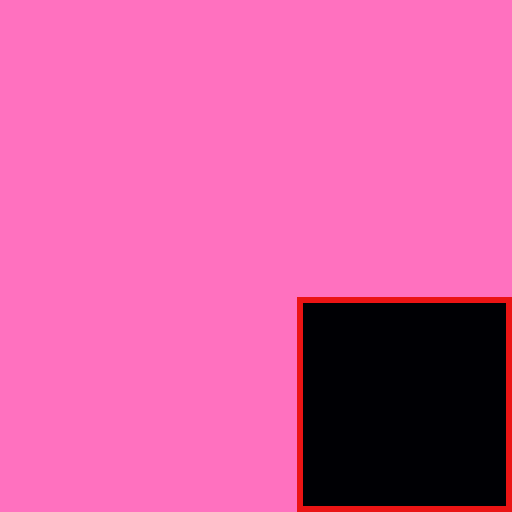} \\[\mipQualRowGap]
    \mipQualStream{ARM} & \mipQualMethod{Image-GS}{0.404 bppc} & \mipQualImg{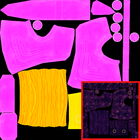} & \mipQualImg{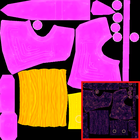} & \mipQualImg{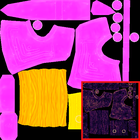} & \mipQualImg{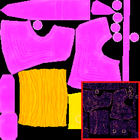} & \mipQualImg{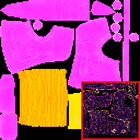} & \mipQualImg{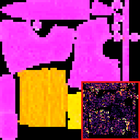} & \mipQualImg{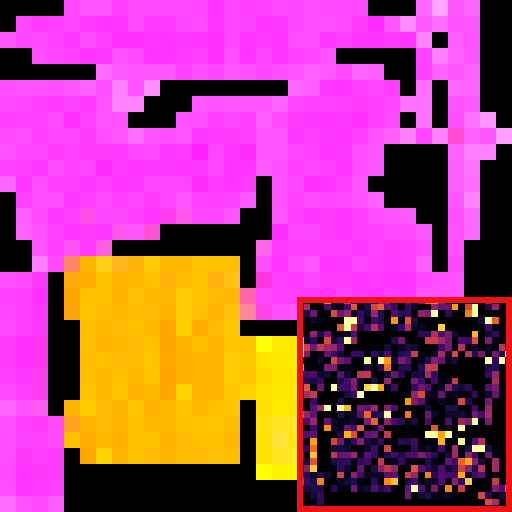} & \mipQualImg{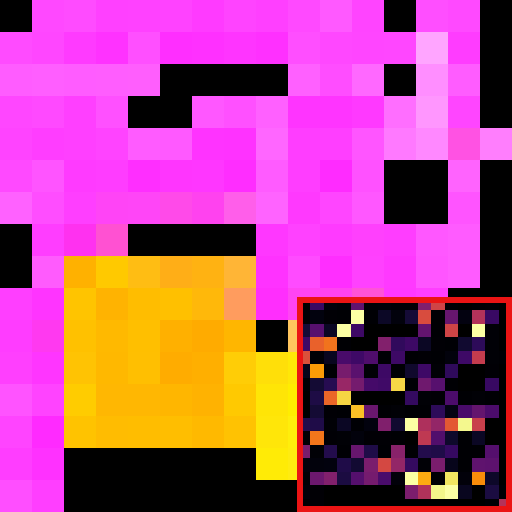} & \mipQualImg{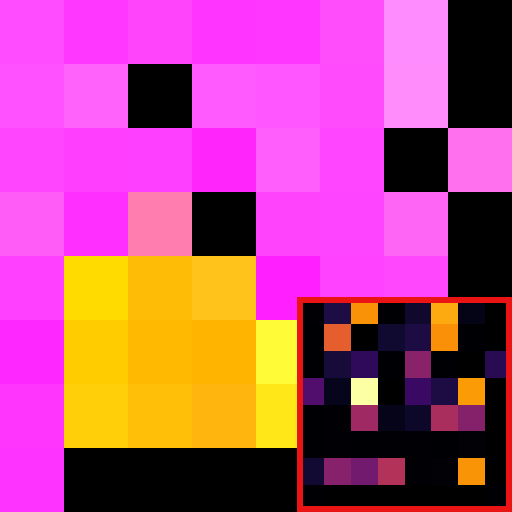} & \mipQualImg{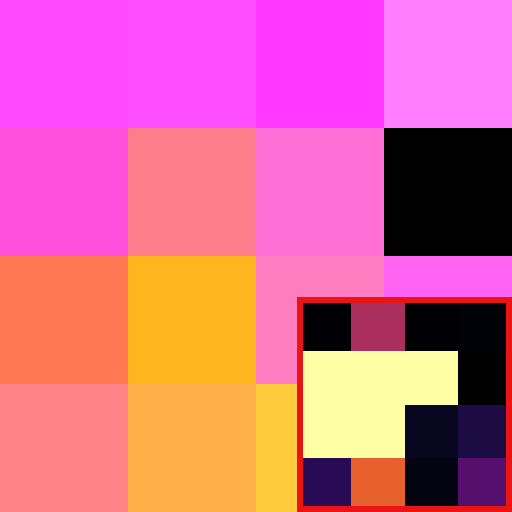} & \mipQualImg{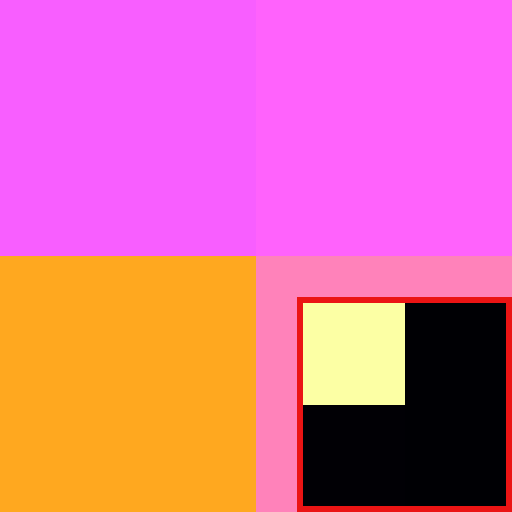} & \mipQualImg{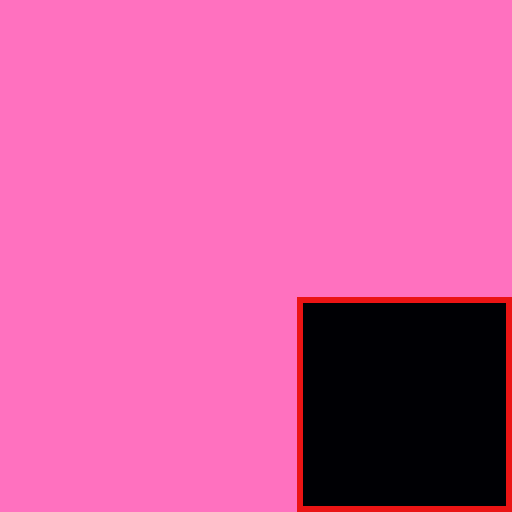} \\[\mipQualRowGap]
      & \mipQualMethod{ASTC}{0.429 bppc} & \mipQualImg{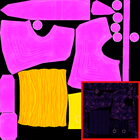} & \mipQualImg{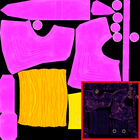} & \mipQualImg{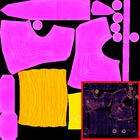} & \mipQualImg{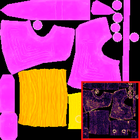} & \mipQualImg{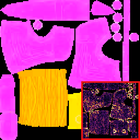} & \mipQualImg{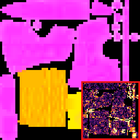} & \mipQualImg{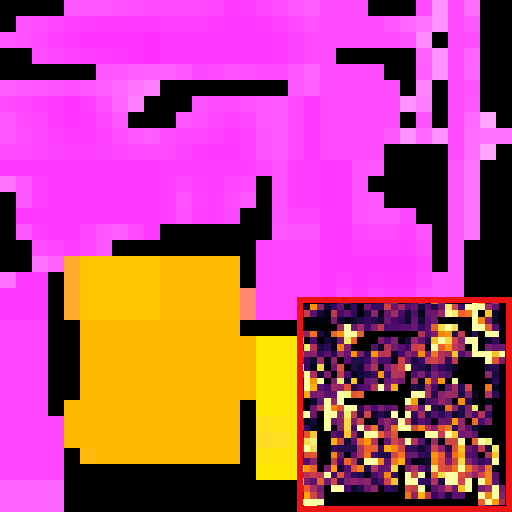} & \mipQualImg{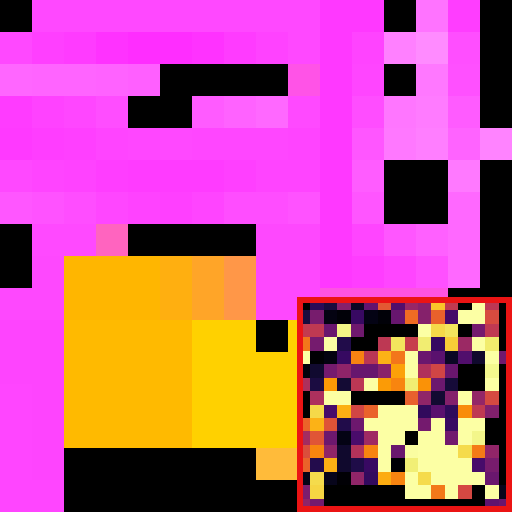} & \mipQualImg{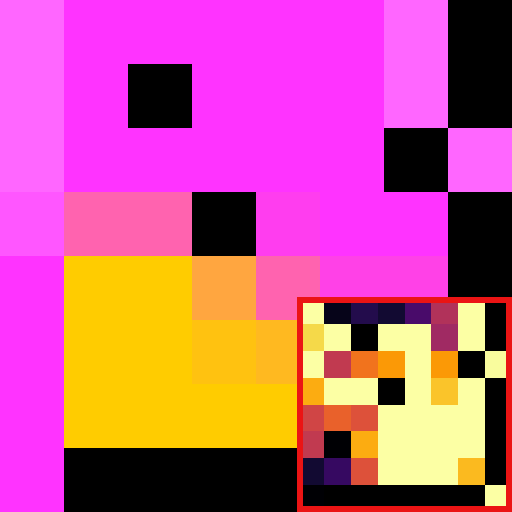} & \mipQualImg{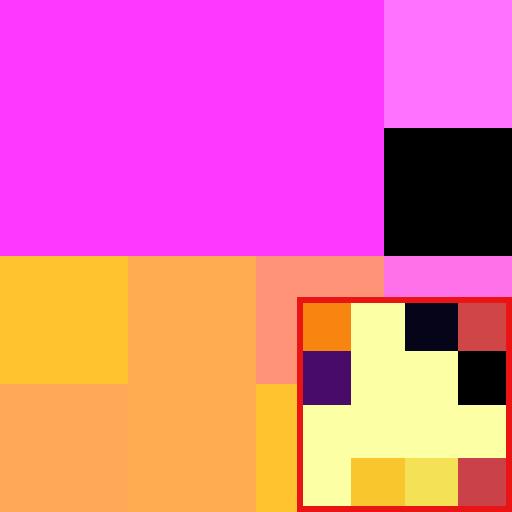} & \mipQualImg{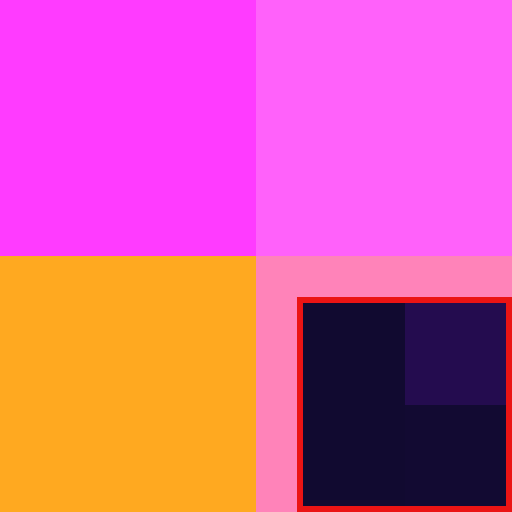} & \mipQualImg{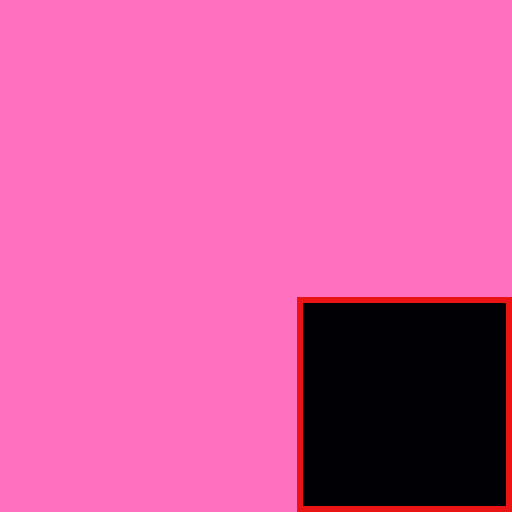} \\[\mipQualRowGap]
      & \mipQualMethod{NTC}{0.226 bppc} & \mipQualImg{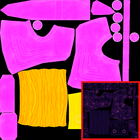} & \mipQualImg{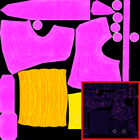} & \mipQualImg{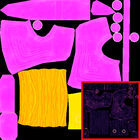} & \mipQualImg{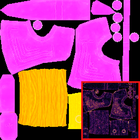} & \mipQualImg{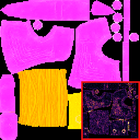} & \mipQualImg{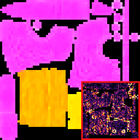} & \mipQualImg{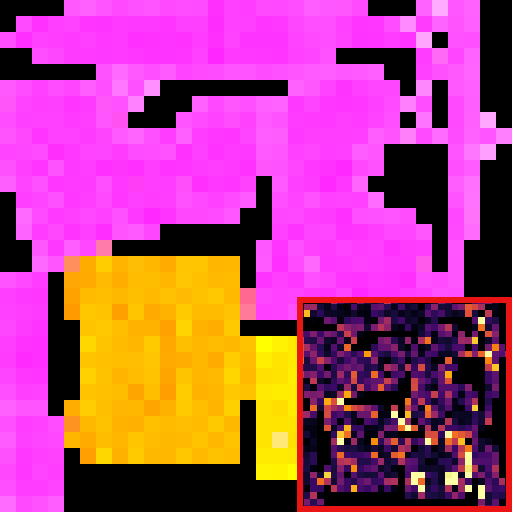} & \mipQualImg{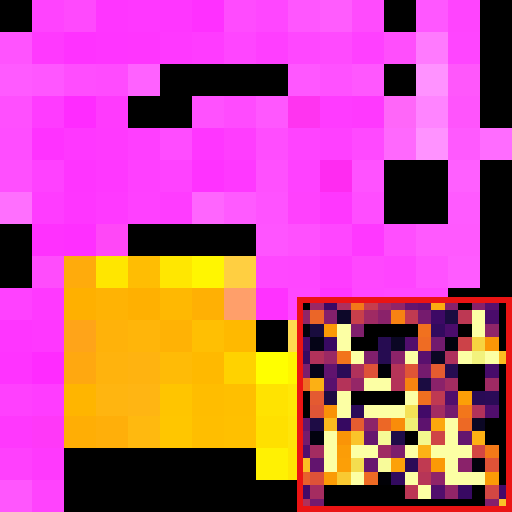} & \mipQualImg{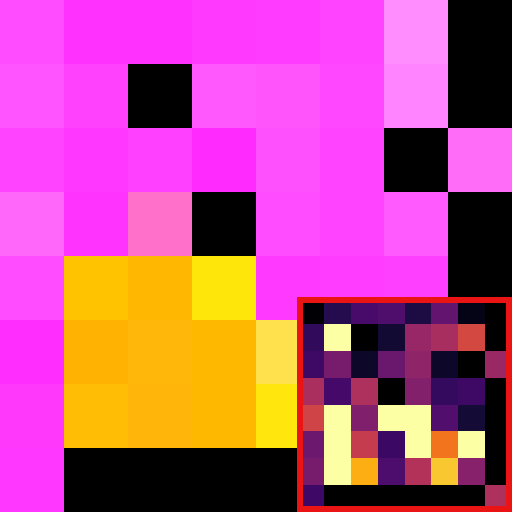} & \mipQualImg{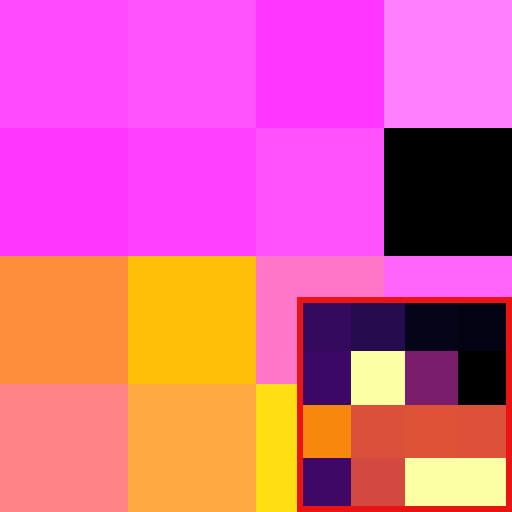} & \mipQualImg{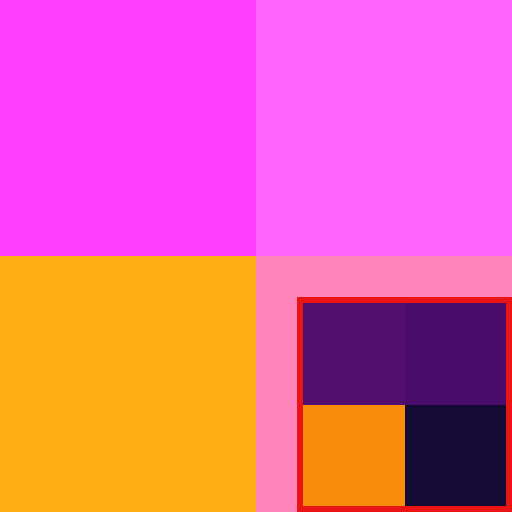} & \mipQualImg{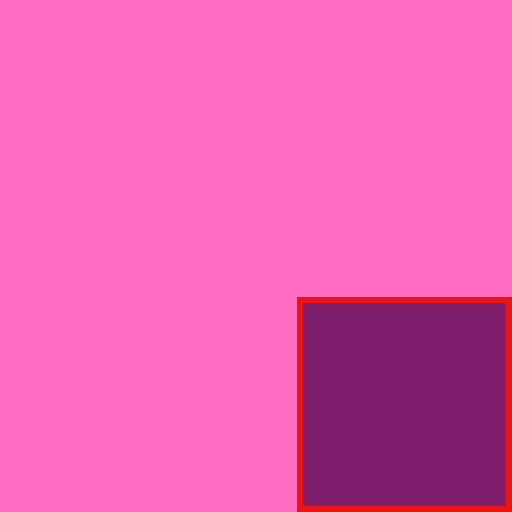} \\

  \end{tabular}%
  }
  \caption{Mipmap reconstruction comparison on the \textit{Hatchet} SVBRDF stack. Rows show the reference and reconstructed mipmap pyramids, with insets showing per-pixel error maps against the reference. ARM denotes ambient occlusion, roughness, and metallic channels.}
  \label{fig:mipmap-error-inset-hatchet}
\end{figure*}

\clearpage
\begin{figure*}[p!]
  \centering
  \mipQualFontSize
  \setlength{\tabcolsep}{0pt}
  \renewcommand{\arraystretch}{1.0}
  \resizebox{\mipQualTableTargetW}{!}{%
  \begin{tabular}{@{}c@{\hspace{\mipQualStreamMethodGap}}c@{\hspace{\mipQualMethodImageGap}}*{11}{c@{\hspace{\mipQualColGap}}}c@{}}
    \multicolumn{2}{c}{} & \mipQualMip{mip 0}{2048$^2$} & \mipQualMip{mip 1}{1024$^2$} & \mipQualMip{mip 2}{512$^2$} & \mipQualMip{mip 3}{256$^2$} & \mipQualMip{mip 4}{128$^2$} & \mipQualMip{mip 5}{64$^2$} & \mipQualMip{mip 6}{32$^2$} & \mipQualMip{mip 7}{16$^2$} & \mipQualMip{mip 8}{8$^2$} & \mipQualMip{mip 9}{4$^2$} & \mipQualMip{mip 10}{2$^2$} & \mipQualMip{mip 11}{1$^2$} \\[\mipQualHeaderGap]
      & \mipQualMethod{GT}{} & \mipQualImg{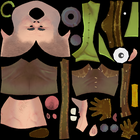} & \mipQualImg{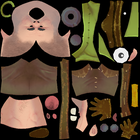} & \mipQualImg{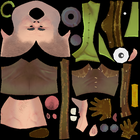} & \mipQualImg{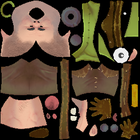} & \mipQualImg{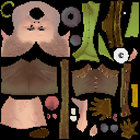} & \mipQualImg{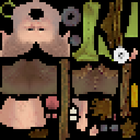} & \mipQualImg{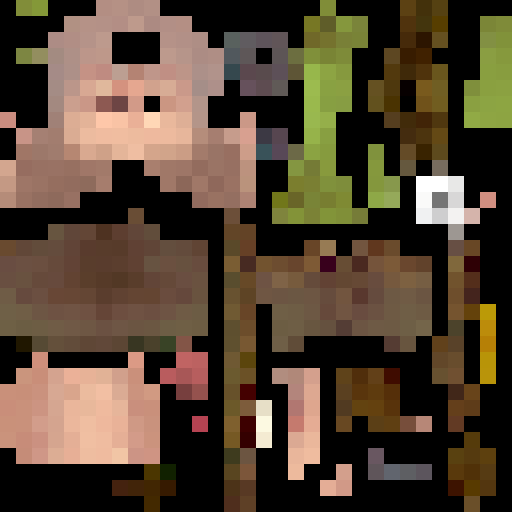} & \mipQualImg{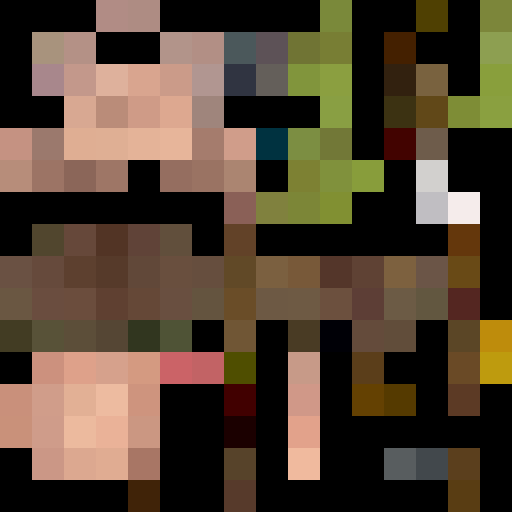} & \mipQualImg{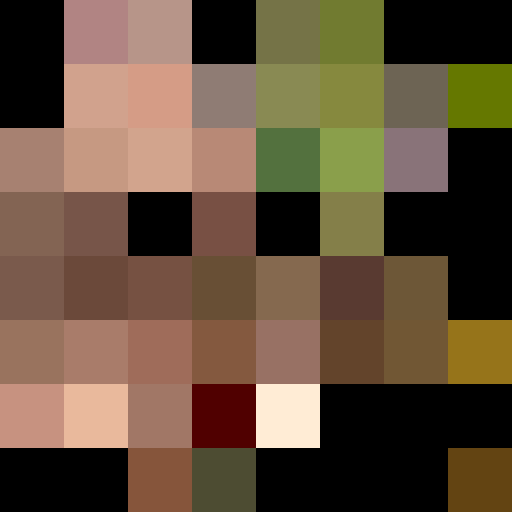} & \mipQualImg{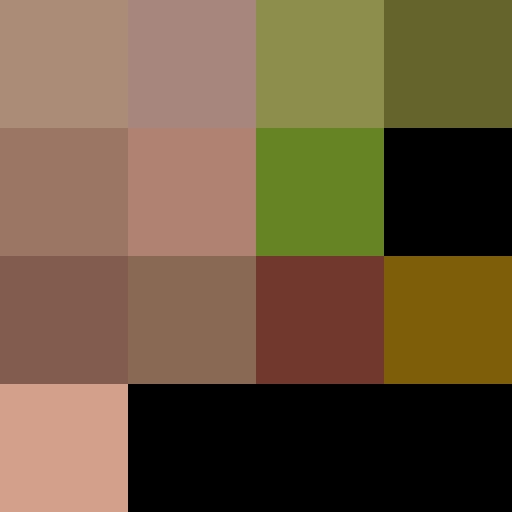} & \mipQualImg{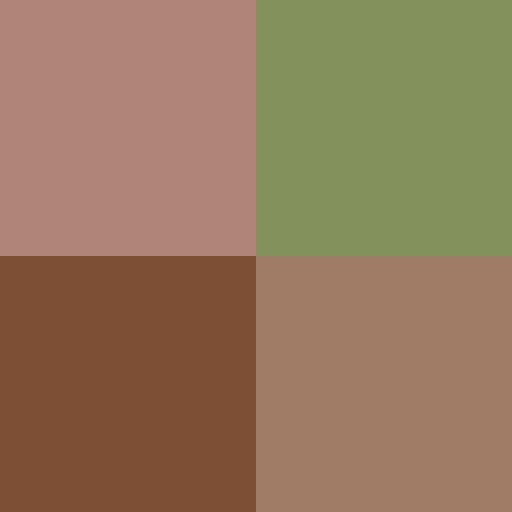} & \mipQualImg{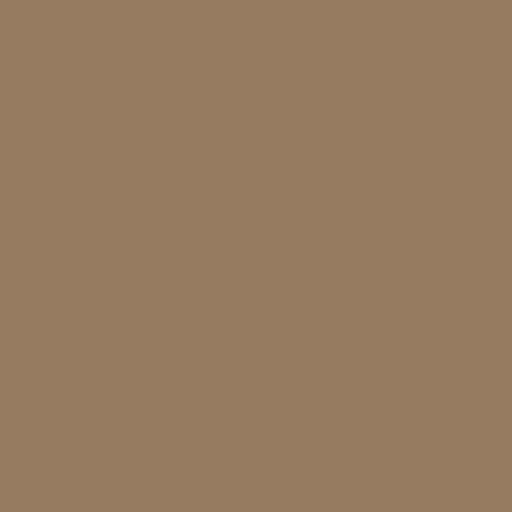} \\[\mipQualRowGap]
      & \mipQualMethod{Ours}{0.097 bppc} & \mipQualImg{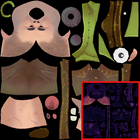} & \mipQualImg{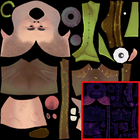} & \mipQualImg{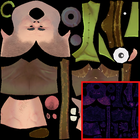} & \mipQualImg{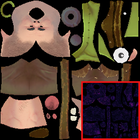} & \mipQualImg{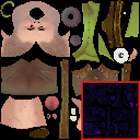} & \mipQualImg{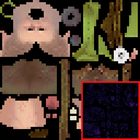} & \mipQualImg{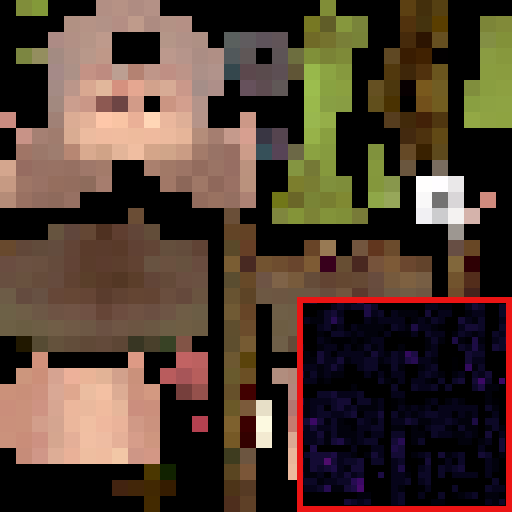} & \mipQualImg{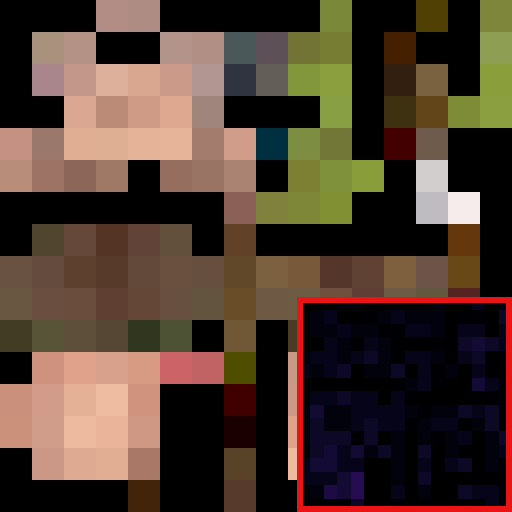} & \mipQualImg{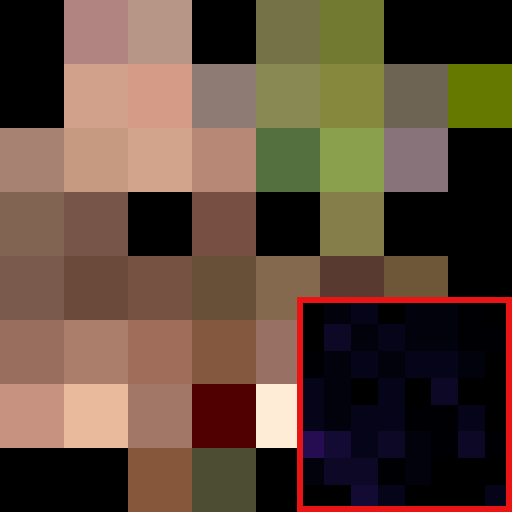} & \mipQualImg{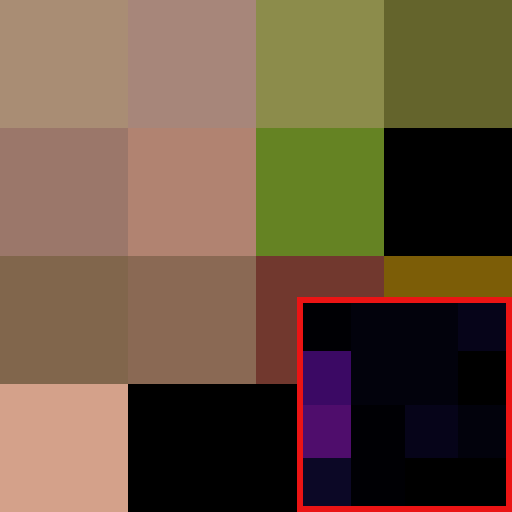} & \mipQualImg{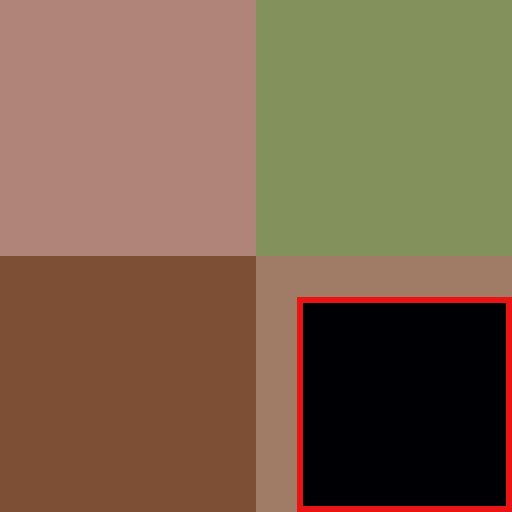} & \mipQualImg{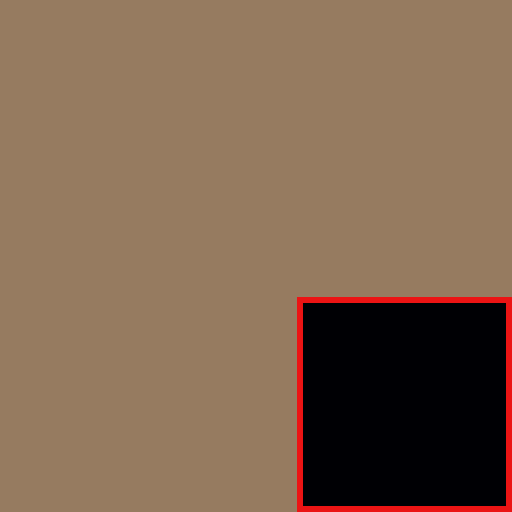} \\[\mipQualRowGap]
    \mipQualStream{Base color} & \mipQualMethod{Image-GS}{0.404 bppc} & \mipQualImg{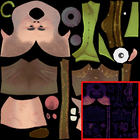} & \mipQualImg{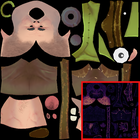} & \mipQualImg{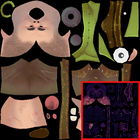} & \mipQualImg{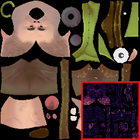} & \mipQualImg{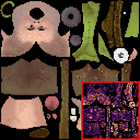} & \mipQualImg{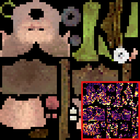} & \mipQualImg{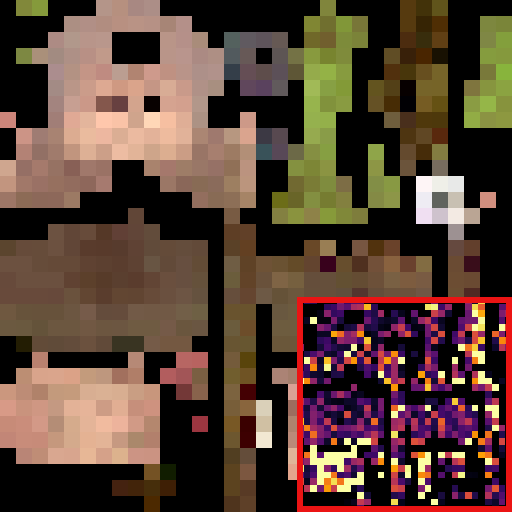} & \mipQualImg{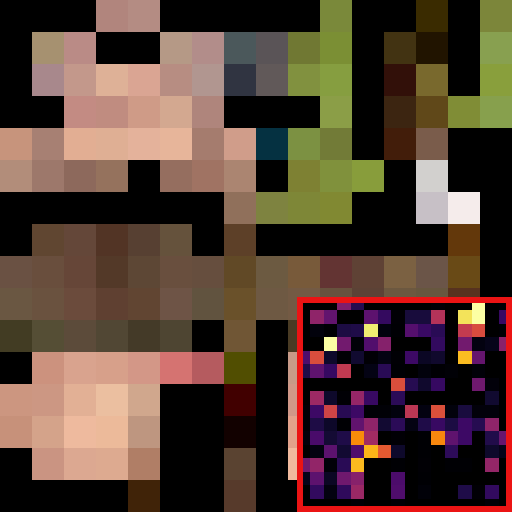} & \mipQualImg{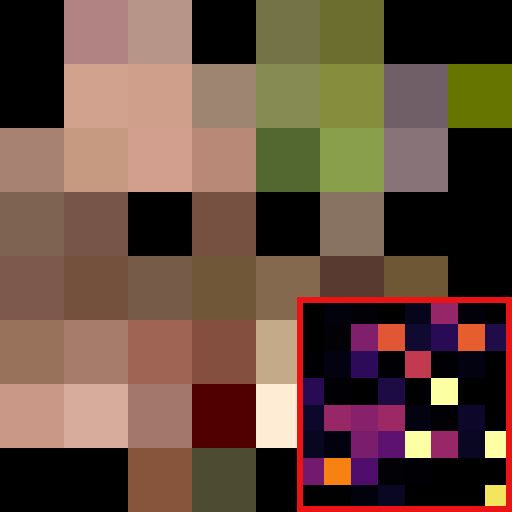} & \mipQualImg{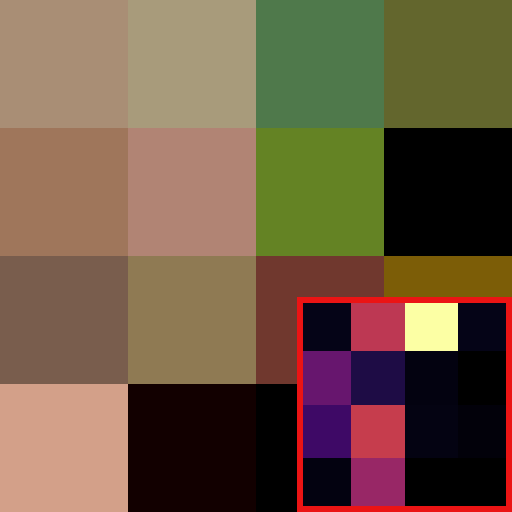} & \mipQualImg{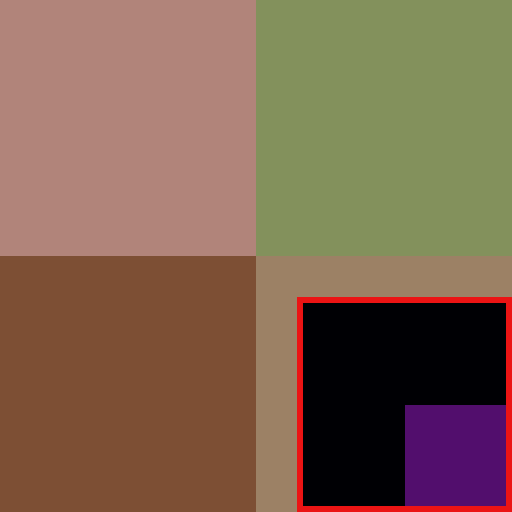} & \mipQualImg{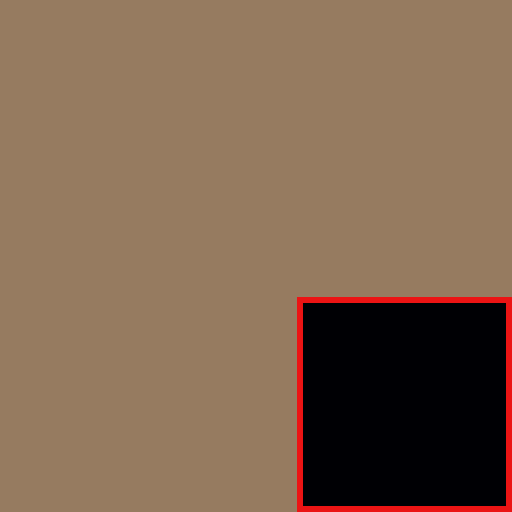} \\[\mipQualRowGap]
      & \mipQualMethod{ASTC}{0.429 bppc} & \mipQualImg{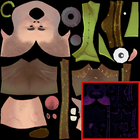} & \mipQualImg{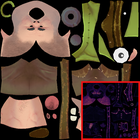} & \mipQualImg{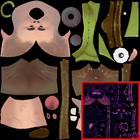} & \mipQualImg{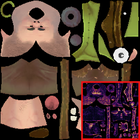} & \mipQualImg{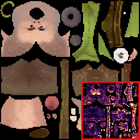} & \mipQualImg{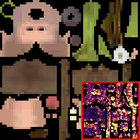} & \mipQualImg{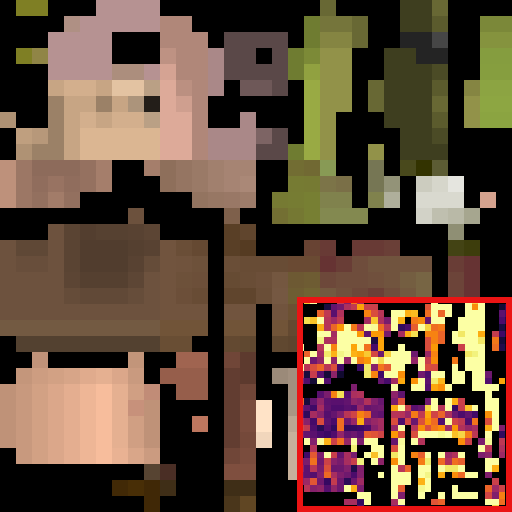} & \mipQualImg{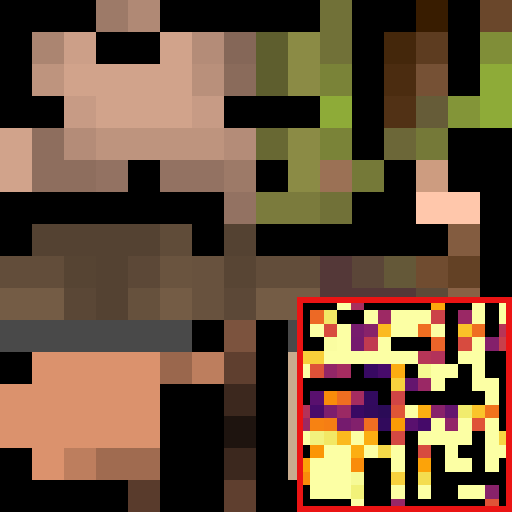} & \mipQualImg{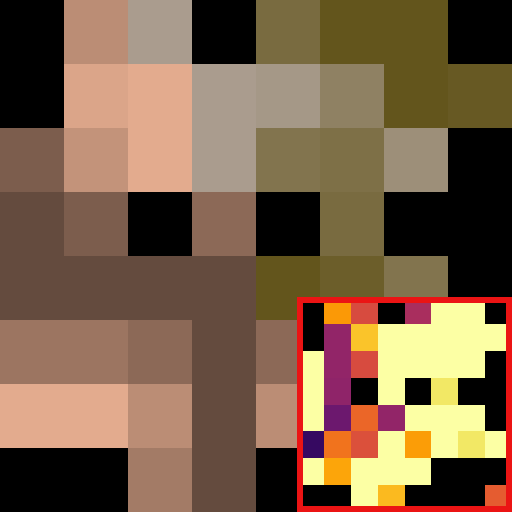} & \mipQualImg{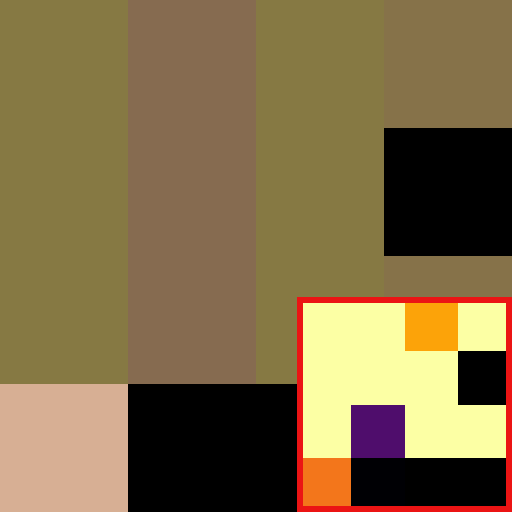} & \mipQualImg{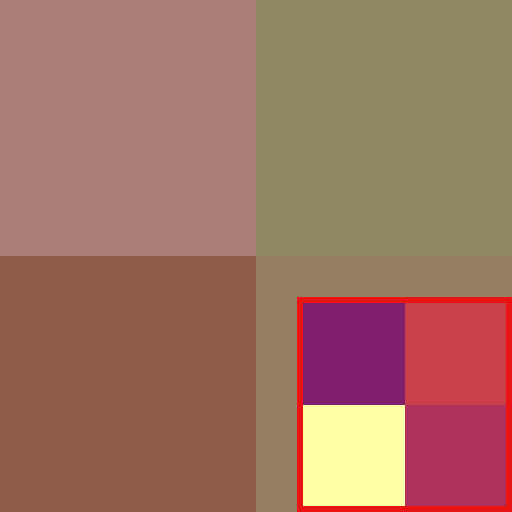} & \mipQualImg{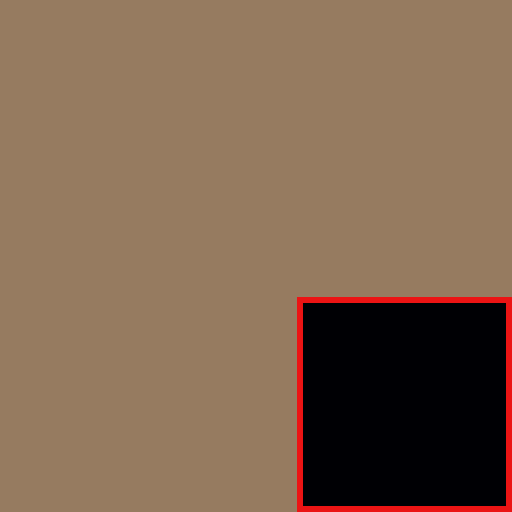} \\[\mipQualRowGap]
      & \mipQualMethod{NTC}{0.102 bppc} & \mipQualImg{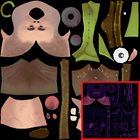} & \mipQualImg{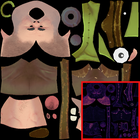} & \mipQualImg{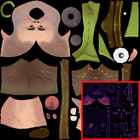} & \mipQualImg{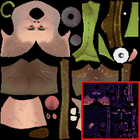} & \mipQualImg{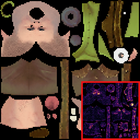} & \mipQualImg{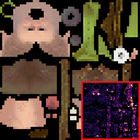} & \mipQualImg{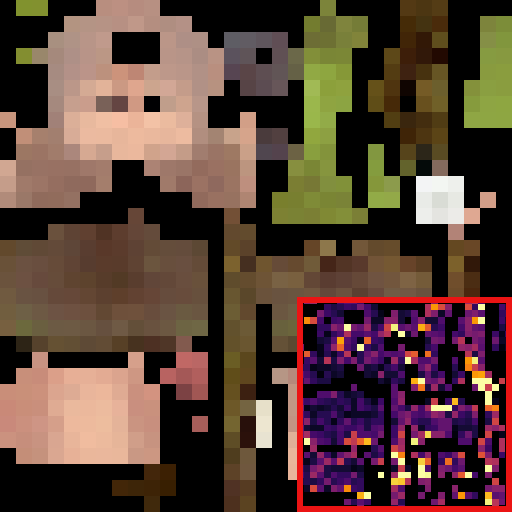} & \mipQualImg{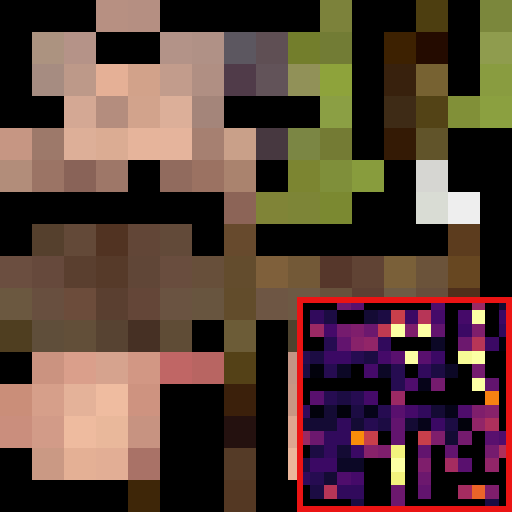} & \mipQualImg{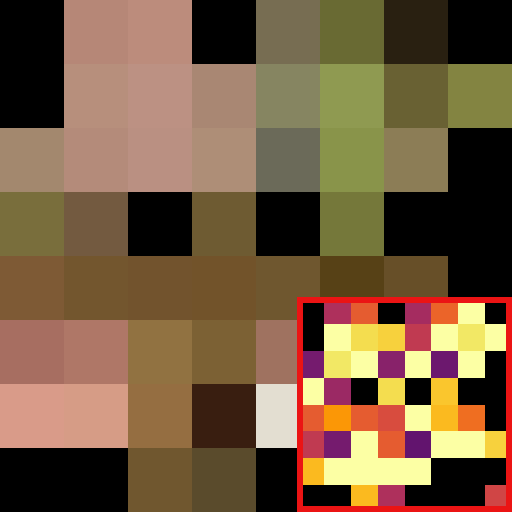} & \mipQualImg{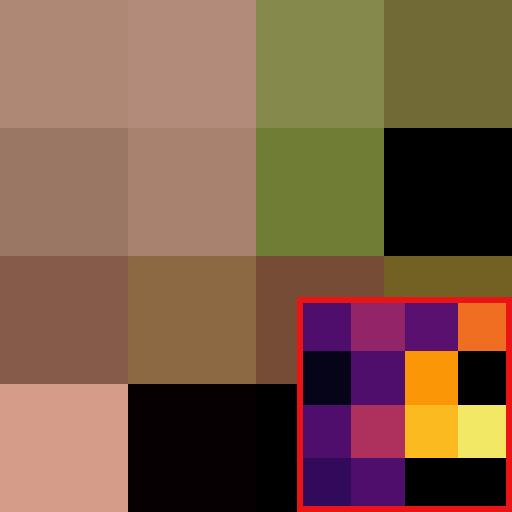} & \mipQualImg{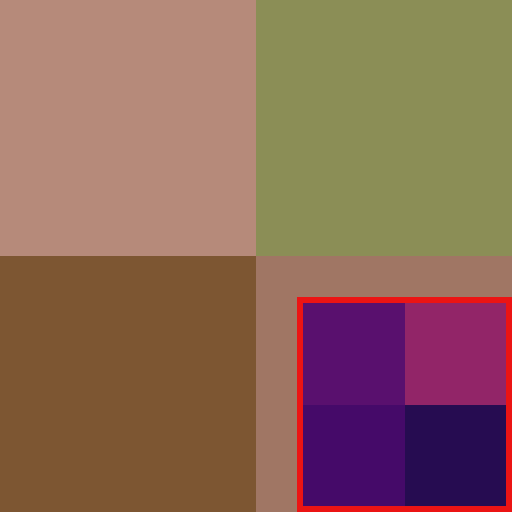} & \mipQualImg{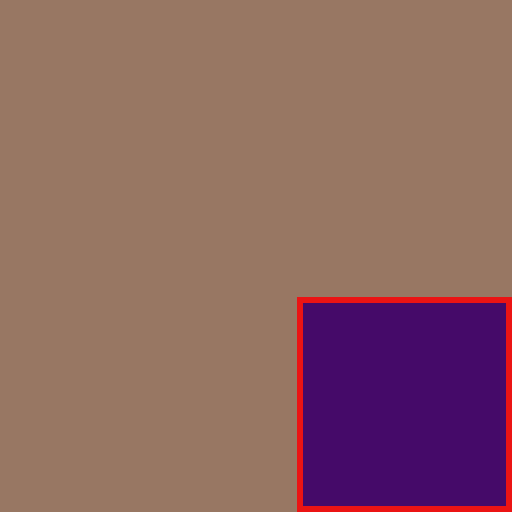} \\
    \mipQualSep
      & \mipQualMethod{GT}{} & \mipQualImg{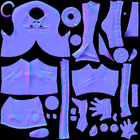} & \mipQualImg{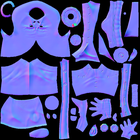} & \mipQualImg{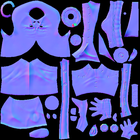} & \mipQualImg{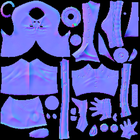} & \mipQualImg{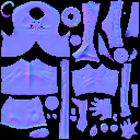} & \mipQualImg{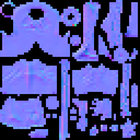} & \mipQualImg{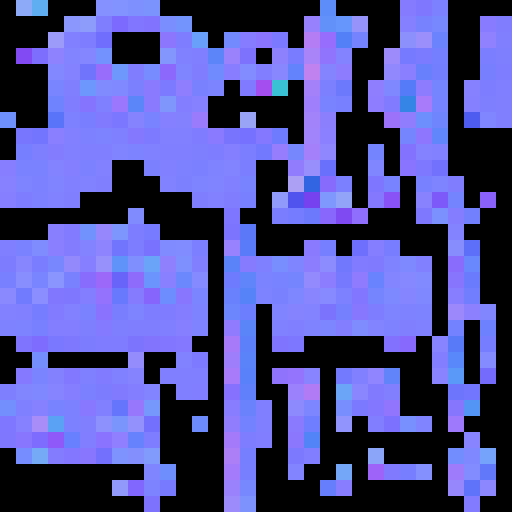} & \mipQualImg{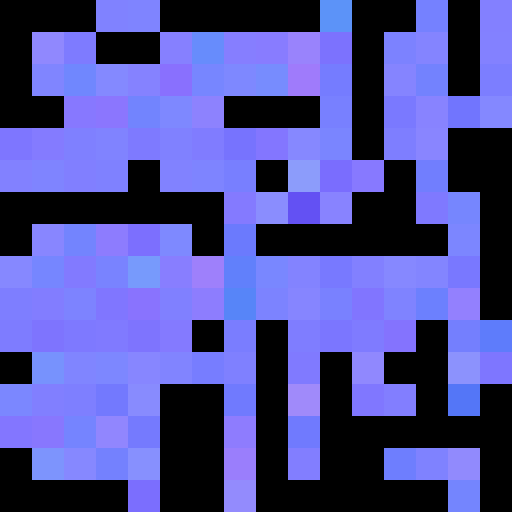} & \mipQualImg{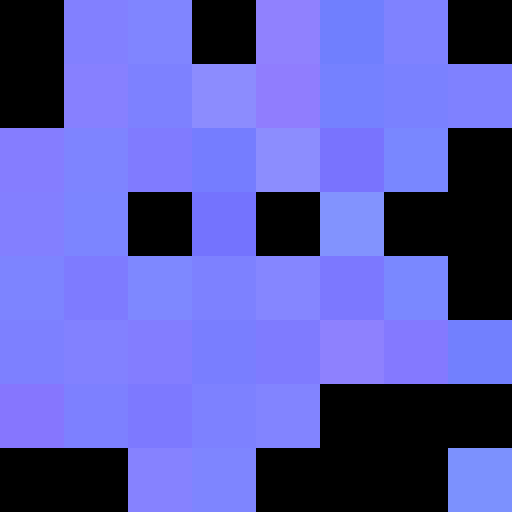} & \mipQualImg{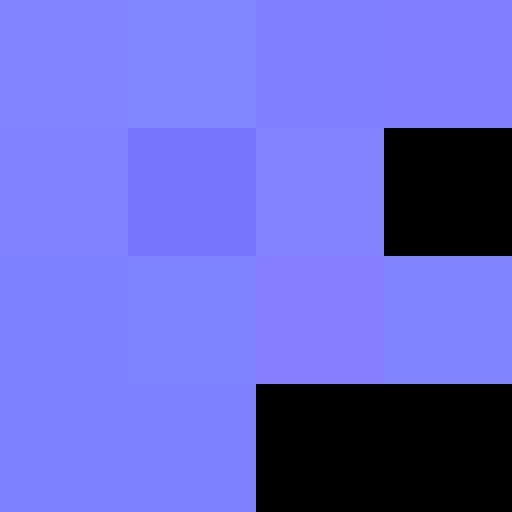} & \mipQualImg{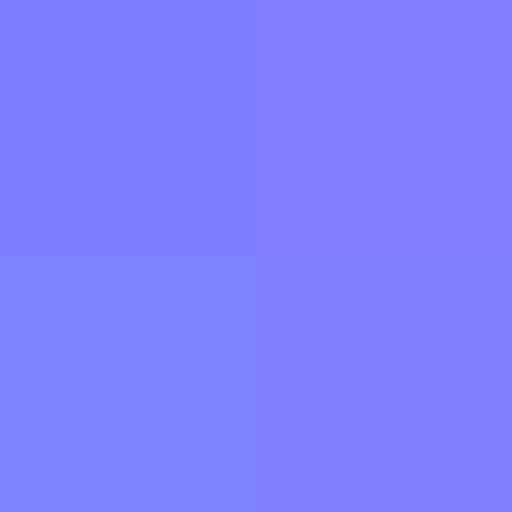} & \mipQualImg{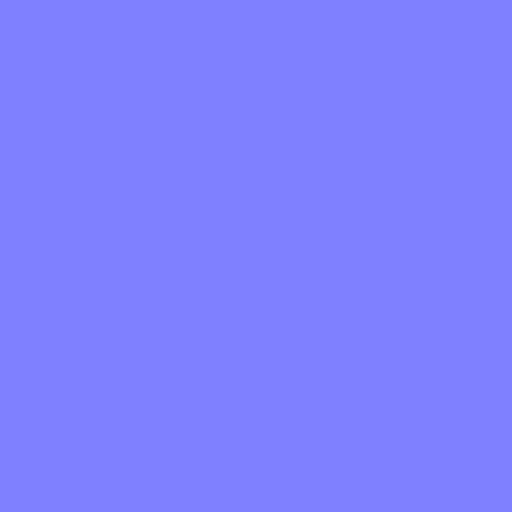} \\[\mipQualRowGap]
      & \mipQualMethod{Ours}{0.097 bppc} & \mipQualImg{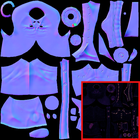} & \mipQualImg{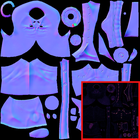} & \mipQualImg{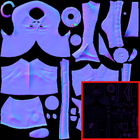} & \mipQualImg{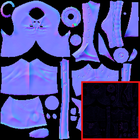} & \mipQualImg{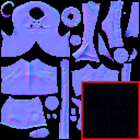} & \mipQualImg{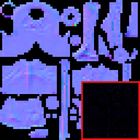} & \mipQualImg{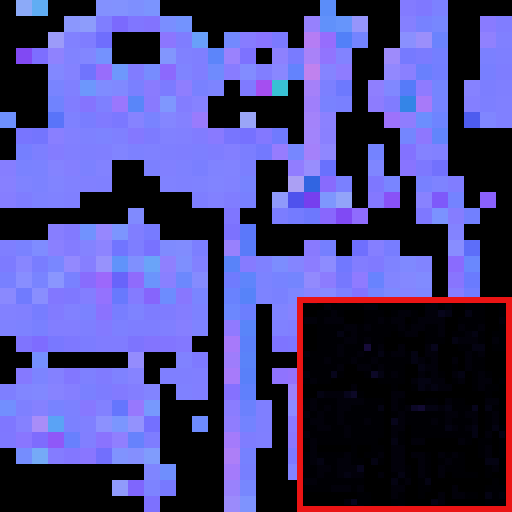} & \mipQualImg{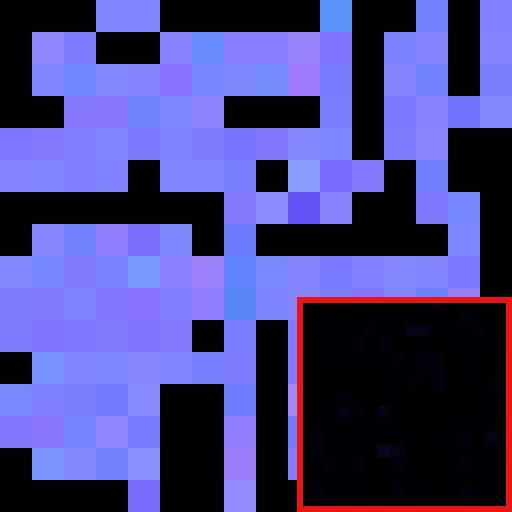} & \mipQualImg{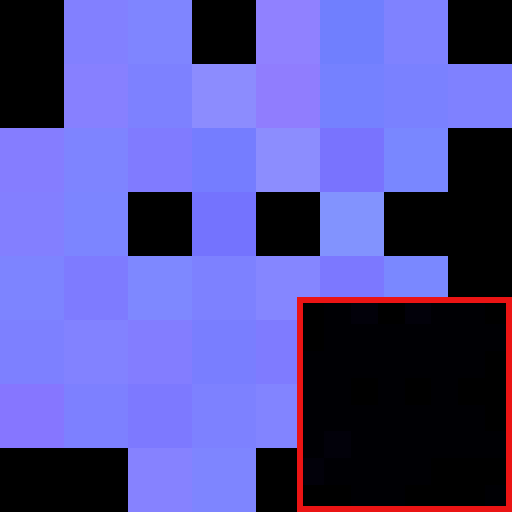} & \mipQualImg{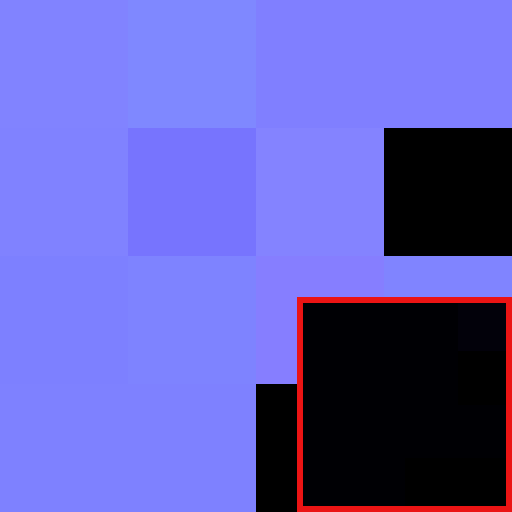} & \mipQualImg{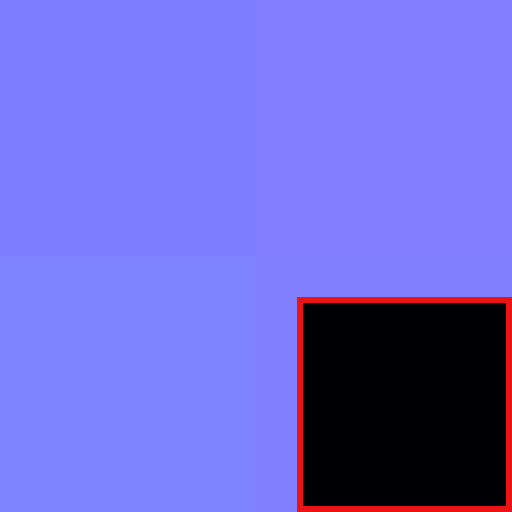} & \mipQualImg{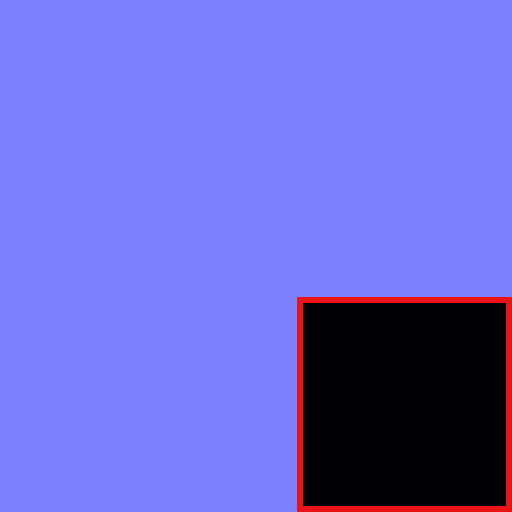} \\[\mipQualRowGap]
    \mipQualStream{Normal map} & \mipQualMethod{Image-GS}{0.404 bppc} & \mipQualImg{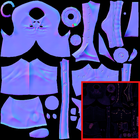} & \mipQualImg{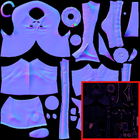} & \mipQualImg{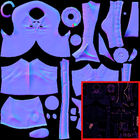} & \mipQualImg{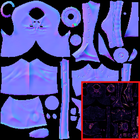} & \mipQualImg{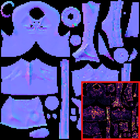} & \mipQualImg{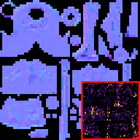} & \mipQualImg{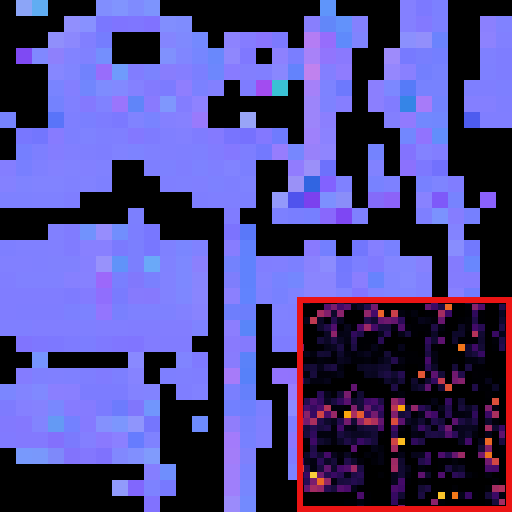} & \mipQualImg{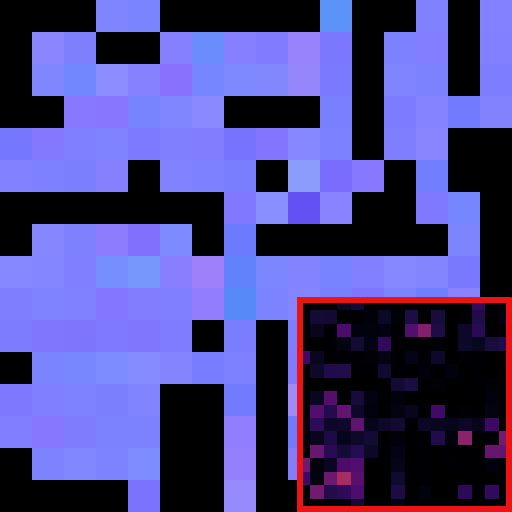} & \mipQualImg{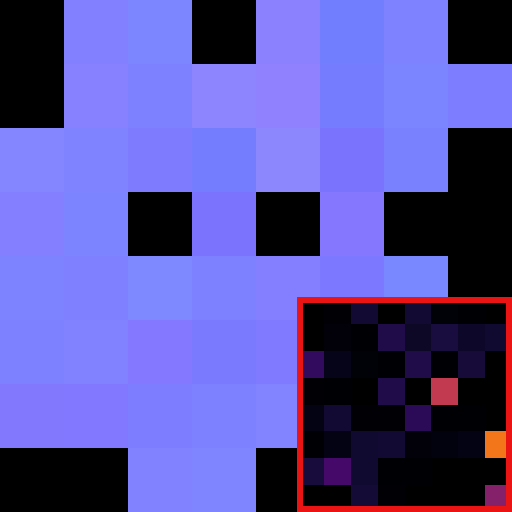} & \mipQualImg{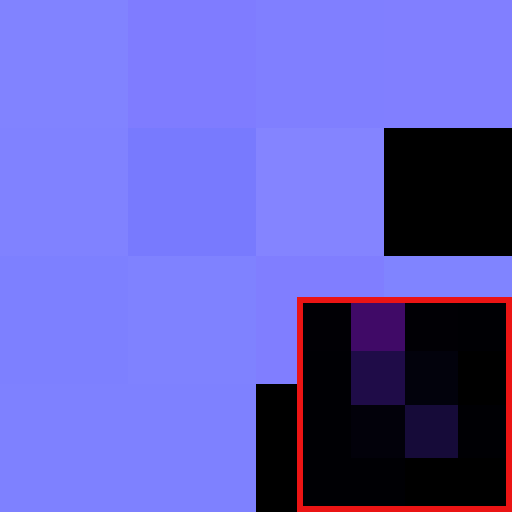} & \mipQualImg{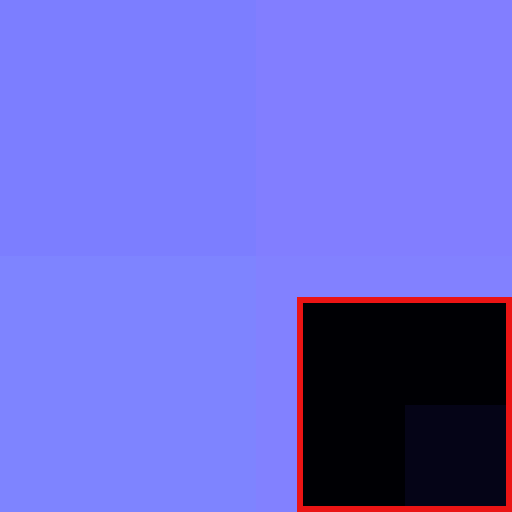} & \mipQualImg{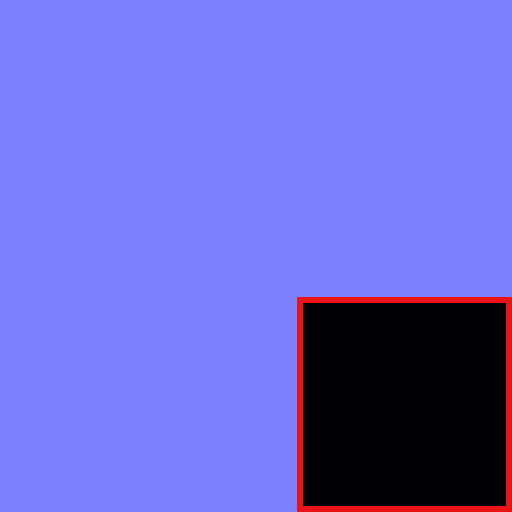} \\[\mipQualRowGap]
      & \mipQualMethod{ASTC}{0.429 bppc} & \mipQualImg{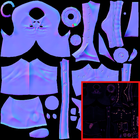} & \mipQualImg{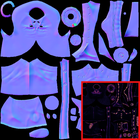} & \mipQualImg{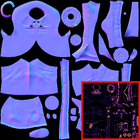} & \mipQualImg{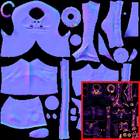} & \mipQualImg{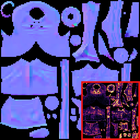} & \mipQualImg{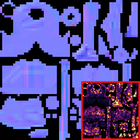} & \mipQualImg{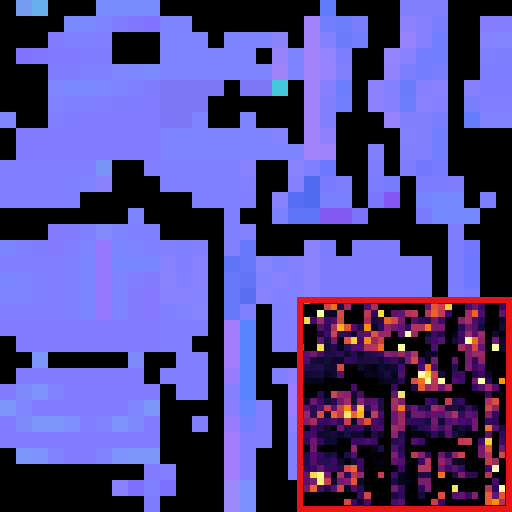} & \mipQualImg{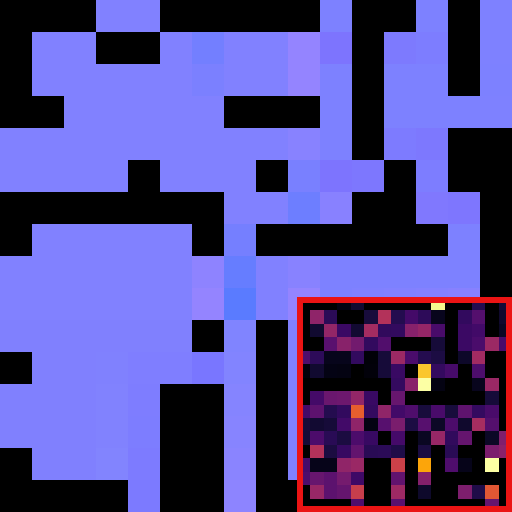} & \mipQualImg{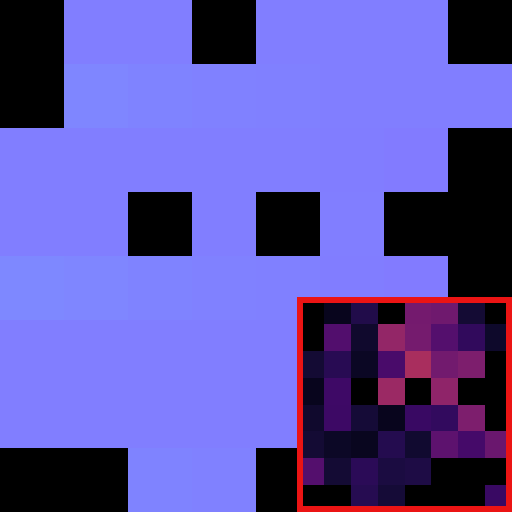} & \mipQualImg{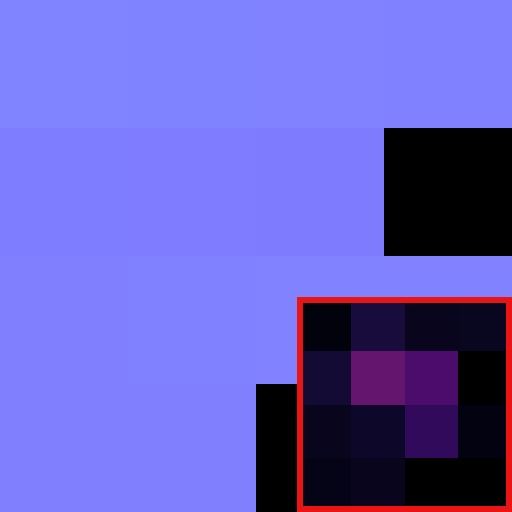} & \mipQualImg{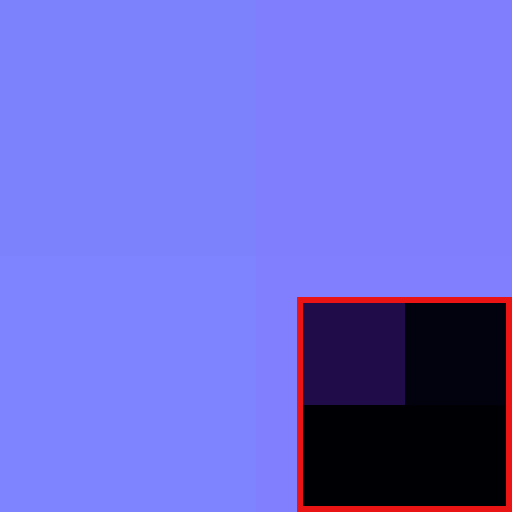} & \mipQualImg{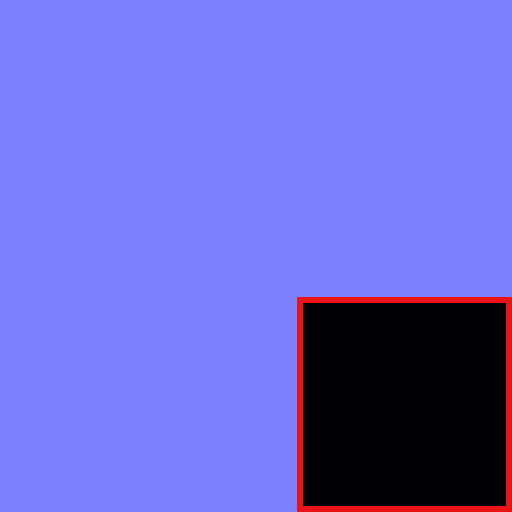} \\[\mipQualRowGap]
      & \mipQualMethod{NTC}{0.102 bppc} & \mipQualImg{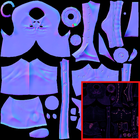} & \mipQualImg{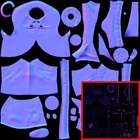} & \mipQualImg{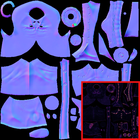} & \mipQualImg{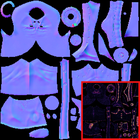} & \mipQualImg{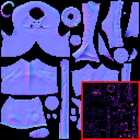} & \mipQualImg{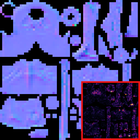} & \mipQualImg{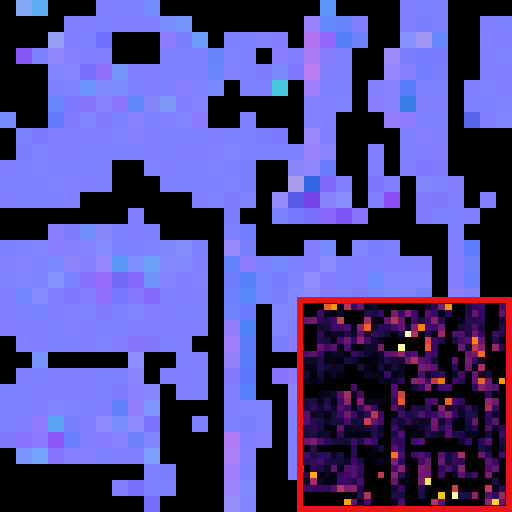} & \mipQualImg{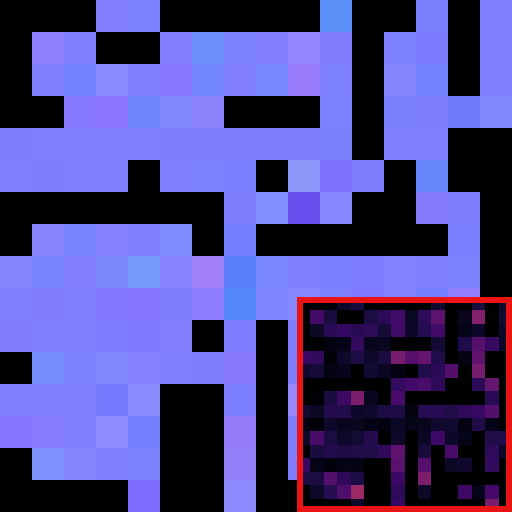} & \mipQualImg{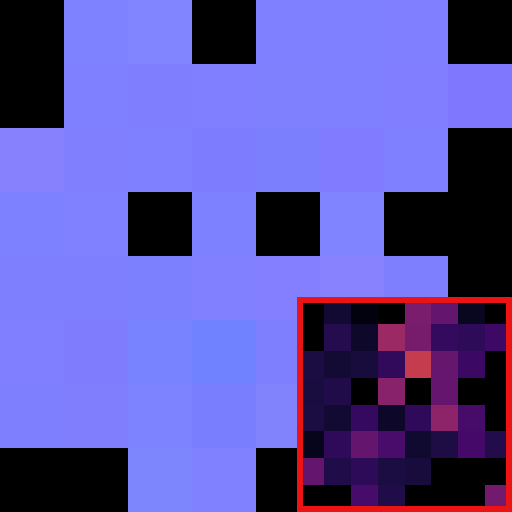} & \mipQualImg{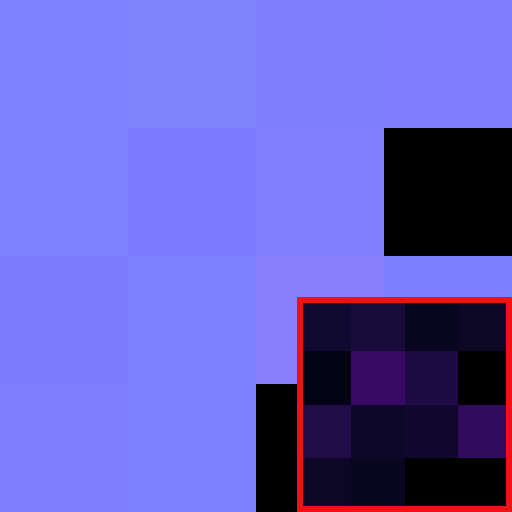} & \mipQualImg{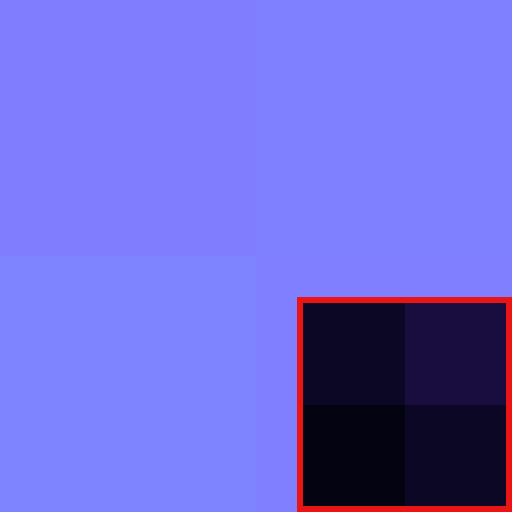} & \mipQualImg{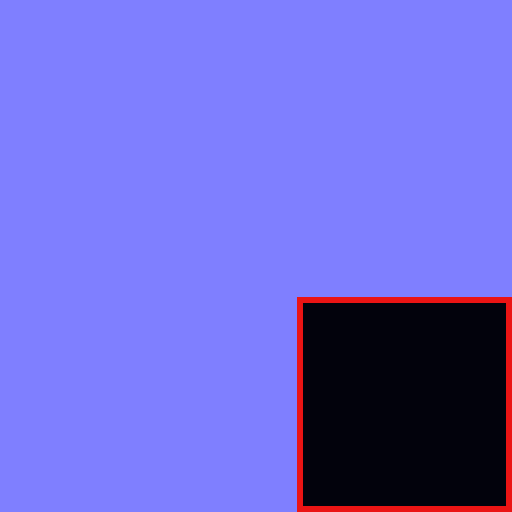} \\
    \mipQualSep
      & \mipQualMethod{GT}{} & \mipQualImg{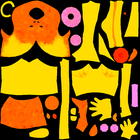} & \mipQualImg{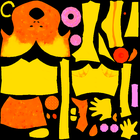} & \mipQualImg{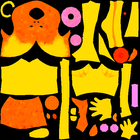} & \mipQualImg{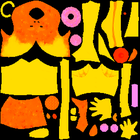} & \mipQualImg{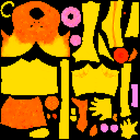} & \mipQualImg{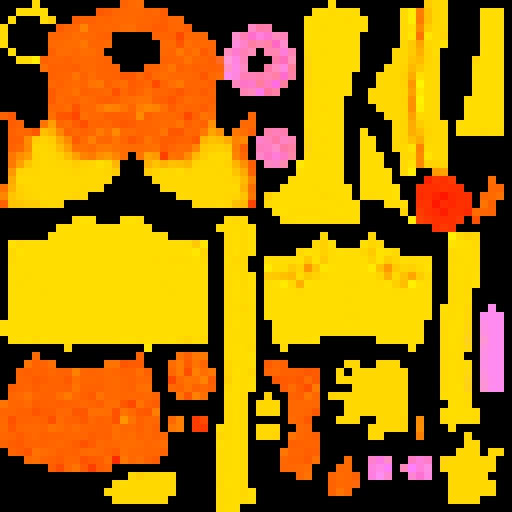} & \mipQualImg{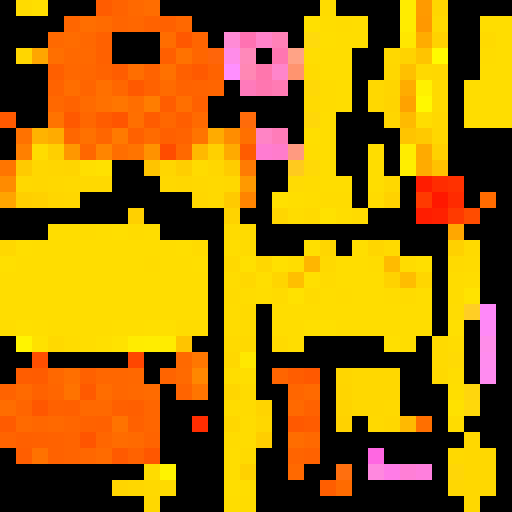} & \mipQualImg{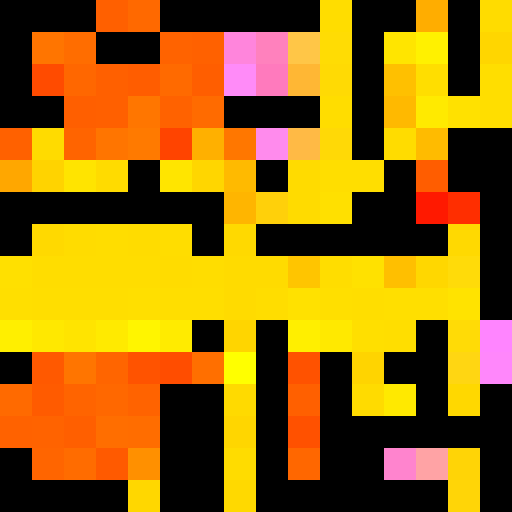} & \mipQualImg{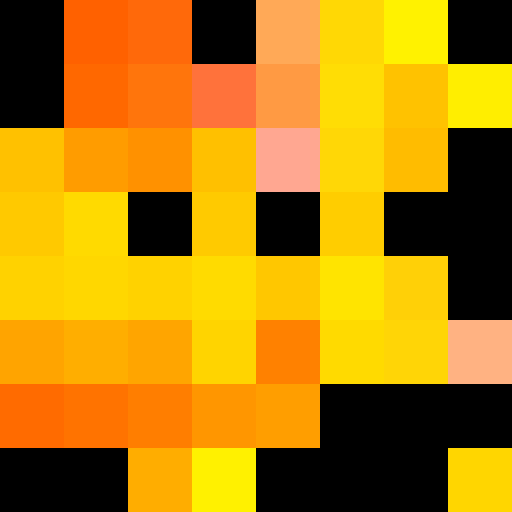} & \mipQualImg{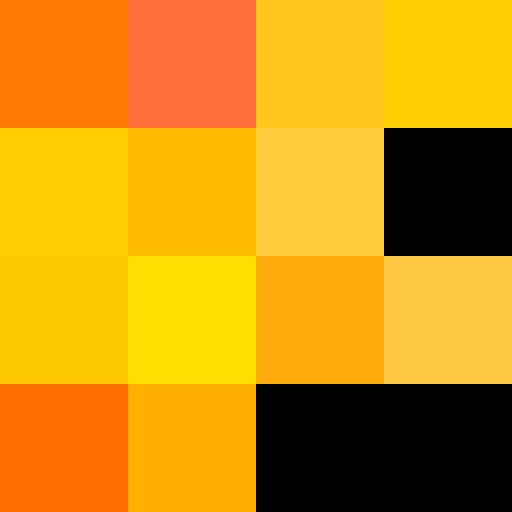} & \mipQualImg{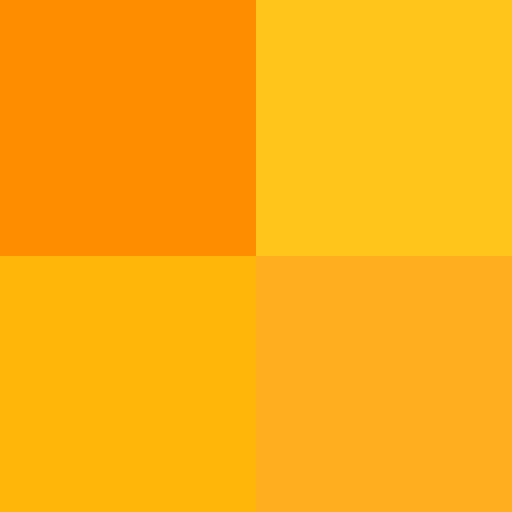} & \mipQualImg{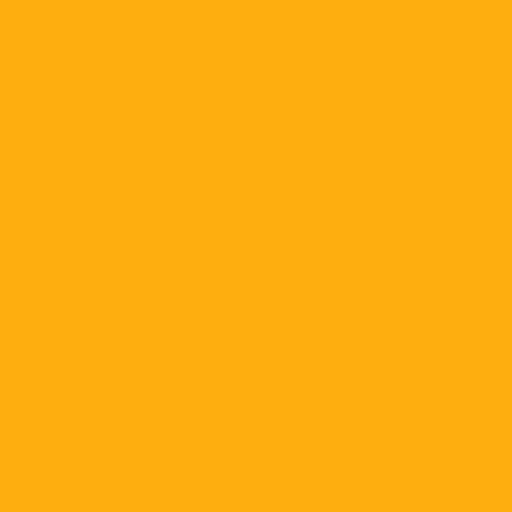} \\[\mipQualRowGap]
      & \mipQualMethod{Ours}{0.097 bppc} & \mipQualImg{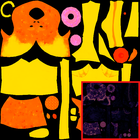} & \mipQualImg{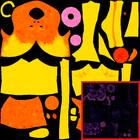} & \mipQualImg{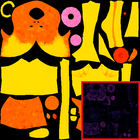} & \mipQualImg{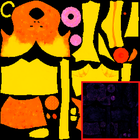} & \mipQualImg{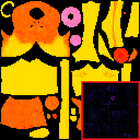} & \mipQualImg{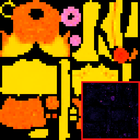} & \mipQualImg{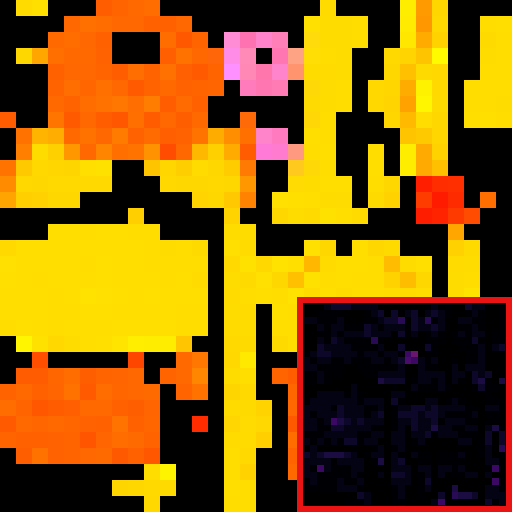} & \mipQualImg{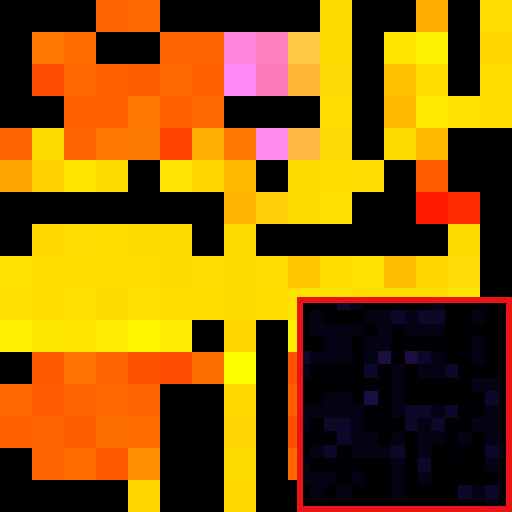} & \mipQualImg{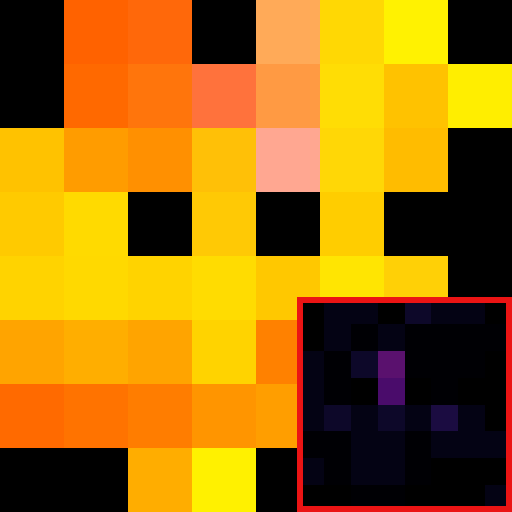} & \mipQualImg{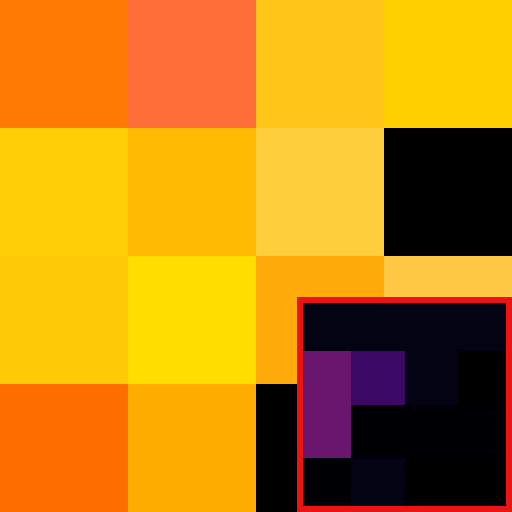} & \mipQualImg{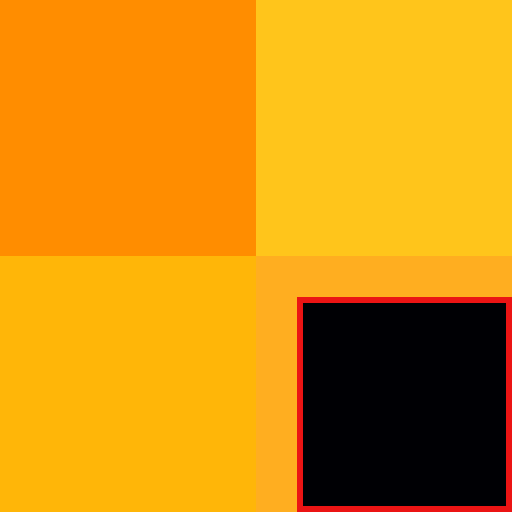} & \mipQualImg{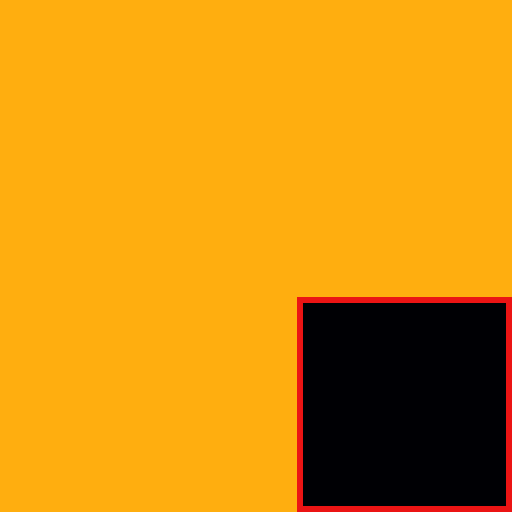} \\[\mipQualRowGap]
    \mipQualStream{ARM} & \mipQualMethod{Image-GS}{0.404 bppc} & \mipQualImg{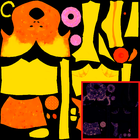} & \mipQualImg{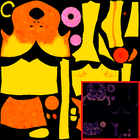} & \mipQualImg{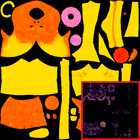} & \mipQualImg{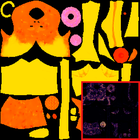} & \mipQualImg{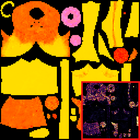} & \mipQualImg{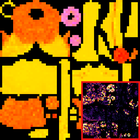} & \mipQualImg{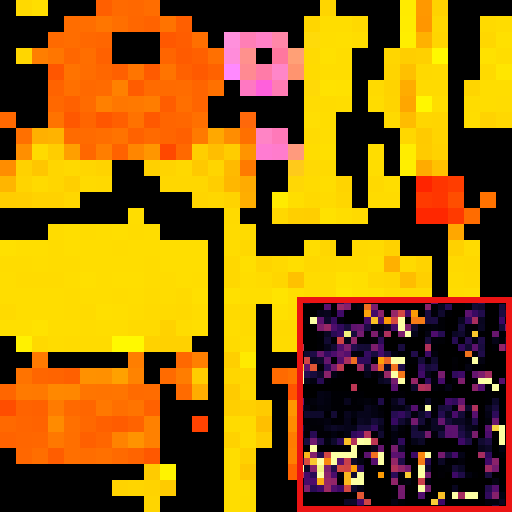} & \mipQualImg{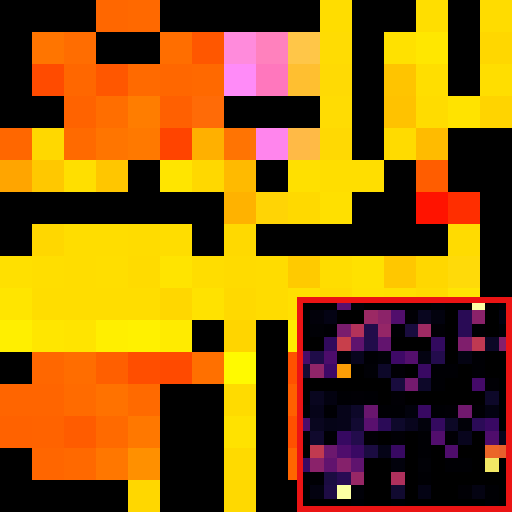} & \mipQualImg{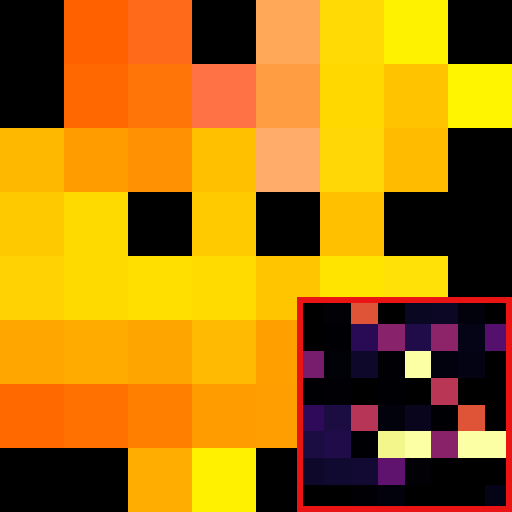} & \mipQualImg{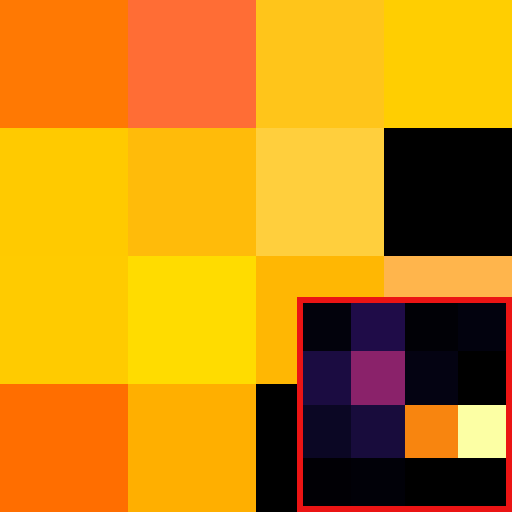} & \mipQualImg{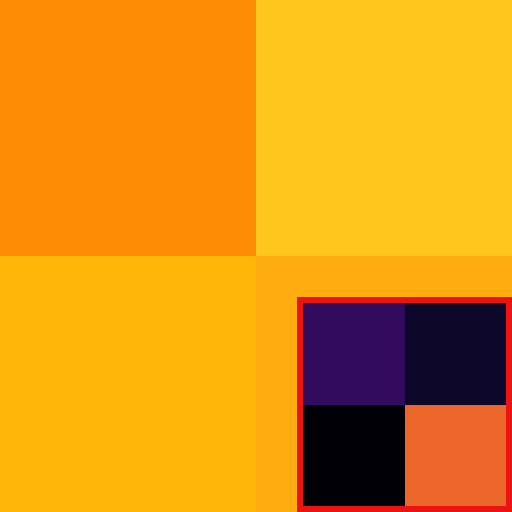} & \mipQualImg{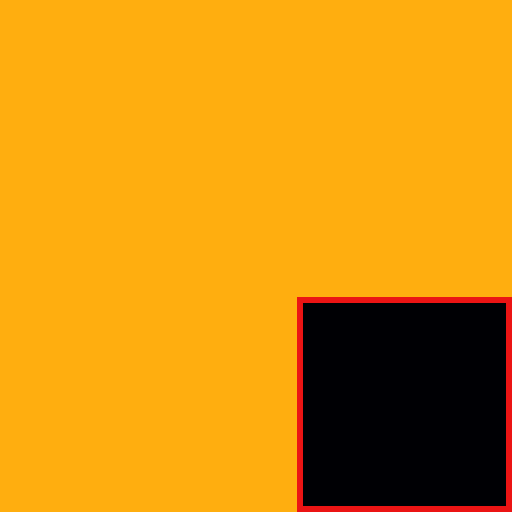} \\[\mipQualRowGap]
      & \mipQualMethod{ASTC}{0.429 bppc} & \mipQualImg{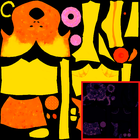} & \mipQualImg{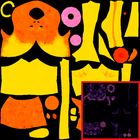} & \mipQualImg{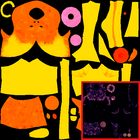} & \mipQualImg{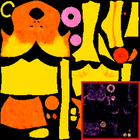} & \mipQualImg{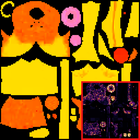} & \mipQualImg{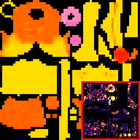} & \mipQualImg{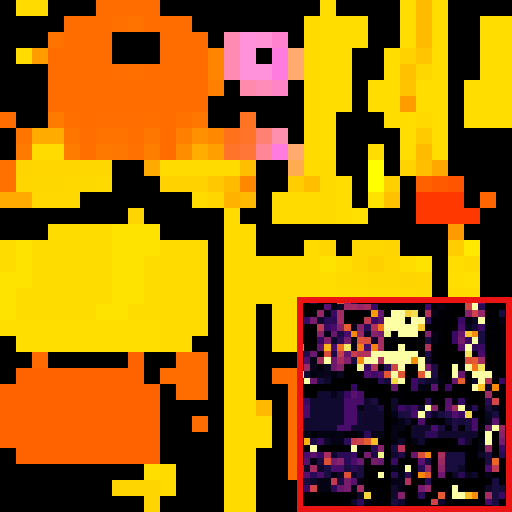} & \mipQualImg{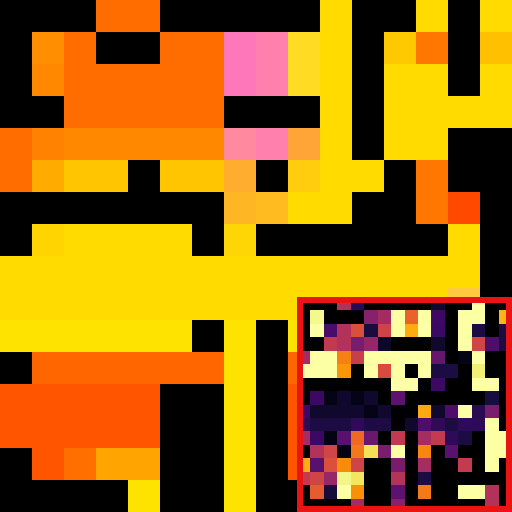} & \mipQualImg{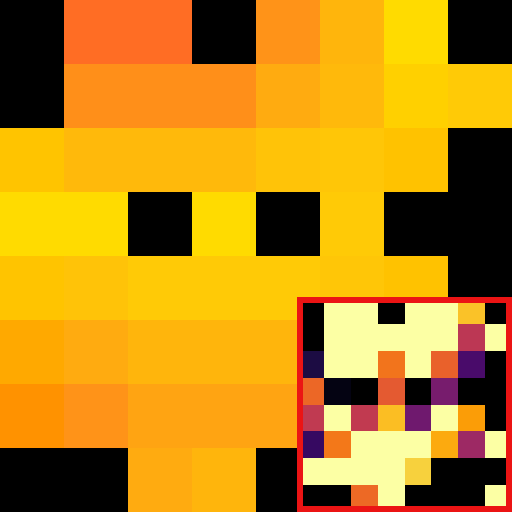} & \mipQualImg{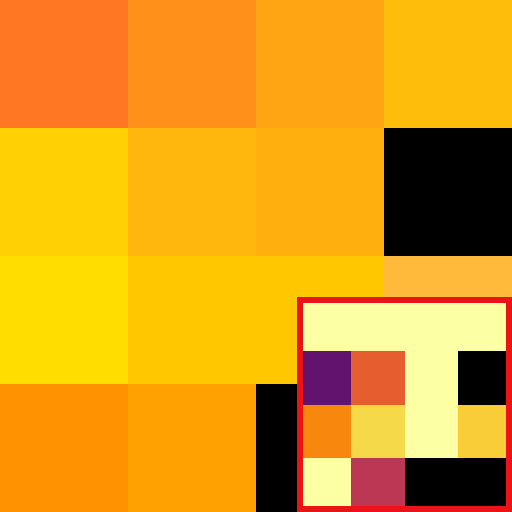} & \mipQualImg{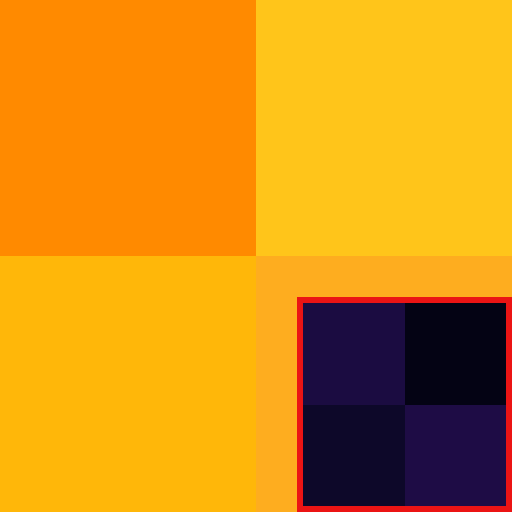} & \mipQualImg{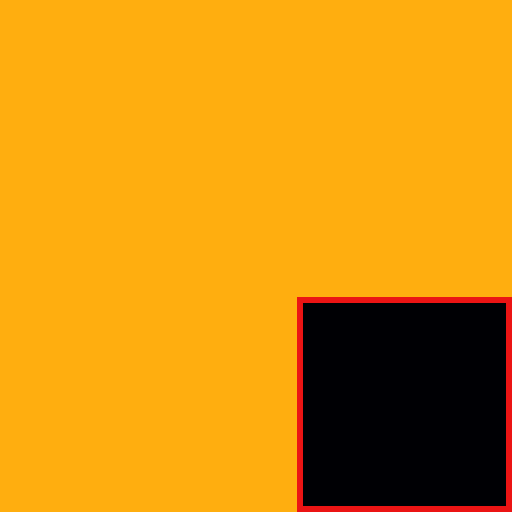} \\[\mipQualRowGap]
      & \mipQualMethod{NTC}{0.102 bppc} & \mipQualImg{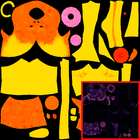} & \mipQualImg{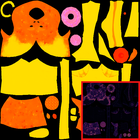} & \mipQualImg{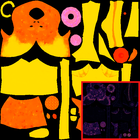} & \mipQualImg{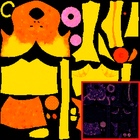} & \mipQualImg{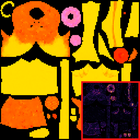} & \mipQualImg{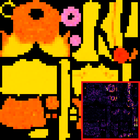} & \mipQualImg{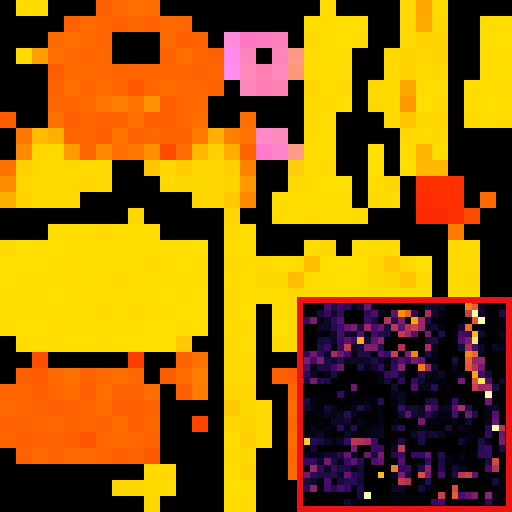} & \mipQualImg{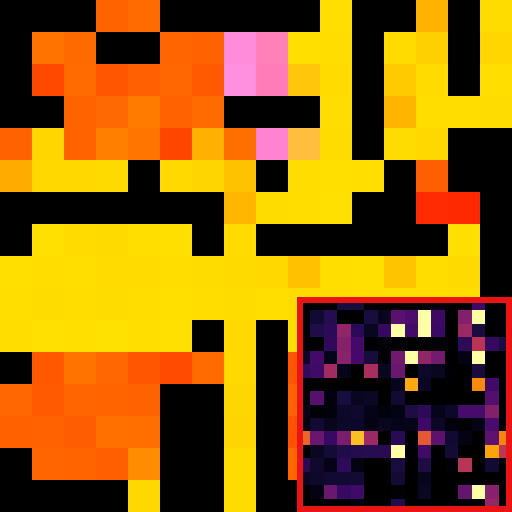} & \mipQualImg{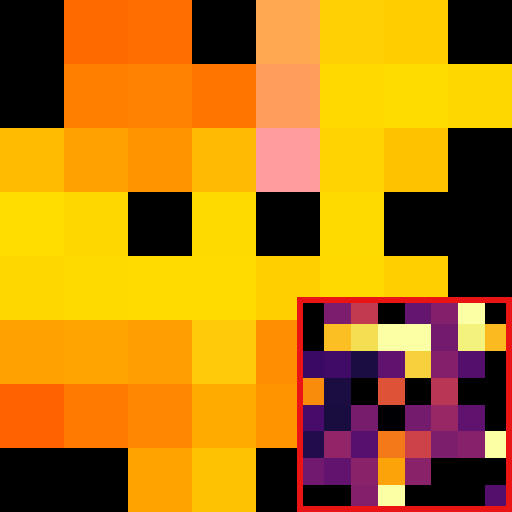} & \mipQualImg{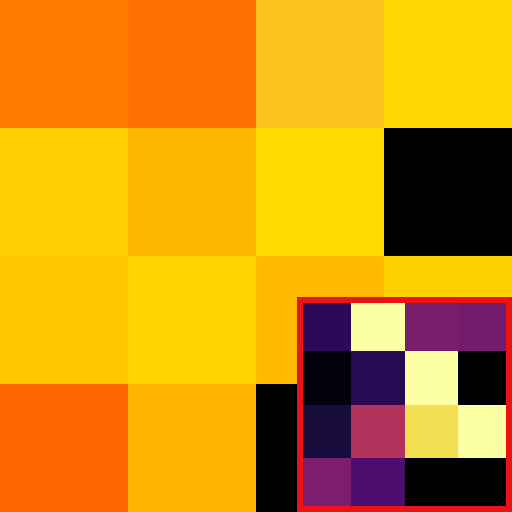} & \mipQualImg{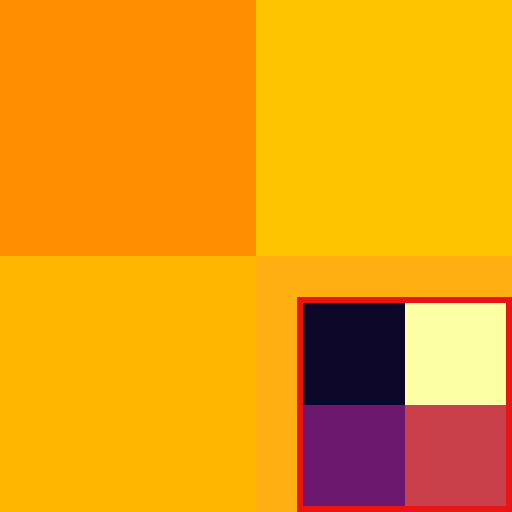} & \mipQualImg{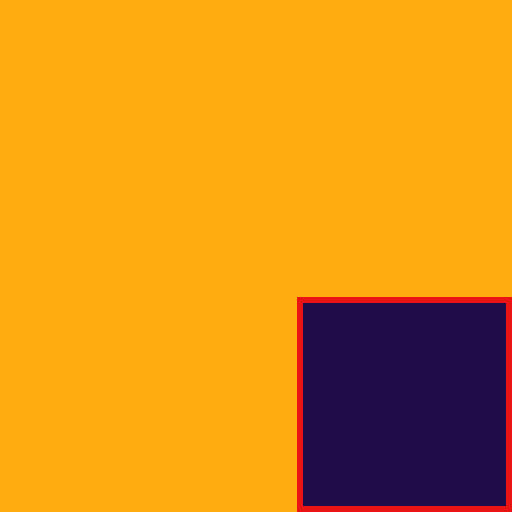} \\

  \end{tabular}%
  }
  \caption{Mipmap reconstruction comparison on the \textit{Outlaw} SVBRDF stack. Rows show the reference and reconstructed mipmap pyramids, with insets showing per-pixel error maps against the reference. ARM denotes ambient occlusion, roughness, and metallic channels.}
  \label{fig:mipmap-error-inset-outlaw}
\end{figure*}

\clearpage
\begin{figure*}[p!]
  \centering
  \mipQualFontSize
  \setlength{\tabcolsep}{0pt}
  \renewcommand{\arraystretch}{1.0}
  \resizebox{\mipQualTableTargetW}{!}{%
  \begin{tabular}{@{}c@{\hspace{\mipQualStreamMethodGap}}c@{\hspace{\mipQualMethodImageGap}}*{11}{c@{\hspace{\mipQualColGap}}}c@{}}
    \multicolumn{2}{c}{} & \mipQualMip{mip 0}{2048$^2$} & \mipQualMip{mip 1}{1024$^2$} & \mipQualMip{mip 2}{512$^2$} & \mipQualMip{mip 3}{256$^2$} & \mipQualMip{mip 4}{128$^2$} & \mipQualMip{mip 5}{64$^2$} & \mipQualMip{mip 6}{32$^2$} & \mipQualMip{mip 7}{16$^2$} & \mipQualMip{mip 8}{8$^2$} & \mipQualMip{mip 9}{4$^2$} & \mipQualMip{mip 10}{2$^2$} & \mipQualMip{mip 11}{1$^2$} \\[\mipQualHeaderGap]
      & \mipQualMethod{GT}{} & \mipQualImg{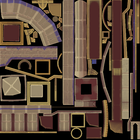} & \mipQualImg{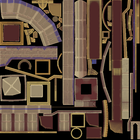} & \mipQualImg{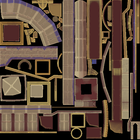} & \mipQualImg{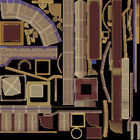} & \mipQualImg{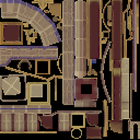} & \mipQualImg{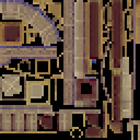} & \mipQualImg{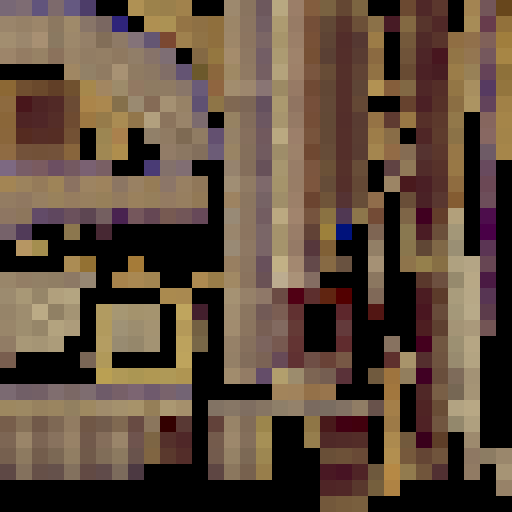} & \mipQualImg{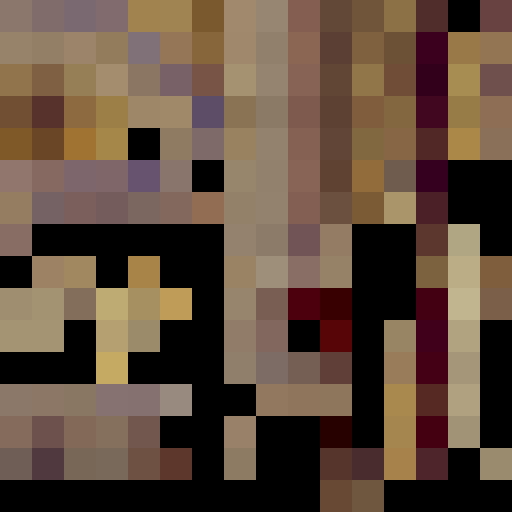} & \mipQualImg{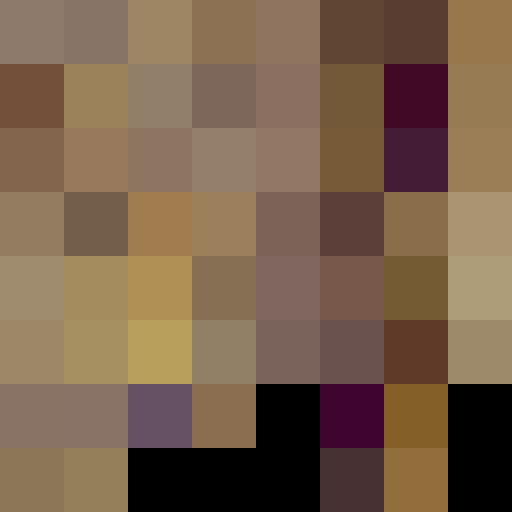} & \mipQualImg{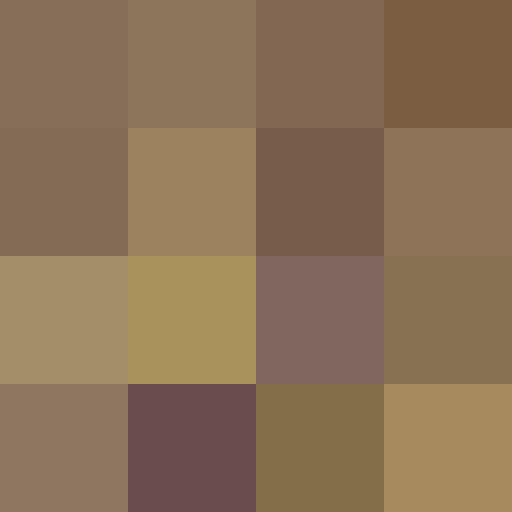} & \mipQualImg{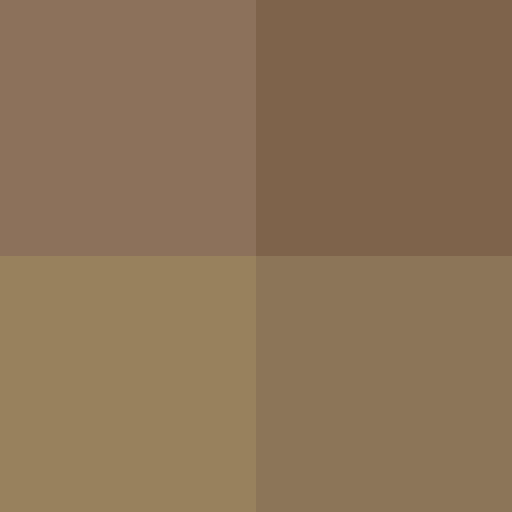} & \mipQualImg{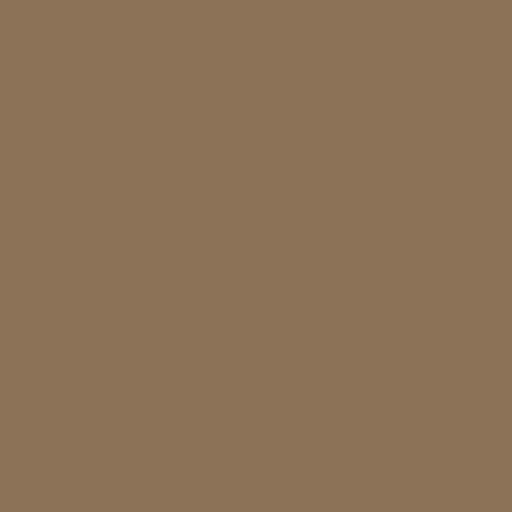} \\[\mipQualRowGap]
      & \mipQualMethod{Ours}{0.172 bppc} & \mipQualImg{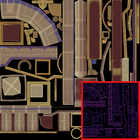} & \mipQualImg{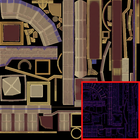} & \mipQualImg{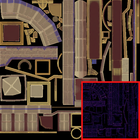} & \mipQualImg{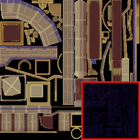} & \mipQualImg{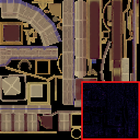} & \mipQualImg{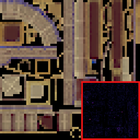} & \mipQualImg{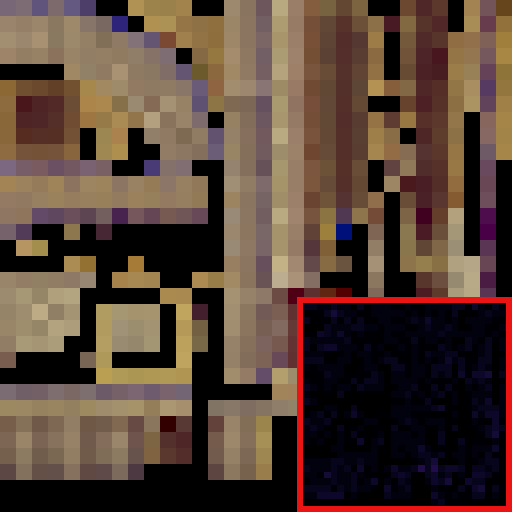} & \mipQualImg{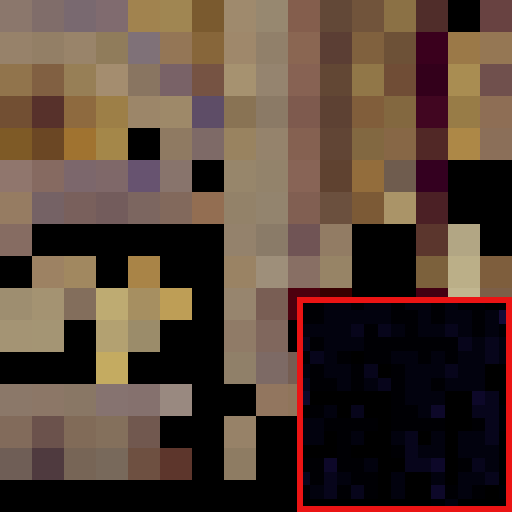} & \mipQualImg{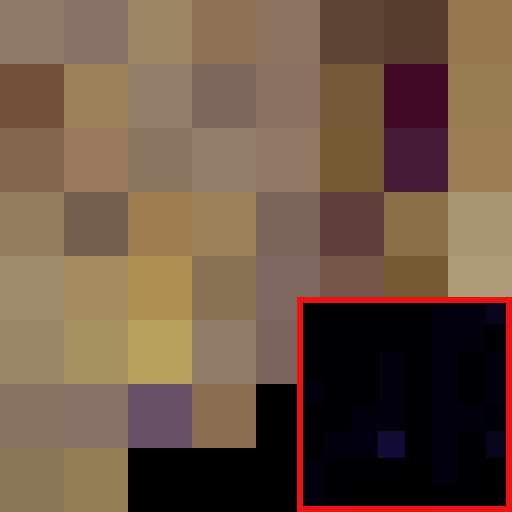} & \mipQualImg{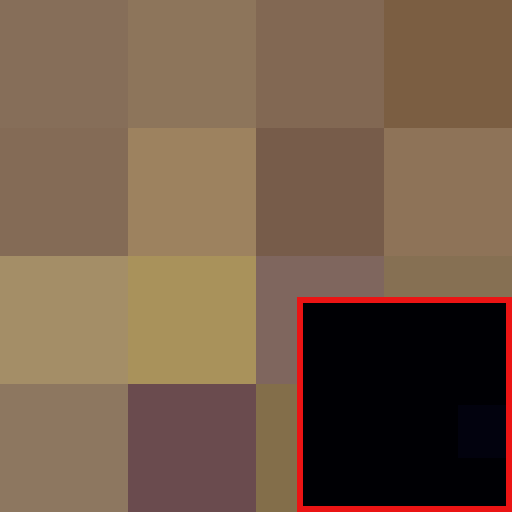} & \mipQualImg{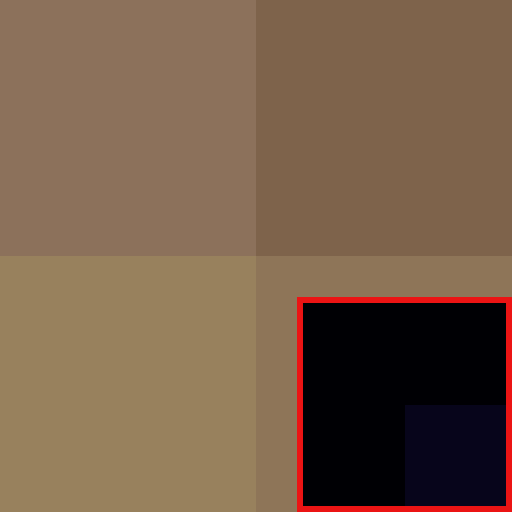} & \mipQualImg{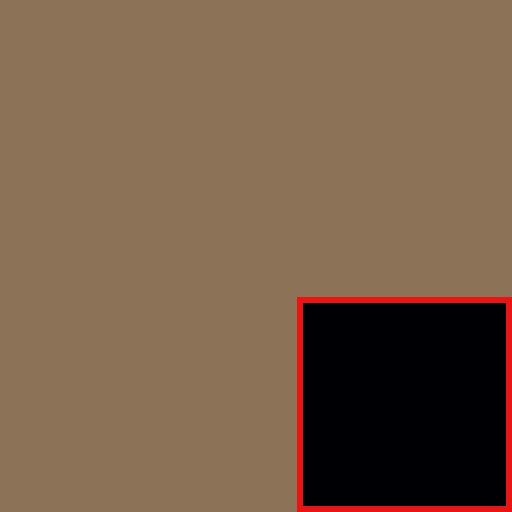} \\[\mipQualRowGap]
    \mipQualStream{Base color} & \mipQualMethod{Image-GS}{0.404 bppc} & \mipQualImg{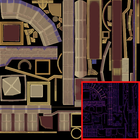} & \mipQualImg{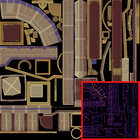} & \mipQualImg{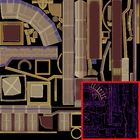} & \mipQualImg{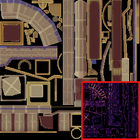} & \mipQualImg{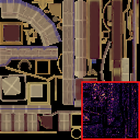} & \mipQualImg{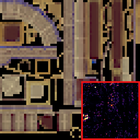} & \mipQualImg{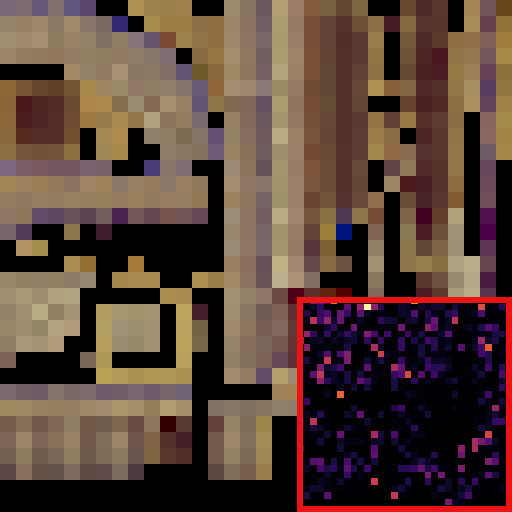} & \mipQualImg{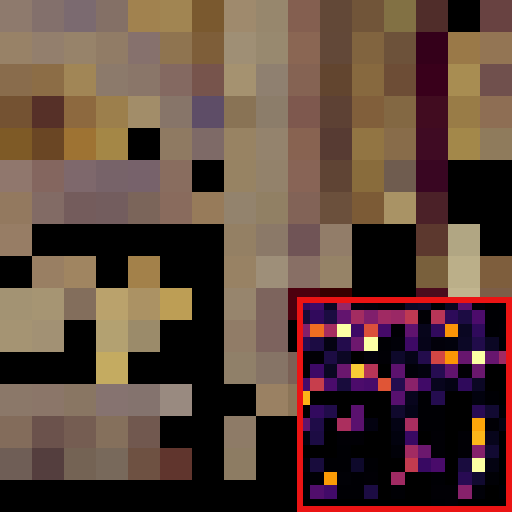} & \mipQualImg{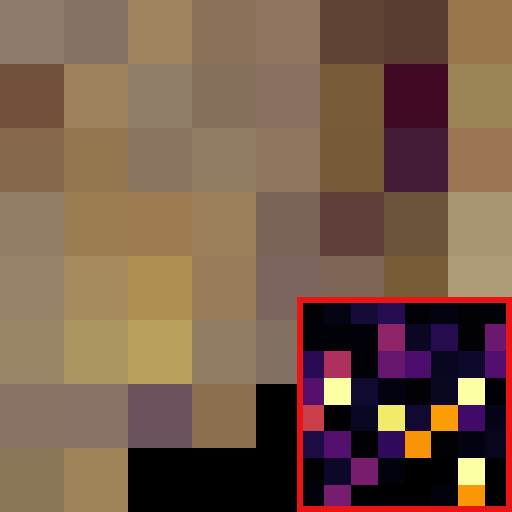} & \mipQualImg{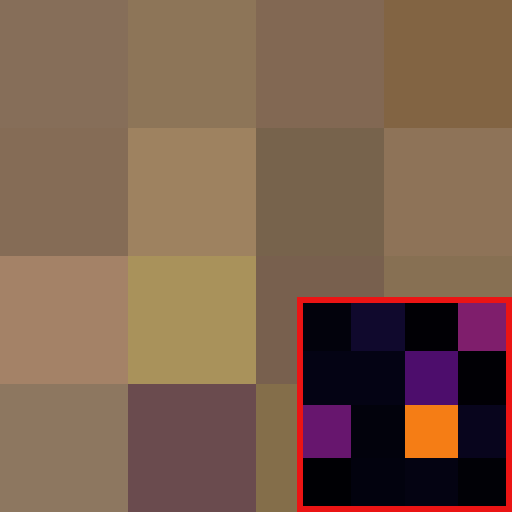} & \mipQualImg{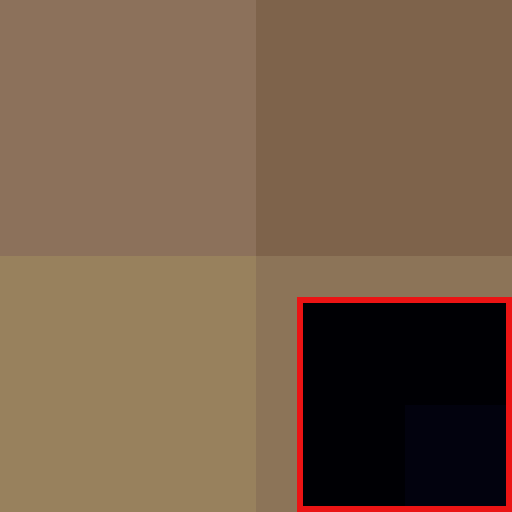} & \mipQualImg{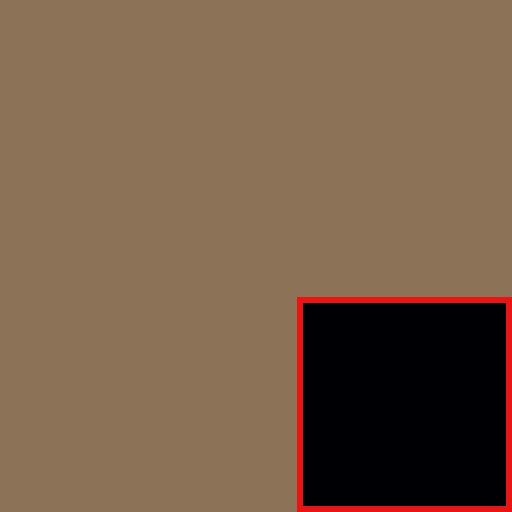} \\[\mipQualRowGap]
      & \mipQualMethod{ASTC}{0.429 bppc} & \mipQualImg{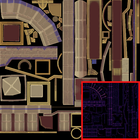} & \mipQualImg{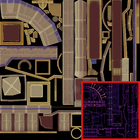} & \mipQualImg{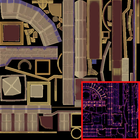} & \mipQualImg{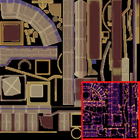} & \mipQualImg{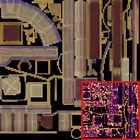} & \mipQualImg{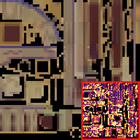} & \mipQualImg{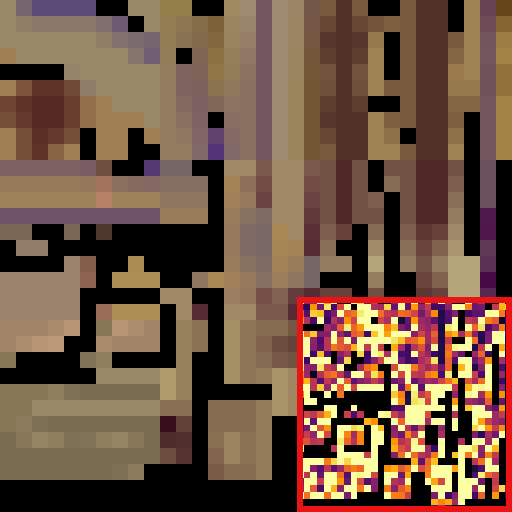} & \mipQualImg{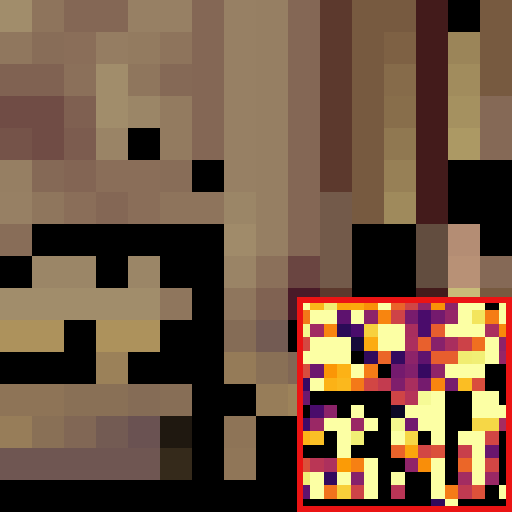} & \mipQualImg{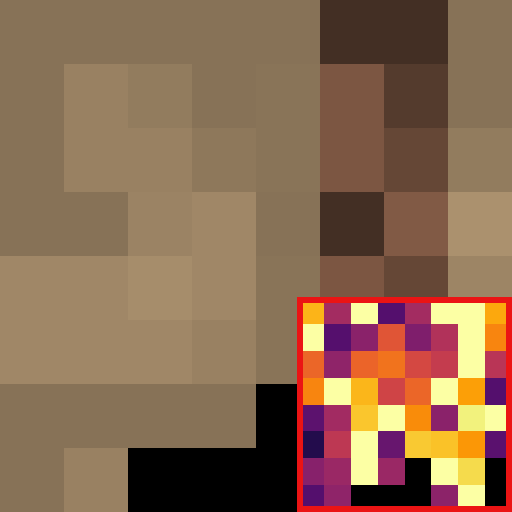} & \mipQualImg{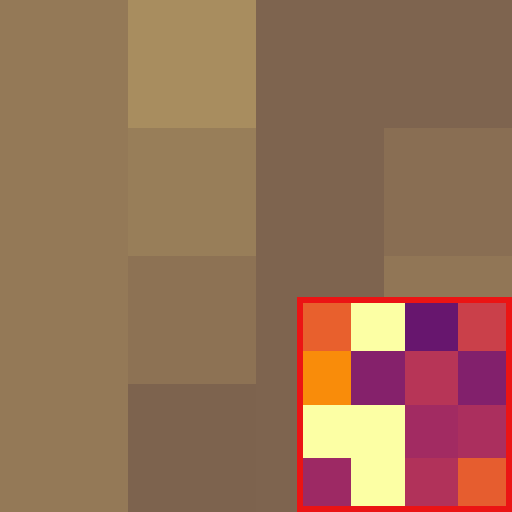} & \mipQualImg{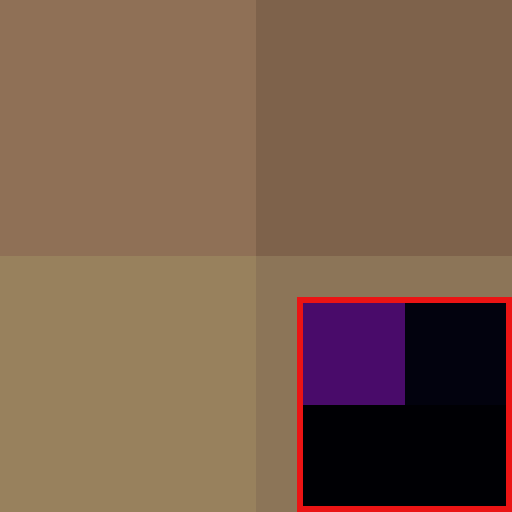} & \mipQualImg{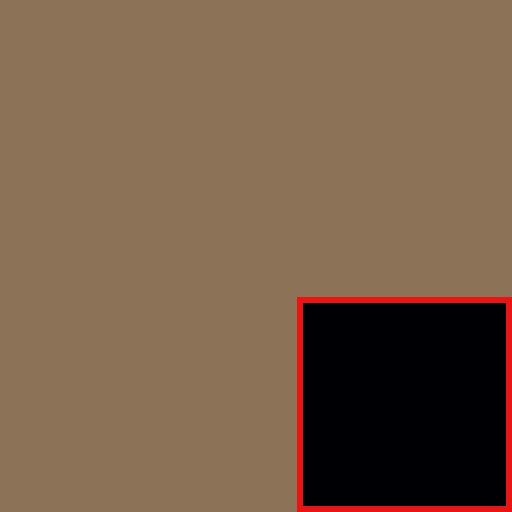} \\[\mipQualRowGap]
      & \mipQualMethod{NTC}{0.220 bppc} & \mipQualImg{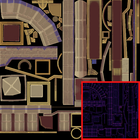} & \mipQualImg{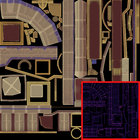} & \mipQualImg{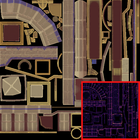} & \mipQualImg{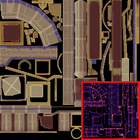} & \mipQualImg{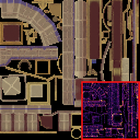} & \mipQualImg{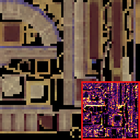} & \mipQualImg{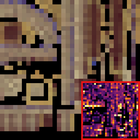} & \mipQualImg{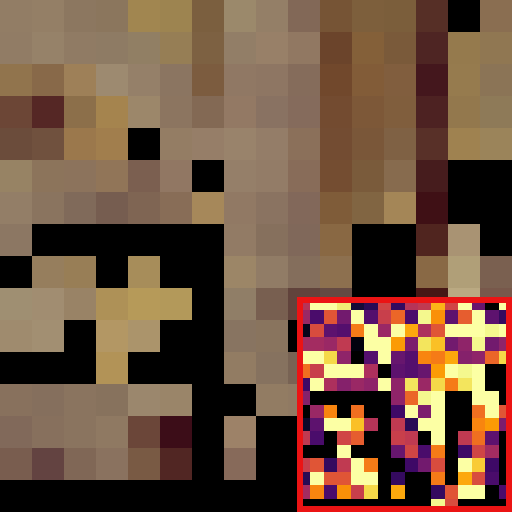} & \mipQualImg{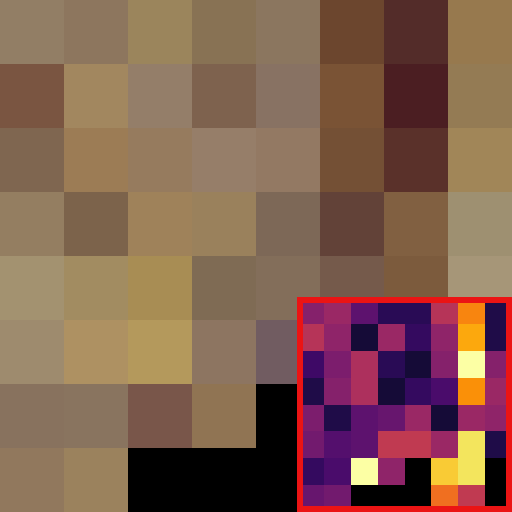} & \mipQualImg{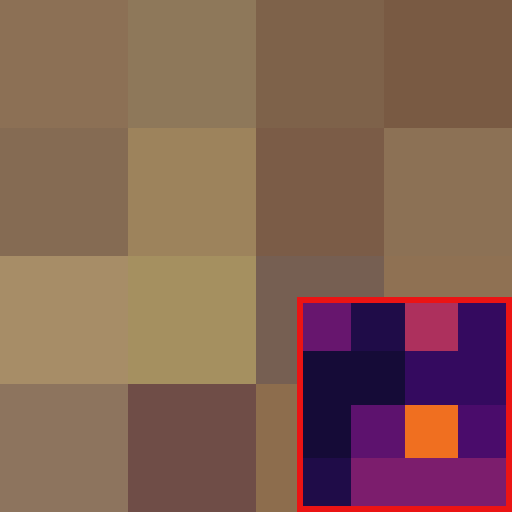} & \mipQualImg{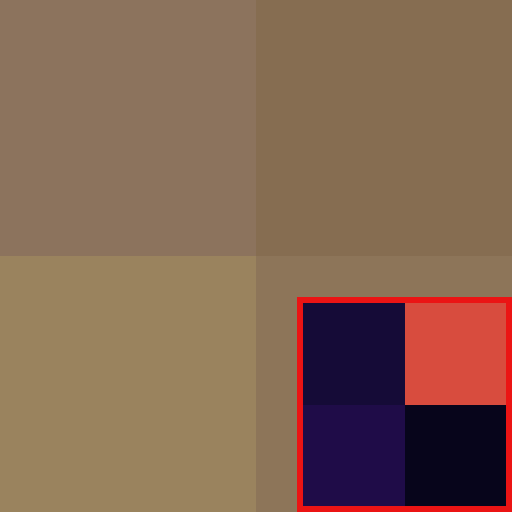} & \mipQualImg{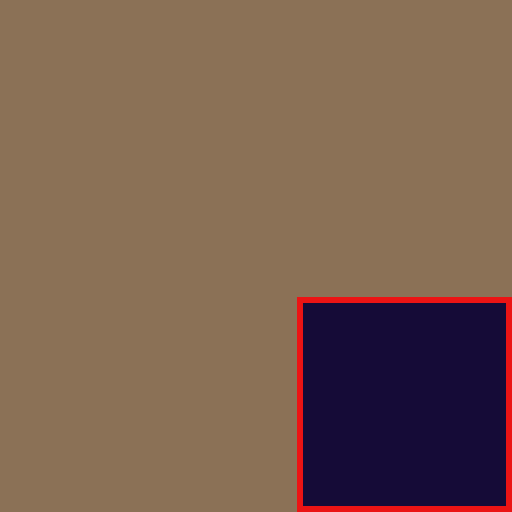} \\
    \mipQualSep
      & \mipQualMethod{GT}{} & \mipQualImg{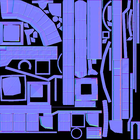} & \mipQualImg{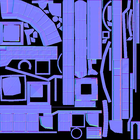} & \mipQualImg{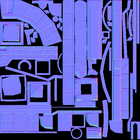} & \mipQualImg{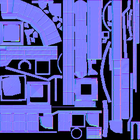} & \mipQualImg{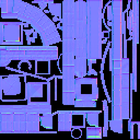} & \mipQualImg{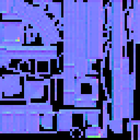} & \mipQualImg{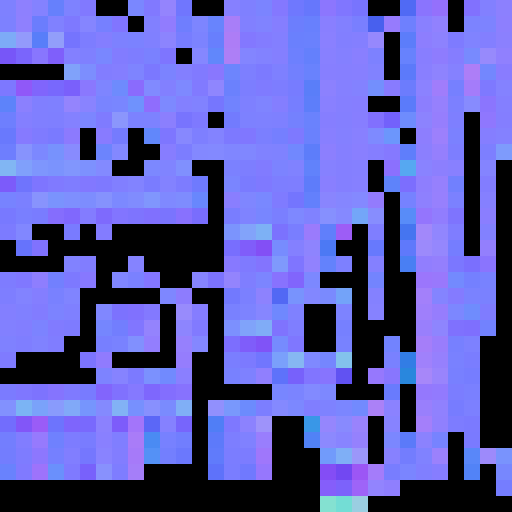} & \mipQualImg{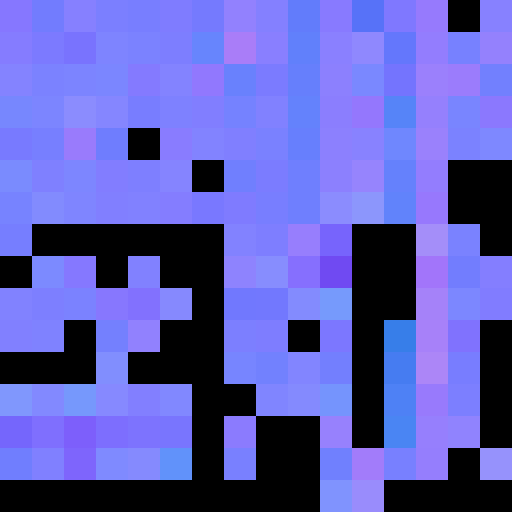} & \mipQualImg{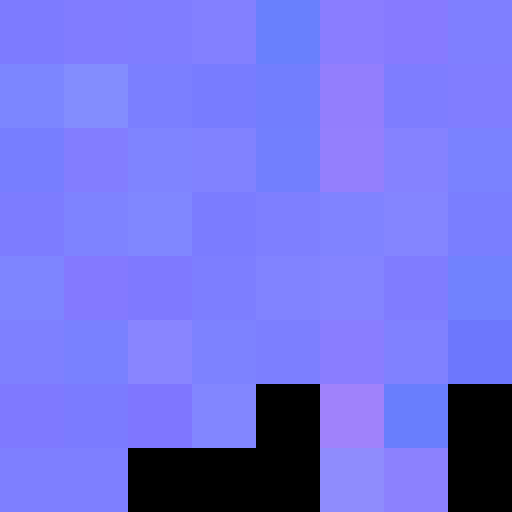} & \mipQualImg{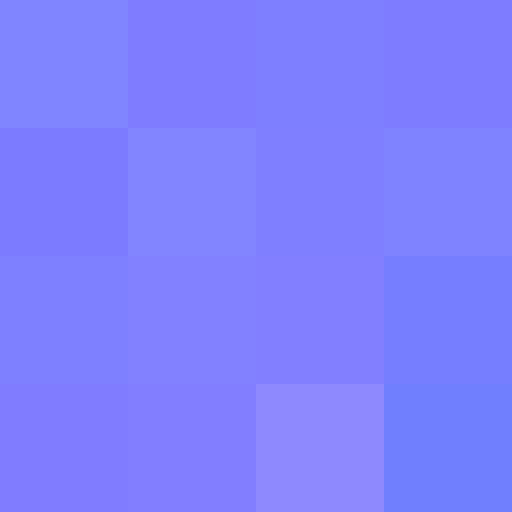} & \mipQualImg{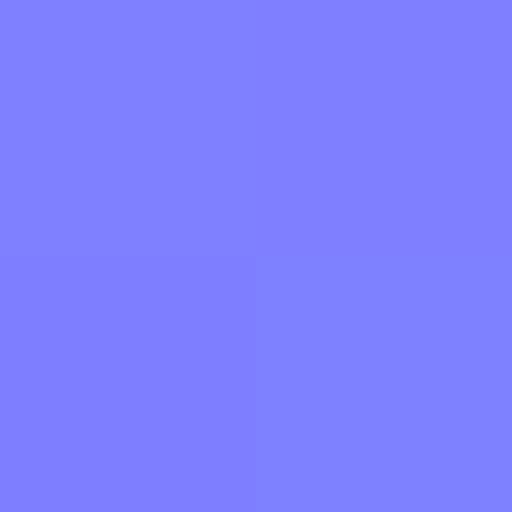} & \mipQualImg{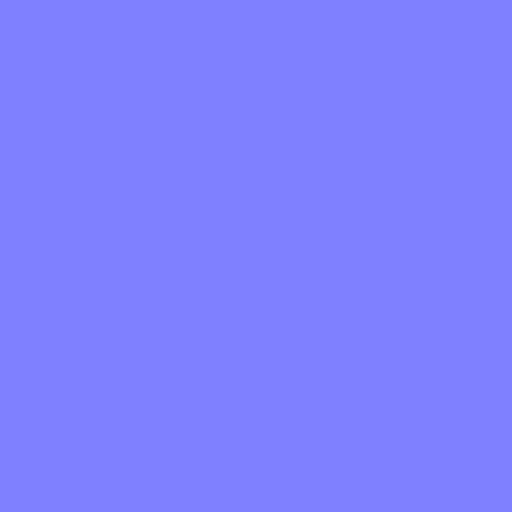} \\[\mipQualRowGap]
      & \mipQualMethod{Ours}{0.172 bppc} & \mipQualImg{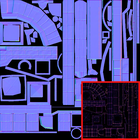} & \mipQualImg{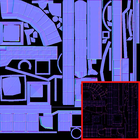} & \mipQualImg{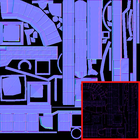} & \mipQualImg{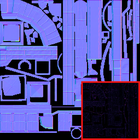} & \mipQualImg{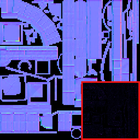} & \mipQualImg{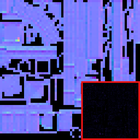} & \mipQualImg{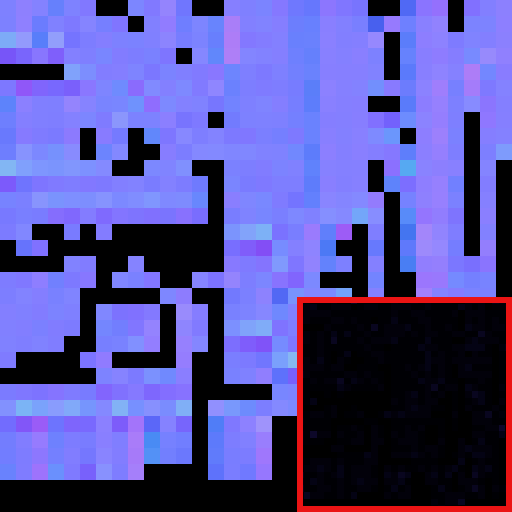} & \mipQualImg{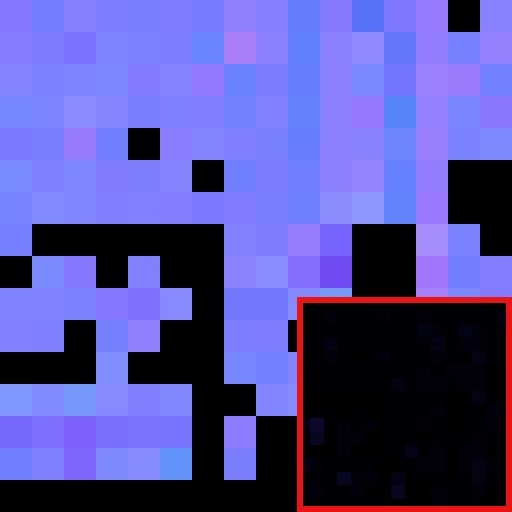} & \mipQualImg{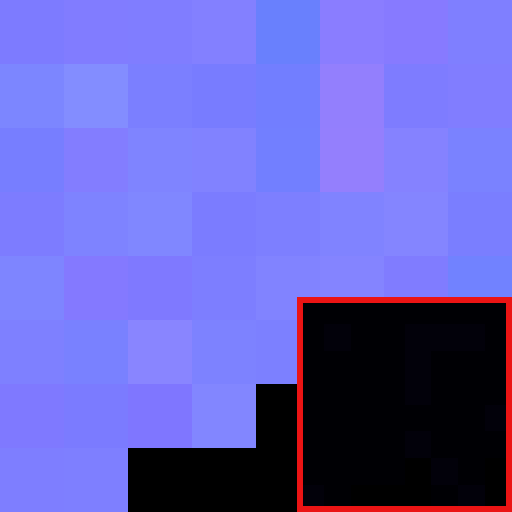} & \mipQualImg{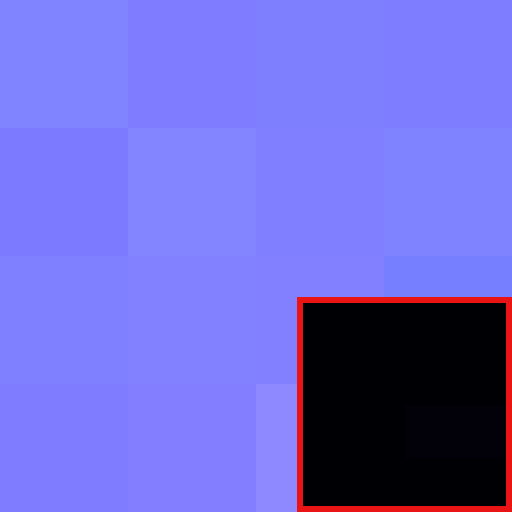} & \mipQualImg{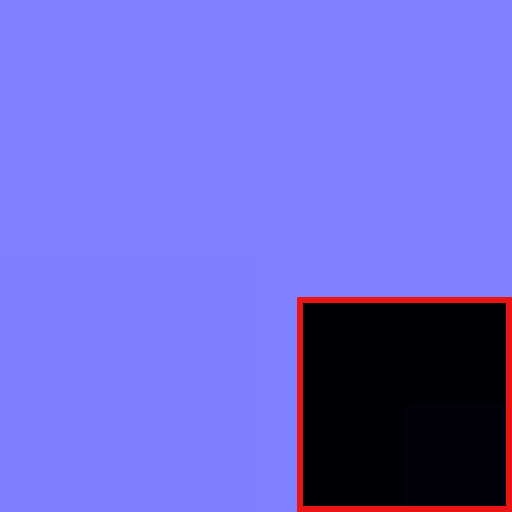} & \mipQualImg{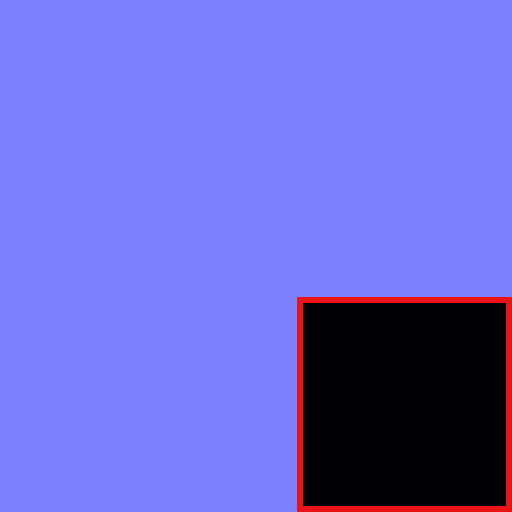} \\[\mipQualRowGap]
    \mipQualStream{Normal map} & \mipQualMethod{Image-GS}{0.404 bppc} & \mipQualImg{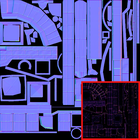} & \mipQualImg{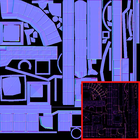} & \mipQualImg{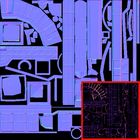} & \mipQualImg{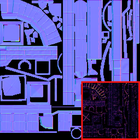} & \mipQualImg{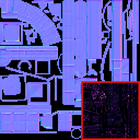} & \mipQualImg{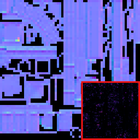} & \mipQualImg{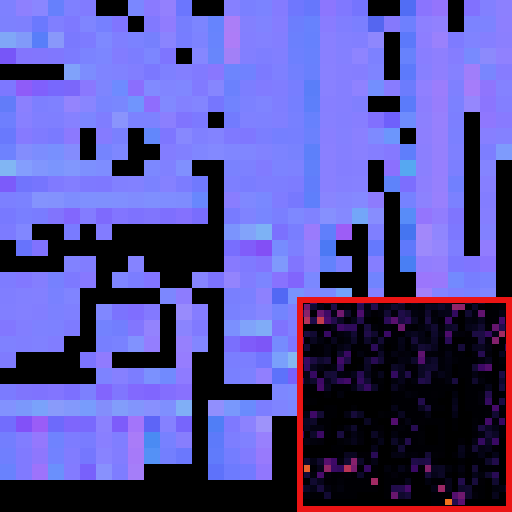} & \mipQualImg{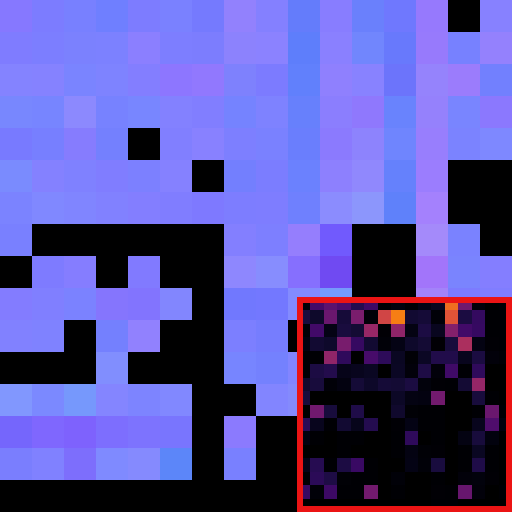} & \mipQualImg{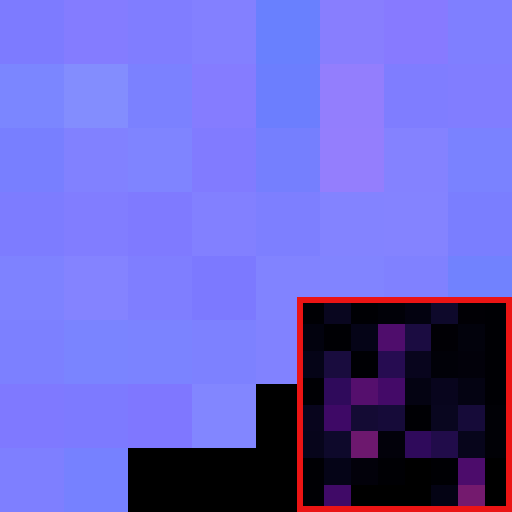} & \mipQualImg{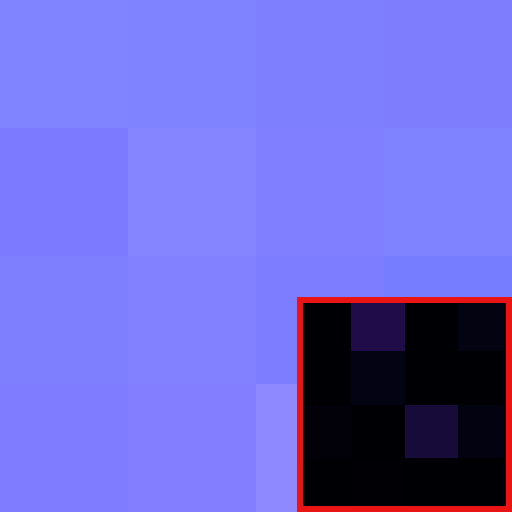} & \mipQualImg{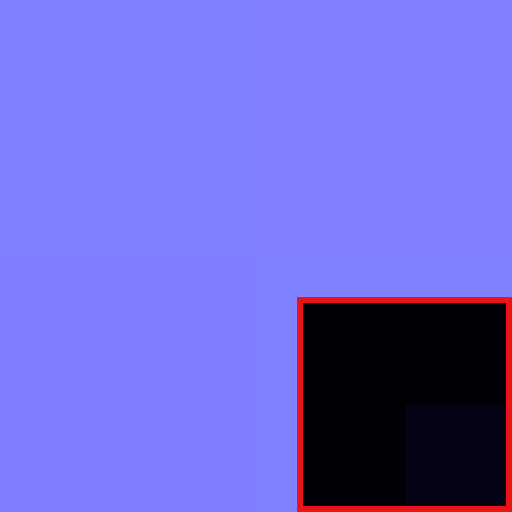} & \mipQualImg{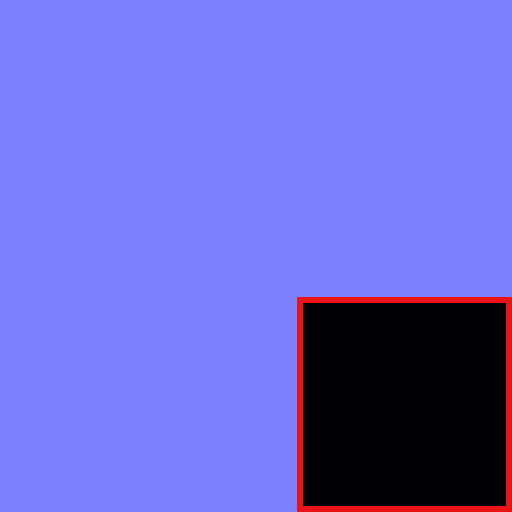} \\[\mipQualRowGap]
      & \mipQualMethod{ASTC}{0.429 bppc} & \mipQualImg{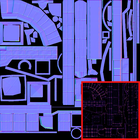} & \mipQualImg{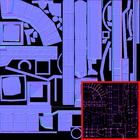} & \mipQualImg{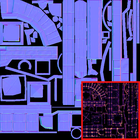} & \mipQualImg{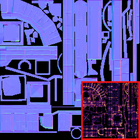} & \mipQualImg{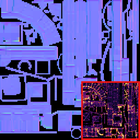} & \mipQualImg{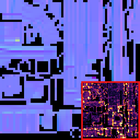} & \mipQualImg{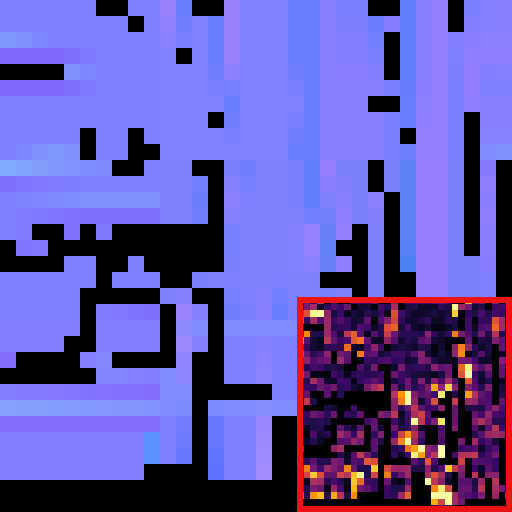} & \mipQualImg{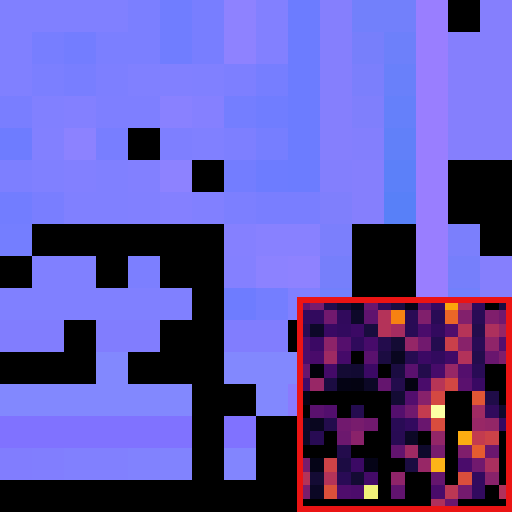} & \mipQualImg{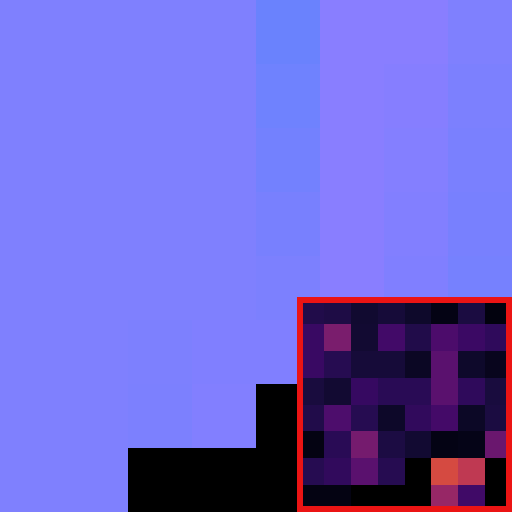} & \mipQualImg{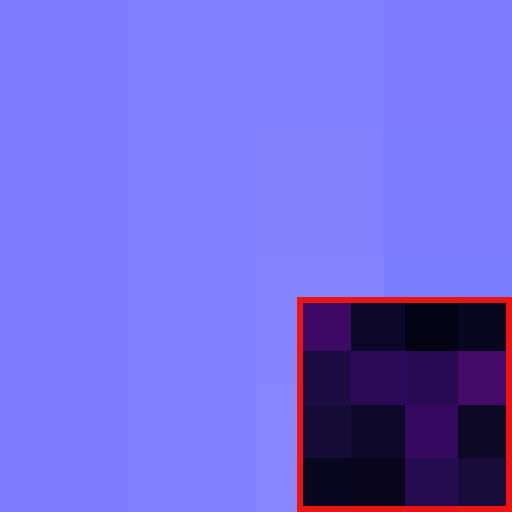} & \mipQualImg{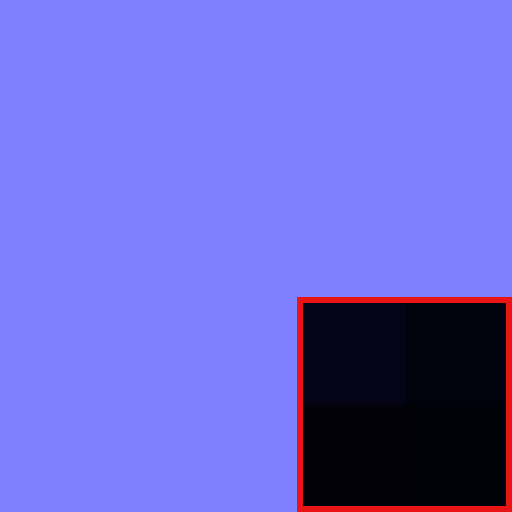} & \mipQualImg{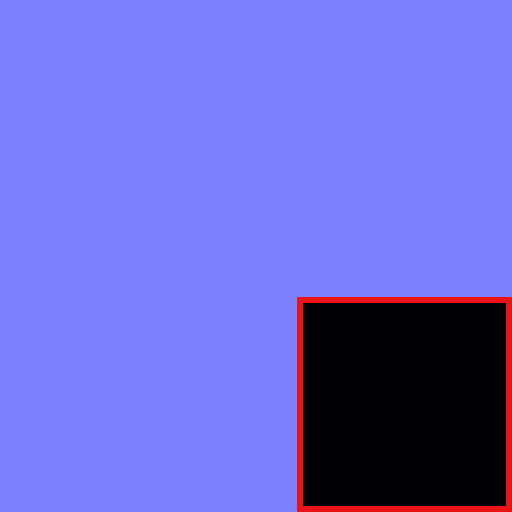} \\[\mipQualRowGap]
      & \mipQualMethod{NTC}{0.220 bppc} & \mipQualImg{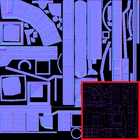} & \mipQualImg{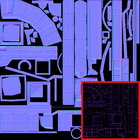} & \mipQualImg{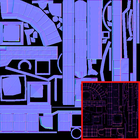} & \mipQualImg{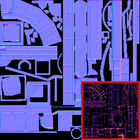} & \mipQualImg{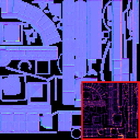} & \mipQualImg{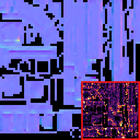} & \mipQualImg{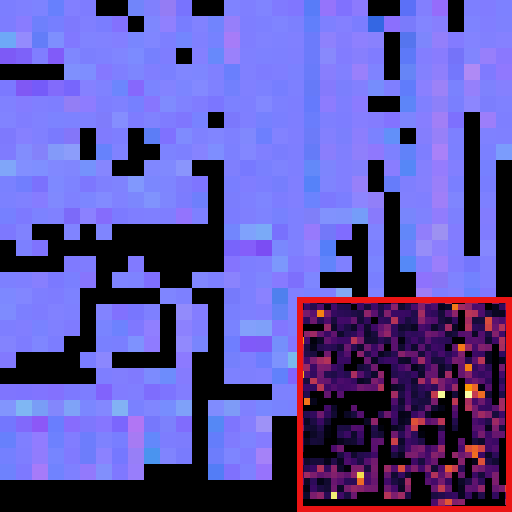} & \mipQualImg{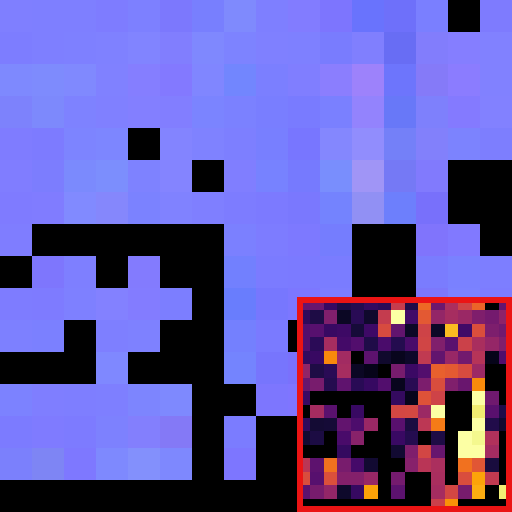} & \mipQualImg{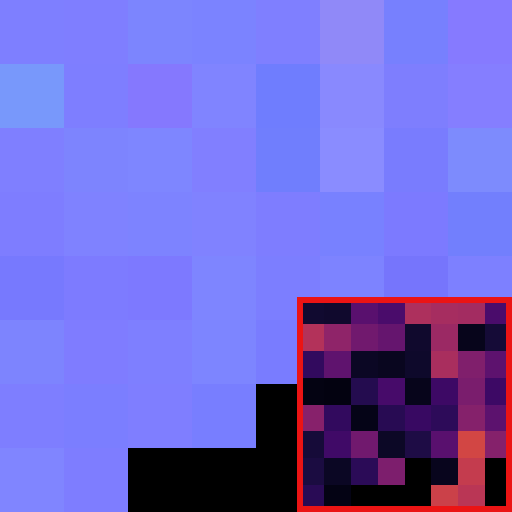} & \mipQualImg{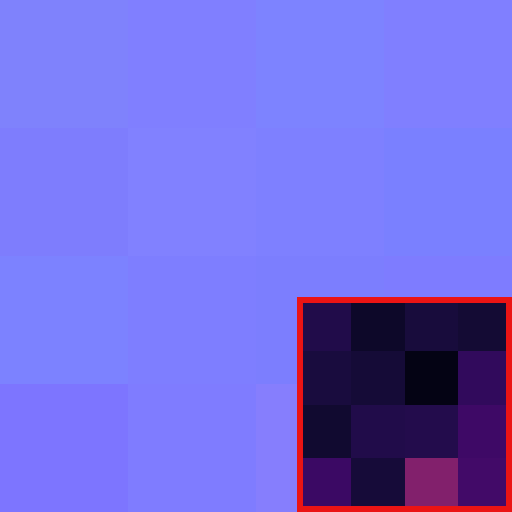} & \mipQualImg{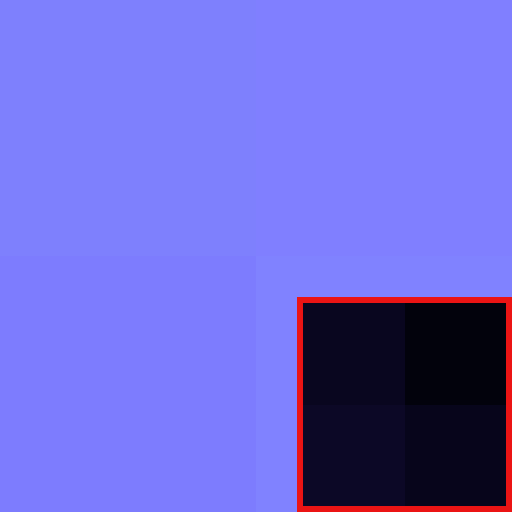} & \mipQualImg{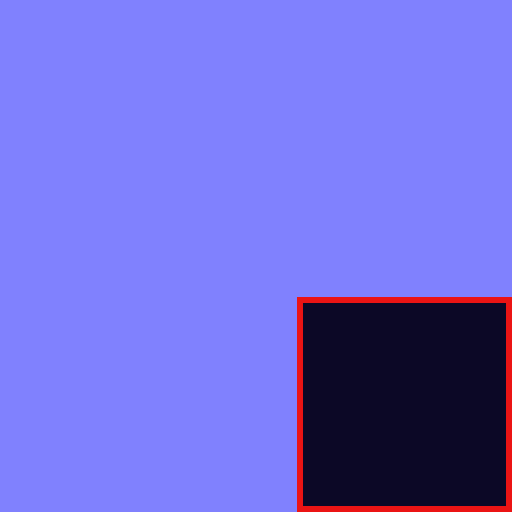} \\
    \mipQualSep
      & \mipQualMethod{GT}{} & \mipQualImg{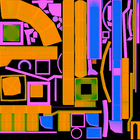} & \mipQualImg{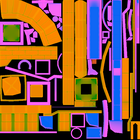} & \mipQualImg{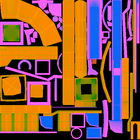} & \mipQualImg{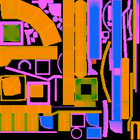} & \mipQualImg{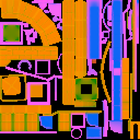} & \mipQualImg{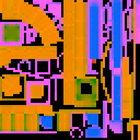} & \mipQualImg{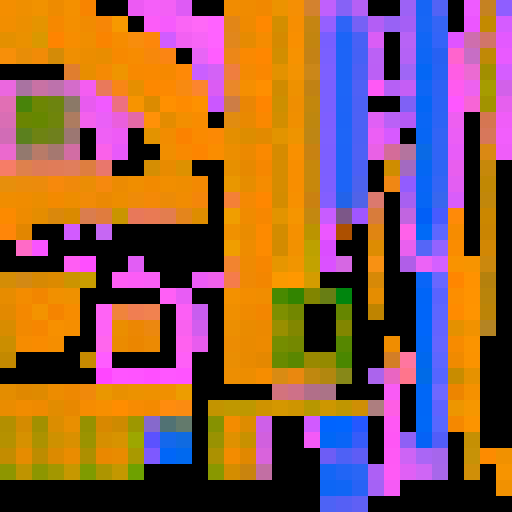} & \mipQualImg{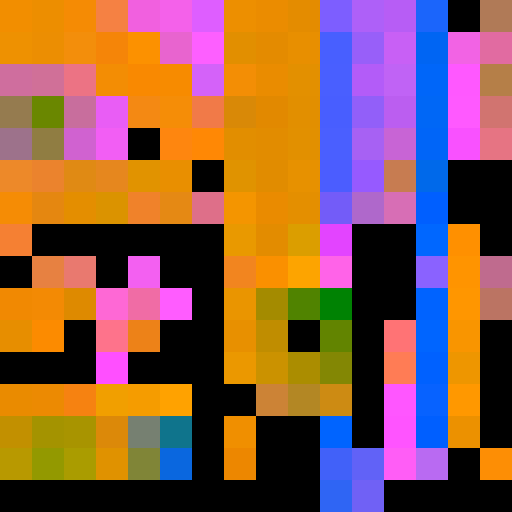} & \mipQualImg{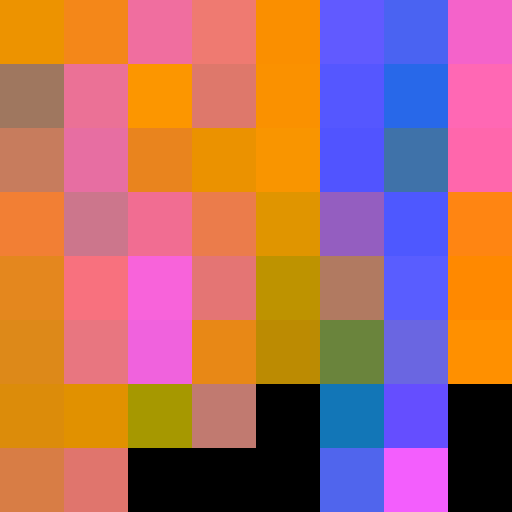} & \mipQualImg{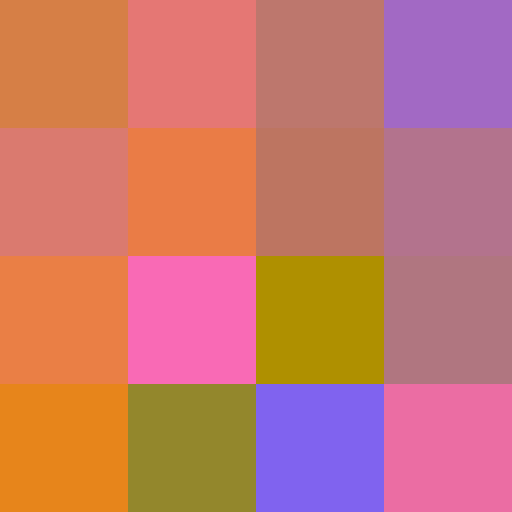} & \mipQualImg{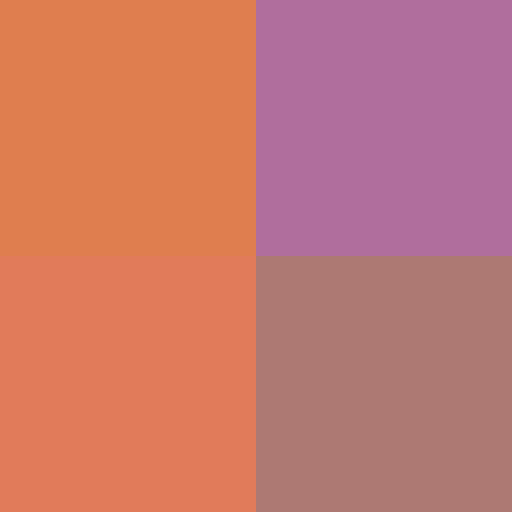} & \mipQualImg{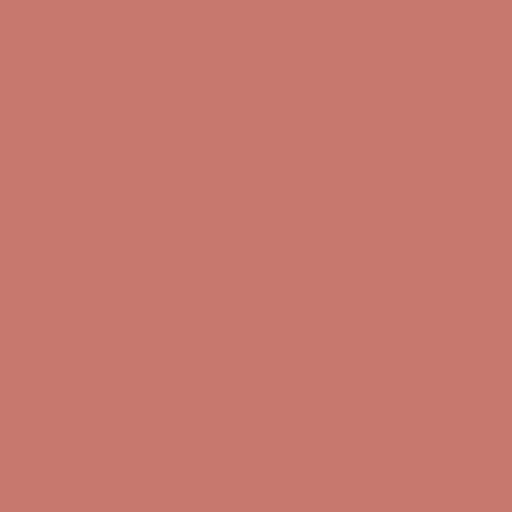} \\[\mipQualRowGap]
      & \mipQualMethod{Ours}{0.172 bppc} & \mipQualImg{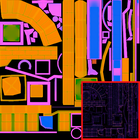} & \mipQualImg{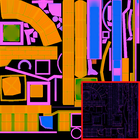} & \mipQualImg{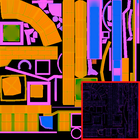} & \mipQualImg{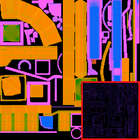} & \mipQualImg{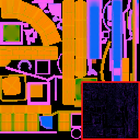} & \mipQualImg{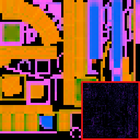} & \mipQualImg{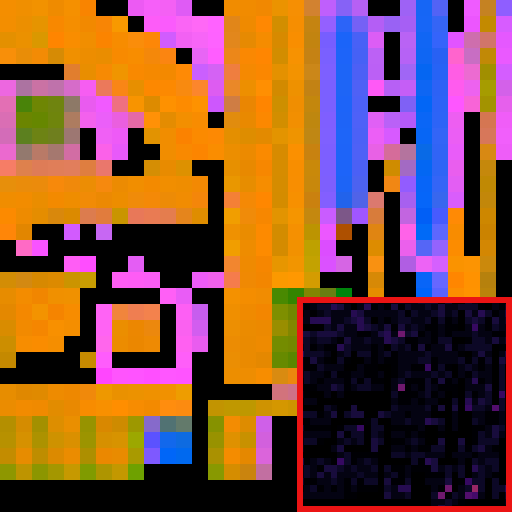} & \mipQualImg{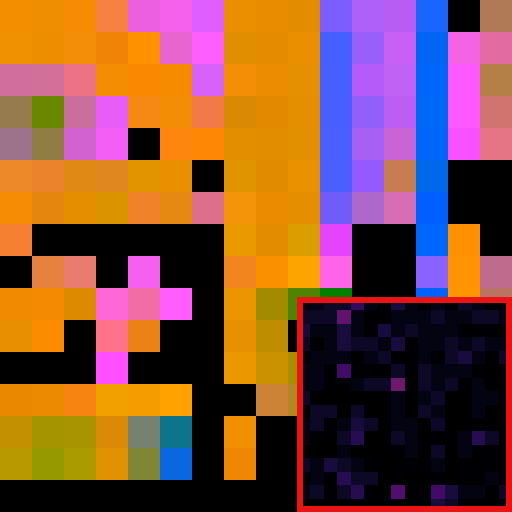} & \mipQualImg{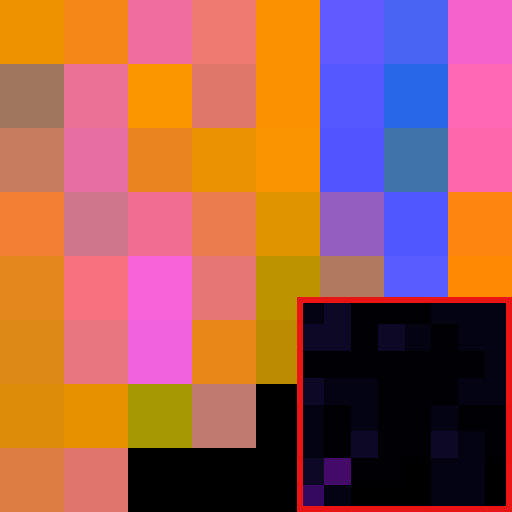} & \mipQualImg{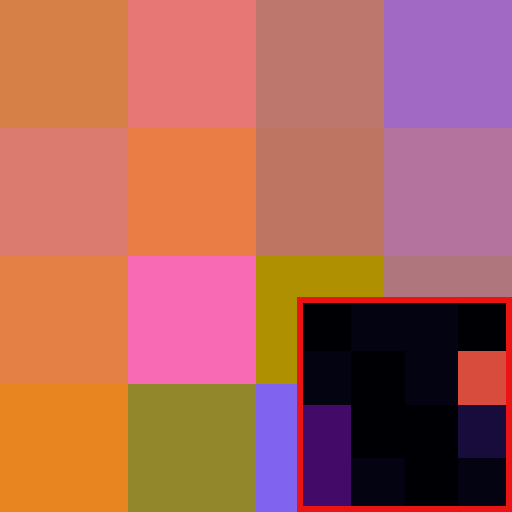} & \mipQualImg{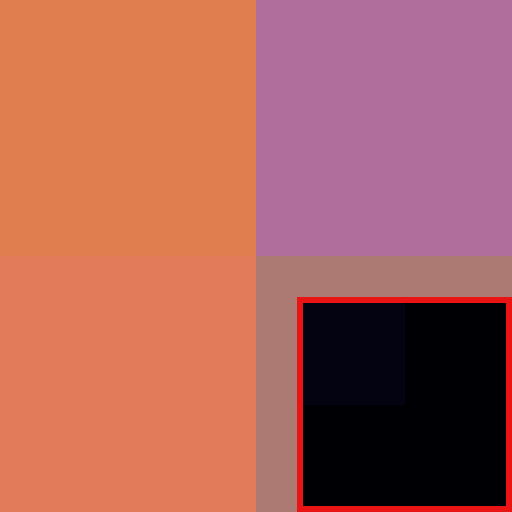} & \mipQualImg{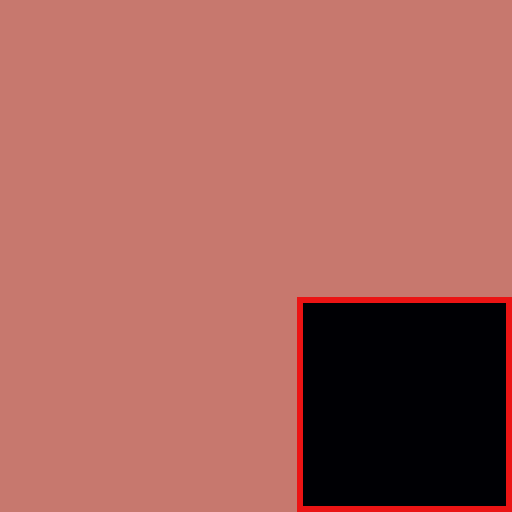} \\[\mipQualRowGap]
    \mipQualStream{ARM} & \mipQualMethod{Image-GS}{0.404 bppc} & \mipQualImg{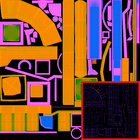} & \mipQualImg{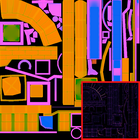} & \mipQualImg{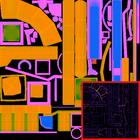} & \mipQualImg{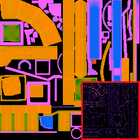} & \mipQualImg{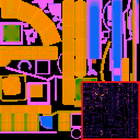} & \mipQualImg{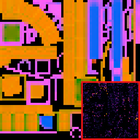} & \mipQualImg{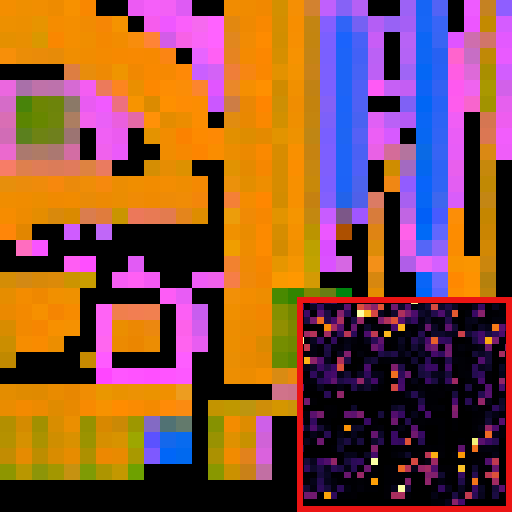} & \mipQualImg{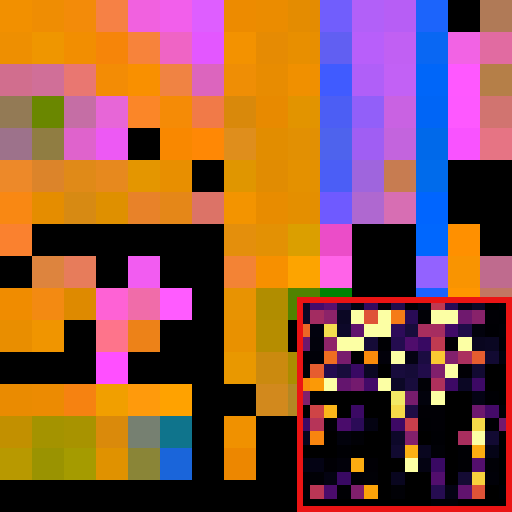} & \mipQualImg{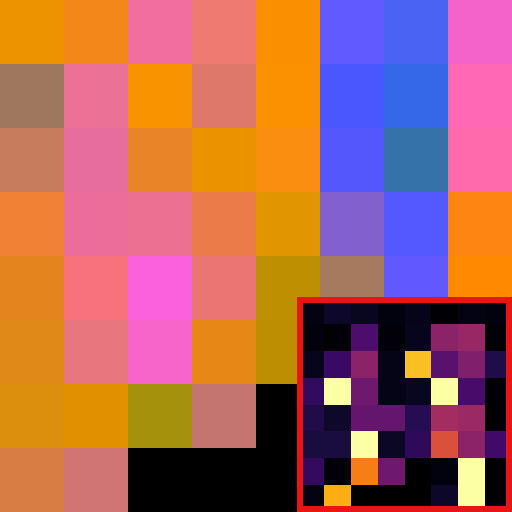} & \mipQualImg{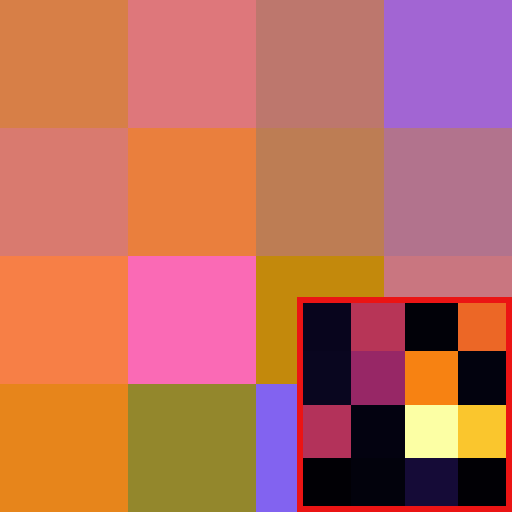} & \mipQualImg{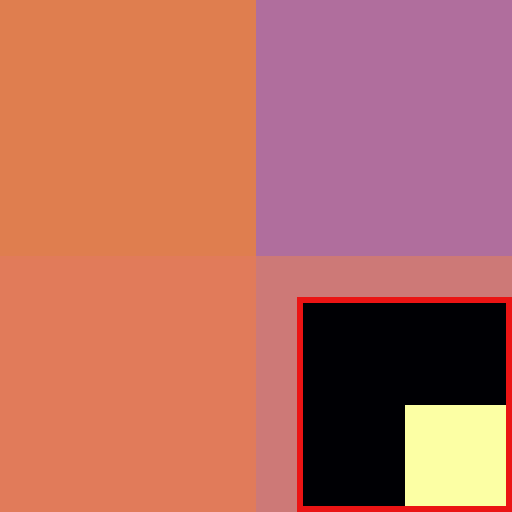} & \mipQualImg{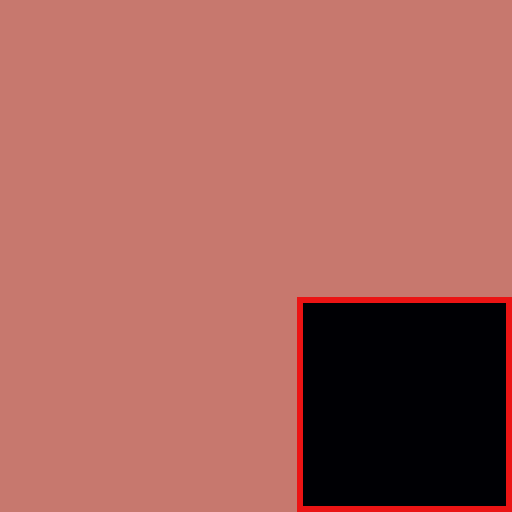} \\[\mipQualRowGap]
      & \mipQualMethod{ASTC}{0.429 bppc} & \mipQualImg{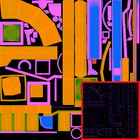} & \mipQualImg{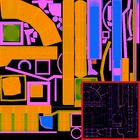} & \mipQualImg{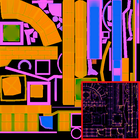} & \mipQualImg{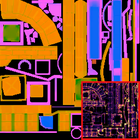} & \mipQualImg{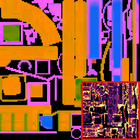} & \mipQualImg{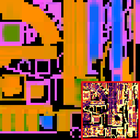} & \mipQualImg{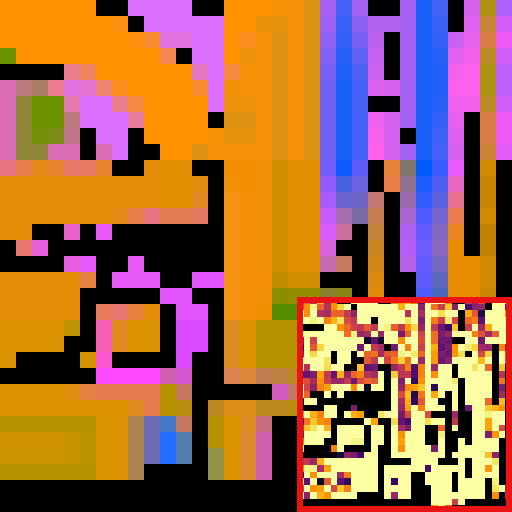} & \mipQualImg{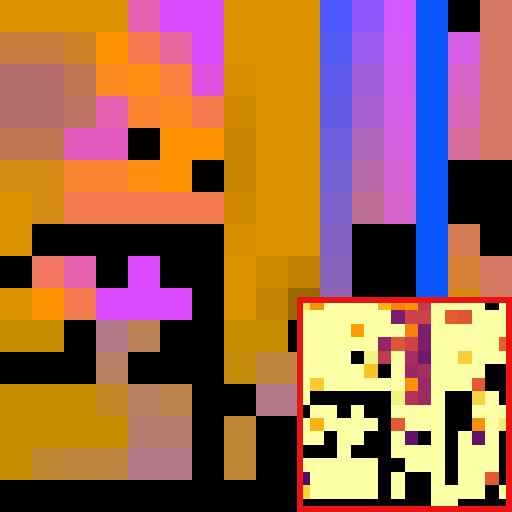} & \mipQualImg{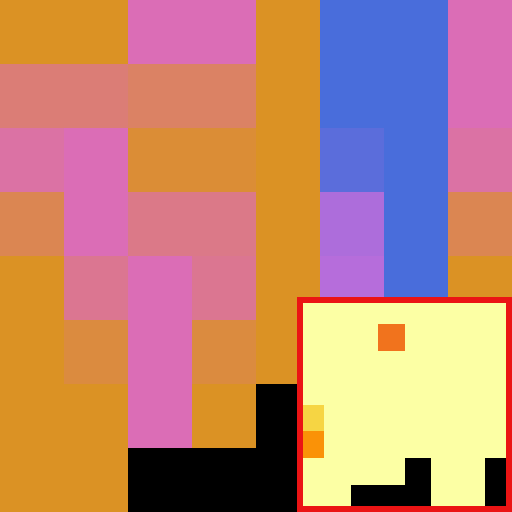} & \mipQualImg{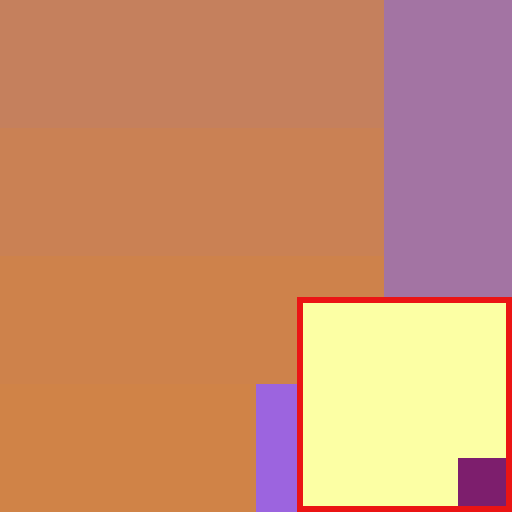} & \mipQualImg{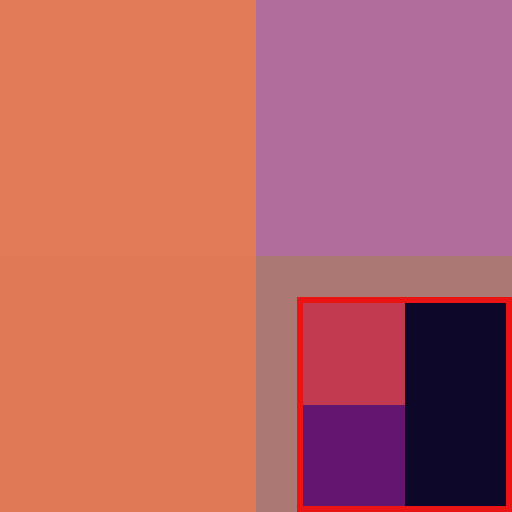} & \mipQualImg{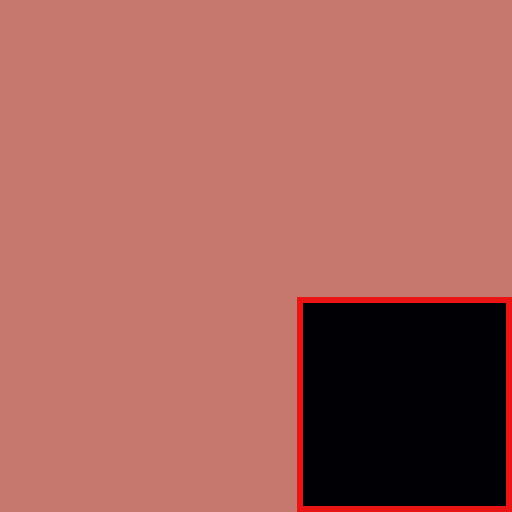} \\[\mipQualRowGap]
      & \mipQualMethod{NTC}{0.220 bppc} & \mipQualImg{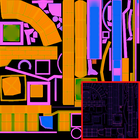} & \mipQualImg{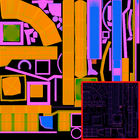} & \mipQualImg{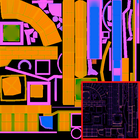} & \mipQualImg{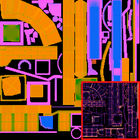} & \mipQualImg{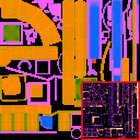} & \mipQualImg{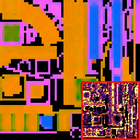} & \mipQualImg{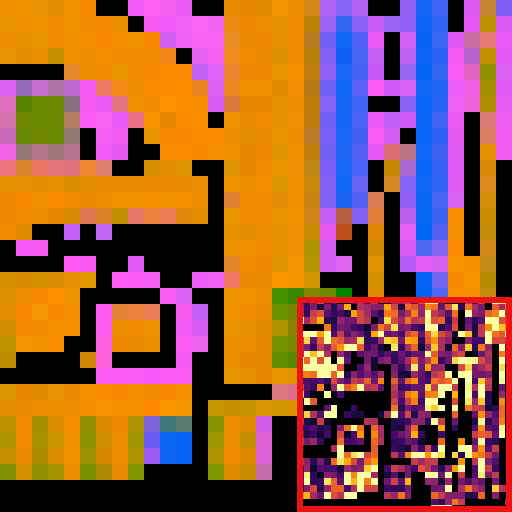} & \mipQualImg{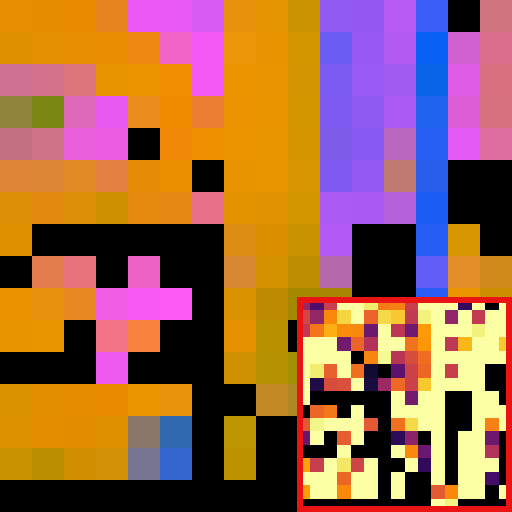} & \mipQualImg{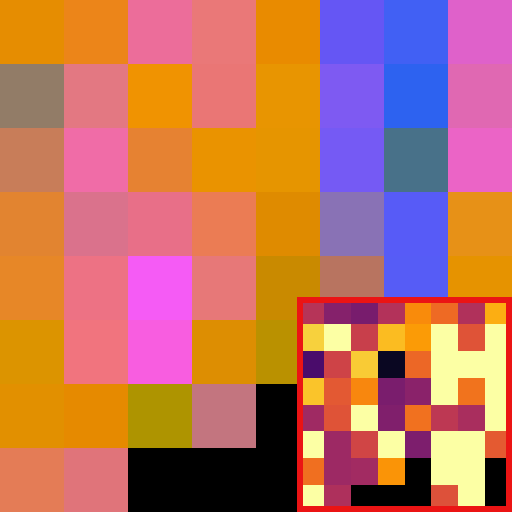} & \mipQualImg{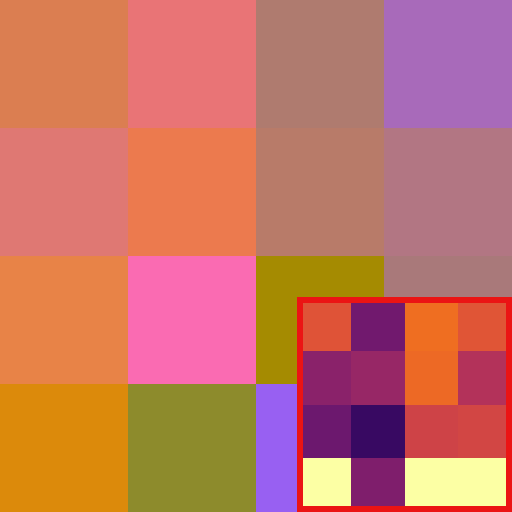} & \mipQualImg{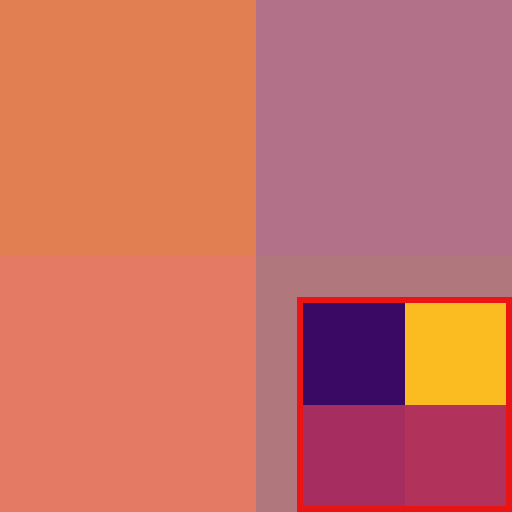} & \mipQualImg{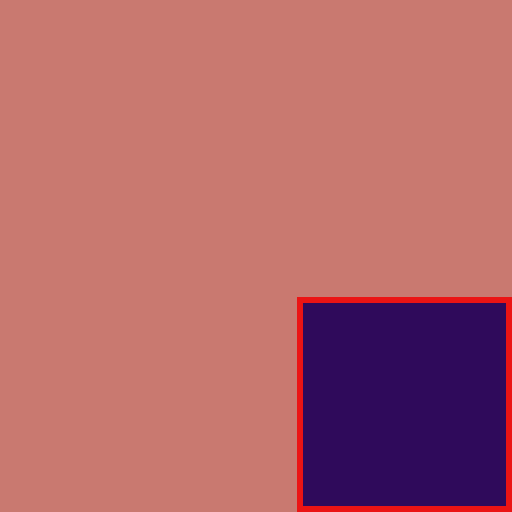} \\

  \end{tabular}%
  }
  \caption{Mipmap reconstruction comparison on the \textit{Temple Wall} SVBRDF stack. Rows show the reference and reconstructed mipmap pyramids, with insets showing per-pixel error maps against the reference. ARM denotes ambient occlusion, roughness, and metallic channels.}
  \label{fig:mipmap-error-inset-temple-wall}
\end{figure*}

\clearpage
\setkeys{Gin}{interpolate=false}

\newcommand{\oursSweepFontSize}{\scriptsize}
\newcommand{\oursSweepMetricGap}{1pt}
\newcommand{\oursSweepMetricTopGap}{27pt}
\newcommand{\oursSweepRowGap}{27pt}
\newcommand{\oursSweepMipLabelW}{0.032\textwidth}
\newcommand{\oursSweepCellW}{0.105\textwidth}
\newcommand{\oursSweepTableTargetW}{\linewidth}
\newcommand{\oursSweepVCenter}[1]{\raisebox{-0.5\height}{#1}}
\newcommand{\oursSweepImg}[1]{\oursSweepVCenter{\includegraphics[width=\oursSweepCellW]{#1}}}
\newcommand{\oursSweepMip}[2]{\oursSweepVCenter{\makebox[\oursSweepMipLabelW][r]{\begin{tabular}{@{}r@{}}\textbf{#1}\\#2\end{tabular}}}}
\newcommand{\oursSweepStream}[1]{\oursSweepVCenter{\rotatebox{90}{\textbf{#1}}}}
\newcommand{\oursSweepMetric}[2]{%
  \vtop{\hbox{\begin{tabular}{@{}c@{}}
    \makebox[\oursSweepCellW][c]{\textbf{#1}}\\[\oursSweepMetricGap]
    \makebox[\oursSweepCellW][c]{#2}
  \end{tabular}}}%
}
\newcommand{\oursSweepMetricName}[1]{%
  \vtop{\hbox{\begin{tabular}{@{}c@{}}
    \makebox[\oursSweepCellW][c]{\textbf{#1}}\\[\oursSweepMetricGap]
    \makebox[\oursSweepCellW][c]{}
  \end{tabular}}}%
}
\newcommand{\oursSweepSep}{\noalign{\vskip 2pt}\hline\noalign{\vskip 2pt}}

\clearpage
\begin{figure*}[p!]
  \centering
  \oursSweepFontSize
  \setlength{\tabcolsep}{0.7pt}
  \renewcommand{\arraystretch}{0.78}
  \resizebox{\oursSweepTableTargetW}{!}{%
\begin{tabular}{@{}c@{\hspace{2pt}}c@{\hspace{3pt}}*{6}{c@{\hspace{2pt}}}c@{}}
     & \oursSweepMip{mip 0}{2048$^2$} & \oursSweepImg{figs/ours_bppc_sweep_qual_subfigs/brick_wall/basecolor_m0_gt.png} & \oursSweepImg{figs/ours_bppc_sweep_qual_subfigs/brick_wall/basecolor_m0_ours_001_bppc0p056.png} & \oursSweepImg{figs/ours_bppc_sweep_qual_subfigs/brick_wall/basecolor_m0_ours_002_bppc0p111.png} & \oursSweepImg{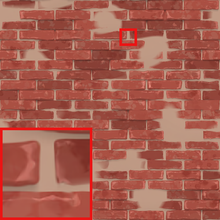} & \oursSweepImg{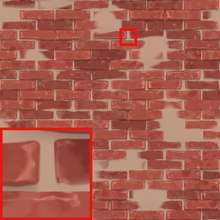} & \oursSweepImg{figs/ours_bppc_sweep_qual_subfigs/brick_wall/basecolor_m0_ours_005_bppc0p248.png} & \oursSweepImg{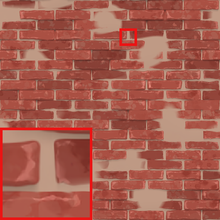} \\[\oursSweepRowGap]
    \oursSweepStream{Base color} & \oursSweepMip{mip 4}{128$^2$} & \oursSweepImg{figs/ours_bppc_sweep_qual_subfigs/brick_wall/basecolor_m4_gt.png} & \oursSweepImg{figs/ours_bppc_sweep_qual_subfigs/brick_wall/basecolor_m4_ours_001_bppc0p056.png} & \oursSweepImg{figs/ours_bppc_sweep_qual_subfigs/brick_wall/basecolor_m4_ours_002_bppc0p111.png} & \oursSweepImg{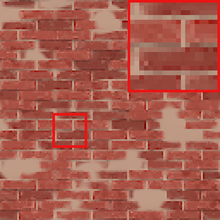} & \oursSweepImg{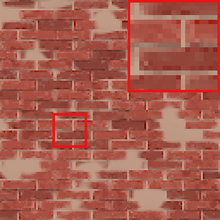} & \oursSweepImg{figs/ours_bppc_sweep_qual_subfigs/brick_wall/basecolor_m4_ours_005_bppc0p248.png} & \oursSweepImg{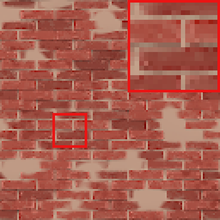} \\
\end{tabular}%
  }
  \caption{Bitrate sweep for our method on the \textit{Brick Wall} SVBRDF stack. ARH denotes ambient occlusion, roughness, and height channels.}
  \label{fig:ours-sweep-brick-wall}
\end{figure*}

\clearpage
\begin{figure*}[p!]
  \centering
  \oursSweepFontSize
  \setlength{\tabcolsep}{0.7pt}
  \renewcommand{\arraystretch}{0.78}
  \resizebox{\oursSweepTableTargetW}{!}{%
  \begin{tabular}{@{}c@{\hspace{2pt}}c@{\hspace{3pt}}*{6}{c@{\hspace{2pt}}}c@{}}
     & \oursSweepMip{mip 0}{2048$^2$} & \oursSweepImg{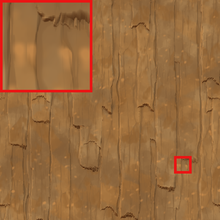} & \oursSweepImg{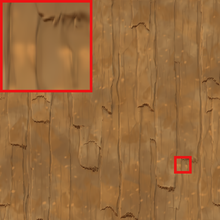} & \oursSweepImg{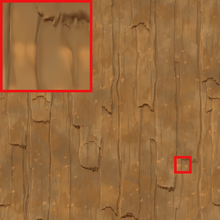} & \oursSweepImg{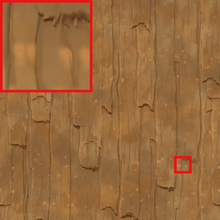} & \oursSweepImg{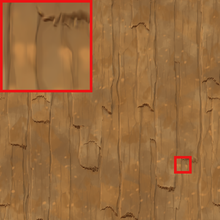} & \oursSweepImg{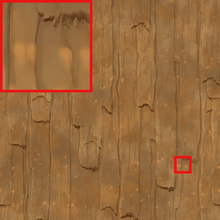} & \oursSweepImg{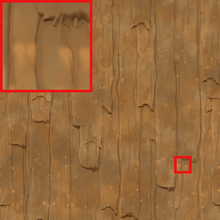} \\[\oursSweepRowGap]
    \oursSweepStream{Base color} & \oursSweepMip{mip 4}{128$^2$} & \oursSweepImg{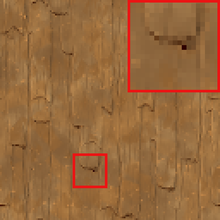} & \oursSweepImg{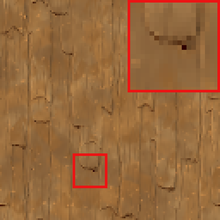} & \oursSweepImg{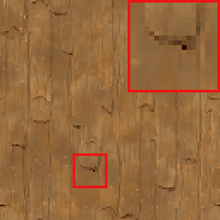} & \oursSweepImg{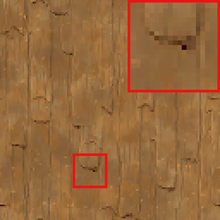} & \oursSweepImg{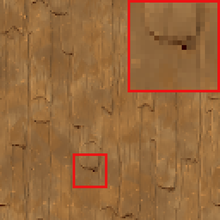} & \oursSweepImg{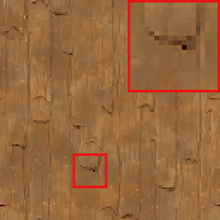} & \oursSweepImg{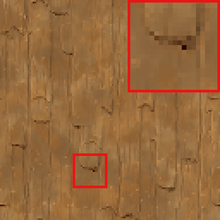} \\
\end{tabular}%
  }
  \caption{Bitrate sweep for our method on the \textit{Bark} SVBRDF stack. ARH denotes ambient occlusion, roughness, and height channels.}
  \label{fig:ours-sweep-bark}
\end{figure*}

\clearpage
\begin{figure*}[p!]
  \centering
  \oursSweepFontSize
  \setlength{\tabcolsep}{0.7pt}
  \renewcommand{\arraystretch}{0.78}
  \resizebox{\oursSweepTableTargetW}{!}{%
  \begin{tabular}{@{}c@{\hspace{2pt}}c@{\hspace{3pt}}*{6}{c@{\hspace{2pt}}}c@{}}
     & \oursSweepMip{mip 0}{2048$^2$} & \oursSweepImg{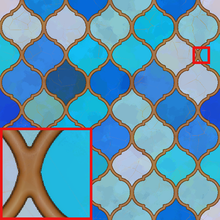} & \oursSweepImg{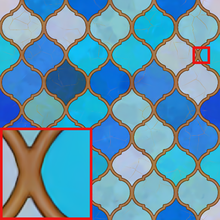} & \oursSweepImg{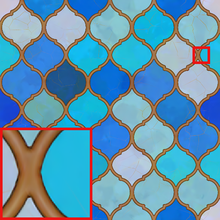} & \oursSweepImg{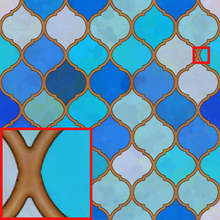} & \oursSweepImg{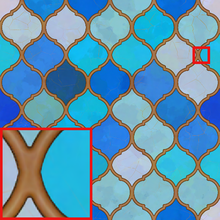} & \oursSweepImg{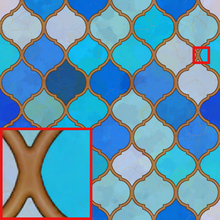} & \oursSweepImg{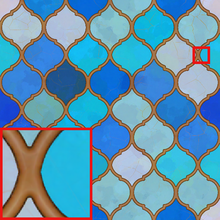} \\[\oursSweepRowGap]
    \oursSweepStream{Base color} & \oursSweepMip{mip 4}{128$^2$} & \oursSweepImg{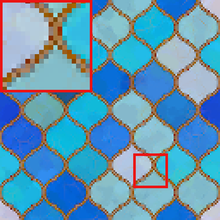} & \oursSweepImg{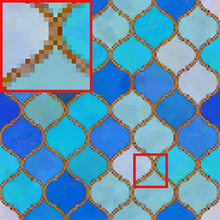} & \oursSweepImg{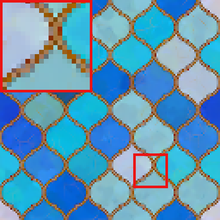} & \oursSweepImg{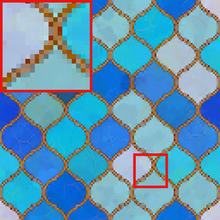} & \oursSweepImg{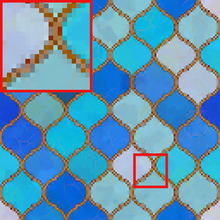} & \oursSweepImg{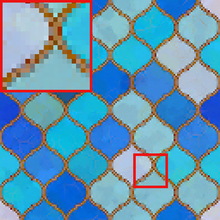} & \oursSweepImg{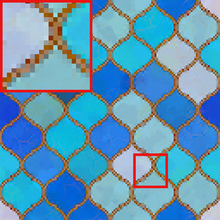} \\
\end{tabular}%
  }
  \caption{Bitrate sweep for our method on the \textit{Window} SVBRDF stack. ARH denotes ambient occlusion, roughness, and height channels.}
  \label{fig:ours-sweep-window}
\end{figure*}

\clearpage
\begin{figure*}[p!]
  \centering
  \oursSweepFontSize
  \setlength{\tabcolsep}{0.7pt}
  \renewcommand{\arraystretch}{0.78}
  \resizebox{\oursSweepTableTargetW}{!}{%
  \begin{tabular}{@{}c@{\hspace{2pt}}c@{\hspace{3pt}}*{6}{c@{\hspace{2pt}}}c@{}}
     & \oursSweepMip{mip 0}{2048$^2$} & \oursSweepImg{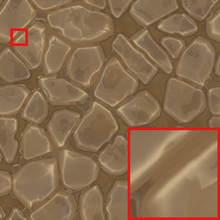} & \oursSweepImg{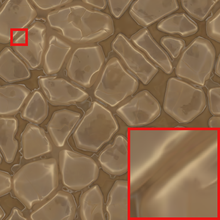} & \oursSweepImg{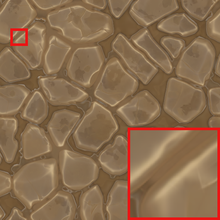} & \oursSweepImg{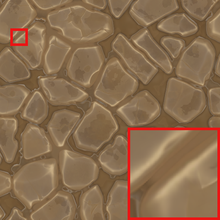} & \oursSweepImg{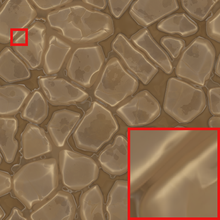} & \oursSweepImg{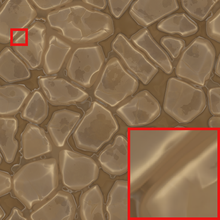} & \oursSweepImg{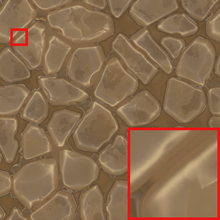} \\[\oursSweepRowGap]
    \oursSweepStream{Base color} & \oursSweepMip{mip 4}{128$^2$} & \oursSweepImg{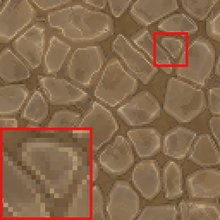} & \oursSweepImg{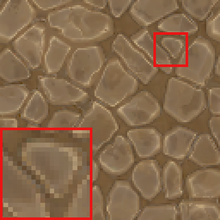} & \oursSweepImg{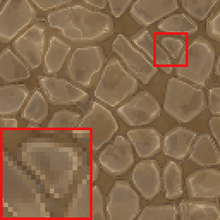} & \oursSweepImg{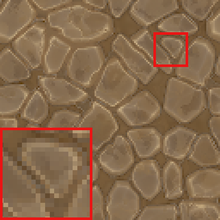} & \oursSweepImg{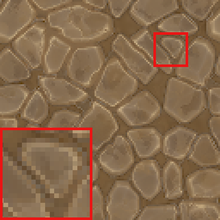} & \oursSweepImg{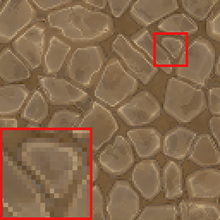} & \oursSweepImg{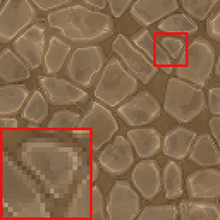} \\
\end{tabular}%
  }
  \caption{Bitrate sweep for our method on the \textit{Wet Rocks} SVBRDF stack. ARH denotes ambient occlusion, roughness, and height channels.}
  \label{fig:ours-sweep-wet-rocks}
\end{figure*}

\clearpage
\begin{figure*}[p!]
  \centering
  \oursSweepFontSize
  \setlength{\tabcolsep}{0.7pt}
  \renewcommand{\arraystretch}{0.78}
  \resizebox{\oursSweepTableTargetW}{!}{%
  \begin{tabular}{@{}c@{\hspace{2pt}}c@{\hspace{3pt}}*{6}{c@{\hspace{2pt}}}c@{}}
     & \oursSweepMip{mip 0}{2048$^2$} & \oursSweepImg{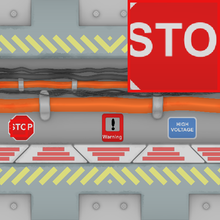} & \oursSweepImg{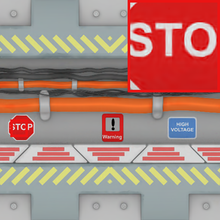} & \oursSweepImg{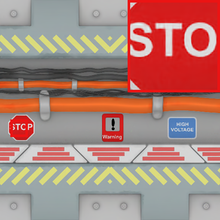} & \oursSweepImg{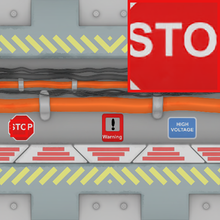} & \oursSweepImg{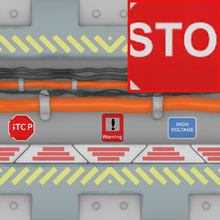} & \oursSweepImg{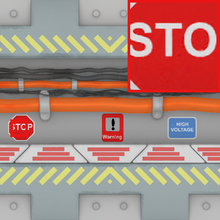} & \oursSweepImg{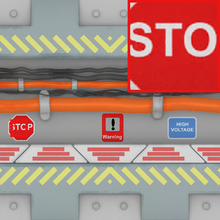} \\[\oursSweepRowGap]
    \oursSweepStream{Base color} & \oursSweepMip{mip 4}{128$^2$} & \oursSweepImg{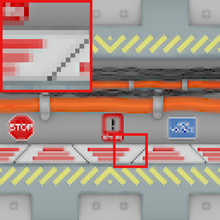} & \oursSweepImg{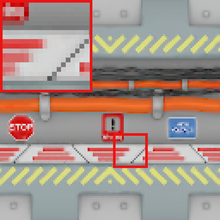} & \oursSweepImg{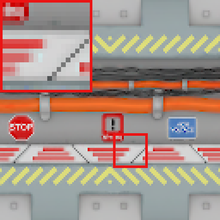} & \oursSweepImg{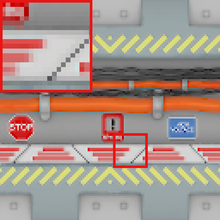} & \oursSweepImg{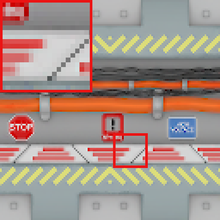} & \oursSweepImg{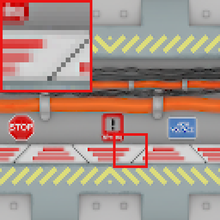} & \oursSweepImg{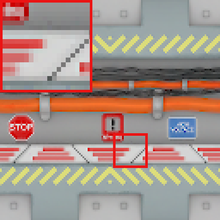} \\
\end{tabular}%
  }
  \caption{Bitrate sweep for our method on the \textit{Industrial Wall} SVBRDF stack. ARH denotes ambient occlusion, roughness, and height channels.}
  \label{fig:ours-sweep-industrial-wall}
\end{figure*}

\clearpage
\begin{figure*}[p!]
  \centering
  \oursSweepFontSize
  \setlength{\tabcolsep}{0.7pt}
  \renewcommand{\arraystretch}{0.78}
  \resizebox{\oursSweepTableTargetW}{!}{%
  \begin{tabular}{@{}c@{\hspace{2pt}}c@{\hspace{3pt}}*{6}{c@{\hspace{2pt}}}c@{}}
     & \oursSweepMip{mip 0}{2048$^2$} & \oursSweepImg{figs/ours_bppc_sweep_qual_subfigs/battle_axe/basecolor_m0_gt.png} & \oursSweepImg{figs/ours_bppc_sweep_qual_subfigs/battle_axe/basecolor_m0_ours_001_bppc0p050.png} & \oursSweepImg{figs/ours_bppc_sweep_qual_subfigs/battle_axe/basecolor_m0_ours_002_bppc0p104.png} & \oursSweepImg{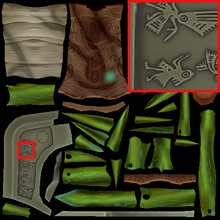} & \oursSweepImg{figs/ours_bppc_sweep_qual_subfigs/battle_axe/basecolor_m0_ours_004_bppc0p135.png} & \oursSweepImg{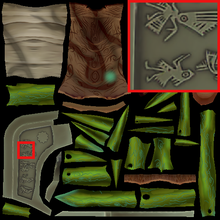} & \oursSweepImg{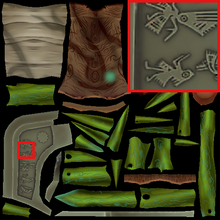} \\[\oursSweepRowGap]
    \oursSweepStream{Base color} & \oursSweepMip{mip 4}{128$^2$} & \oursSweepImg{figs/ours_bppc_sweep_qual_subfigs/battle_axe/basecolor_m4_gt.png} & \oursSweepImg{figs/ours_bppc_sweep_qual_subfigs/battle_axe/basecolor_m4_ours_001_bppc0p050.png} & \oursSweepImg{figs/ours_bppc_sweep_qual_subfigs/battle_axe/basecolor_m4_ours_002_bppc0p104.png} & \oursSweepImg{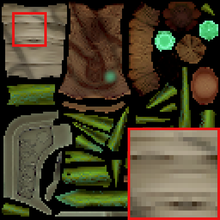} & \oursSweepImg{figs/ours_bppc_sweep_qual_subfigs/battle_axe/basecolor_m4_ours_004_bppc0p135.png} & \oursSweepImg{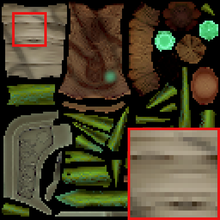} & \oursSweepImg{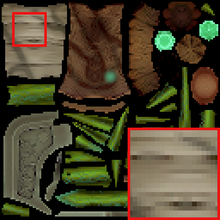} \\
\end{tabular}%
  }
  \caption{Bitrate sweep for our method on the \textit{Battle Axe} SVBRDF stack. This asset contains an emissive map instead of a normal map; ARM denotes ambient occlusion, roughness, and metallic channels.}
  \label{fig:ours-sweep-battle-axe}
\end{figure*}

\clearpage
\begin{figure*}[p!]
  \centering
  \oursSweepFontSize
  \setlength{\tabcolsep}{0.7pt}
  \renewcommand{\arraystretch}{0.78}
  \resizebox{\oursSweepTableTargetW}{!}{%
  \begin{tabular}{@{}c@{\hspace{2pt}}c@{\hspace{3pt}}*{6}{c@{\hspace{2pt}}}c@{}}
     & \oursSweepMip{mip 0}{2048$^2$} & \oursSweepImg{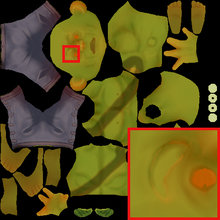} & \oursSweepImg{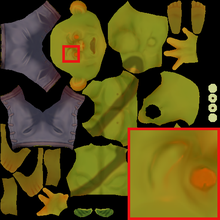} & \oursSweepImg{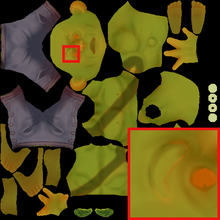} & \oursSweepImg{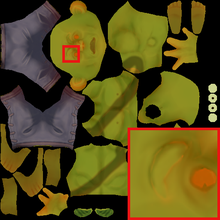} & \oursSweepImg{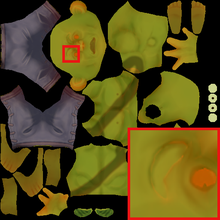} & \oursSweepImg{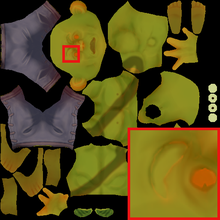} & \oursSweepImg{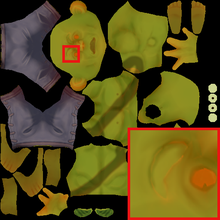} \\[\oursSweepRowGap]
    \oursSweepStream{Base color} & \oursSweepMip{mip 4}{128$^2$} & \oursSweepImg{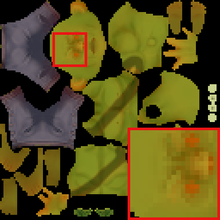} & \oursSweepImg{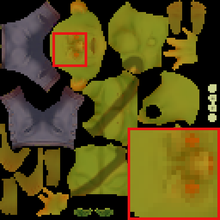} & \oursSweepImg{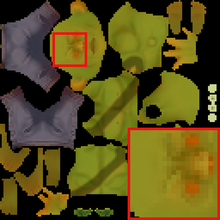} & \oursSweepImg{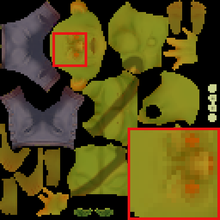} & \oursSweepImg{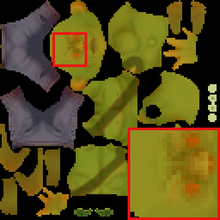} & \oursSweepImg{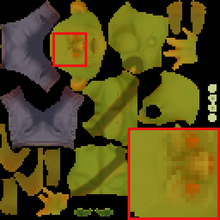} & \oursSweepImg{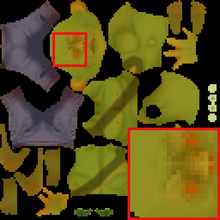} \\
\end{tabular}%
  }
  \caption{Bitrate sweep for our method on the \textit{Orc Warrior} SVBRDF stack. ARM denotes ambient occlusion, roughness, and metallic channels.}
  \label{fig:ours-sweep-orc-warrior}
\end{figure*}

\clearpage
\begin{figure*}[p!]
  \centering
  \oursSweepFontSize
  \setlength{\tabcolsep}{0.7pt}
  \renewcommand{\arraystretch}{0.78}
  \resizebox{\oursSweepTableTargetW}{!}{%
  \begin{tabular}{@{}c@{\hspace{2pt}}c@{\hspace{3pt}}*{6}{c@{\hspace{2pt}}}c@{}}
     & \oursSweepMip{mip 0}{2048$^2$} & \oursSweepImg{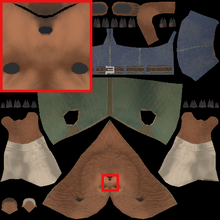} & \oursSweepImg{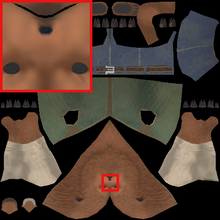} & \oursSweepImg{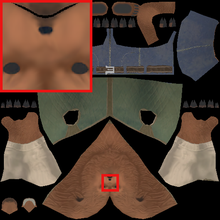} & \oursSweepImg{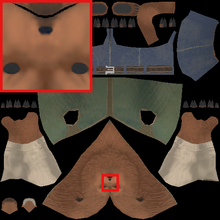} & \oursSweepImg{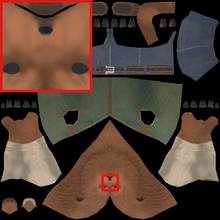} & \oursSweepImg{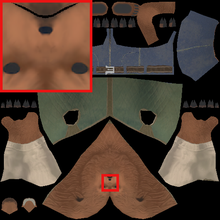} & \oursSweepImg{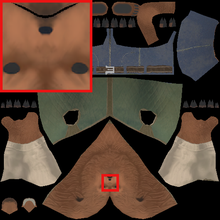} \\[\oursSweepRowGap]
    \oursSweepStream{Base color} & \oursSweepMip{mip 4}{128$^2$} & \oursSweepImg{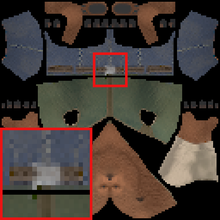} & \oursSweepImg{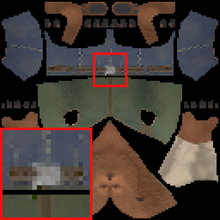} & \oursSweepImg{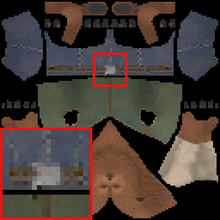} & \oursSweepImg{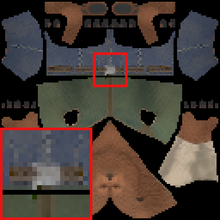} & \oursSweepImg{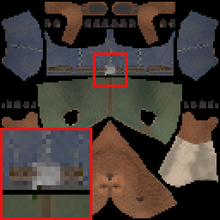} & \oursSweepImg{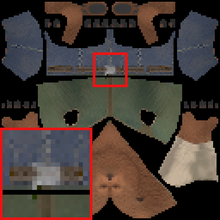} & \oursSweepImg{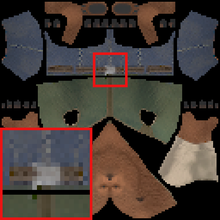} \\
\end{tabular}%
  }
  \caption{Bitrate sweep for our method on the \textit{Bear Character} SVBRDF stack. ARM denotes ambient occlusion, roughness, and metallic channels.}
  \label{fig:ours-sweep-bear-character}
\end{figure*}

\clearpage

\end{document}